# Towards A Novel Unified Framework for Developing Formal, Network and Validated Agent-Based Simulation Models of Complex Adaptive Systems

**Muaz Ahmed Khan Niazi**

**Computing Science and Mathematics**
**School of Natural Sciences**
**University of Stirling**
**Scotland UK**

**This thesis has been submitted to the University of Stirling**
**In partial fulfillment for the degree of Doctor of Philosophy**

**2011**

# Abstract

Literature on the modeling and simulation of complex adaptive systems (cas) has primarily advanced vertically in different scientific domains with scientists developing a variety of domain-specific approaches and applications. However, while cas researchers are inherently interested in an interdisciplinary comparison of models, to the best of our knowledge, there is currently no single unified framework for facilitating the development, comparison, communication and validation of models across different scientific domains. In this thesis, we propose first steps towards such a unified framework using a combination of agent-based and complex network-based modeling approaches and guidelines formulated in the form of a set of four levels of usage, which allow multidisciplinary researchers to adopt a suitable framework level on the basis of available data types, their research study objectives and expected outcomes, thus allowing them to better plan and conduct their respective research case studies.

Firstly, the complex network modeling level of the proposed framework entails the development of appropriate complex network models for the case where interaction data of cas components is available, with the aim of detecting emergent patterns in the cas under study. The exploratory agent-based modeling level of the proposed framework allows for the development of proof-of-concept models for the cas system, primarily for purposes of exploring feasibility of further research. Descriptive agent-based modeling level of the proposed framework allows for the use of a formal step-by-step approach for developing agent-based models coupled with a quantitative complex network and pseudocode-based specification of the model, which will, in turn, facilitate interdisciplinary cas model comparison and knowledge transfer. Finally, the validated agent-based modeling level of the proposed framework is concerned with the building of in-simulation verification and validation of agent-based models using a proposed Virtual Overlay Multiagent System approach for use in a systematic team-oriented approach to developing models. The proposed framework is evaluated and validated using seven detailed case study examples selected from various scientific domains including ecology, social sciences and a range of complex adaptive communication networks. The successful case studies demonstrate the potential of the framework in appealing to multidisciplinary researchers as a methodological approach to the modeling and simulation of cas by facilitating effective communication and knowledge transfer across scientific disciplines without the requirement of extensive learning curves.

# Declaration

I, Muaz Niazi hereby declare that this work has not been submitted for any other degree at this University or any other institution and that, except where reference is made to the work of other authors, the material presented is original. Some portions of the thesis chapters have been published as follows:

- Parts of Chapter 3 in [2], [3].
- Parts of Chapter 4 in [4].
- Parts of Chapter 5 in [5], [6].
- Parts of Chapter 6 in [7-9]. Qasim Siddique helped in some of the simulation experiments. Some ideas of research simulation were formulated based on advice from Dr. Saeed Bhatti and Dr. Abdul Rauf Baig.

**Muaz Niazi**

# Acknowledgements

This thesis would not have been possible without the help and support of a large number of individuals. First and foremost, I would like to thank my family members, especially my beloved parents, who have endured my absence during my research and helped me tremendously in all ways possible. Without your continued help, support and guidance, this would never have been possible. Thank you.

I am deeply indebted to my principal supervisor and founding Head of our COSIPRA Lab, Dr. Amir Hussain who provided stimulating advice, guidance and encouragement to me every step of the way. I would like to thank Kamran, Dr. Erfu, Thomas, Rozniza, Erik and Andrew from the COSIPRA lab. I would also like to thank Kerstin Rosee who, on behalf of Dr. Amir, would keep me on my toes by requiring regular updates. Writing updates always pushed me, by forcing me to quantify my progress and obtain critical feedback on a periodic basis. I would also like to thank Ian MacLellan from the International Office. Finally, I would especially like to thank my Examiners Dr. Keshav Dahal (from Bradford University) and Dr. Savitri Maharaj for their constructive criticism and helpful comments, which helped me considerably in shaping up the thesis in a much better way than I had originally planned.

# Table of Contents

# List of Figures

# List of Tables

# List of Abbreviations

| | |
|---|---|
| ABM | Agent-based Model |
| CNA | Complex Network Analysis |
| ABC | Agent-based Computing |
| MAS | Multiagent System |
| VOMAS | Virtual Overlay Multiagent System |
| SECAS | Sensing Emergence in Complex Adaptive Systems |
| CAS or cas | Complex Adaptive System |
| SME | Subject Matter Expert |
| SS | Simulation Specialist |
| UML | Unified Modeling Language |
| V&V | Verification and Validation |
| WSN | Wireless Sensor Network |
| CSN | Cognitive Sensor Network |
| P2P | Peer to Peer |
| MANET | Mobile Ad-hoc Networks |
| MASON | Multi-agent Simulator of neighborhoods/ networks |
| JADE | Java Agent Development Framework |
| WOS | Web of Science |
| JCR | Journal Citation Reports |
| LISP | List Processing (Language) |

# 1 Introduction

*"Framework: A structure composed of parts framed together, esp. one designed for inclosing or supporting anything; a frame or skeleton."* - Oxford English Dictionary

Complex adaptive systems (cas) [10] is a set of special type of natural and artificial complex systems. It is representative of the notion of a system where "The whole is more than the parts". In other words, these are systems where numerous and perhaps quite simple components interact in a nonlinear fashion to give rise to global unprecedented and often unpredictable behaviors visible and detectable at a higher level of abstraction. Phenomena and mechanisms associated with cas include "emergence"[1], self-organization[11] and self-assembly[12]. These phenomena are typically observable and comprehensible by intelligent observers at higher levels of abstraction[13] but not easily quantifiable at the micro level. At times, while emergent patterns have been noted to be both interesting as well as unprecedented, it has been observed by cas researchers that there is no known way of predicting these patterns based solely on an understanding of the individual components. Thus, an understanding of the individual components of a cas is not enough to develop an understanding of the entire system [14] because it is the actual interactions of the components which plays a key role in the observed global behaviors.

With a strong presence and association with the natural sciences from the physical world, cas are well-known to transgress disciplinary boundaries. Therefore, it is quite common to find an interest in different types of cas as demonstrated by the existence of a large number and variety of models in literature. Multidisciplinary researchers from a wide

[1] According to the Oxford Dictionary, the earliest use of the term emergence dates back to 1755 by Brooke University "Beauty i. 10 From the deep thy [Venus'] bright emergence sprung." And formally defined as "The process of coming forth, issuing from concealment, obscurity, or confinement. lit. and fig. (Cf. emerge v. 3, 4.) Also said of the result of an evolutionary process"

and diverse set of backgrounds ranging from social sciences, ecological sciences, computational economics, system biology and others have been studying cas and related phenomena. However, a study of this complex behavior can typically turn out to be a non-trivial exercise; firstly because a research study can involve a set of extensive data collection exercises coupled with model development exercises for the cas components as well as interactions. Subsequently these models need to be correlated with their observed data to develop an understanding of the underlying dynamics of the cas system under study.

Technically, developing an understanding of a cas is associated with the formation of “explicit” models of the system. While “implicit” models are of a mental cognitive nature[15], “explicit” models are more effective for communication[16].

In terms of effort and commitment, these models can range from exploratory, requiring a small time to develop, to inquisitory, requiring perhaps man-years of effort and resources. These projects can also range from smaller to larger teams with different team structures of personnel for gathering real-world data and for developing agent-based or complex network-based models of various aspects of the system of interest. While cas researchers conduct research in parallel disciplines and have an inherent interest in examining, comparing and contrasting models across disciplines, currently such comparisons are performed informally. As such, to the best of our knowledge, there is no unified framework for the multidisciplinary modeling and simulation of cas. As such, most researchers resort to either performing informal comparisons[17] or else evolving domain-specific methods of developing models[18-22].

## 1.1 Modeling cas

Epstein defines modeling as the development of a “simplified” representation of “something”. Epstein also clarifies misconceptions about modeling and simulation giving a list of different reasons for modeling [16]. He notes that modeling can be "implicit" (a mental model), in which case, "assumptions are hidden, internal consistency is untested, logical consequences and relation to data are unknown". Thus these types of models are not effective in the communication of ideas. The alternative is "explicit" models, which are more effective in communicating ideas and concepts associated with the model. In "explicit" models, the assumptions are carefully laid out and are coupled with details such as the outcomes of the modeling exercise. Another important feature of developing explicit models is that these models facilitate the replication of results obtained from the model by the scientific community.

For any cas, the ability to come up with explicit models thus demonstrates the attainment of a level of understanding. Although cas are extremely commonplace since all living systems themselves tend to be cas at several levels of hierarchy, they do not surrender themselves easily to modeling. As such, in the absence of a single comprehensive framework governing various aspects of modeling and simulation of cas, these systems have more or less been loosely modeled using a number of different paradigms with closely related roots. While most initial modeling of complex systems has traditionally been used with the goal of simplification[23] such as using differential equations, system dynamics or Monte Carlo simulations, more recently multidisciplinary literature has noted[24] a preference of cas researchers to the use of one of the two modeling approaches as follows :

1. Agent-based (or Individual-based) modeling approaches[25].
2. Complex network-based approaches[26].

To specifically comprehend the cas modeling problem, a closer examination reveals that the development of a unified framework for modeling cas would entail answering several open research questions. These questions, ordered from the less abstract to the more abstract, can be phrased as follows:

1. How to better develop models of interaction data?
2. How to describe cas to facilitate communication across scientific and disciplinary boundaries?
3. In the case of a dearth of real-world data, how can cas simulation models be validated primarily using meta-data or concepts?
4. In general, how can multidisciplinary cas research projects be structured and executed based on availability of resources and commitment?

Next, these four questions are examined in further detail.

1. How to better develop models of interaction data?

As part of a search for suitable data for cas modeling research, it is common to first discover that while there are a considerable number of existing data sources, not many of them might be useful for modeling. One key problem lies in the fact that most available data typically consists of statistical summarized data, which is often not very helpful in developing models of the exact nature of the underlying dynamics in cas [27]. This approach does not give sufficient details to develop more advanced data-driven models. As such, cas researchers need to figure out exactly how to structure their data collection exercise during their case study to be able to come up with models which actually give some useful information.

A particularly useful paradigm for developing models of interaction data is the Complex Networks Analysis (CNA) approach. Complex networks are an extensions of the traditional graphs[28]. The idea is to use interactions and other relationships of various as-pects and components in cas to develop network models. Once the network models have been developed, cas researchers can use network manipulation to extract useful quantitative as well as visual information about the topological structure of the system. Networks can be trimmed and particular types of nodes or links can be manipulated to highlight useful visual information about the topological aspects of the cas. In addition, various quantitative measures such as network diameter, clustering coefficients, matching indices and various centrality measures such as eccentricity, betweenness, degree etc. can be calculated for the network. These measurements allow making recommendations such as classification of a particular type of network (Such as small-World, Scale-Free or Random network models) or else for highlighting important components or interactions inside the cas.

While there are numerous available software tools for developing complex network models, selecting the right type and quantity of data can pose as a difficult research challenge. Currently this is not an automatic process and thus existing data mining techniques cannot be directly applied to implicitly select important data from large number of data types and columns. Specifically what cas researchers might be looking for, are specific aspects of cas, tied closely with emergence and other complex phenomena.

Taking a specific example of biological networks, currently a large amount of data sources are available in online repositories such as nucleic acid sequences and bioinformatics analysis clusters which can be used to execute complex algorithms. However, to actually develop networks, cas researchers have to first understand and develop an implicit mental model of exactly what should be a suitable node (e.g. a gene) and what would constitute a suitable link (e.g. co-expression) to ensure that the resulting network is useful for their domain. Some of these networks can themselves be quite complex and hierarchically structured. An example of these is the gene regulatory networks, which are themselves well-known as a system consisting of many sub-networks[29]. Other examples include protein interaction networks which are developed based on protein molecules in undirected graphs, signal transduction networks which are directed networks demonstrating various biochemical reactions and social networks such as friendship networks where the weighted links depend upon the perceived level of friendship of the subjects (persons) involved in the study[30] etc.

2. How to describe cas to facilitate communication across scientific and disciplinary boundaries?

Multidisciplinary researchers exploring cas can come from a variety of possible backgrounds. While they are experts in their subject and can have significant domain knowledge, it is possible that they might not be comfortable with the nomenclature followed by other disciplines. As such, to describe cas, there needs to be a common easily accessible description format which ties in closely with the cas model but is not specific to a particular cas scientific discipline. While it should not be highly technical, it should still allow the construction of descriptive models which should be comparable across case studies and disciplines. There are many inherent benefits of having such a description; firstly it would allow for a comparative study of cas models across scientific disciplines and domains. Secondly, it would allow high fidelity of simulation models with the specification models. Thirdly, it can be used for learning cas concepts from models of other domains.

3. In the case of a dearth of real-world data, how can cas simulation models be validated primarily using meta-data or concepts?

Being able to develop simulation models of cas is one thing and being able to validate them is another. Unlike traditional simulation models, the concepts related to cas are typically quite abstract in nature and not easy to describe verbally. While researchers have attempted to describe some aspects of a cas, it has traditionally been quite difficult to give generalized definitions of cas concepts such as emergence or self-organization etc. using terms globally acceptable by all disciplines. As such, the acceptability of the results of any cas simulation model is tied closely with how valid these results appear to the researchers. While for social scientists, models might be considered valid if the results of a simulation appear similar to what they observe from population studies, for biologists, simulation models might need a much higher level of fidelity with the actual components of the system such as bio-chemical molecules etc. As such, what might seem valid to scientists from one discipline might not be an acceptable validation for another set of scientists from another discipline. As such, validation needs to be customizable. The problem with this approach is that currently developing validation can be a fairly nontrivial exercise and there is currently no way of structure the efforts of the Subject Matter Experts (SME) and the Simulation Specialists (SS). As such, validation of cas simulation models varies from case study to another without any basic common validation techniques unlike more traditional simulation models of complicated systems. These engineered systems are often better describable in terms of mathematical equations and mostly lend themselves better to generating suitable data for validation of models.

Another problem associated with validation of cas models is that at times, instead of being able to validate using data, the validation of complex concepts and emergent behavior of the cas under study is required so as to ensure that the simulated system behaves close enough to the cas being modeled. Such behavior is typically hard to put in an explicit model. Traditional validation techniques being typically data-driven either use various attributes from the results of a simulation model and compare them with "actual" results from the systems of interest or compare the results with another model. In the absence of either of these, SS need to rely on "Turing or face validation"[31], where the observer compares output of the model visually based on previous experience with the system. In a cas modeling scenario, at times, it can be difficult to acquire the exact data required for validation and thus, even though there might be some data, it might not always be a good candidate for traditional validation schemes. In other words, some points regarding validation of simulation models of cas can be noted as follows:

a. Cas are extremely interactive in nature. A simple change in the composition can massively change the global behavior. Having aggregate data in the form of purely statistical tables might not be a real representation of the actual cas and might only represent a certain propagation of states of the cas. If validation were performed similar to traditional simulation models, it might not be valid itself technically as it is similar to taking a zoomed picture of a bird's yellow beak in a simulation and validating it with a yellow life jacket simply because the colors match. Thus, even if the "graphs" or "plots" appear to coincide, the actual models may be totally different and technically speaking, the validation would not be valid by itself.

b. Secondly, validation of cas is tied closely with the interaction (run-time behavior) of the various "agents" inside the model and not just the output data. For a good cas validation scheme, this important fact needs to be taken into consideration.

4. In general, how can multidisciplinary cas research projects be structured and executed based on availability of resources and commitment?

Cas researchers develop models in a number of different ways, with numerous goals and for various cas aspects [32]. One problem very often faced by researchers (as can be observed on agent-based modeling mailing lists such as netlogo-users and repast) is when researchers are planning research on cas but there is currently no clear way of identifying which types of models to develop or how to move further, based on the expected outcomes of the research. As such, researchers can find it difficult to structure their time and resource commitments to better develop cas research projects. Occasionally, researchers might have a few days to develop models to demonstrate future feasibility. Subsequently, if they succeed in this phase and are able to secure funding for further research, they might have access to more resources to perform modeling and data collection exercises with a larger team.

To the best of our knowledge, there is no existing multidisciplinary framework allowing the structuring of cas case studies and model selection based on commitment levels or goals in addition to allowing a combination of complex network and agent-based modeling methods for multidisciplinary cas researchers. Having such a single unified framework for multidisciplinary cas research would thus assist considerably in developing and communicating cas models across scientific disciplines and structuring cas research projects.

## 1.2 Motivation

As noted in the previous section, there are several open questions associated with developing various types of “explicit” cas models. The key problem here is the multidisciplinary nature of cas systems. Researchers from diverse backgrounds such as Life Sciences, Social Sciences and Communication Engineers need to work with cas. As such, there is neither a common methodology nor a set of concrete guidelines for developing cas models spanning multiple scientific disciplines. Researchers can be unaware of exactly how to proceed and what to expect when developing cas models. In addition, this problem is compounded by the fact that most cas researchers are non-specialist in Computer Sciences and therefore in spite of being experts in their particular domains, they can tend to be neither interested in nor are often able to develop advanced highly technical models. However, it can be noted from an examination of multidisciplinary cas literature that they still feel comfortable developing various types of explicit models with visual components (such as agent-based and complex network-based models).

This thesis has been motivated by the lack of a single unified framework for the modeling of complex adaptive systems. Although many different models have previously been developed for various types of cas, in general, cas modeling has evolved vertically with little cross-flow of ideas between various application domains. As such, although there are numerous examples of applied modeling and simulation in literature, to the best of our knowledge, there does not exist a common guiding framework for interested multidisciplinary cas researchers providing concrete guidance on how to approach cas problems, how to develop different types of models and how to decide which type of data would be most suitable for developing models, how to describe simulation models with a high fidelity to the actual model, how to develop model descriptions allowing for visual and quantitative comparisons of models and last, but not the least, how to structure validation studies of cas simulations. Such a framework might also assist in the removal of ambiguities in the usage of terms associated with cas nomenclature[2] prevalent due to a parallel evolution of modeling practices in different scientific disciplines, enabling different cas models from different case studies and scientific disciplines to be comparable with each other.

## 1.3 Aims and Objectives

The aim of this research is to work towards a set of unified framework guidelines allowing researchers interested in multidisciplinary and inter-disciplinary cas studies to explore the development of explicit models of cas by using a combination of complex network and agent-based modeling approaches based on their research goals and level of commitment.

[2] Examples include the terms individual-based modeling, agent-based modeling and multiagent systems. More details are given in Chapters 2 and 3.

## 1.4 Original Contributions

Original research conducted and reported in this thesis is aimed at developing first steps towards a unified framework in the form of concrete guidelines coupled with detailed case study examples for use by multidisciplinary cas researchers. Some of the key original contributions in the modeling and simulation of cas are summarized as follows:

A. The key contribution is a unified framework allowing multidisciplinary researchers to plan their cas research case studies formulated based on levels of research goals and commitment. The proposed framework uses a combination of agent-based and complex network models to allow for interaction-based, exploratory, descriptive and validated models of cas.

B. The proposed framework is structured in the form of different levels composed of a set of methodological guidelines. Each of the framework levels allows the planning and execution of a specific type of cas research study based on the availability and type of data, the objectives of the case study and the expected levels of commitment.

C. The first two levels of the proposed framework are structured specifically to encompass existing modeling and simulation research which mainly uses complex network and agent-based simulation modeling techniques whereas the rest of the two framework levels allow for more advanced model development using a combination of these approaches.

D. The complex network modeling level of the proposed framework is structured and linked with the availability of suitable interaction data from cas components. Using interaction data, complex network models can thus be developed for cas exploration. These models can be manipulated and subsequently visualized using various mathematical and software tools giving qualitative as well as quantitative inference capability to the cas researchers.

E. Using the exploratory agent-based modeling level, researchers can use agent-based modeling as an exploratory tool to develop proof-of-concept cas models to explore feasibility of future research thus paving the way for more sophisticated techniques.

F. The descriptive agent-based modeling level of the proposed framework is useful for researchers who are primarily interested in cross-disciplinary communication and comparison of models. Descriptive modeling approach is being proposed which uses a combination of pseduocode-based specification and complex network modeling as a means of modeling agent-based models. There are several benefits of this approach; firstly it allows the description of cas models in a way such that there is a high degree of fidelity of the model with the ABM. Secondly, it allows for quantitative, visual and non-code based comparison of cas models developed in multiple disciplines. Thirdly, it allows the exploitation of learning opportunities for researchers by allowing the examination of models across scientific disciplines thus facilitating the creation of heterogeneous multi-domain ABMs.

G. The validated agent-based modeling level of the proposed framework is based on a step-by-step methodology for the development of in-simulation validation for agent-based models by means of an interactive collaborative effort involving both Subject Matter Experts (SME) as well as the Simulation Specialists (SS). This approach is based on concepts from multiagent systems, software engineering and social sciences. Using a systematic approach, the outcome of the methodology is an agent-based model of the cas validated by means of design-by-contract invariants in the simulation model where the contract is enforced by means of in-simulation cooperative agents termed as the Virtual Overlay Multiagent System (VOMAS). Building a VOMAS allows SME and SS to collaboratively develop custom in-simulation verification and validation schemes for the cas application case study.

H. The viability of the proposed framework is demonstrated with the help of various case studies spanning different individual scientific disciplines and some case studies spanning multiple scientific disciplines.

## 1.5 Proposed framework for the Modeling and Simulation of cas

In this section, firstly an overview of the proposed framework is presented. Next the framework is described from two different perspectives firstly in terms of study objectives of conducting a cas research case study and the expected level of commitment. Secondly, the framework is described in correlation with the availability and access to specific data types.

### 1.5.1 Overview of the proposed framework

For the sake of practicality, the framework guidelines have been developed in the form of levels of abstraction. Thus, cas researchers can opt for modeling at a particular level depending on factors such as availability of data, meta-data as well as the level of interest and how much time cas researchers can invest in pursuing a research project. In addition, example case studies are presented to demonstrate the usage of various framework levels. Based on the proposed framework, research in cas can be conducted by choosing and subsequently following one of the following four proposed levels for developing cas models:

- The complex network modeling level of the framework is useful if interaction data is readily available. In this level, complex network models of cas can be developed using this interaction data and subsequently Complex Network Analysis (CNA) can be per-formed for network classification as well as determination of various global and local quantitative measures from the network for the extraction of useful information. Such information can give details of emergent behavior and patterns which would otherwise have not been evident using statistical or other more traditional mathematical methods.

- The exploratory agent-based modeling level of the framework extends existing ideas of agent-based modeling prevalent in multidisciplinary literature which focus on the development of exploratory agent-based models of cas to examine and extricate possible emergent trends in the cas. Building exploratory models allows cas researchers to develop experimental models which help lay foundation for further research. These proof-of-concept models also assist researchers in determining the feasibility of future research in the domain using the selected model design.

- Developing models at the descriptive agent-based modeling level of the framework entails developing concrete DescRiptivE Agent-based Models (DREAM) by using a combination of pseudocode-based specification, a complex network model and a quantitative model "fingerprint" based on centrality measures of the agent-based model which are all associated closely with the ABM. The pseudocode-based specification is developed in the form of non-code template schemas and has several benefits; firstly it is close to an executable specification but is not tied with any single programming language thus allowing use by cas researchers for developing agent-based models based on the specification using tools of their own liking. Secondly, this specific type of specification allows a one-to-one correspondence of ABM concepts with the descriptive model. Thirdly this specification allows communication and comparison of models in multidisciplinary studies by using visual as well as quantitative methods.

- The validated agent-based modeling level of the proposed framework is concerned with developing verified and validated agent-based models. This level allows perform-ing in-simulation verification and validation of the agent-based models using a Virtual Overlay Multiagent System (VOMAS) based on a cooperative set of agents inside the simulation allowing the verification and validation of the cas model by means of design-by-contract invariants. These invariants are developed as a result of collaboration of the Subject Matter Expert (SME) and the Simulation Specialist (SS). In this level, concepts originating from software engineering, multiagent Systems and social sciences are all used in tandem to propose a systematic methodology for ABM verification and validation.

### 1.5.2 Proposed framework levels formulated in terms of cas study objectives

In Figure 1, it can be noted how different framework levels can be used by multidisciplinary cas researchers to develop models based on their particular study objectives and expected outcomes.

If there is sufficient interaction data available then the cas research study can proceed by using the complex network modeling level of the proposed framework. In this level, researchers first analyze the data columns, extract suitable data, develop complex network models and subsequently perform network manipulation and complex network analysis for the discovery of emergent patterns.

However, as is often the case, if such data is unavailable and the goal of the research study is to determine the feasibility of future research, then it might be possible to feasible to proceed in their research study by using the exploratory ABM level of the proposed framework.

These two framework levels essentially can also be used to encompass existing cas research studies which have primarily used either complex network modeling and analysis or else agent-based modeling or both in their analyses.

If however, the goal of the research is to perform an inter-disciplinary comparative case study then the descriptive agent-based modeling level of the proposed framework allowing for developing a DREAM model can be chosen. This particular framework level has the benefit of allowing for inter-disciplinary model comparison, knowledge transfer and learning.

Finally, if the goal of the study is develop simulations with a high degree of correlation with real-world systems such as in the development of decision support systems, then the appropriate framework level for usage would be the validated agent-based modeling level based on the development of an in-simulation validation scheme using the VOMAS approach. This framework level is also more suitable for large-scale team oriented projects and requires adherence to team-oriented protocols with the goal of a verified and validated agent-based model of the cas system under study.

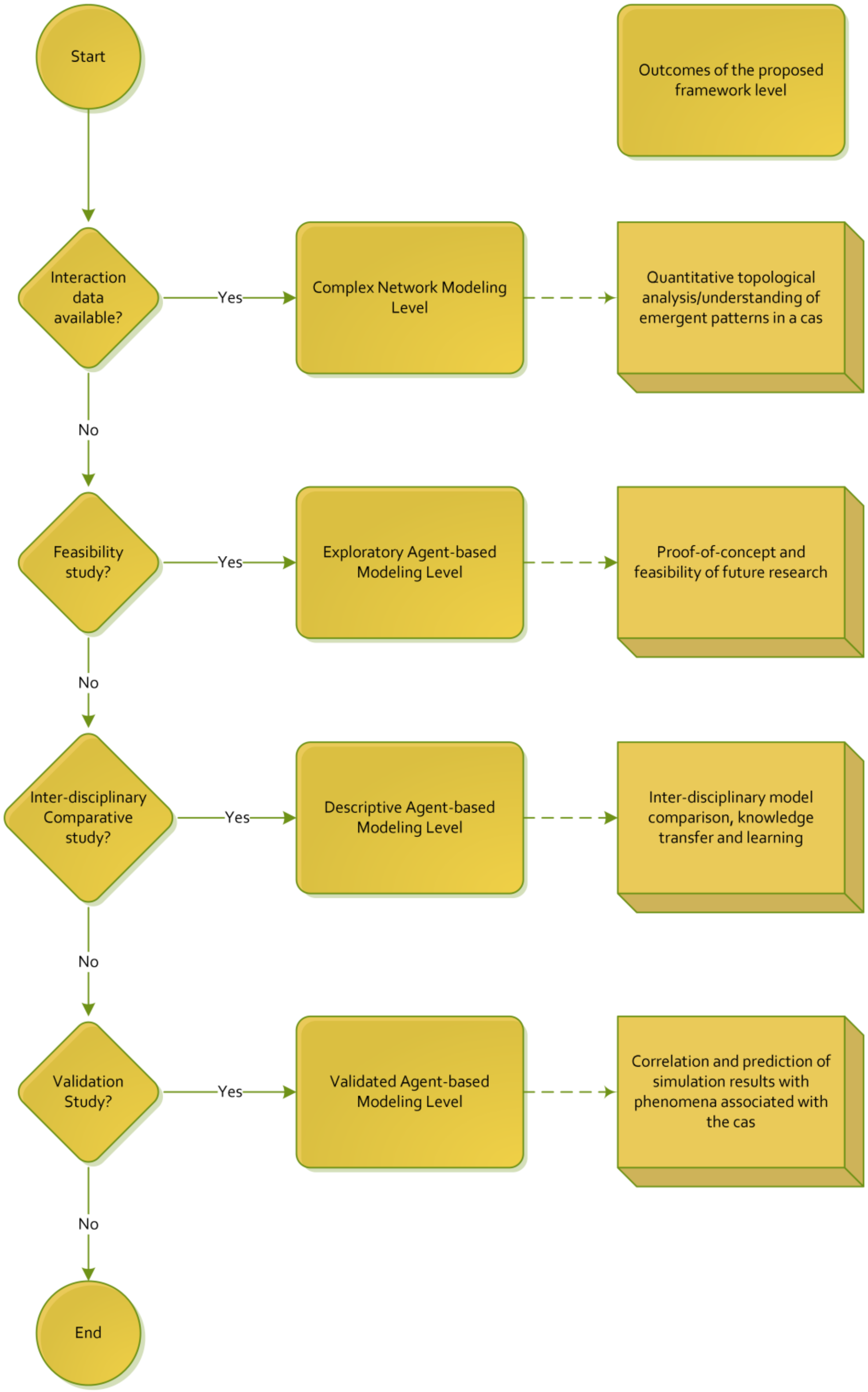

**Figure 1: An overview of the decision-making process for choosing framework levels in relation to cas research study objectives**

### 1.5.3 Proposed framework levels formulated in relation to available data types

In the previous section, an overview of the different framework levels in relation to the cas research study objectives. Here, in this section, we shall describe the framework in terms of available data or knowledge about the cas. As we can note from Figure 2, a descriptive specification of the cas model can be developed based on the metadata or knowledge about the cas. An ABM can be developed either from this specification or directly from the knowledge of cas. The ABM can be verified and validated using in-simulation validation (which has been developed as a result of extensive meetings between SMEs and SSs) that is performed by building a VOMAS model. By the help of invariant constraints enforced by the cooperative agents forming the VOMAS, the simulation can be verified and validated using in-simulation methods. In addition, if interaction data is available for the development of network models, complex network models can be developed and CNA can be used to manipulate and analyze various structural topological features of the interactions using various information visualization-based and mathematical tools.

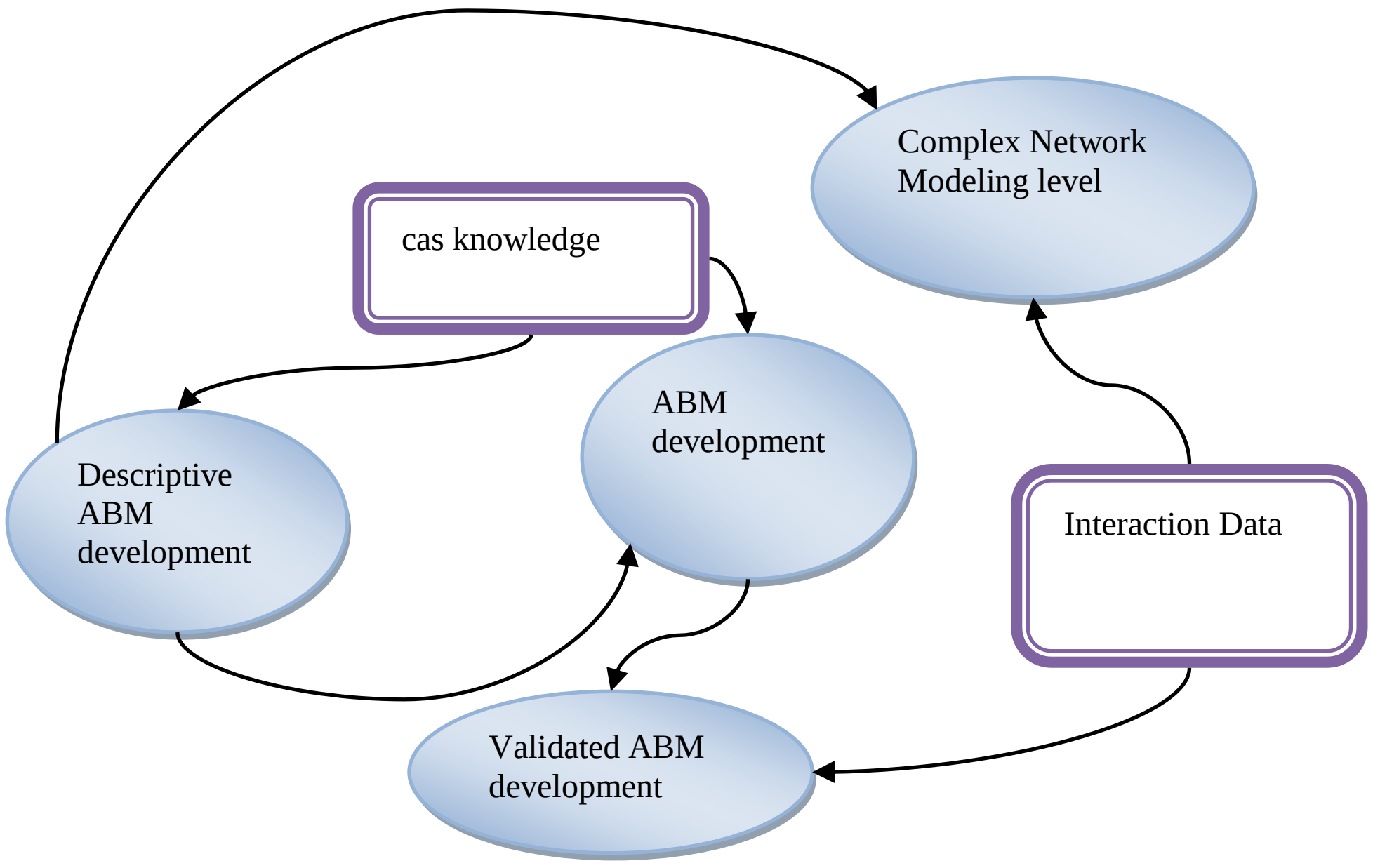

**Figure 2: Data Driven View of Proposed Framework**

## 1.6 Publications

Some parts of the chapters in this thesis have been published in various peer-reviewed venues. Because of the peculiar inter-disciplinary nature of cas research, some of the work performed was in collaboration with different researchers and the specific contributions of the co-authors have been noted earlier in the declaration at the start of the thesis.

### 1.6.1 Refereed Conferences and Workshops

1. Niazi MA, Hussain A. Social Network Analysis of trends in the consumer electronics domain. In: Proc Consumer Electronics (ICCE), 2011 IEEE International Conference on, Las Vegas, NV, 9-12 Jan. 2011, 2011. pp 219-220. (Chapter 3)
2. Niazi MA, Hussain A, Baig AR, Bhatti S. Simulation of the research process. 40th Conference on Winter Simulation. Miami, FL: Winter Simulation Conference; 2008. pp 1326-1334. (Chapter 6)
3. Niazi MA, Hussain A, Kolberg M. Verification &Validation of Agent Based Simulations using the VOMAS (Virtual Overlay Multi-agent System) approach. MAS&S 09 at Multi-Agent Logics, Languages, and Organisations Federated Workshops. Torino, Italy; 2009. pp 1-7. (Chapter 6)
4. Niazi MA, Siddique Q, Hussain A, Kolberg M. Verification and Validation of an Agent-Based Forest Fire Simulation Model. SCS Spring Simulation Conference. Orlando, FL, USA: ACM; 2010. pp 142-149. (Chapter 6)

### 1.6.2 Published Journal papers

1. Niazi MA, Hussain A. Agent-based Computing from Multi-agent Systems to Agent-Based Models: A Visual Survey. Springer Scientometrics 2011;In-press.(Chapter 3)

2. Niazi MA, Hussain A. Agent based Tools for Modeling and Simulation of Self-Organization in Peer-to-Peer, Ad-Hoc and other Complex Networks. IEEE Communications Magazine 2009; 47(3):163 - 173. (Chapter 4)
3. Niazi MA, Hussain A. A Novel Agent-Based Simulation Framework for Sensing in Complex Adaptive Environments. IEEE Sensors Journal 2011; 11(2):404-412. (Chapter 5)
4. Niazi MA, Hussain A. Sensing Emergence in Complex Systems. IEEE Sensors Journal 2011; 11(10): 2479-2480. (Chapter 5)

## 1.7 Overview of the thesis

Here, first an overview of the different explored case studies is provided. This is followed by an overview of the chapters.

### 1.7.1 Overview of case studies

To ensure that the research was in line with the norms of various cas, we have worked in tandem with teams of cas researchers and domain experts from life sciences, social sciences and telecommunications. The following list gives details of some of the case studies discussed in the thesis as a means of examples of the application of the proposed methods associated with various levels of framework along in correlation with the thesis chapters:

- In chapter 3, the proposed framework level is applied on two different case studies in the domain of scientometric data of agent-based computing and consumer electronics domains.
- In chapter 4, a comprehensive case study on the use of unstructured search algorithms from the domain of P2P networks in the domain of "Cyber-physical systems"[33] by the development of an "Internet of things"[34] has been presented.

- In chapter 5 a case study on the development of a heterogeneous ABM of sensing single-hop Wireless Sensor Network for sensing complex behaviors of flocking “boids” is presented.
- In chapter 6, three different case studies are presented. The first case study is in the domain of ecological modeling and models forest fire simulations. The second is in the domain of multi-hop Wireless Sensor Networks modeled in the form of a Quasi Unit Disk Graph (QUDG). The third case study is in the domain of simulation of the evolution of researchers on the basis of their Hirsch index.

### 1.7.2 Outline of the Thesis

The rest of the thesis is organized as follows:

**Chapter 2:** This chapter presents background and related work required to comprehend the rest of the thesis.

**Chapter 3:** This chapter presents complex network modeling level of the proposed framework. These methods are further applied to two different domains; Agent-based Computing and Consumer Electronics. The first case study is given in considerable detail and various Scientometric indicators such as key papers, important authors, highly cited Institutions, key journals and a number of other indicators are identified using complex network modeling. The second case study is given as a means of validation of the proposed methods in another separate domain and identifies key Journals, authors and research papers from the consumer electronics domain.

**Chapter 4:** This chapter presents exploratory Agent-based modeling level of the proposed framework. As a demonstration of the proposed methods, a comprehensive exploratory agent-based model in the domain of Cyber-Physical Systems is developed demonstrating a combination of unstructured P2P search methods to locate content in static and mobile physical computing devices.

**Chapter 5:** In this chapter, descriptive agent-based modeling level of the proposed framework is presented. Descriptive modeling entails the development of a DescRiptivE Agent-Based Modeling (DREAM) model by using a combination of a complex network model, a quantitative centrality-based fingerprint and a pseudocode-based specification model with a high degree of fidelity with the actual agent-based model. As a means of demonstration of the proposed framework level, the DREAM approach is applied in a comprehensive case study of a heterogeneous cas ABM of a WSN observing a set of flocking “boids”.

**Chapter 6:** In this chapter, the validated agent-based modeling level of the proposed framework is proposed. The proposed methodology based a team-oriented approach of in-simulation validation is demonstrated using three different application case studies from three different scientific disciplines of ecology, telecommunications and social simulation allowing for a proof of concept of the generalized and broad applicability of the proposed methods.

**Chapter7:** In this chapter, the thesis is concluded. First the key research contributions are presented and are followed by a critical review of the framework levels in relation with existing approaches in literature. This is followed by limitations and proposed future enhancements proposed in the framework levels. Finally a set of possible future application case studies ideas which can effectively employ the different framework levels is presented.

**Appendix 1:** The appendix contains a description of the exact keywords associated with the Scientometric study of agent-based computing.

# 2 Complex Adaptive Systems: Background and Related Work

In the previous chapter, an overview of the entire thesis and the key contributions were presented. In this chapter, basic background and a literature review of various concepts needed for the understanding of the thesis are presented.

## 2.1 Overview

We start by first giving an overview of cas and their key characteristics. Next we give specific examples of cas from natural and artificial systems. Subsequently we give an overview of modeling of cas. Next, we give a review of agent-based modeling tools. This is followed by a review of verification and validation of simulation models. Finally an overview of different communication network simulators is presented.

## 2.2 Complex Adaptive Systems (cas)

Cas are a special type of complex systems which arise due to nonlinear interactions of smaller components or agents[17]. While it is difficult to exactly define cas, Holland [35] notes that:

*"Even though these complex systems differ in detail, the question of coherence under change is the central enigma for each. This common factor is so important that at the Santa Fe Institute we collect these systems under a common heading, referring to them as complex adaptive systems (cas). This is more than terminology. It signals our intuition that general principles rule cas behavior, principles that point to ways of solving the attendant problems."*

The modern approach of cas has evolved from a series of attempts by researchers to develop a comprehensive and general understanding of the world around us. Our world is burgeoning with interactive entities. Most times, such interactions manifest themselves as change of some type either internal to the entities (change of internal state) or else external (change of external state or behavior). In other words, these entities adapt in networks of interactions spread nonlinearly across the entire system spatially as well as temporally[36].

In the science of complexity, it is the “small” and “numerous” that govern changes in the “large” and “few”. The interaction of small (micro) and perhaps simple components gives rise to structures and behaviors, which are amazingly complex when observed from the macro perspective[37]. Comprehension of the intricacies of these systems is so hard that it has lead to not just one but a set of theories based on different observations and different types of modeling methodologies targeting different aspects of the system.

A simple example of such systems is that of life. Although not easily quantifiable, living systems exhibit elegance absent in the monotonicity of complex but relatively inadaptable inorganic and chaotic systems. Life surrounds us and embosoms us. Every living system and life form known to us asserts itself with a display of a dualistic nature which on one hand is decrepit and frail and on the other it has its own tinge of sturdiness and resilience, hard to imagine in any engineered systems[38]. The frailty of complex living systems is apparent because all complex life forms seem stamped with a “certain” programmed expiration date because the growth and aging of most multi-cellular living organisms is governed by the complex behavior encoded in the genes. On the other hand, a closer examination reveals the Complexity and resilience working inside the systems. From apparent humble beginnings and small components (DNA, RNA, genomes, amino acids, proteins and cells etc.), the complexity of each life form emerges to give complex features such as muscles, tissues, systems, of which the most important is cognition and “self” for higher life forms.

A human embryo which starts as just a single cell replicates and forms a complete human being which is not just a blob of identical cells but has an extremely intricate and complex structure, with organs, systems and dynamics. Looking at the single cell embryo, it is hard to imagine how it could end up in forming this amazingly complex life form[39].

If we examine the cell closely, it is hard not to think as to how exactly does this one cell, which divides into two and then four and so on, apparently suddenly bloom into complex structures without any apparent set of "physical" guiding forces or advanced sensory abilities such as one which coordinates all cells to form global structures. And when the structures do mature into systems, it makes one wonder how really do the multiple cells synchronize to gradually start the dynamic phenomenon associated with complex life forms, such as the blood flow, breathing, digestion, clotting, immune systems and emotions, thoughts, family, social systems and so on? The interesting part of all this is that the guiding principles of the entire life of the organism or life form are coded inside the genes, part of the cell's nucleus.

Another example of similar complex systems is human and animal social systems, where interactions are once again the key to understanding the larger perspective of things. The concept of "country", "social or ethnic groups", "families", "clans", "universities", "civilizations", "religion" and so on are extremely powerful and govern the life of every human being. However, these phenomena are in one way or the other, rooted in the concepts of interactions. All this makes one thing very obvious. Our current science is known to miss the big picture of complexity. Developing this understanding further would assist us in understanding life, the universe and everything that exists[37].

This absence of the "complete picture" is at the heart of research in cas and the closely related movements such as the "Complexity theory"[17] and the "Systems Biology theory"[14]. In contrast to reductionism which is rooted in simplifications and thus gives a misleading confidence that understanding the parts will somehow "ensure" that we will be able to understand the whole, the complex systems approach focuses instead on a specific meaning of the phrase “The whole is more than the parts”. Here, literature notes that the key factor in understanding cas is to understand the “interactions” of the parts which cannot be quantified easily.

### 2.2.1 The Seven Basics of cas

Holland maintains that there are seven *basics* of cas. Four of these are properties and three are mechanisms. These have been termed basics because he notes that other cas-related ideas can be considered as derivates of these. The following discussion is based on Holland’s description of properties and mechanisms of cas [35]:

1. Aggregation (property)

The word “Aggregation” follows from “aggregate”, which, according to the Oxford English Dictionary, means “Collected into one body”. Aggregation is useful in two ways, firstly as a generalization where different items can be categorized in a single big and oft-reused umbrella, e.g. animals, plants, bags, baskets etc. In other words, this is a way of abstraction as well since it allows focus on the important details to be modeled and ignores those which can be ignored. This type of aggregation is perhaps classification and is more related to cognition rather than actual physical containment, to which the second type of aggregation may be attributed. Holland points out the formation of complex agents called meta-agents in living and social systems (whether natural or artificial) based on complex behavior of smaller, simpler agents.

An example can be seen in Figure 3, where we notice that seemingly without any global intelligence, the shape of the melon is an example of emergence based on the complex interactions of its constituent components, such as skin, seeds and other parts, which are made up of other structures, eventually coming down to cells and sub-cellular structures.

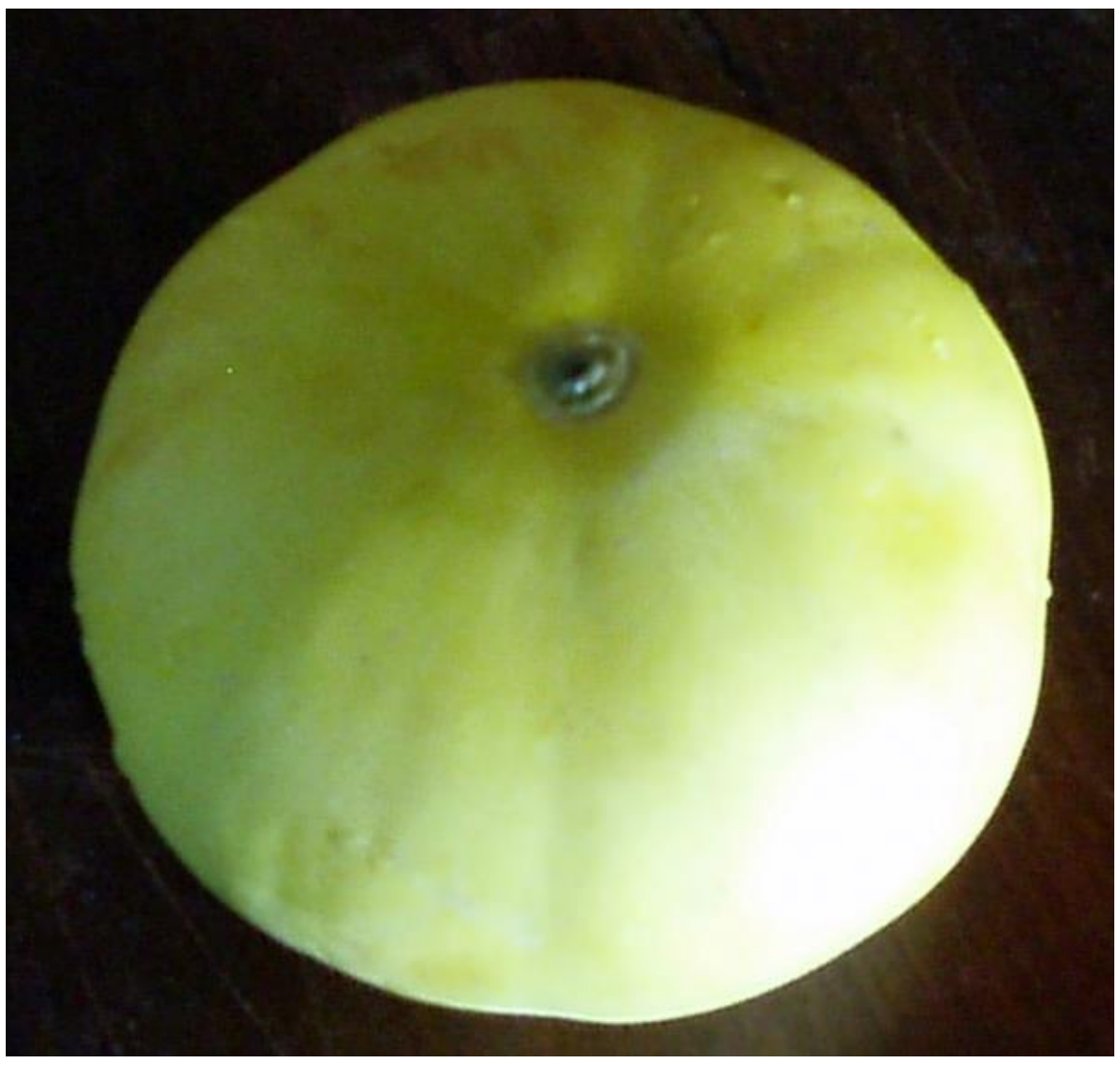

**Figure 3: Peculiar shape and striations in a melon**

2. Tagging (Mechanism)

Tagging is a mechanism which is frequently observed in cas. Tagging allows for the formation of aggregates. Tagging is exhibited in cas in the same manner as flags are used to identify troops or messages are given IP addresses to reach the correct destination in a network etc.

3. Nonlinearity (Property)

Nonlinear interactions are the norm in cas and are one of the reasons in the emergent global behaviors which indicate that the system is a cas.

4. Flows (Property)

Another property of cas is the formation of dynamic networks and flows. As such there are numerous attributes such as seen in Economics e.g. The Multiplier Effect when an extra amount of a resource is added to a flow causing it to transform as well as transmit between different nodes. Another important behavioral property that can be observed in flows is the recycling effect where resources are recycled over the flows via the network nodes thus enriching the emergent behavior. An example of the effects of recycling is the emergence of large number of insect and animal species in undisturbed forests as discussed by Nair [40]. Similar example is in the case of plants; Lowman notes the patterns of emergence, growth, mortality and herbivory of five rain-forest tree species[41].

5. Diversity (Property)

The species of a rainforest, people living in large towns, vendors in malls, structure of the mammalian brain all exhibit extreme diversity. So, it is rare to see the same type of components if any of these cas are explored in depth.

Being a dynamic property, diversity also has self-repairing and self-healing characteristics. Disturbing the population of a particular predator in a forest can result in an increase of numbers of several prey species which can then result in an increase of numbers of the same or other predators. Lima[42] notes that continuous predation can lead to major impact on entire ecosystems. Hein and Gillooly [43] have recently demonstrated that dispersal and resource limitation can jointly regulate assembly dynamics of predators and preys.

6. Internal Models (Mechanism)

Internal models are a mechanism by which agents inside a cas keep models of other individuals. On first looks, such a model would seem to be the property of only intelligent mammals. However, a closer look reveals that even the seemingly extremely simple life forms such as bacteria and viruses need to keep models of a certain kind of their environment and their hosts. These models are important for the survival of the species in general and thus the fact that a particular species still exists is a testimony that the species can defend its own in the particular environment (as has been happening for perhaps millions of years). Berg[44] notes that E. coli exhibit model behavior because they exhibit motility towards regions in their environment that are deemed favorable.

7. Building Blocks (Mechanism)

Internal models as described in the previous section need to be enhanced by realistic mechanisms. These realistic mechanisms are termed as the “Building blocks”. As an example, a diverse numbers of human beings form the 6+ Billion human population and they can each be basically differentiated from each other based on only a few set of building blocks such as eyes, nose, lips, hair, face etc. Variations in these blocks include changes in constituent attributes such as their color, shape etc. and eventually form the basis of the internal models.

In this section, we have noted how complex adaptive systems are structured and how they have a set of complex features which need to be examined closely to see why the systems behave in the peculiar manner at the global scale. We next talk a bit about “emergence”.

### 2.2.2 Emergence

Emergent behavior in cas has traditionally been considered difficult to define and hard to detect. Yaneer Bar Yam[45] defines emergence as follows:

“Emergence refers to the existence or formation of collective behaviors — what parts of a system do together that they would not do alone.”

Yaneer also notes that emergence can be considered as a means of explaining collective behavior or properties of a group of components or agents or else it can also be used to describe a system in relation to the environment. Boschetti and Gray [13] describe three levels of emergence:

1. Pattern Formation and detection such as oscillating reactions etc.
2. Intrinsic Emergence such as flocking behavior
3. Causal Emergence such as human behavior using messaging.

## 2.3 Examples of cas

Cas can be both natural as well as artificial. Here we give some examples of different types of cas. Next, we give specific examples of natural and artificial cas.

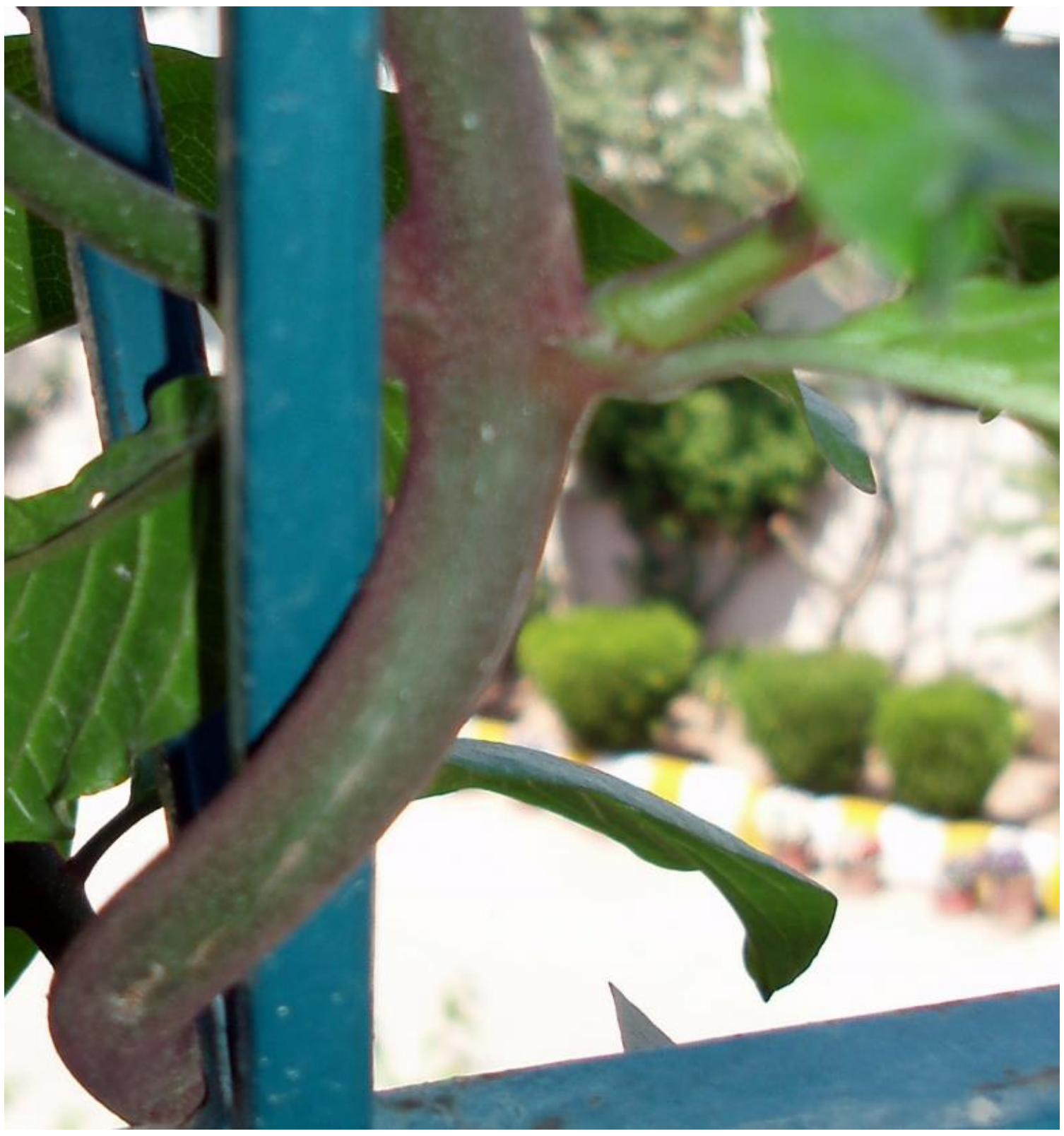

**Figure 4: Tree stem adapting structure temporally**

### 2.3.1 Natural cas example 1: Cas in plants

An example of adaptation in a natural system can be seen in Figure 4 where we see a tree stem adapting according to the peculiar shape of an artificial metallic structure. One cas which can be noted here is the complete plant. The plant itself is made up of numerous cells interacting and performing tasks based on programs from their genomes. However each cell is unaware of the large scale features of the plant or the way it interacts with the environment such as the emergent behavior of adaptation that can be observed in the figure in response to the position of the metallic structure as well as the path followed by sun for a better chance at getting sunlight. This emergent behavior can be seen as essentially a trait which has helped plants and animal species to survive over millions of years.

### 2.3.2 Natural cas example 2: Cas in social systems

While humans themselves are made up of several cas originating from living cells and a variety of bacteria, viruses and other biochemical organic molecules. All these play an important role in the individual as well as social lives of living beings. An example is given here in the domain of the research process, which can, by far, be considered as one of the most complex social interaction, based on intelligent behavior.

A research paper published in a peer-reviewed venue represents the culmination of efforts of a large number of interactions. Some of the entities involved in a research paper are as follows:

- Authors and their Papers which were read during the study by the researcher
- Discussions and meetings with colleagues
- Advice by advisors
- Advice by conference or journal referees during the peer-review process
- And so on . . .

One way of understanding this all is to notice that this entire process reflects a highly dynamic cas with multiple intelligent and inorganic (or even cyber-) entities such as papers, reviews and Journals. Several emergent behaviors can be observed here such as the propagation of research, formation of research groups, rise and fall of academic journals and their impact factors, emerging trends and these are all based on an enormous community of people spread globally, which do not even understand or even, at times, do not need to understand the exact structure and dynamics of how research leads to emergent behavior.

A quantitative study of scientific communication is technically termed as scientometrics [46]. Scientometrics requires the use of a multitude of sophisticated techniques including Citation Analysis, Social Network Analysis and other quantitative techniques for mapping and measurement of relationships and flows between people, groups, organizations, computers, research papers or any other knowledge-based entities.

Scientometrics sometimes involves the use of domain visualization, a relatively newer research front. The idea is to use information visualization to represent large amounts of data in research fronts[47]. This allows the viewer to look at a large corpus and develop deeper insights based on a high level view of the map[48]. Visualization using various network modeling tools has been performed considerably for social network analysis of citation and other complex networks[49]. Various types of Scientometric analyses have previously been performed for domains such as HIV/AIDS [50], Scientometrics [51], Mexican Astronomers[52], scientific collaborations [53] and engineers in South Africa [54]. Extensive work on research policy has been performed by Leydesdroff [46]. Some of the recent studies in this direction include visualization of the HCI domain[55], identification of the proximity clusters in dentistry research[56], visualization of the pervasive computing domain[57], visualization of international innovation Journals[58] as well as identification of trends in the Consumer Electronics Domain[2].

Scientometric studies which combine co-citation analysis with visualizations greatly enhance the utility of the study. They allow the readers to quickly delve into the deeper aspects of the analysis. Co-citation analysis measures the proximity of documents in various scientific categories and domain. These analyses include primarily the author and the journal co-citation analyses. Journal co-citation analysis identifies the role of the key journals in a scientific domain. In contrast, the author co-citation analysis[59] especially by using visualization[60] offers a view of the structures inside a domain. The idea is based on the hypothesis that the higher the frequency of co-citation of two authors, the stronger the scientific relation between them. Whereas document co-citation maps co-citations of individual documents [61-63], author co-citation focuses on body of literature of individual authors. In addition, co-citation cluster analysis of documents is useful for the display of the examination of scientific knowledge evolution structure at the macro level. In initial cluster analysis, techniques involved clustering highly cited documents using clustering based on single-links. Subsequently clustering is also performed on the resultant clusters numerous times[64]. Recent techniques involve the use of computational algorithms to perform this clustering. These clusters are then color coded to reflect that.

If we were to take some of these entities, and analyze the data, we can come up with quite interesting networks e.g. as shown in Figure 5. This network represents top papers, in terms of citations, with topics related to "agent-based modeling" from the years 1990 – 2010. As can be seen here, the network visualization is clearly allowing the separation of the top journals in this field in terms of "centrality" measures such as degree, betweenness, eccentricity or indices or clustering coefficients as described in previous section. Complex Network are not only formed from social interactions. Rather they can be developed from any cas system interaction such as systems inside living beings[29].

Journal Citation Reports (JCR) has been considered in literature as the key indicator of a Journal's scientific repute[65]. It is pertinent to note here that previously researchers such as Amin and Mabe [66] have critically reviewed the way Journal impact factor might be used by authors and journals. While alternatively other researchers have proposed new alternative impact factors such as by Braun [67] and Fersht [68]. However, the fact remains that, in general, the scientific contributions listed in Thomson Reuters Web of Science are considered highly authentic by the overall scientific community[69], [70].

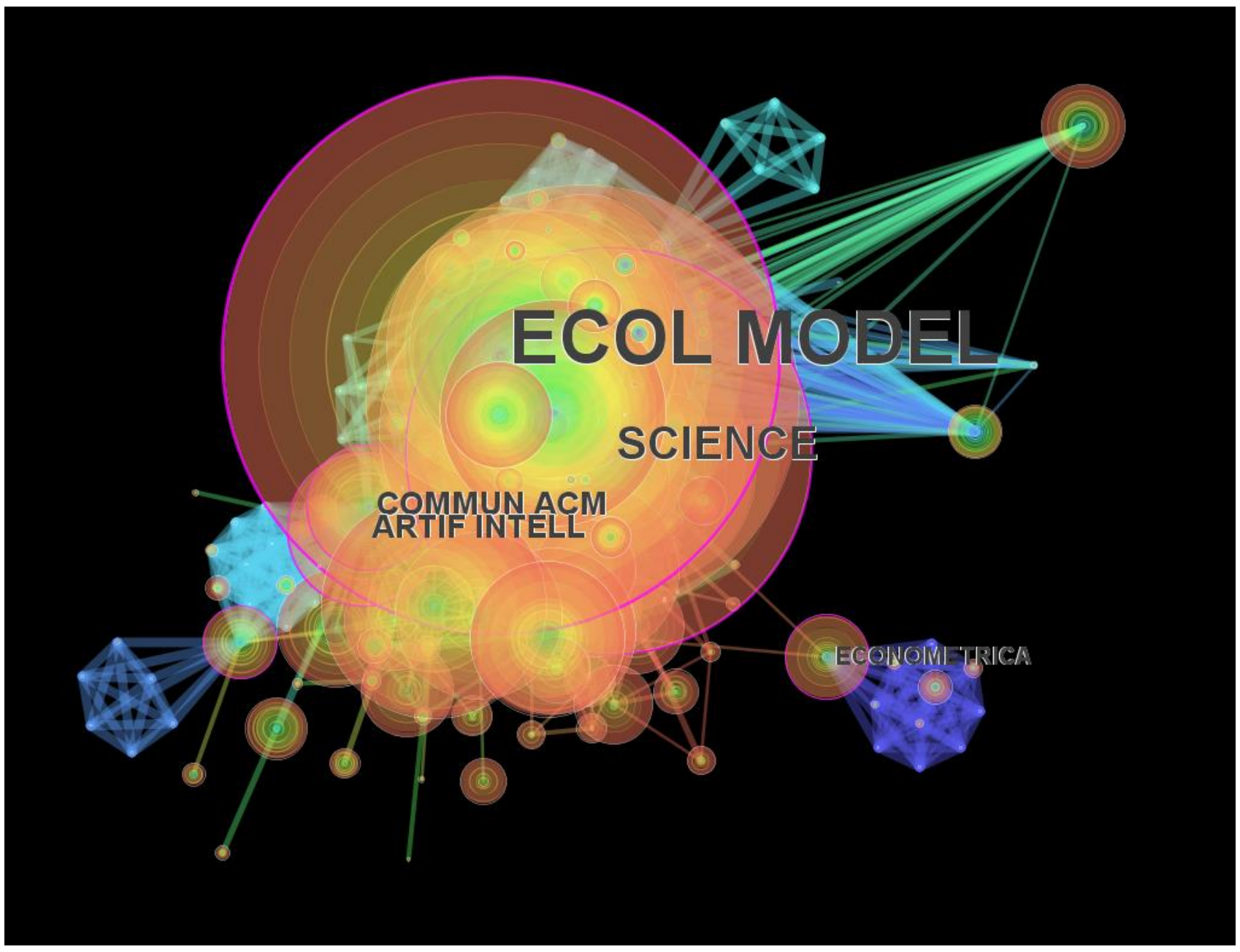

Figure 5: Complex Network model showing different Journals publishing articles related to agent-based modeling. Bigger caption represents a higher centrality measure explained later in the chapter.

### 2.3.3 Artificial cas example 1: Complex adaptive communication networks

In this section, we introduce complex adaptive communication networks. Complex adaptive communication networks are a recent advancement in artificial cas. Primarily these are communication networks which have a large number of components resulting in behavior associated with cas. These networks arise due to a recent rapid advancements in communication technology because today's communication networks such as those formed by wireless sensor, ad-hoc, Peer-Peer (P2P), multiagent, nano-Communication and mobile robot communication networks, are all expected to grow larger and more complex than ever previously anticipated. Thus, these networks, at times, can possibly give rise to complex global behaviors similar to natural cas. As a result, network designers can expect to observe unprecedented emergent patterns. Such patterns can be important to understand since, at times, they can have considerable effect on various aspects in a communication network such as unanticipated traffic congestion, unprecedented increase in communica-tion cost or perhaps a complete network/grid shutdown as a result of emergent behavior. Some well-known examples include the emergence of cascading faults in Message Queue-based financial transactions after New Year holidays[71], recent cascading failures reported in the Amazon.com cloud[72], effects of viral and worm infections in large networks, effects of torrent and other complex traffic in Internet Service Providers and corporate networks, multi-player gaming and other similar P2P traffic in company intranets, self-organization and self-assembly related effects in sensor and robotic networks[73].

The torrent protocol is an example of an artificial cas since it exhibits a number of interesting properties associated with cas. It relies on software called the “torrent clients” which, as can be seen in Figure 6 allow autonomous and almost anonymous interaction with other clients around the globe. The different peers autonomously self-organize and adapt to both download as well as upload files. Here, the emergent phenomenon is the downloading of the file. The interactions, allocations of bandwidth, uploading and downloading of chunks are part of the nonlinear interactions.

The way this entire process the process of uploading and downloading (sharing) files works is outlined by Cohen in [74] as follows:

1. The peer computer which starts to share a file breaks the file into a number of identically sized pieces with byte sizes of a power of 2 anywhere from 32 KB to 16 MB each. Each piece is hashed using SHA-1 hash function and this information is recorded in the .torrent file.
2. Peers with a complete file copy are called seeders and those providing the initial copy are called the initial seeders.
3. The .torrent is a collective set of information about the file coupled with the information about tracker servers allowing peer discovery of downloaders.
4. The trackers give random lists of peers to other peers.

5. The different peers automatically self-adapt to download the complete file from distributed chunks of file from all over the globe.
6. Peers who only download and do not upload are punished by lesser bandwidth while peers who upload as well are given higher bandwidths.
7. The emergent behaviors here include the file sharing (upload and download) and reduction of peers who only download.

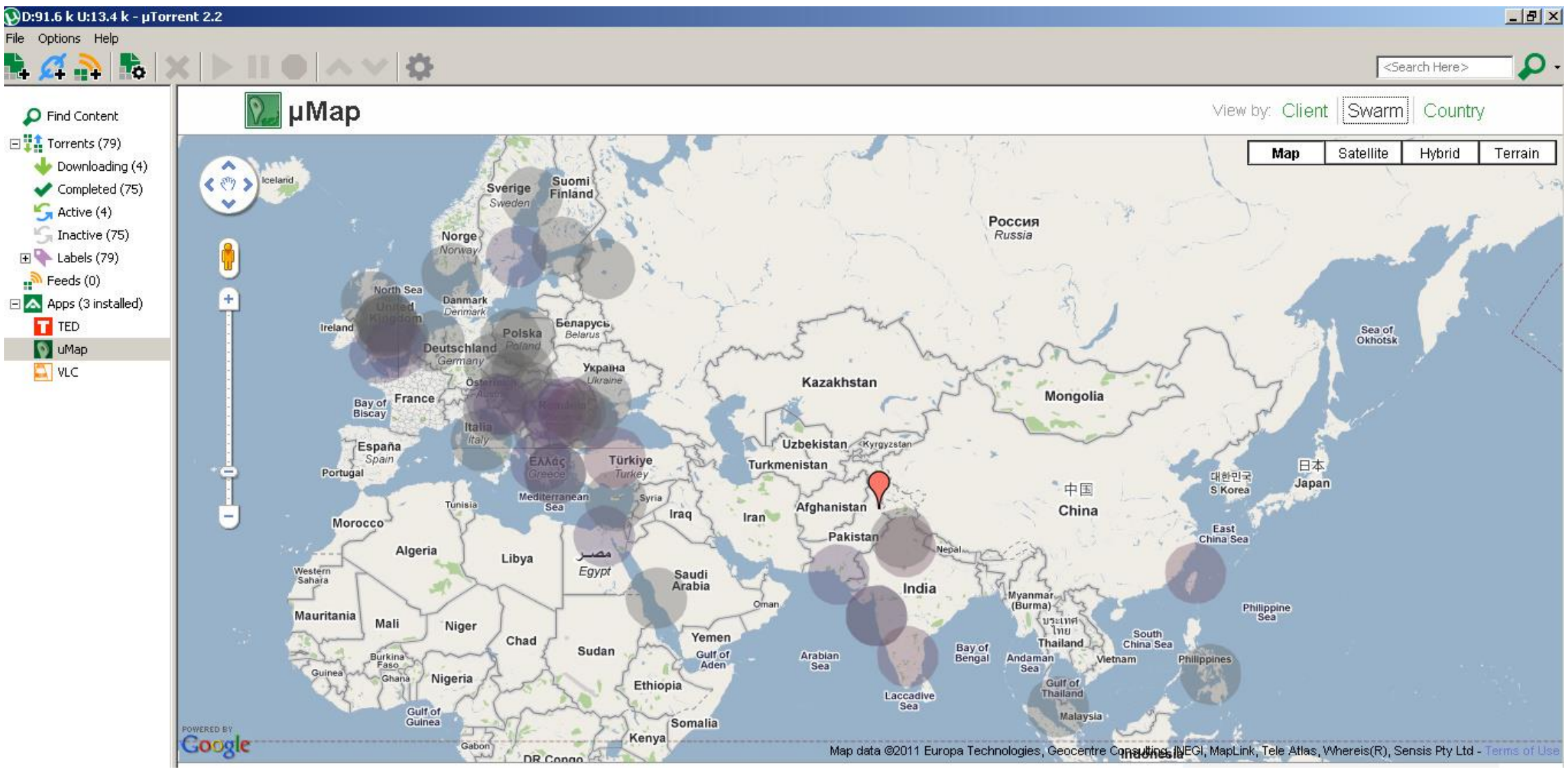

**Figure 6: Artificial Global cas formed by P2P clients observed using utorrent software**

### 2.3.4 Artificial cas example 2: Simulation of flocking boids

In 1986, a computational model of flocking animal motion was presented by Craig Reynolds [75] called the "boids". The model is based on three separate but simple rules. The first rule is "separation" which involves steering to avoid crowding. The second rule is "alignment", which steers the boid towards the average heading of local flockmates. And the third rule is "cohesion", which steers towards the average position of local flockmates. The "boids" model has been considered a model of realistic simulations ranging from SIGGRAPH presentations to Hollywood. This model offers an example of complex adaptive behavior based on apparently local rules of the mobile agents or boids.

## 2.4 Modeling cas

The word "Model" is defined by Merriam-Webster in 13 different ways:

*1 obsolete : a set of plans for a building*

*2 dialect British : copy, image*

*3 : structural design *a home on the model of an old farmhouse**

*4 : a usually miniature representation of something; also : a pattern of something to be made*

*5 : an example for imitation or emulation*

*6 : a person or thing that serves as a pattern for an artist; especially : one who poses for an artist*

*7 : Archetype*

*8 : an organism whose appearance a mimic imitates*

*9 : one who is employed to display clothes or other merchandise : Mannequin*

*10 a : a type or design of clothing b : a type or design of product (as a car)*

*11 : a description or analogy used to help visualize something (as an atom) that cannot be directly observed*

*12 : a system of postulates, data, and inferences presented as a mathematical description of an entity or state of affairs*

*13 : Version*

Critically speaking, modeling a system is important for the survival of the human race. The entire concept of cognition appears to be based on modeling. We start our learning by experiencing this world, which leads us to the development of cognitive "mental" models. Then we use these models to associate with other experiences or knowledge to keep learning. As an example a baby develops a first implicit "model" of heat if s/he touches a hot bottle of milk. Later on, in her life, she keeps learning more and more about temperature and heat while whether she consciously realizes it or not, all new models are rooted essentially with the basic cognitive model. This kind of models have been termed as "implicit" models[76].

Implicit models, however, are hard to express and are not very communicable. What one person perceives of a certain concept, might mean a totally different thing to another.

This is the reason for the development of what Epstein, an authority in modeling, mentions as "explicit models" [16].

Modeling implies a certain level of understanding and prowess over the system and requires a deep study of certain aspects of the system. When it is not just one system, rather a group of systems, modeling implies developing abstractions, which can cover different domains.

Previous techniques have included focus on physics as a means of modeling. Such formal models typically are simplified models representing the system as e.g. Differential equations. Obviously by limiting an entire system to only a few equations implies that most aspects of the system including the components have been reduced to simple numbers, e.g. quantities such as change of numbers etc. However, more recently modeling cas often entails using one of two basic methods of developing computational models i.e. Either use of Agent based modeling to develop simulation models or else development of complex network models for analysis of interactions using real world or simulation data as shall be discussed in detail in the next section. Although these models prevail in literature, to the best of our knowledge, there is no unified framework which outlines this set of ideas and couples them along with agent-based models.

### 2.4.1 Complex Network Modeling

In this section, an overview of complex network methods is presented. The basic idea of Complex Network Analysis originates from graph theory. A complex network is essentially an advanced graph, where unlike theoretical graphs, each node and edge is loaded with more information. As an example, if we were to develop a social network of friends, the network edges could represent the level of emotions which the friends have for each other. So, if a person Alice considers three people (say Bob, Cassandra and David) as friends, then this could be represented as a network. Suppose Alice considers Cassandra as her best friend so this could be modeled in the network by giving a higher attribute value to the link/edge between Alice and Cassandra. To make things more interesting, the network could be directed so it is possible that Bob might not even consider Alice as a friend. So, each friendship relation should be represented by a directed edge.

Using various Mathematical tools and models, complex networks can be subsequently analyzed and compared with other models from the same or different domains. As an example Barabasi [77] compares Biological networks with models of the world wide web. Besides classification, networks can also be used to discover important features of the numerous nodes (or components) of the cas. While Complex Networks are a general set of methods, the growth of literature of complex network usage can be identified across various cas literature ranging from Social Network Analysis[30] to Biological Networks [29] and in Citation Networks[78].

#### 2.4.1.1 Complex Network Methods

In a cas, interactions can be modeled as networks. The key difficulty in network design is to understand which data needs to be captured to develop the network model(s). Once the data has been converted to a network, there exist a large number of complex network analysis Mathematical and software tools for performing the analysis. However, the hard part here is the selection of data, the design and extraction of the complex network models. While application of the actual centrality measures appears to be deceptively easy due to the existence of a large set of software tools for assistance, it can be very difficult to actually come up with data which is relevant as well as generates useful networks, which can represent the entire cas properly. Development of partial networks can result in completely incorrect results as complete networks are required for the accurate calculation of centrality measures (as discussed in the next section). Calculation of centrality measures requires the use of all possible nodes in a connected network.

#### 2.4.1.2 Theoretical basis

Graph theory has a long history. One of the earliest known representation of graph problem is the Leonard Euler's "Konigsberg bridge problem" in 1735which involved representing a map of the river Pregel in the form of a graph[79].

A *graph* G = (V, E) is essentially formed by a set of *vertices* or *nodes* V and a set of *edges* or *arcs* E. Each edge is assigned to two vertices, where the vertices might not be disjunct. Graph nodes as well as edges are given other alpha-numeric attributes including the physical coordinates for display purposes. Networks are typically analyzed using the following methods:

1. Global network properties: Since real-world cas differ from purely randomly generated networks, they can be distinctly categorized using global properties. Three distinct types of network models have been used in literature. The first one is the Erdős-Rényi random network model[80]. The second model was proposed by Watts and Strogatz by analysis of many real-world networks which they discovered to have a smaller average shortest path length coupled with a high clustering coefficient as compared with random networks. These networks are labeled the Watts and Strogatz small-world model[81] after their discoverers. Another model, called the scale-free network model was proposed by Barabasi and Albert [82] based on ideas of complex networks from real-world cas domains. These networks exhibit some degree of a power law distribution in the "degree centrality". This implies that they contain very few nodes with a higher degree and a large number of nodes with a lower degree.

2. Network centralities: Unlike global network properties, centralities deal with particular nodes. As such, using a centrality analysis, different nodes can be ranked according to their importance in *Centrality indices*. It is pertinent to note that not all centrality measures might give substantial results for a particular network. As such, it is typical to develop numerous networks as well as perform considerable analysis by trial and error before a particular extraction of network as well as a centrality measure is discovered showing significant results.

3. Network motifs: Motifs are commonly used network architectural patterns. They are often used in cas modeling to discover particular patterns of local interactions. Examples include signal transduction[83] and gene regulatory biological networks[84]. Recent work includes Davidson [85] who has demonstrated emerging properties in Gene regulatory networks.

4. Network clustering: Biological networks have been studied in the form of hierarchical structures with motifs and pathways at lower levels to functional molecules for large scale organization of networks by Oltvai and Barabasi[86].

#### 2.4.1.3 Centralities and other Network measures

In this section, a description of some of the commonly used quantitative measures in complex networks is provided.

##### *2.4.1.3.1 Clustering Coefficient*

The clustering coefficient is a means of quantitative measurement of the local cohesiveness in a complex network. It is a measure of the probability that two vertices with a common neighbor are connected.

The formula for calculation of the clustering coefficient is given as follows:

$$C_i = \frac{2E_i}{k_i(k_i - 1)} \qquad (2.1)$$

Where

$E_i$ = number of edges between neighbors of a node i

$k_i$ = degree of the node i

The global/mean clustering coefficient C = <$C_i$> is found by taking an average of all the clustering coefficients.

#### *2.4.1.3.2 Matching Index*

Matching index gives a quantitative means of measuring the number of common neighbors between any two vertices. In other words, for empirical networks, two functionally similar vertices need not be directly connected with each other. It is formally defined as following:

$$M_{ij} = \frac{N_{k,l} A_{ik} A_{jl}}{k_i + k_j - N_{k,l} A_{ik} A_{jl}} \qquad (2.2)$$

Where the numerator represents the common neighbors while the denominator represents the total number of neighbors of the two nodes $i$ and $j$.

#### *2.4.1.3.3 Centrality measures*

A centrality is formally defined as a function $C : V \mapsto R$ for a directed or undirected graph $G = (V, E)$ [29]. A centrality is a real number assigned to each vertex allowing for a pair-wise comparison across the entire network.

#### 2.4.1.3.3.1 Degree Centrality

Degree centrality or simply “degree” of a node is a basic centrality measure which identifies the connections of a node with other nodes. For directed networks, there can be two kinds of degree centrality measures. The in-degrees and the out-degrees can be found by the summation of the total in-links and the total out-links respectively. Degree centrality is regarded as a local centrality measure however nodes having a higher degree can be more important as compared with nodes with a lower degree. As such, deletion of high degree nodes can disrupt the network structure and flow of interactions.

#### 2.4.1.3.3.2 Eccentricity Centrality

Eccentricity of a node v is defined as the maximum distance between v and u for all nodes u. Eccentricity centrality assists in computation of the maximum distance, which can be defined as the length/distance of the longest shortest path to all other vertices present in the network. Nodes which are easily reachable from other nodes receive the highest value. Eccentricity identifies the node that can be reached within equal distance from all other nodes in the network. Mathematically, eccentricity can be defined as:

$$C_{ecc}(s) = \frac{1}{\max\{dist(s,t)\}} \tag{2.3}$$

Where

*s* and *t* are the nodes belonging to the set of vertices V

#### 2.4.1.3.3.3 Closeness Centrality

Closeness centrality is similar to eccentricity but utilizes the minimum distance from a target node to all other nodes within the network.

The formula for the calculation of the closeness centrality is of a node *s* for all nodes *t* is given as follows:

$$C_{clo}(s) = \Sigma \frac{1}{dist(s,t)} \tag{2.4}$$

**2.4.1.3.3.4 Shortest Path Betweenness Centrality**

It is calculated by measuring the number of times a particular node comes in the shortest path between any two nodes. In other words, it identifies the ability of a node to monitor communication across the network. A higher value implies a better ability of a particular node to monitor network communication. Mathematically it can be defined as:

$$C_{spb}(v) = \Sigma \frac{\sigma_{st}(v)}{\sigma_{st}} \tag{2.5}$$

Where

$\sigma_{st}(v)$ is the number of the shortest paths between nodes s and t containing the vertex v between the nodes

$\sigma_{st}$ is the total number of the shortest paths present between two nodes s and t

2.4.1.4 Software tools for Complex Networks

While Complex Network Analysis uses a number of mathematical tools, some of which have been described in this chapter, for practical reasons, network scientists use a number of software tools for performing network operations. These include network construction, extraction, simulation, visualization and analysis tools such as Network Workbench[87], Pajek[88] and Citespace[89], Cytoscape[90], Visone[91] and others.

### 2.4.2 Agent-based Modeling and Agent-based Computing

Agent-based Computing ranges from building intelligent agents to multiagent systems and agent-based simulation models. There has been considerable focus in the Artificial Intelligence and other research communities on something which has been termed agent-based computing[92], [93]. Although prevalent across a number of domains, there is considerable confusion in the literature regarding clearly a differentiation between its various flavors. Here we shall attempt to disambiguate its various levels by providing a hopefully clear set of definitions of the various terms and their uses.

#### 2.4.2.1 Agent-Oriented Programming

The first of the agent-oriented paradigm which we see is focused more on the increased use of artificially intelligent agents such as proposed by Russell and Norvig [94]. The focus here is on developing more complexity in an individual agent rather than on a large set of agents.

#### 2.4.2.2 Multi-agent oriented Programming

While agent-oriented programming focused on individual agents, it was difficult to actually design, implement and run multiple agents with such intricate internals because of the need for massively parallel computational resources for even simple problems. However, multiagent-oriented programming typically attempts to add at least some intelligence to agents and observe their interactions. One thing to note here is that typically such models have either few agents with a focus on certain aspects only, or else do not implement a whole lot of complex behaviors[94], [95]. The primary reason for not implementing all intelligent paradigms is the inability to perform all such calculations in real-time in any currently known artificial computational system. That is one reason why online algorithms in AI typically use simplistic mechanisms for searching such as heuristic functions other than purely reactive simple or table-driven agents (which can just perform an if else type of actions). Another key problem is the knowledge-engineering bottleneck[96], where the agents had difficulty communicating with each other. Although partially solved by languages such as Knowledge Query Manipulation Language (KQML)[97] and DARPA Agent Markup Language (DAML)[98] and DAML+OIL[99] language released in 2000-01 time frame, practically large-scale intelligent agents have not seen the daylight of most commercial or real-life scenarios (such as earlier predicted by agent-technology evangelists).

#### 2.4.2.3 Agent-based Or Massively multiagent Modeling

One particular application of agents, which actually has found extensive use is what is termed as individual or agent-based models. The key idea in these models is the focus on use of "simple" simulated agents interacting with a large population of other agents with emphasis on observing the global behavior. Here instead of the individual agents themselves, the interest of researchers is on the use of the interaction at the micro-level to observe features at the global values. The focus of main-stream agent-based modeling and simulation has been either social systems [20], [100] or else ecosystems[101], [102] .

. In this section, we shall review these paradigms further. Agent-based models [103] (or individual-based models as termed in some scientific disciplines) have been used for the modeling of a wide variety of complex systems[104]. Their use ranges from as diverse as Biological Systems[105-120] to Social systems[121], [122], from Financial systems [123] to supply chains [124], from the modeling of honey bees[125] to modeling of traffic lights[126]. With such an impressive set of applications the strength of agent-based modeling[127] is quite apparent. However, with such a parallel evolution of ideas, there also exists a set of domain-specific and perhaps conflicting concepts. These emerge primarily due to a lack of communication or standardization between different scientific domains.

One possible explanation of the prevalence of agent-based modeling in such a diverse set of scientific domains could be the models is their similarity to the human Cognition because perhaps this is how most humans perceive the surrounding world. In general, agent-based models are particularly useful for developing models of Complex Adaptive Systems. Typical examples of software used to develop agent-based modeling and simulation are NetLogo[128], StarLogo[129], Repast [19], MASON[95], Swarm [130] and others .

#### 2.4.2.4 Benefits of Agent-based Thinking

For system modeling involving a large number of interacting entities, at times it is more appropriate to be able to access parameters of each individual "agent". In emerging communication paradigms, network modelers need to model a variety of concepts such as life forms, sensors and robots. Agent-based thinking allows for direct addressability of individual entities or agents. The concept of breeds allows system designer to freely address various system entities. In agent-based modeling and simulation, agents are designed to be addressed for performing a certain action.

Using a type name, agents or "turtles" can be asked to perform actions or change their attributes. Agent-based models are developed with a high level of abstraction in mind. In contrast to programming abstracts such as loops, designers ask agents without worrying about the low-level animation or interaction details necessary for execution. As such, this results in the production of compact programs with a high degree of functionality.

An important benefit of small programs is their ability to greatly reduce the tweak-test-analyze cycle. Thus it is more likely to model complex paradigms within a short time without worrying about the lower layers of other parameters unless they are important for the particular application. As we shall see in this chapter, it also has great expressive power and most of all, is fun to use. Although being used very frequently to model complex and self-organizing systems it has not previously been used extensively to model computer networks.

## 2.5 A Review of Agent-based Tools

In the arena of agent-based tools, a number of popular tools are available. These range from Java based tools such as Mason to Logo based tools such as: StarLogo, NetLogo [128] etc. Each of these tools has different strengths and weaknesses. In the rest of the chapter, we focus on just one of these tools: NetLogo as a representative of this set. Building on the experience of previous tools based on the Logo language, NetLogo has been developed from grounds up for complex systems research. Historically, NetLogo has been used for modeling and simulation of complex systems including multi-agent systems, social simulations, biological systems etc, on account of its ability to model problems based on human abstractions rather than purely technical aspects. However, it has not been widely used to model computer networks, to the best of our knowledge.

### 2.5.1 NetLogo Simulation: An overview

In this section, we introduce NetLogo and demonstrate its usefulness using a number of modeling and simulation experiments. NetLogo is a popular tool based on the Logo language with a strong user base and an active publicly supported mailing list. It provides visual simulation and is freely available for download and has been used considerably in multi-agent systems literature. It has also been used considerably in social simulation and complex adaptive networks[131]. One thing which distinguishes NetLogo from other tools is its very strong user support community. Most times, you can get a reply from someone in the community in less than a day. NetLogo also contains a considerable number of code samples and examples. Most of the time, it is rare to find a problem for which there is no similar sample freely available either within NetLogo's model library or elsewhere from the NetLogo M&S community.

Based on the Logo language, the NetLogo world consists of an interface which is made up of "patches". The inhabitants of this world can range from turtles to links. In addition, one can have user interface elements such as buttons, sliders, monitors, plots and so on.

NetLogo is a visual tool and is extremely suitable for interactive simulations. When one first opens up a NetLogo screen, an interface with a black screen is visible. There are three main tabs and a box called the command center. Briefly, the interface tab is used to work on the user interface manually and the "Information" tab is used to write the documentation for the model. And finally the "procedures" tab is where the code is actually written. The "command center" is a sort of an interactive tool for working directly with the simulation. It can be used for debugging as well as trying out commands similar to the interpreter model which, if successful, can be incorporated in one's program. The commands of a NetLogo procedure can be executed in the following main contexts:

The key inhabitants of the Logo world are the turtles which can be used to easily model network nodes. The concept of agents/turtles is to provide a layer of abstraction. In short, the simulation can address much more complex paradigms which include pervasive models, environment or terrain models or indeed any model the M&S specialist can conceive of without requiring much additional add-on modules. However, the tool is extensible and can be directly connected to Java based modules. By writing modules using Java, the tool can potentially be used as the front end of a real-time monitoring or interacting simulation. For example, we could have a java based distributed file synchronization system, which reports results to the NetLogo interface and vice versa, the NetLogo interface could be used by the user to setup the simulation at the backend (e.g. how many machines, how many files to synchronize and subsequently with the help of the simulation, the user could simply monitor the results). Although the same can be done with a lot of other tools and technologies, the difference is that NetLogo offers these facilities almost out of the box and requires minimal coding besides being non-commercial, free and easy-to-install. A single place where the turtle exists is a patch. Observer is a context, which can be used in general without relating to either a patch or a turtle. The NetLogo user manual, which comes pre-packaged with NetLogo, says: “Only the observer can ask all turtles or all patches. This prevents you from inadvertently having all turtles ask all turtles or all patches ask all patches, which is a common mistake to make if you're not careful about which agents will run the code you are writing.”

Inside the NetLogo world, we have the concept of agents. Coming from the domain of complex systems, all agents inside the world can be addressed in any conceivable way, which the user can think of, e.g., if we want to change the color of all nodes with communication power less than 0.5W, a user can simply say: ask nodes with [power < 0.5] [set color green] or if a user wants to check nodes with two link neighbors only, this can be done easily too and so on.

The context of each command is one of three. Observer object is the context, when the context is neither turtle nor the patch. It is called the observer because this can be used in an interactive simulation where the simulation user can interact in this or other context using the command window.

Although there are no real rules to creating a NetLogo program, one could design a program to have a set of procedures which can be called directly from the command center. However, in most cases, it suffices to have user interface buttons to call procedures. For the sake of this chapter, we shall use the standard technique of buttons.

In this program there will be two key buttons; Setup and the Go buttons. The idea is that the “setup” button is called only once and the “go” button is to be called multiple times (automatically). These can be inserted easily by right clicking anywhere on the interface screen and selecting buttons. So, just to start with NetLogo, the user will need to insert these two buttons in his or her model, remembering to write the names of the buttons in the commands. For the go, we shall make it a forever button. A forever button is a button which calls the code behind itself repeatedly. Now, the buttons show up with a red text. This is actually NetLogo’s way of telling us that the commands here do not yet have any code associated with them. So, let us create two procedures by the name of “setup” and “go”. The procedures are written, as shown in Figure 7, in the procedures tab and the comments (which come after a semi-colon on any line in NetLogo) explain the actions that are performed by the commands. This code creates 100 turtles (nodes in this case) on the screen. However the shape is a peculiar triangle by default and colors are assigned at random. Note that we have written code here to have the patches colored randomly.

```
to setup
ca ; Clears everything so if we call setup again, it won't make a mess
crt 100 ;This  means we are creating 100 turtles
[
setxy random-pxcor random-pycor   ;These  100  turtles,  we want them to be spaced out at random
; patch x and y co-ordinates
]
let mycolor random 140 ; Randomly select a color value from 0 to 139
ask patches
[
```

**Figure 7 Code for setup**

To create the procedure for the go, the code can be written as listed in Figure 8.

```
to go
ask turtles
[
fd 0.001; ask each turtle to move a small step
]
end
```

**Figure 8 Code for go**

Now, if we press setup followed by go, we see turtles walking slowly in a forward direction on the screen, a snapshot of which is shown in Figure 9.

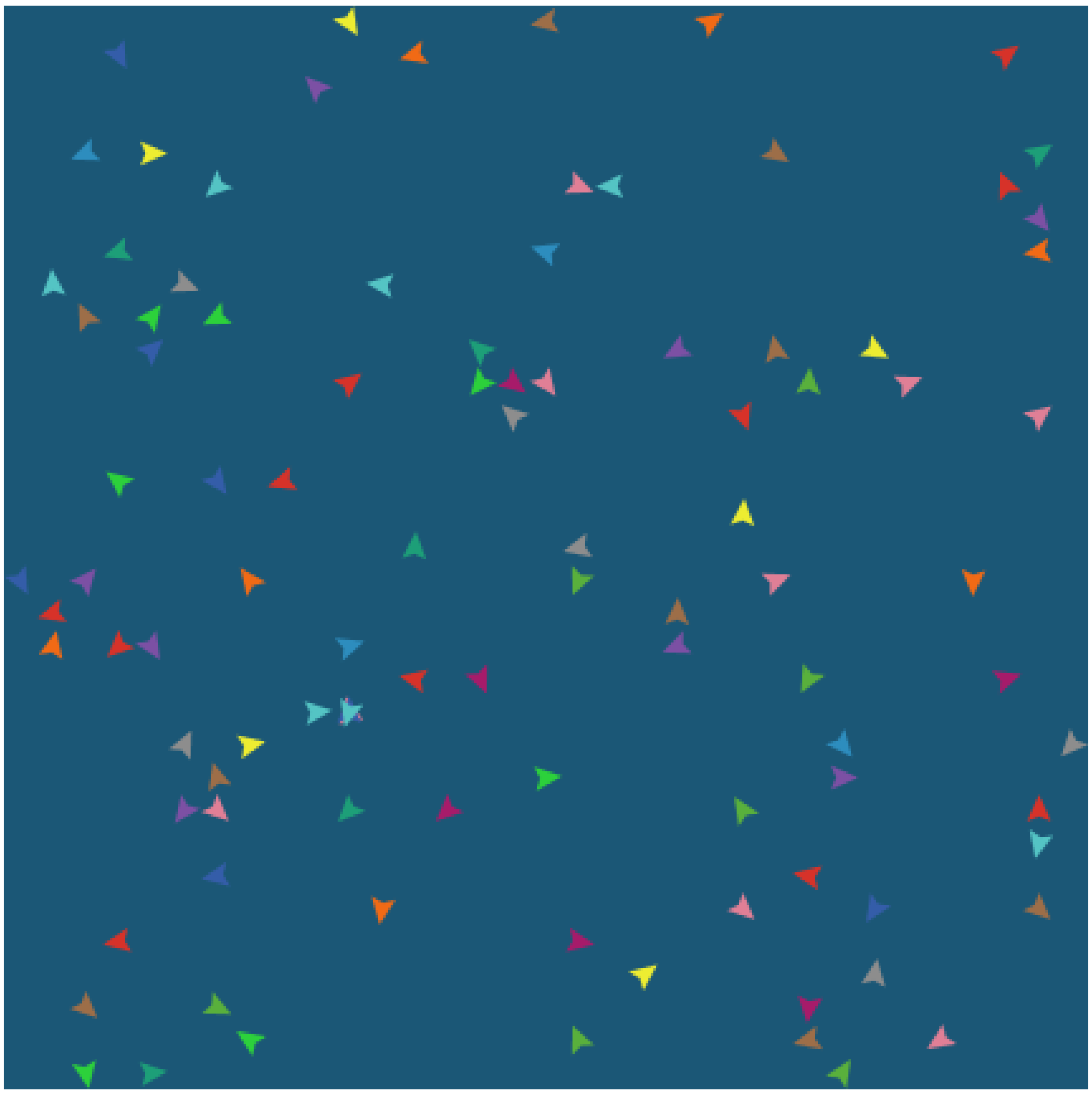

**Figure 9: Mobile Nodes**

### 2.5.2 Overview of NetLogo for Modeling Complex interaction protocols

NetLogo allows for the modeling of various complex protocols. The model, however, does not have to be limited to the simulation of only networks; it can readily be used to model human users, intelligent agents, and mobile robots interacting with the system or virtually any concept that the M&S designer feels worthwhile having in the model. NetLogo, in particular, has the advantage of LISP-like[132] list-processing features. Thus modeled entities can interact with computer networks. Alternatively, the simulation specialist can interact and create run-time agents to interact with the network to experiment with complex protocols, which are not otherwise straightforward to conceive in terms of programs.

As an example, let us suppose, if we were to model the number of human network managers (e.g. from 10 to 100) attempting to manage a network of 10000 nodes by working on workgroups the size of n nodes (e.g. ranging from 5 to 100) at one time while giving a total of 8 hour shifts with network attacks coming in as a Poisson distribution; this can be modeled in less than a few hours with only a little experience in NetLogo per se. The simulation can then be used to evaluate policies of shifts to minimize network attacks.

Another example could be the modeling and simulation of link backup policies in case of communication link failures in a complex network of 10,000 nodes along with network management policies based on part-time network managers carrying mobile phones for notification and management vs. full-time network managers working in shifts etc. all in one simulation model. And to really make things interesting, we could try these out in reality by connecting the NetLogo model to an actual J2ME based application in Symbian phones using a Java extension; so the J2ME device sends updates using GPRS to a web server which is polled by the NetLogo program to get updates while the simulation is updated in a user interface provided by NetLogo. Again, although the same could be done by a team of developers in a man-year or so of effort using different technologies, NetLogo provides for coding these almost right out of the box and the learning curve is not steep either.

This expressive nature of NetLogo allows modeling non-network concepts such as pervasive computing alongside human mobile users (e.g. in the formation of ad-hoc networks for location of injured humans) or Body Area Networks come in play along with the network. Now, it is important to note here that simulation would have been incomplete without effective modeling of all related concepts which come into play. Depending upon the application, these could vary from ambulances, doctors, nurses to concepts such as lap-tops, injured humans etc. in addition to readily available connectivity to GIS data provided by NetLogo extensions.

### 2.5.3 Capabilities in handling a range of input values

Being a general purpose tool, by design, the abstraction level of NetLogo is considered higher. As such, the concepts of nodes, antenna patterns and propagation modeling are all user-dependent. On one hand, this may look burdensome to the user accustomed to using these on a regular basis, as it might appear that he or she will be working a little extra to code these in NetLogo modeling. On the other hand, NetLogo allows for the creation of completely new paradigms of network modeling, wherein the M&S specialist can focus on, for example, purely self-organization aspects or on developing antenna patterns and propagation modeling directly.

### 2.5.4 Range of statistics and complex metrics

NetLogo is a flexible tool in terms of using statistics and measurements. Any variable of interest can be added as a global variable and statistics can be generated based on single or multiple-run. Plots can be automatically generated for these variables as well.

Measurements of complex terms in NetLogo programs are very easy to perform. As an example, if it is required to have complex statistics such as network assembly time, global counters can be used easily for this. Similarly, statistics such as network throughput, network configuration time, throughput delay can be easily modeled by means of similar counters (which need not be integral). By default, NetLogo provides for real-time analysis. Variables or reporters (functions which return values) can be used to measure real-time parameters and the analyst can actually have an interactive session with the modeled system without modifying the code using the “Command Window”.

## 2.6 Verification and Validation of simulation models

### 2.6.1 Overview

Validation of a simulation model is a crucial task[31], [133]. Simulations, however well-designed, are always only an approximation of the system and if it was so easy to build the actual system, the simulation approach would never have been used [134]. In traditional modeling and simulation practices, a standard approach to verification and validation (V&V) is the three step approach given by Naylor et al. in [135] as follows:

1. The first step is to develop a model that can intrinsically be tested for a higher level of face validity.
2. The next step is to validate the assumptions made in the model
3. Finally, the transformations of the input to output from the model need to be compared with those for the actual real-world system.

### 2.6.2 Verification and Validation of ABMs

While verification and validation are important issues in any simulation model development, in this section, we discuss the peculiarities associated with ABMs in the domain of modeling and simulation of cas.

Firstly ABMs can be difficult to validate because there is a high tendency of errors and un-wanted artifacts to appear during the development of an ABM as noted by Galan et al. [136]. In addition, since the number of parameters in an ABM model can tend to be quite high, Lucas et al. have noted that it is possible to fall into the trap of tweaking the variables [137].

Bianchi et al. note that validation of agent based models can be quite challenging [138]. Another problem noted by Hodges and Dewar is as to how to ensure that the observed be-havior is a true representative of the actual system[139]. Fagiolo et al. mention four set of issues with Agent-Based models [140]:

1. A "lack of robustness".
2. Absence of a high degree of comparability between developed agent-based models.
3. A dire shortage of standard techniques for the construction and analysis of agent-based models.
4. Difficulties in correlation of the models with empirical data.

Keeping these issues in mind, while validation techniques of agent-based models have been mentioned in some domains such as computational economics and social simulation, there are still a number of open issues:

A. There is no standard way of building Agent-based Models.
B. There is no standard and formal way of validation of Agent-based Models.
C. Agent-Based Modeling and Agent-Based Simulations are considered in the same manner because there is no formal methodology of agent-based modeling [141].
D. Agent-Based Models are primarily pieces of Software however no software process is available for development of such models.
E. All validation paradigms for agent-based models are based on quantifying and measurable values but none caters for emergent behavior [142], [143] such as traffic jams, structure formation or self-assembly as these complex behaviors cannot be quantified easily in the form of single or a vector of numbers.
F. Agent-based models are occasionally confused with multi-agent systems (MAS) even though they are developed for completely different objectives; ABM are primarily built as simulations of cas, whereas MAS are typically actual software or at time, robotic systems. Although MASs may themselves be simulated in the form of an ABM but that does not change their inherent nature. We would like to note here that differences between MAS and ABMs have been discussed earlier in the chapter and a further exposition to this effect will be performed in Chapter 3.

### 2.6.3 Related work on V&V of ABM

In social sciences literature such as in the case of Agents in Computation Economics (ACE), empirical validation of agent-based models has been described by Fagiolo et al. in [140]. Alternate approaches to empirical validation have been noted by Moss in [144]. Validation of models is closely related to model replication as noted by Wilensky and Rand in [145]. An approach of validation based on philosophical truth theories in simulations has been discussed by Schmid in [146]. Another approach called "companion modeling", an iterative participatory approach where multidisciplinary researchers and stakeholders work together throughout a four-stage cycle has been proposed by Barreteau et al. in [147].

A different point of view also exists in literature which uses agent-based simulation as a means of validation and calibration of other models such as by Makowsky in [148]. In addition, agent-based simulation has also been shown to be useful in the validation of multi-agent systems by Cossentino et al. in [149].

## 2.7 Overview of communication network simulators

In this section, an overview of different types of simulators used for the simulation of communication networks is provided.

### 2.7.1 Simulation of WSNs

Current sensor network simulation toolkits are typically based on simulators such as NS-2, OPNET[150], J-Sim[151], TOSSIM[152] etc. SensorSim is a simulation framework [153]. Amongst traditional network simulations, J-Sim and NS2 have been compared in [151] and J-Sim has been shown to be more scalable. Atemu [154] on the other hand, focuses on simulation of particular sensors. Likewise, TOSSIM [152] is a TinyOS Simulator.

### 2.7.2 Simulation of P2P networks

In terms of simulation of P2P networks, on one hand, there are actual implementations and on the other, there are simulators of the P2P protocols. These include Oversim [155] which is based on the OMNET++[156] simulator. Another simulator for P2P networks is the PeerSim[157] simulator.

### 2.7.3 Simulation of Robotic Swarms

In the case of swarm robotics, simulators include Swarm-Bot simulator [158] and WebotsTM robot simulator[159]. Other simulators include LaRoSim which allows for robotic simulations of large scales[160].

### 2.7.4 ABM for complex communication networks simulation

While we have noted above that individual sub-areas of this domain such as WSNs, Swarm robots as well as P2P networks all have dedicated as well as general purpose simulators but all of them are limited to their specific domain, the benefit of using ABM in the modeling and simulation of complex communication networks is that it is a general purpose methodology and thus it can be used to simulate any combination of these in addition to being able to simulate other entities such as humans, plants and animals in the simulation.

## 2.8 Conclusions

In this chapter, we have discussed the background and related work required for an understanding of the subsequent chapters. We have discussed modeling and simulation and its relation to cas research. In the next chapter, we present complex network modeling level of the proposed framework, which is concerned with developing Complex Network models and performing Complex Network Analysis of large amounts of cas interaction data.

# 3 Proposed Framework: Complex Network Modeling Level

In this chapter, we propose the complex network modeling level of the unified framework. This idea is to develop a level which encompasses existing complex network modeling studies as well as allow other studies to be conducted. This level is applicable when suitable interaction data is available from a given cas. The basic idea is to develop graphs in the form of complex networks using interaction data of different cas components and characteristics. Subsequently these complex networks can be used to perform a CNA for the discovery of emergent patterns in the data; patterns which would otherwise not have been apparent using either legacy statistical or other mathematical methods. While this chapter can be considered a generalized methodology, it can also be considered as an extension of the previous chapter since the data used for analysis of the case studies includes a case study about of agent-based computing. Whereas the second case study is in the domain of consumer electronics selected specifically to demonstrate the generalized applicability of the proposed framework level.

## 3.1 Overview

This chapter demonstrates the application of CNA on two separate case studies. The idea is to use Information visualization and Complex Network Analysis to discern complex patterns in large corpora of reliable scientific interaction data.

Both case studies use different corpora but are based on the well-defined structured data from the standard source of The Thomson Reuters Web of Science (WOS) which is well-known as a standardized indexing database for scientific research papers. The corpora was selected as the WOS since it is a stable index and has a well-defined methodology for adding new data. While it is always possible to find some indexing errors in any database, this particular dataset presents a suitable candidate better than most other manually collected datasets. The first case study of the application of the proposed methods is on "Agent-based Computing" while the second is on "Consumer Electronics".

## 3.2 Generalized Methodology

In Figure 10, we give details of how complex network modeling level of the proposed framework can be useful in developing complex network models of a cas and subsequently performing CNA to determine emergent behavior and patterns in the data.

The first step is the acquisition or retrieval of interaction data. In other words, only data which allows the development of interaction models would be needed in this framework level. Typical summarized statistical data or records would not be sufficient for the development of these models. Typically such data must demonstrate important relations of some kind between various components to prove useful for complex network development.

After data retrieval, the next step is to develop complex network models. While it might appear a trivial exercise to develop these models, it can actually take significant time and effort before suitable networks can be developed. One especially difficult task is the selection of the most suitable columns for network extraction and construction. This exercise can require extensive experimentation using trial and error before the most suitable columns can be discovered for network construction. Network manipulation is often applied to develop and extract suitable networks in an iterative manner. This can involve selective removal, insertion, merging and splitting of various nodes and edges followed by refinement of the network using various colors and sizes of the nodes and edges based on loaded data from the cas components.

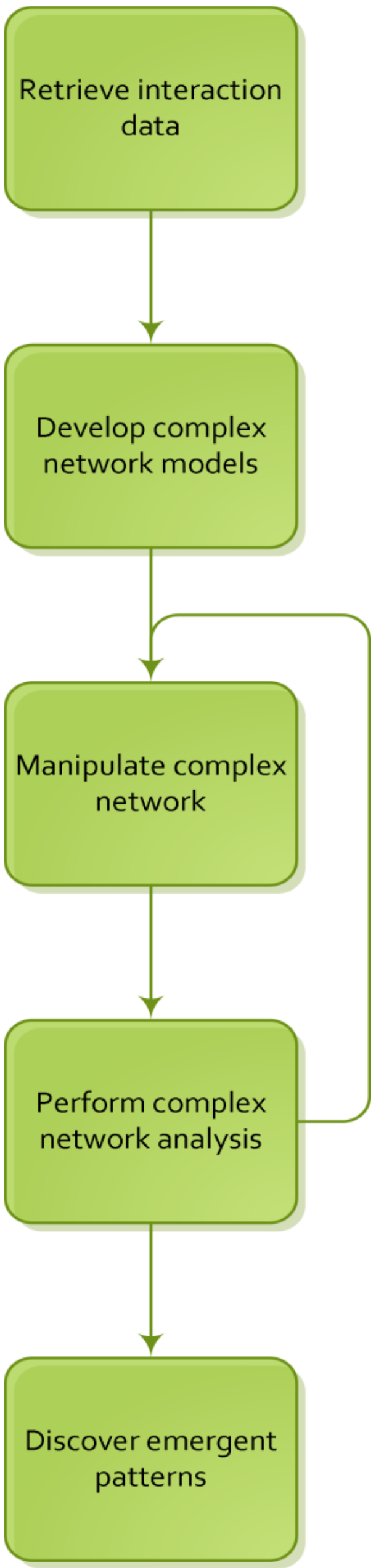

**Figure 10 Proposed Methodology for the complex network modeling level (for the discovery of emergent patterns based on available interaction data of cas components)**

Gradually the networks start to shape up. Eventually this process can result in giving useful information about the topological structure of the entire domain. This approach is particularly useful when other basic approaches to modeling fail (such as using statistical tests, machine learning, data mining etc.). The next step in this level is to perform CNA. There are essentially two different types of CNA, which have been used in previous literature (and described earlier in Chapter 2). One type of CNA is useful for classification of the entire network using global properties such as degree distribution. The other type of CNA can often prove to be more useful and involves the calculation of various quantitative measures such as Centralities, Clustering coefficients and indices. Once these measures have been calculated, quite often, the analysis is still far from being over. Next, we demonstrate the application of the complex network methods on two different case studies.

### 3.3 Case Study I: Agent-based Computing

The area of interest of the first case study is agent-based computing, a diverse research domain concerned with the building of intelligent software based on the concept of "agents". In the first case study, we perform Scientometric analysis to analyze all sub-domains of agent-based computing. The data consists of 1,064 journal articles indexed in the ISI web of knowledge published during a twenty year period: 1990-2010. These were retrieved using a topic search with various keywords commonly used in sub-domains of agent-based computing. The proposed approach employs a combination of two applications for analysis, namely Network Workbench and CiteSpace as mentioned earlier in chapter 2. The idea is that Network Workbench allowed for the analysis of complex network aspects of the domain whereas a detailed visualization-based analysis of the bibliographic data was performed using CiteSpace. Results include the identification of the largest cluster based on keywords, the timeline of publication of index terms, the core journals and key subject categories. We also identify the core authors, top countries of origin of the manuscripts along with core research institutes. Finally, the results have revealed the strong presence of agent-based computing in a number of non-computing related scientific domains including life sciences, ecological sciences and social sciences.

### 3.3.1 Problem Statement:

As discussed earlier in the previous chapter, agent-based computing encompasses both multiagent systems as well as agent-based modeling. Agent design and simulation go hand in hand but in completely different ways in different sub-domains of agent-based computing. So, e.g. on one hand, there are researchers whose research goals revolve around the design of various types of agents where the role of simulation is closely linked to validation of the future operation of actual or physical agents[161]. On the other, there is another group of researchers whose goal is not agent-design but rather the agent-design is a means of developing simulations which can lead to better understanding of global or emergent phenomena associated with complex adaptive systems[6], [141]. This broad base of applications of this research area thus often leads to confusions regarding the exact semantics of various terms in the literature. This is tied closely to the evolution of "agent-based computing" into a wide assortment of communities. These communities have at times, perhaps nothing other than the notion of an "agent" in common with each other.

The goal of the CNA here is to use network development, analysis and visualization to give a Scientometric survey of the diversity and spread of the domain across its various sub-domains, in general, and to identify key concepts, which are mutual to the various sub-domains, in particular. These includes identifying such visual and Scientometric indicators as the core journals, the key subject categories, some of the most highly cited authors, most central authors, institutes of highly cited authors and the top countries of manuscript origin.

Our goal is to use a combination of two different tools based on their relative strengths; Network Work Bench [87], [162] for performing a Complex Network Analysis using Network Analysis Toolkit and CiteSpace[89].

The objectives of this study can be summarized as follows:

- To identify the largest clusters of inter-connected papers for the discovery of complex interrelations of various sub-areas of agent-based computing.
- To discover "bursts"[163], [164] of topics along with a timeline of publication of the index terms used.
- To identify the core journals for the whole of agent-based computing ranging from agents, multi-agent systems to agent-based and individual-based simulations, in terms of citations of articles.
- To identify the key subject categories.
- To identify and study the most productive authors, institutes and countries of manuscript origins.

### 3.3.2 Methodology

In this section, we give an overview of the research methodology for the first case study. While the overall methodology of any CNA is similar, there are always peculiarities of application in each case study as well as each cas domain. As such this involved data retrieval as well as subsequent CNA performed using the two separate tools. It is pertinent to note here that this analysis might appear trivial and a simple application of the complex network methods and software, however, the actual case is far from that. Simply finding and understanding the correct data tags for development of complex networks requires extensive understanding of the data types and transformations. Thus the analyses in this chapter reflect the last iteration of the application of these methods rather than a simple and naïve application of these tools. In addition, the datasets were time and again verified manually and the data was collected initially by hand and consisted of several tens of thousands of records. Subsequently the networks which are demonstrated here were developed however again these are not the first networks which were developed. Network manipulation was also applied extensively. In short, developing complex network models and pruning net-works for custom application of Complex Network methods is a nontrivial task and takes considerable time, effort and learning for each application. It is best learnt by practicing and observing existing case studies, which is the reason we are demonstrating two different case studies as a means of how other case studies can be applied using the proposed complex network modeling level of the framework.

#### 3.3.2.1 Data collection

The input data was retrieved using searches from the WOS [165]. A thorough topic search for data was devised to cater for various aspects and keywords used in agent-based computing in the following three sub-domains:

1. Agent-based, multi-agent based or individual-based models
2. Agent-oriented software engineering
3. Agents and multi-agent systems in AI

The search was performed in all four databases of Web of Science namely SCI-EXPANDED, SSCI, A&HCI, CPCI-S for all years. Details of the search have been provided in the Appendix 1. For the sake of analysis, the range of years 1990-2010 was selected with search limited to only Journal articles. Bibliographic records retrieved from SCI include information such as authors, titles, abstracts as well as references. The addition of cited references resulted in a total of 32, 576 nodes. The details of the search keywords along with reasoning for the selection are given in the Appendix 1.

### 3.3.3 Results

The results are presented starting with a basic look at the overall picture of articles retrieved from the Web of Science. As can be noted in Figure 11, the articles in this domain start primarily during the early 1990's and gradually keep rising and as such reach a total of 148 articles published merely in the last year 2009.

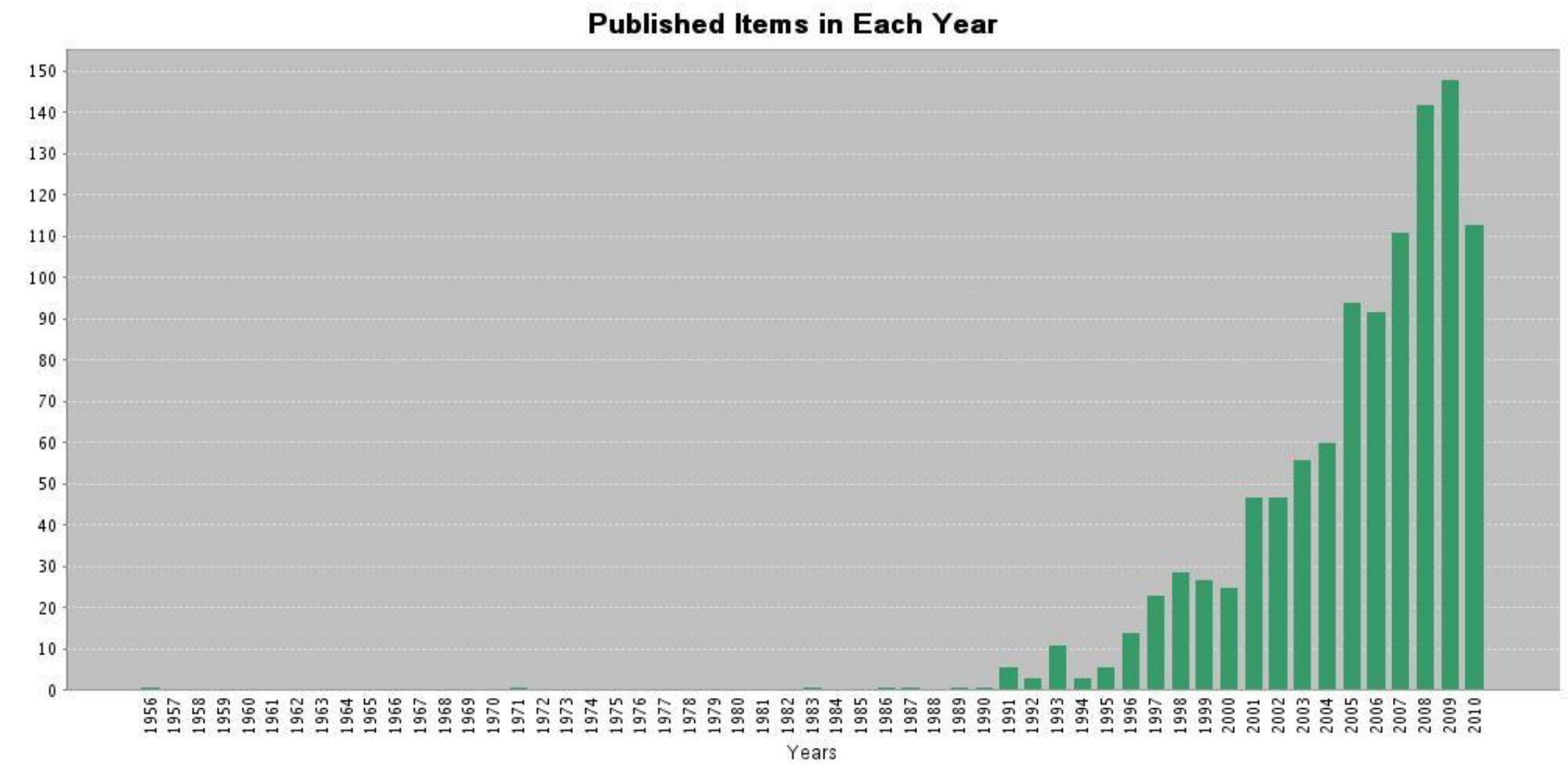

**Figure 11: Articles published yearly**

In addition, since the popularity of a domain is known to be based on its number of citations, we need to observe this phenomenon closely. It can be observed in the graph using data from the Web of Science in Figure 12. Thus, starting from a very small number of citations, Agent-based computing has risen to almost 1630 citations alone during the year 2009.

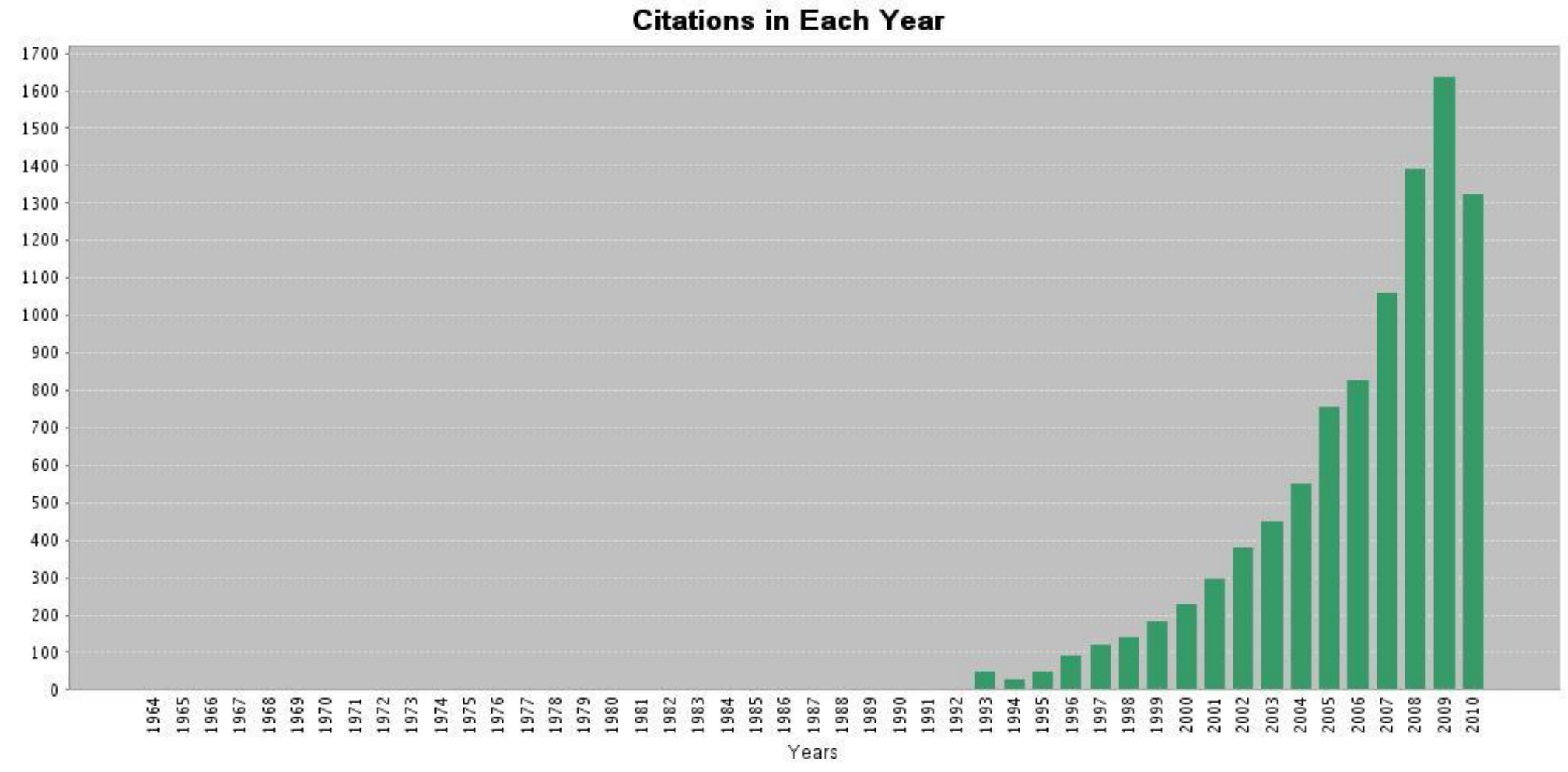

**Figure 12: Citations per year**

#### 3.3.3.1 Network Workbench (NWB)

Next, to get the big picture of the citation network, a network (paper co-citation network) was extracted from the ISI records using Network Workbench tool. In addition, analysis was performed for the extracted network using NWB based Network Analysis Toolkit [166]. The extracted global properties of the network are shown in Table 1. The first thing to note here is the large number of nodes i.e. 32, 576 which appears much larger than reported above from the Web of Science data. This is primarily because it includes the cited references as well as the original nodes. Now, we see here that there are no isolated nodes, which is obvious because every paper will at least cite some other papers at the very least. This is followed by the 39, 096 edges. Another interesting feature is the average degree, which we see is 2.4. The fact that the average degree is not significantly higher shows actually that a large number of papers have been co-cited in this corpus otherwise the degree would have been much higher.

The graph itself is not weakly connected but it consists of 78 weakly connected components with a largest component of size 31,104 nodes. The weakly connected components are formed based on articles connected with a small number of other articles identifying the importance of these papers (Which shall be investigated in depth later on in this chapter). Finally, we note that the density is 0.00004 however knowing the density does not give much structural information about the Scientific domain per se. These properties are a characteristic of the retrieved empirical data. However, we note here that while interesting, they do not provide in depth insights of the data. So we perform a further set of analysis using NWB

**Table 1 Basic Network Analysis of Extracted Paper Citation Network**

| Attribute | Value |
|---|---|
| Nodes: | 32, 576 |
| Isolated nodes: | None |
| Edges: | 39, 096 |
| Self Loops | None |
| Parallel Edges | None |
| Edge Attributes | None |
| Valued Network | No |
| Average total degree: | 2.400 |
| Weakly Connected | No |
| Weakly connected components | 78 |
| Largest connected component Size | 31, 104 nodes |
| Density (disregarding weights): | 0.00004 |

Here the structure of the overall agent-based computing domain can be examined further using NWB. To allow for the examination of the domain using the specific strengths, central to the NWB tool, the top nodes were extracted using a local citation count > 20. Subsequently the network was visualized using GUESS[167] and nodes were resized according to the values of the citation count. The result can be observed in Figure 13. Here we can note the peculiar relation of some of the top papers connected with Volker Grimm's paper in the center. Apart from Grimm's own papers, these include Robert Axelrod's as well as Parker's and Deangelis's works. In addition, we can note that here Kreft's and Huston's work in Microbiology and Biosciences as well as Wooldridge's and Ferber's work in the domain of multiagent systems are also showing up. For further analysis, we move on to the more advanced Information Visualization tool, namely CiteScape.

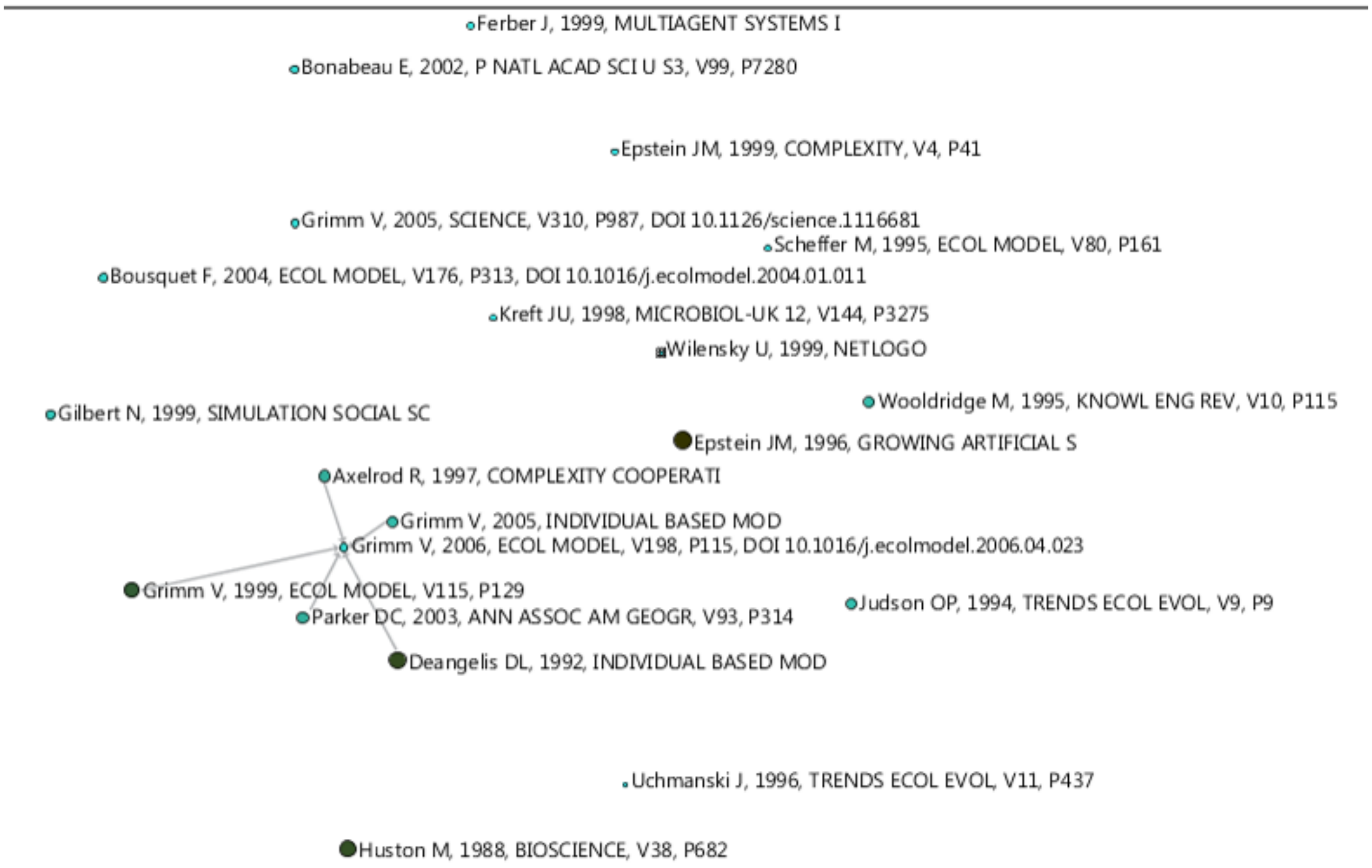

**Figure 13: Top papers with citations > 20**

3.3.3.2 CiteScape

The CiteScape tool allows for a variety of different analysis. CiteScape directly operates on the downloaded ISI data and builds various networks using time slicing. Subsequently using the various options selected by the user, the network can then be viewed in different ways and parameters can be analyzed based on centrality (betweenness) as well as frequency.

### *3.3.3.2.1 Identification of the largest cluster*

The goal of the first analysis was to observe the big picture and identify the most important indexing terms. Based on a time slice of one year, here in Figure 14, we see the largest cluster.

These clusters are formed using CiteSpace, which analyzes clusters based on time slices. Here the links between items show the particular time slices.

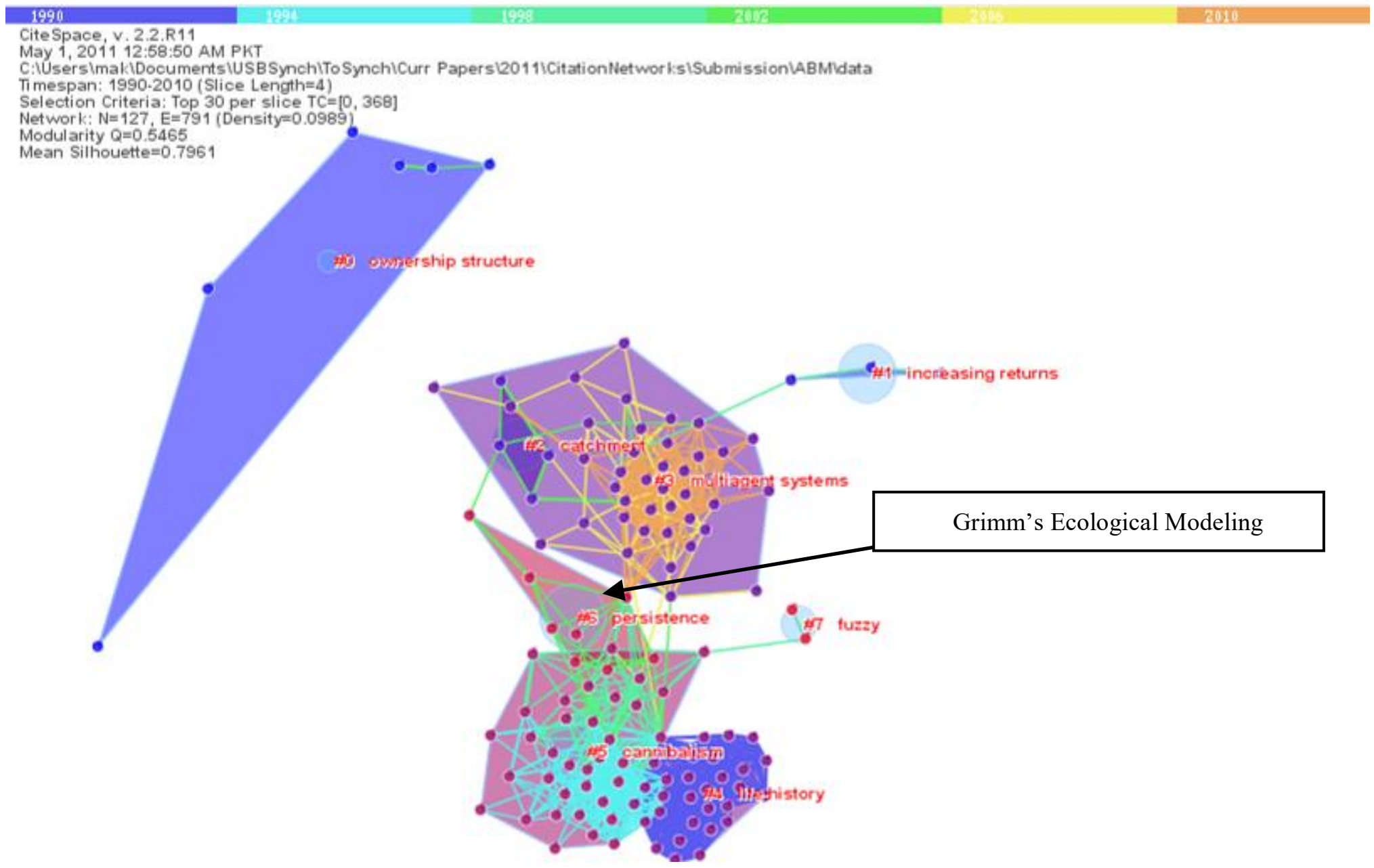

**Figure 14: Largest Cluster based on indexing terms**

In this figure, we first note the top where the year slices from 1990 to 2010 show up in 4 year slices. Using different colors in CiteSpace allows us to clearly identify turning points and other interesting phenomena in literature. We notice here that one of Grimm's papers is actually the key turning point from Agent-based Modeling to Multiagent-systems cluster. Another thing that can be noted here is the role of this paper in connecting a large number of papers in 2010 to the other papers in 2002-2006 era showing a somewhat delayed acceptance of the concepts.

### *3.3.3.2.2 Timeline and bursts of index terms*

The next analysis performed was to observe the clusters on a timeline as shown in Figure 15.

In complex networks, various types of centrality measures such as degree centrality, eccentricity, closeness and shortest path betweenness centralities etc. Citespace, in particular, uses betweenness centrality[89]. As discussed in previous chapter, this particular centrality is known to note the communication monitoring ability of a vertex for other network verti-ces. In other words, a higher centrality ensures that the vertex is between more of the shortest paths between other nodes relative to other nodes with lower centrality.

Using a time line especially helps identify the growth of the field. Please note here that these include papers which are based on agent-based computing as well as papers which are cited by these papers. Here, the red items are the “bursts”. Bursts identify sudden interest in a domain exhibited by the number of citations.

We can see that there are a lot of bursts in the domain of agent-based model. In addition, even the preliminary analytics and visualization here confirms the hypothesis that agent-based computing has a very wide spread article base across numerous sub-domains. This is obvious here as we see clusters ranging from “drosophila” and “yellow perch bioenergetics” to those based on “group decision support systems”, all ranging from different domains. Further analysis strengthens this initial point of view based on an examination of the various clusters.

Here, the results demonstrate the effects of the semantic vagueness discussion earlier. So, e.g. Where concepts such as group decision support system, rule-based reasoning and coordination are concepts tied closely with developing intelligent agents, they also show up in the domain right alongside agent-based modeling.

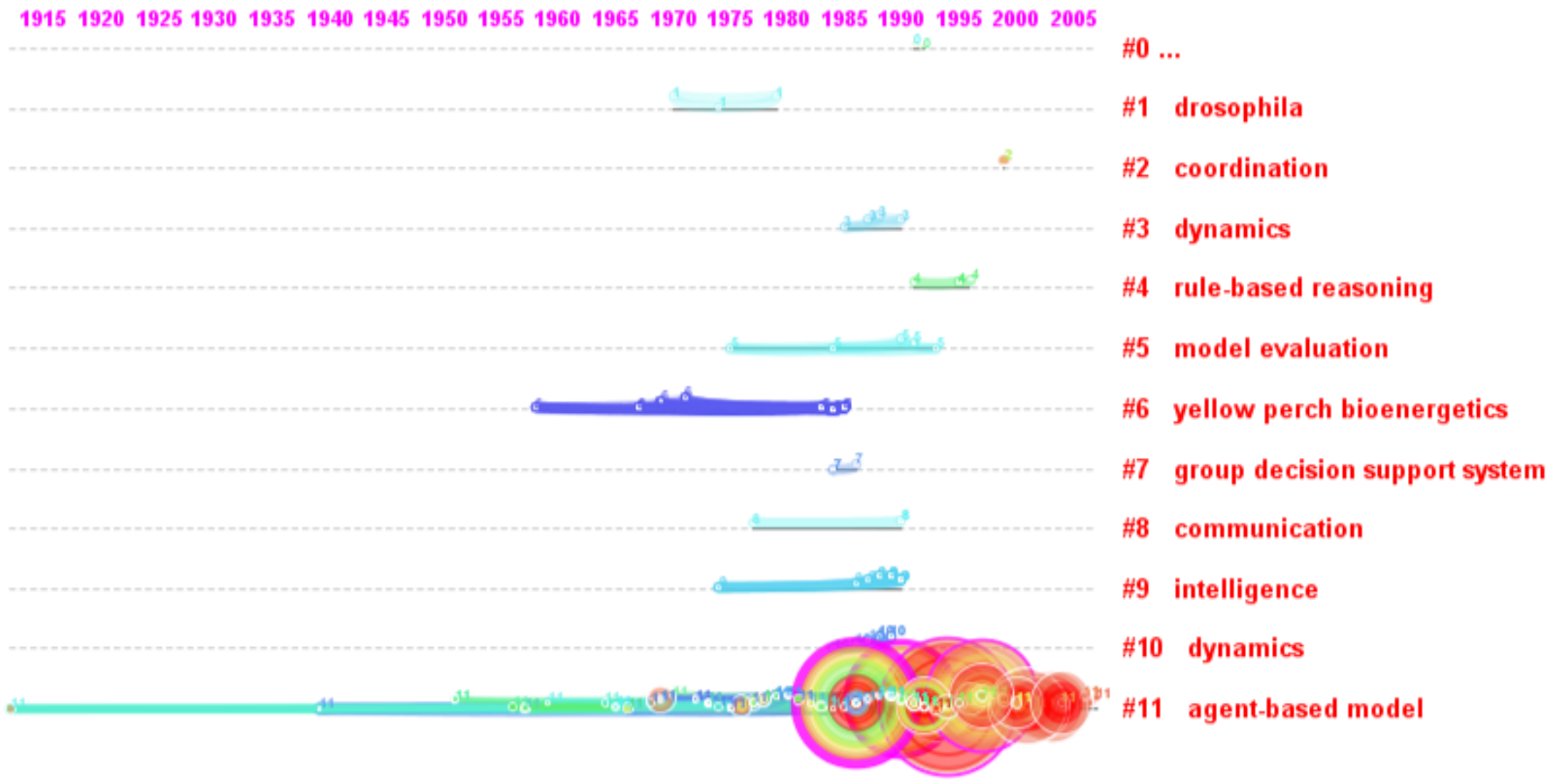

**Figure 15: Timeline view of terms and clusters of index terms (based on centrality) also showing citation bursts**

### *3.3.3.2.3 Analysis of Journals*

Our next analysis was to identify the key publications of the domain[3]. This can be seen in Figure 16. Here the key journals are identified based on their centrality.

Once again, we can note here that the vagueness in the terms of use again shows up in the set of mainstream Journals of the domain with "Artificial Intelligence" and "Communications of the ACM" being relevant mostly to Agents associated with the concepts of "Intelligent agents" and "Multiagent Systems" whereas "Econometrica", "Ecological Modeling" and the journal "Science" representing the "agent-based modeling" perspective.

[3] It is pertinent to note here that we faced one peculiar problem in the analysis of the retrieved ISI data. The Web of Science data identified a Journal named "Individual-based model". However extensive searches online did not find any such journal.

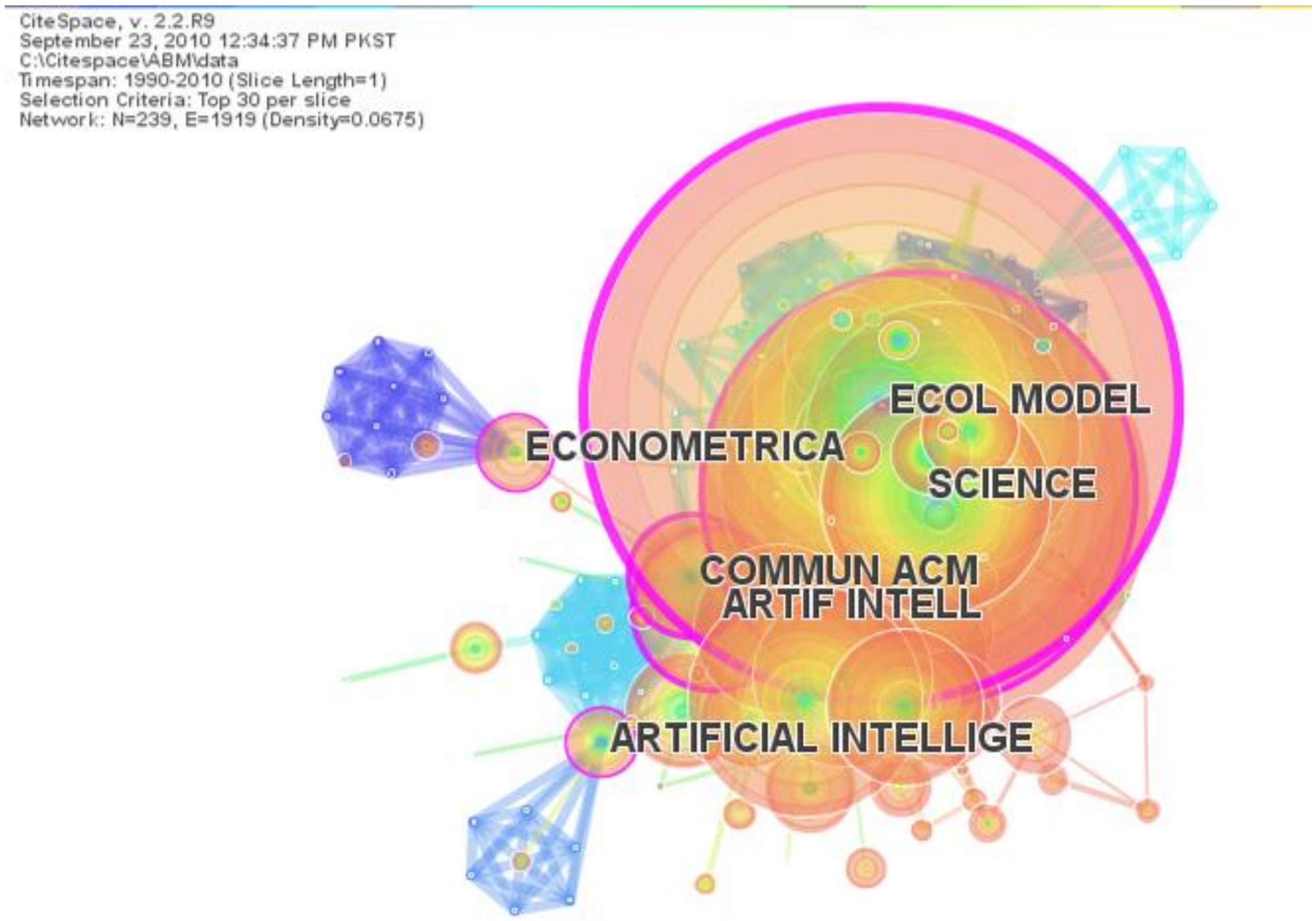

**Figure 16: Key Journals based on centrality**

In Table 2, we give details of these top journals based on centrality shown in the figure. This represents the centrality of the top ten key journals. In terms of centrality, the "ECOL MODEL" Journal has the highest value of centrality among all the journals. In addition, here we observe that "ANN NY ACAD SCI", "CAN J FISH AQUAT SCI", "NATURE" and the "ANN NY ACAD SCI" are also some of the top Journals of this domain in terms of Centrality.

**Table 2 Top Journals based on Centrality**

| Centrality | Title | Abbreviated Title |
|---|---|---|
| 0.47 | Ecological Modelling | ECOL MODEL |
| 0.29 | Science | SCIENCE |
| 0.21 | Communications of the ACM | COMMUN ACM |
| 0.32 | Artificial Intelligence | ARTIF INTELL and ARTIFICIAL INTELLIGE |
| 0.15 | Econometrica | ECONOMETRICA |
| 0.09 | Journal of Theoretical Biology | J THEOR BIOL |
| 0.08 | Canadian Journal of Fisheries and Aquatic Sciences | CAN J FISH AQUAT SCI |
| 0.08 | Nature | NATURE |
| 0.08 | Annals of the New York Academy of Sciences | ANN NY ACAD SCI |

Next, we analyze the publications in terms of their frequencies of publication as given in Table 3. Now, interestingly, the table sorted in terms of article frequency, gives a slightly different set of core journals. Through frequency analysis of the title words of 240 journals, it can be seen that “ECOL MODEL” is still at the top with a frequency of 295 articles. “NATURE” and “SCIENCE” follow with 231 and 216 published articles respectively. “J THEO BIO” is next with 167 articles. Next “ECOLOGY” has published 145 articles. This is followed by “AM NAT”, “TRENDS ECOL EVOL”, “LECT NOTES ARTIF INT” and “P NATL ACAD SCI USA”.

**Table 3 Core Journals based on frequency**

| Frequency | Title | Abbreviated Title |
|---|---|---|
| 295 | Ecological Modelling | ECOL MODEL |
| 231 | Nature | NATURE |
| 216 | Science | SCIENCE |
| 167 | Journal of Theoretical Biology | J THEOR BIOL |
| 145 | Ecology | ECOLOGY |
| 123 | The American Naturalist | AM NAT |
| 121 | Trends in Ecology & Evolution | TRENDS ECOL EVOL |
| 121 | Lecture Notes in Artificial Intelligence | LECT NOTES ARTIF INT |
| 121 | Proceedings of the National Academy of Sciences | P NATL ACAD SCI USA |

### *3.3.3.2.4 Analysis of Categories*

Our next analysis was to discover the prevalence of various agent-based computing articles in various subject categories. This visualization is shown in Figure 17. The detailed analysis of the subject category based on centrality follows in Table 4.

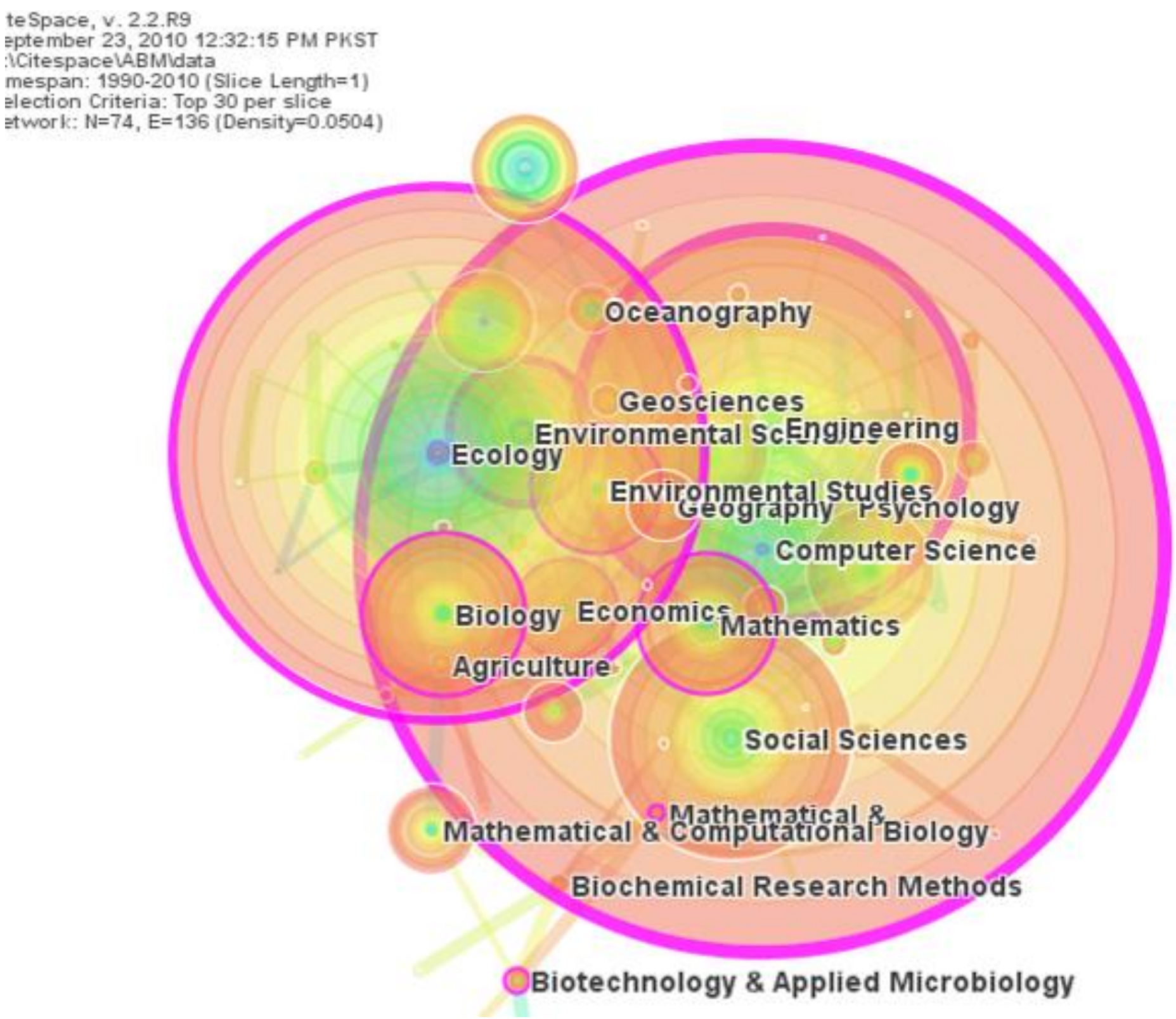

**Figure 17: Complex network model of the category data**

**Table 4 Key categories based on Centrality**

| Centrality | Category |
|---|---|
| 0.44 | Engineering |
| 0.4 | Computer Science |
| 0.36 | Mathematics[4] |
| 0.34 | Ecology |
| 0.31 | Environmental Sciences |
| 0.14 | Biotechnology & Applied Microbiology |
| 0.13 | Biology |
| 0.13 | Economics |
| 0.12 | Psychology |

[4] Shown as two categories erroneously by CiteScape

The table represents the centrality based ordering of the key subject categories. It is important to note here is that this table shows top categories from a total of 75 categories. Here, it can be observed that in terms of centrality, the "Engineering" category leads other categories. It is however, closely followed by computer science, mathematics, ecology and environmental sciences. It appears however that the "Psychology" category has the lowest value of centrality among all other categories.

For comparative analysis, we also analyze these categories using the publication frequency of articles. The results of this analysis are presented in Table 5.

**Table 5 Subject Categories according to frequency**

| Frequency | Category |
|---|---|
| 287 | Computer Science |
| 195 | Ecology |
| 145 | Engineering |
| 92 | Social Sciences |
| 66 | Biology |
| 57 | Environmental Sciences |
| 57 | Mathematics |
| 53 | Environmental Studies |
| 52 | Operations Research & Management Science |
| 52 | Fisheries |

The table represents the frequency of the top ten key categories. Through frequency analysis of the title words of 75 categories, we interestingly come up with a slightly different set of results. Here, "Computer Science" with a frequency of 287 articles leads the rest and is followed closely by ecology and engineering. An interesting observation based on the two tables is that there are certain categories which have relatively low frequency but are still central (in terms of having more citations) such as Mathematics. Amongst the top categories, we can also see categories with a relatively lower frequency such as 53 for "Environmental Sciences" and 52 for "Operations Research & Management Science" as well as "Fisheries". This detailed analysis shows that prevalence of agent-based computing is not based on a few sporadic articles in a variety of subject categories. Instead, there are well-established journals and researchers with interest in and publishing a considerable number of papers in this domain, especially agent-based modeling and simulation.

#### *3.3.3.2.5 Bursts in subject categories*

Next, we analyze how various subject categories have exhibited bursts. This is shown in the Table 6. Here we can see that fisheries have the largest burst associated with the year 1991. Next are two closely related categories "Marine and Freshwater Biology" and "Ecology" in the same time frame. One very interesting finding here is that there are a lot of bursts in non-computational categories.

**Table 6 Key bursts in subject categories**

| Burst | Category | Year |
|---|---|---|
| 13.09 | Fisheries | 1991 |
| 10.3 | Marine & Freshwater Biology | 1991 |
| 9.36 | Ecology | 1991 |
| 5.58 | Economics | 1996 |
| 4.25 | Evolutionary Biology | 1993 |
| 3.76 | Mathematics | 1990 |
| 3.3 | Genetics & Heredity | 1993 |
| 3.2 | Oceanography | 1995 |

#### *3.3.3.2.6 Analysis of Author Networks*

In this section, we analyze the author co-citation networks. Figure 18 shows the visualization of the core authors of this domain. Here red color exhibits burst of articles and concentric circles identify separation of years of publications. The size of these circles based on the centrality values of the author. The blue is color represents the older papers and the green gives not very old papers. Yellowish and reddish colors are for relatively more recent publications. Here, the descriptions are based on the color figures, which can be viewed in the online version of the paper, since Citespace, being a visualization tool, relies extensively on the use of colors to depict styles.

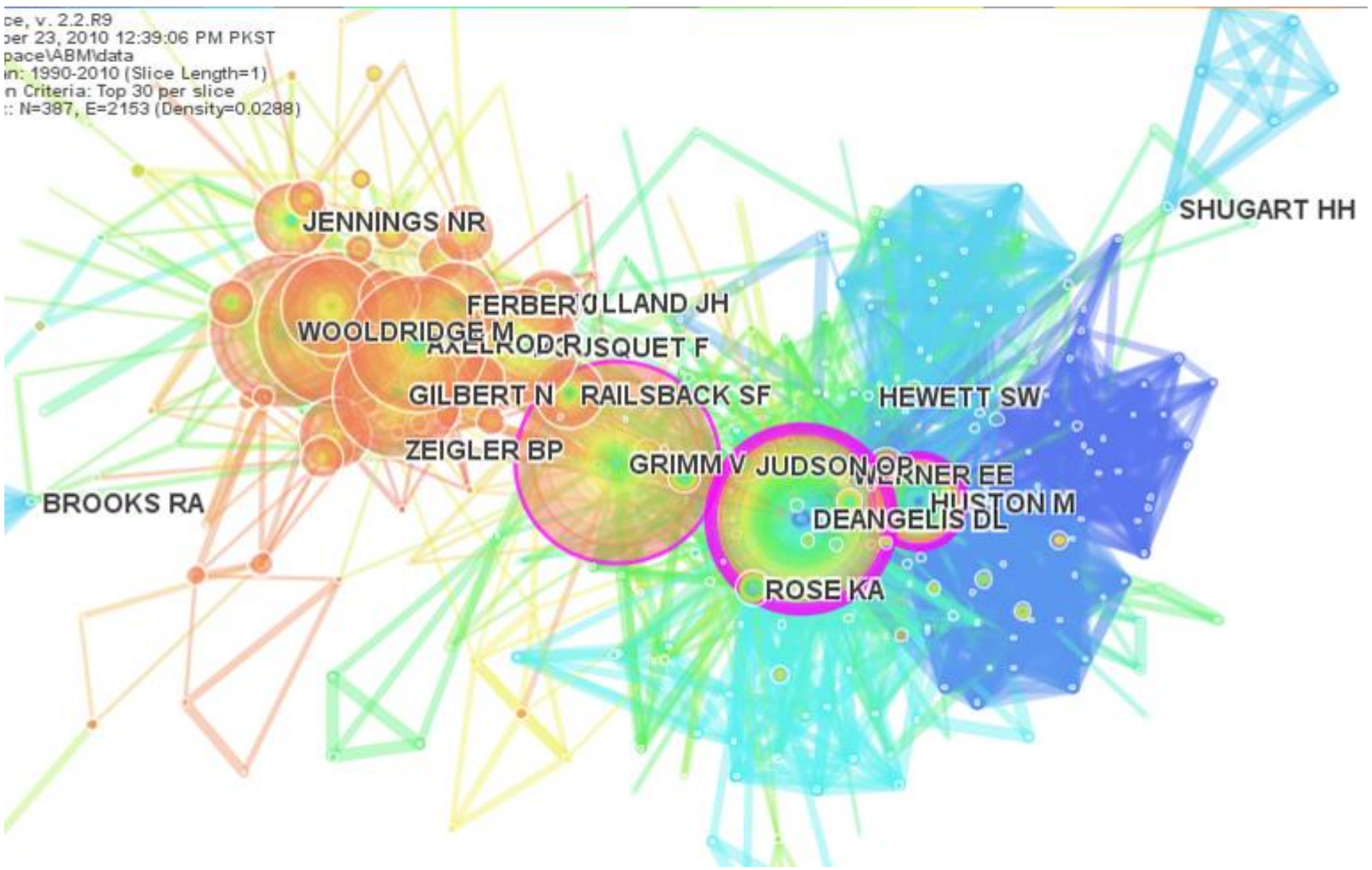

**Figure 18: Co-Author network visualization**

Although this visualization perhaps gives a broad picture of the various authors, we also present a detailed analysis of this data. This can be seen in a tabular form as shown in Table 7. Here, we can observe that the top cited (most central) author is Don DeAngelis, a Biologist. Don is followed by another Biologist Michael Huston. Next is Volker Grimm, an expert in agent-based and individual-based modeling and Kenneth A Rose, an ecologist. Next, we have Robert Axelrod, a Political Scientist. Nigel Gilbert, a Sociologist and Mike Wooldridge, a Computer Scientist is next in the list. Finally, we have François Bousquet from the field of Ecological Modeling (Agriculture) and Steven F. Railsback, an Ecologist. This is quite an interesting result because agents and agent-based computing in general was supposed to be primarily from Computer Science/AI and have very specific meanings. However, the results show that it is actually quite prevalent and flourishing in an uninhibited manner in various other fields in terms of renowned (ISI-indexed) archival Journal articles.

**Table 7 Authors in terms of centrality**

| Centrality | Author |
|---|---|
| 0.5 | DEANGELIS DL |
| 0.33 | HUSTON M |
| 0.16 | GRIMM V |
| 0.08 | ROSE KA |
| 0.08 | AXELROD R |
| 0.07 | GILBERT N |
| 0.07 | WOOLDRIDGE M |
| 0.06 | BOUSQUET F |
| 0.06 | RAILSBACK SF |

For further comparative analysis, we plotted the top authors in terms of frequency of publications. This is shown in Figure 19.

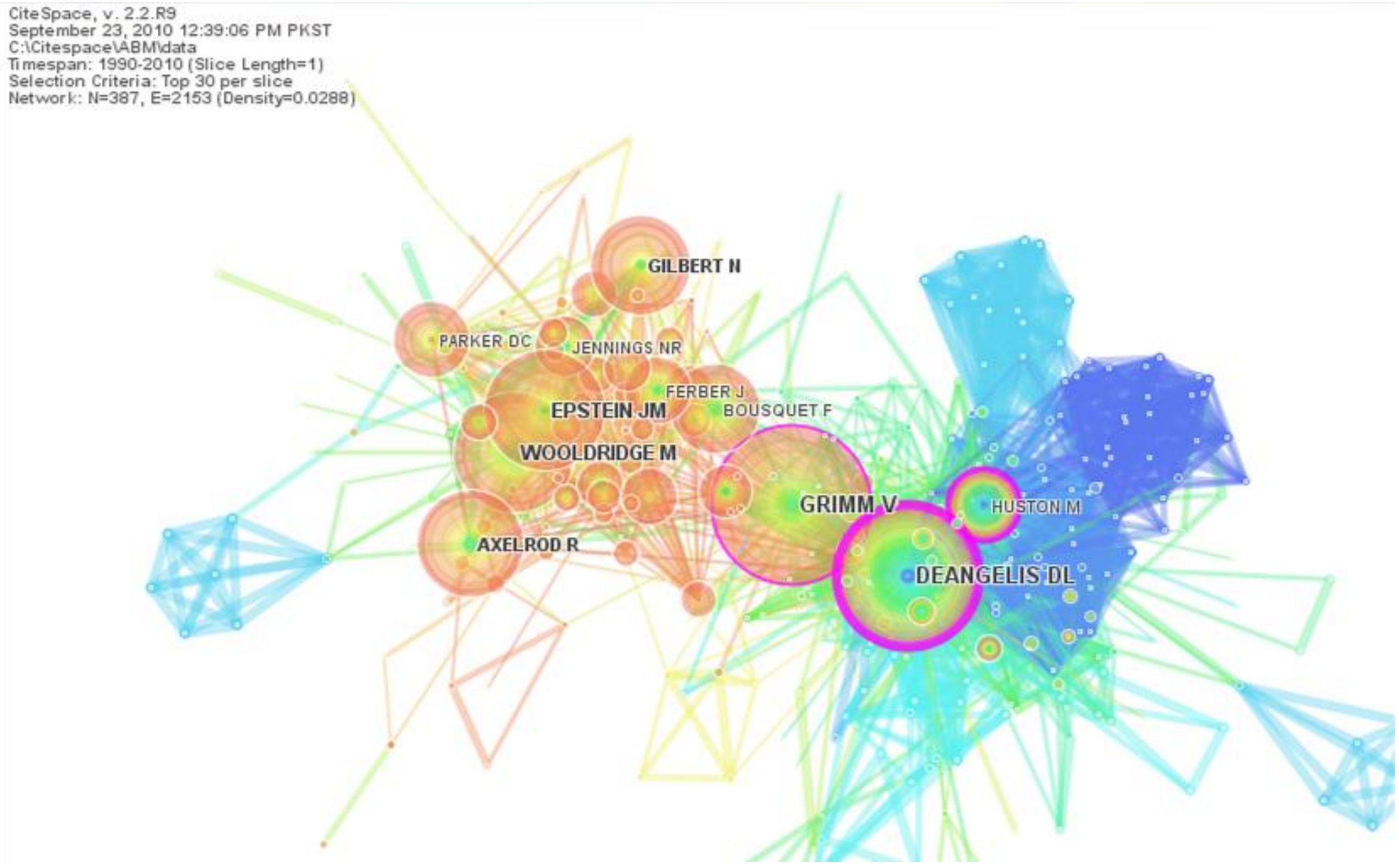

**Figure 19: Authors in terms of frequency**

The detailed analysis of the metrics of the key authors in this domain is shown in a tabular form in Table 8 in terms of their frequency of publication in the selected pruned network.

**Table 8 Top authors based on frequency**

| Frequency | Author |
|---|---|
| 128 | GRIMM V |
| 118 | DEANGELIS DL |
| 101 | EPSTEIN JM |
| 99 | WOOLDRIDGE M |
| 90 | AXELROD R |
| 83 | GILBERT N |
| 76 | BOUSQUET F |
| 68 | HUSTON M |
| 64 | FERBER J |
| 59 | PARKER DC |

It is important to note here that while the results are based on a total of 387 cited authors, new names appear in this table such as Joshua M. Epstein, a Professor of Emergency

Medicine. Prof. Epstein is an expert in human behavior and disease spread modeling. Jacques Ferber, a Computer Scientist is another new name in the list along with Dawn C. Parker, an agricultural Economist.

### *3.3.3.2.7 Country-wise Distribution*

Next, we present an analysis of the spread of research in agent-based computing originating from different countries based on centrality. Here, in Figure 20, we can see the visualization of various countries. Please note here that the concentric circles of different colors/shades here represent papers in various time slices (we have selected one year as one time slice). The diameter of the largest circle thus represents the centrality of the country. Thus the visualization identifies the key publications in the domain to have originated from the United States of America. This is followed by papers originating from countries such as England, China, Germany, France and Canada.

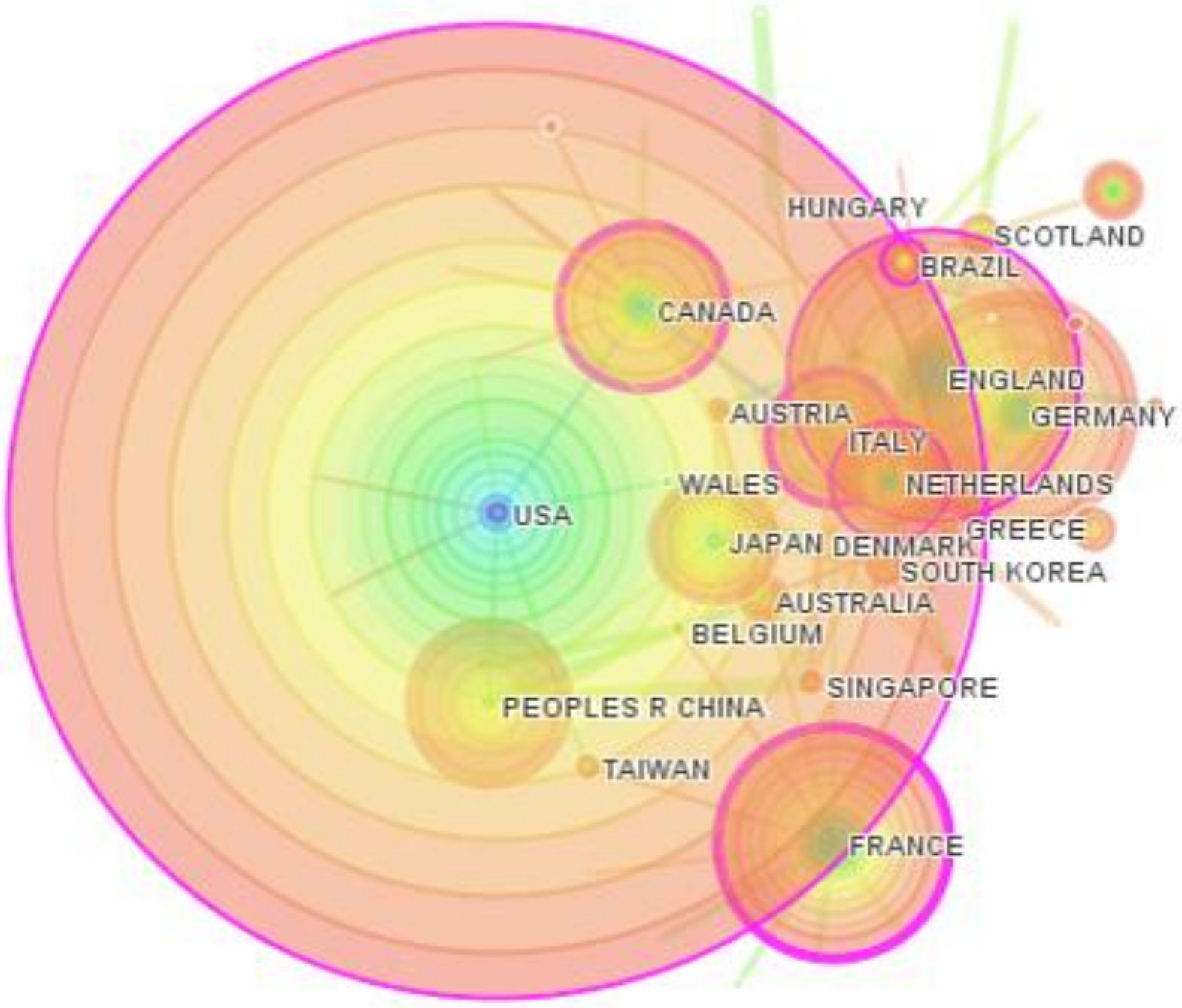

**Figure 20: Countries with respect to centrality**

### *3.3.3.2.8 Analysis of Institutes*

In this sub-section, we present visualization for a detailed analysis of the role of various Institutes. We can see the temporal visualization of various institutes in the domain and the assortment of popular keywords associated with them on the right in Figure 21.

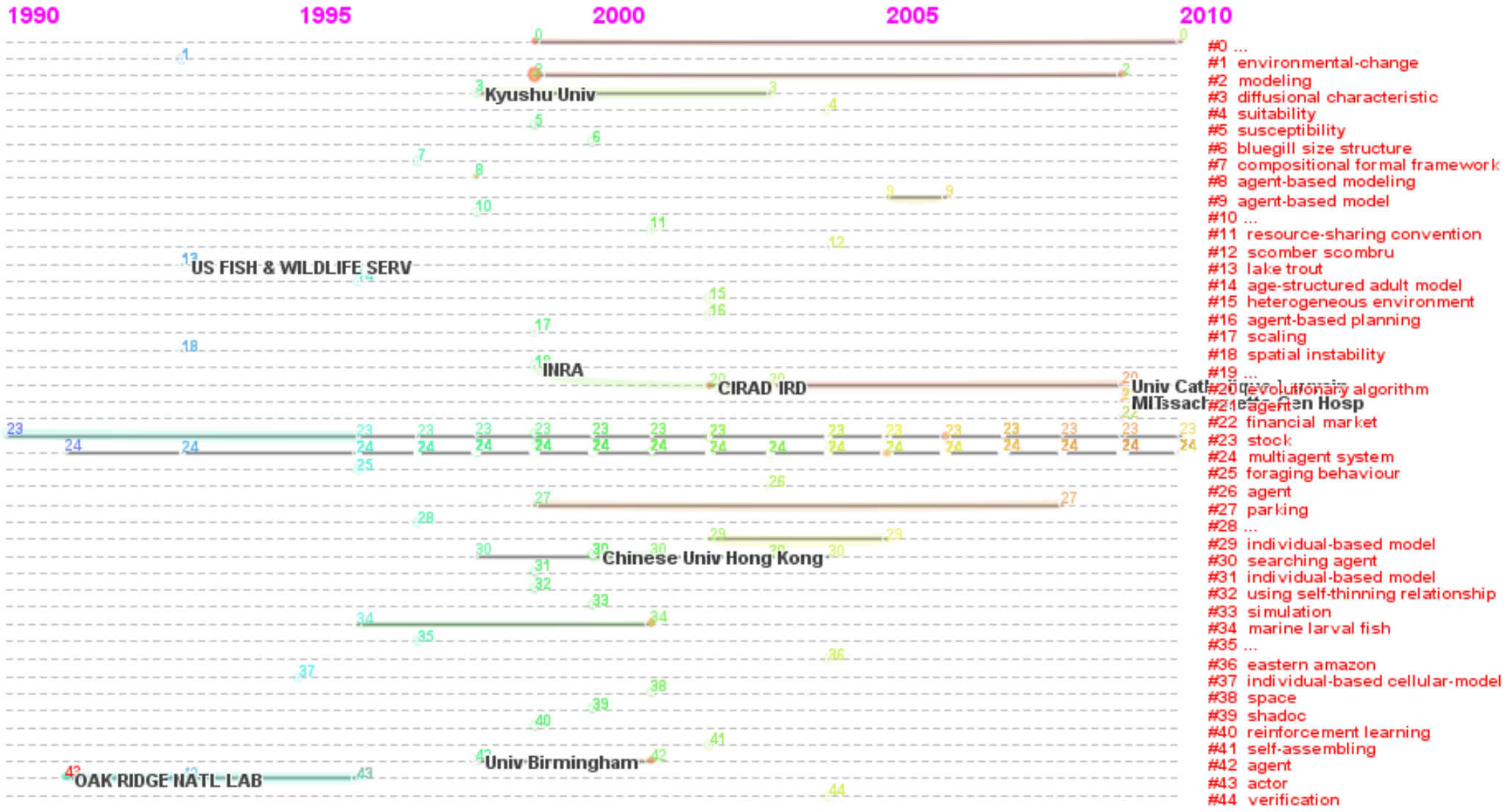

**Figure 21: Timeline based visualization of institutes**

Here, the prevalence of manuscripts originating from the Oak Ridge National Laboratories from the early 1990's can be observed here. Next, we see papers from US Fish and Wildlife services. Then, we can see papers from Kyushu University, Japan, University of Birmingham, UK and French National Institute of Agricultural Research (INRA) close to 1998. From this time to 2000, a prevalent institute is Chinese University, Hong Kong. In around 2002, the French institute Centre de coopération Internationale en recherche agronomique pour le développement (CIRAD), an Institute of Agriculture Research for Development is prevalent followed closely by a sister institute, the Institut de Recherche pour le Développement (IRD). More recent newcomers to the field of agents include the MIT and the Massachusetts General Hospital, associated with the Harvard University.

In the following Table 9, we perform an alternative analysis which is based instead on the frequency of articles.

**Table 9 Core Institutes based on frequency**

| Frequency | Institute |
|---|---|
| 11 | University of Illinois, USA |
| 10 | INRA, France |
| 9 | University of Michigan, USA |
| 9 | University of Minnesota, USA |
| 8 | University of Sheffield, UK |
| 8 | Nanyang Technological University, Singapore |
| 8 | Italian National Research Council (CNR), Italy |
| 8 | Oak Ridge National Labs, USA |
| 7 | Harvard University, USA |
| 7 | MIT, USA |
| 7 | University of Washington, USA |
| 7 | University of Hong Kong, P. R. China |
| 7 | IRD, France |

The Table 9 listed represents the top institutes in terms of frequency. It should be noted that the frequency analysis was performed based on title words of a total of 328 institutes. "University of Illinois"[5] in general has a top ranking with a frequency of 11 articles. It is followed closely by INRA (France) with a frequency of 10 articles. University of Michigan and University of Minnesota, both from the US follow next with 9 articles each. With 8 articles each, next we have University of Sheffield (UK), Nanyang Technological University (Singapore), Italian National Research Council (CNR) and the Oak Ridge National Labs (USA).

[5] ISI data extracted using CiteSpace does not differentiate further as to which exact campus of the University of Illinois is considered here (of the primarily three campuses i.e. UIUC, UIC, UIS.)

## 3.4 Case Study II: Consumer Electronics

The second case study demonstrates the application of the Complex Network methods to scientific literature in the consumer electronics domain. As mentioned earlier at the start of the chapter, while the first case study can be considered as an extension of the background of the previous chapter, the second case study can be considered as a proof-of-concept of the generalized application of the complex network methods.

### 3.4.1 Problem Statement

Consumer electronics as a domain has evolved with the evolution of electronics technology [168]. The guiding principles for this gradual evolution range from consumer demand[169] in general and the drive from the industry in particular[170]. Thus the state of the domain, as it is today, reflects a number of trends from the past. To demonstrate the variety of effects which can dictate trends, trends have been noted to be governed by seemingly unrelated interactions such as a company's internal management policies[171]. The evolution of the domain is perhaps reflected by its growth in terms of sales which if aggregated, have shown to be almost exponential [172]. As evident from the analysis of the Thomson Reuters citation data, a major evolution in the domain occurred during the last 25 years as the citation corpus does not reflect the indexing term of "consumer electronics" in the WOS before this period.

Our research questions include the identification of the highly central journals in consumer electronics. In addition, the analyses give results ranging from the identification of the top authors and the top papers.

### 3.4.2 Methodology

Similar to the previous study on agent-based computing, citation data used here for the identification of trends and interest of the general Consumer Electronics Community was retrieved from the WOS using the topic "Consumer Electronics". All available data sources were used to get a complete picture of trends. These included the SCI-EXPANDED, SSCI, A&HCI and the CPCI-S databases. The data was retrieved based on search on the "Consumer Electronics" topic (which translates to a search including keywords, abstract, title etc.) in addition to all records of the IEEE Transactions on Consumer Electronics. Subsequently, the data obtained from the ISI Web of Knowledge was imported and analyzed primarily using various tools such as CiteSpace. First the data was spliced into yearly data. Next, different co-citation networks were formed such as for Journals, authors and so on. These were then analyzed based on both the frequency of publication as well as centrality (in terms of being cited by others).

### 3.4.3 Results

The citation data was collected from all years including the current year (2010). The data included a total of 5,372 papers and including cited references, it amounted to 26,860 records.

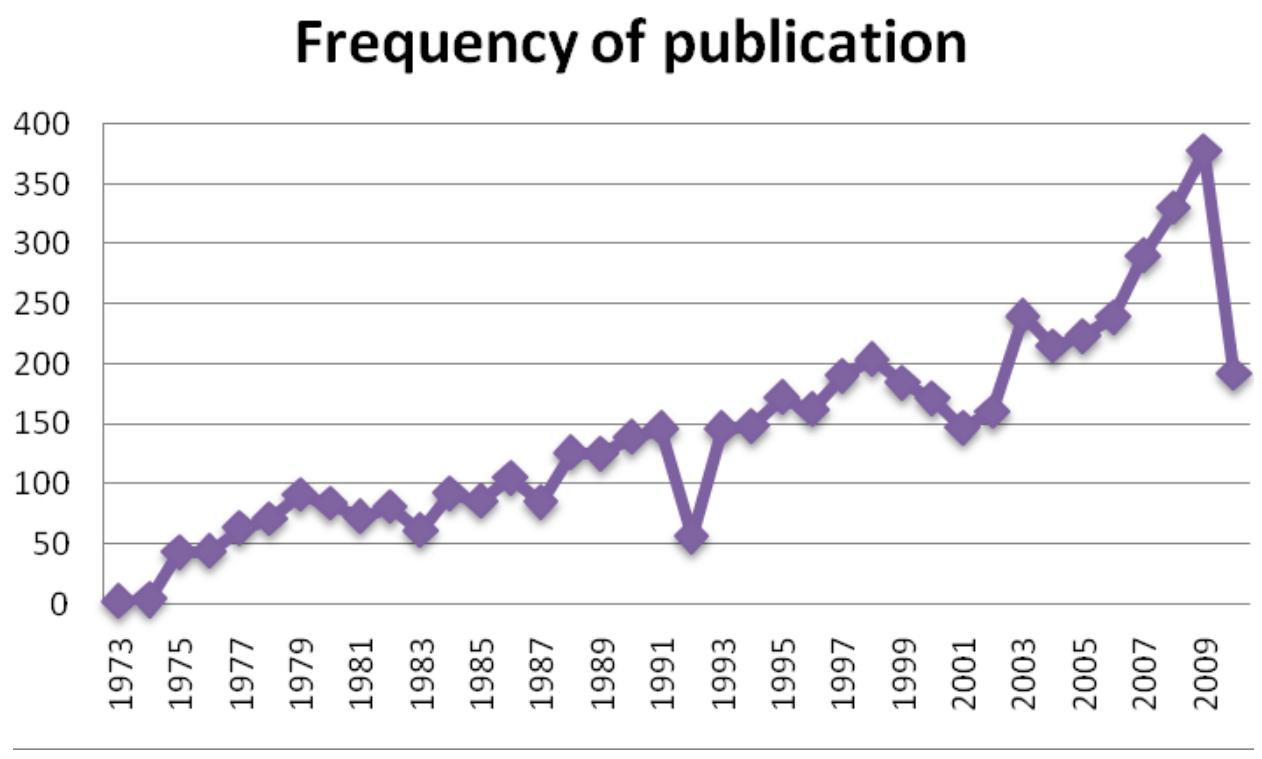

**Figure 22: Frequency of papers in Consumer Electronics domain**

Figure 23 visualizes the top journals of the domain. Different colors (or scales of grey in print) identify the year slices with each slice made up of ten years. The node sizes identify the frequency of articles, which in essence is an indicator of trends or community interest. The number of articles listed hailing from IEEE Transactions on Consumer Electronics (CE) is 1665. The second Journal in the list is IEEE Transactions on Communications with around one third number of the IEEE Transactions on CE articles (i.e. 560). Here the centrality of the IEEE Transactions on Consumer Electronics is apparent here with the yellow color representing the papers published in the nineties. The results for the top 10 Journals are summarized in tabular form in Table 10.

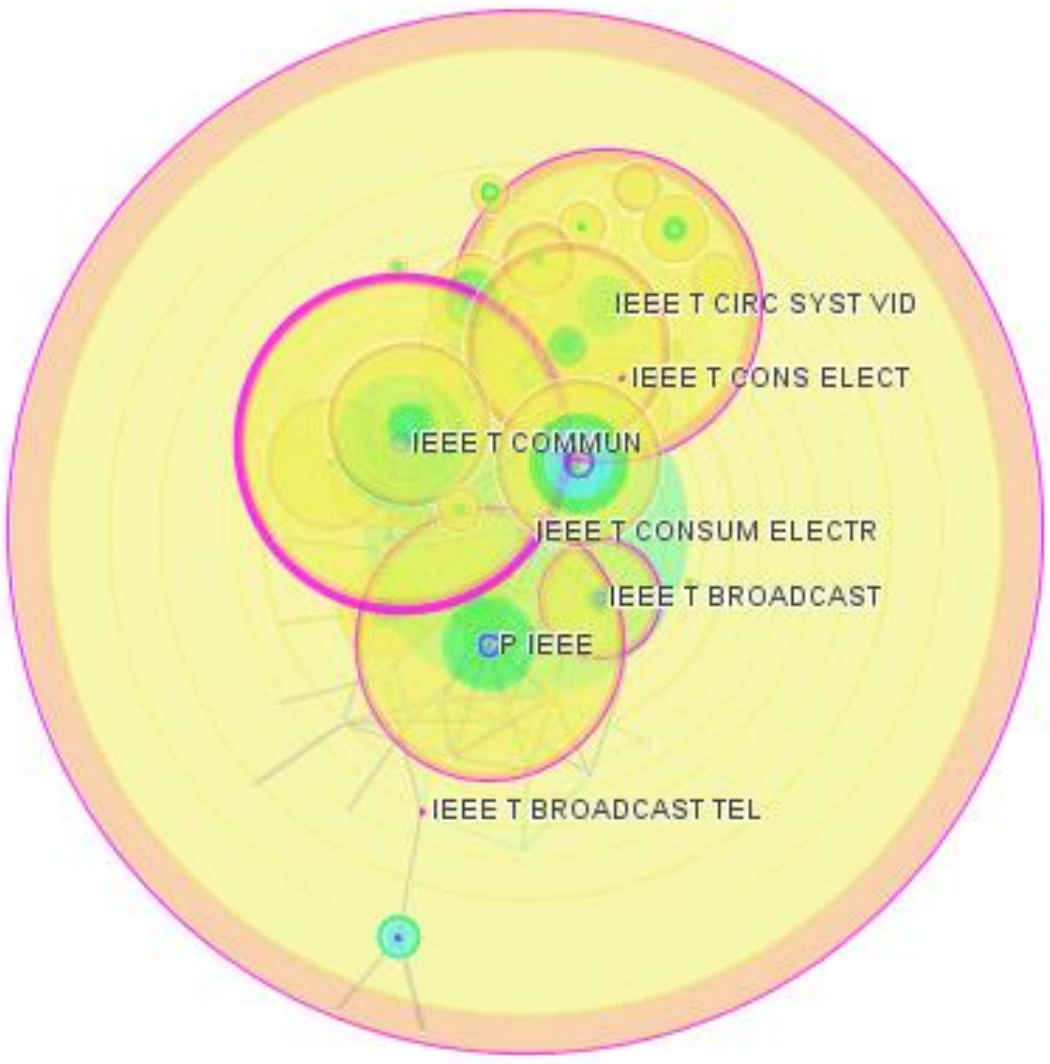

**Figure 23: Visualization of the citation network of the top Journals of the domain**

**Table 10 Top Journals in Consumer Electronics**

| Serial | Journal Name |
|---|---|
| 1. | IEEE T CONSUM ELECTR |
| 2. | IEEE T COMMUN |
| 3. | IEEE T CIRC SYST VID |
| 4. | P IEEE |
| 5. | IEEE T IMAGE PROCESS |
| 6. | IEEE J SEL AREA COMM |
| 7. | IEEE COMMUN MAG |
| 8. | IEEE T CIRCUITS SYST |
| 9. | IEEE T BROADCAST |
| 10. | ELECTRON LETT |

Here, we can note that Figure 24 gives the overall picture of the top papers of the domain/ Here, the size of the nodes and the font is scaled according to the higher value of the centrality of the paper.

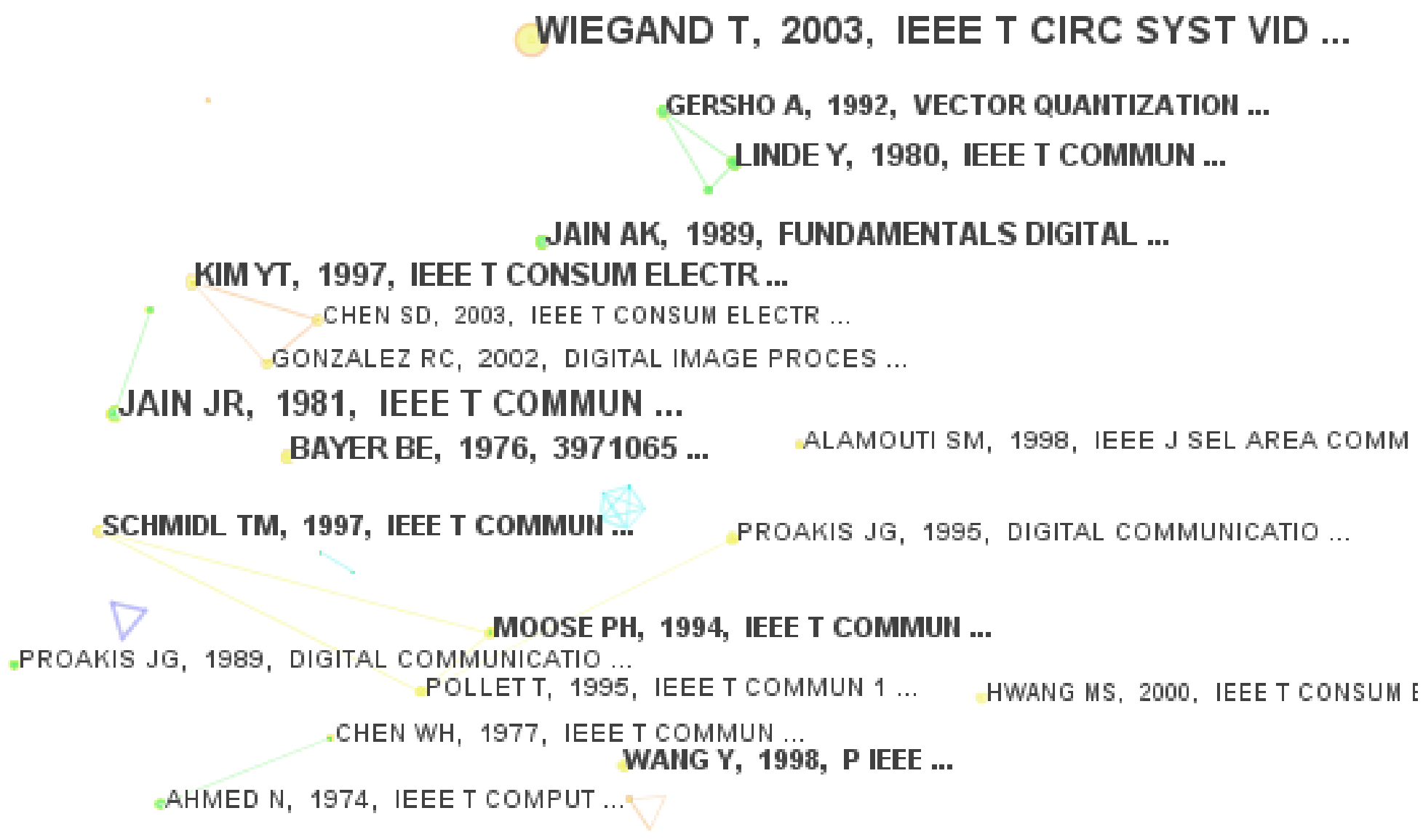

**Figure 24: Network of top papers in Consumer Electronics**

The Table 11 summarizes the top 10 papers of the IEEE Transactions on CE.

**Table 11 Top papers of IEEE Transactions on Consumer Electronics**

| Serial | Paper | Citations |
|---|---|---|
| 1. | The JPEG2000 still image coding system: An overview (2000) | 398 |
| 2. | Pilot tone selection for channel estimation in a mobile OFDM system (1998) | 242 |
| 3. | A new remote user authentication scheme using smart cards (2000) | 192 |
| 4. | An efficient remote use authentication scheme using smart cards (2000) | 147 |
| 5. | Contrast enhancement using brightness preserving bi-histogram equalization (1997) | 127 |
| 6. | The Trustworthy Digital Camera - Restoring Credibility To The Photographic Image (1993) | 118 |
| 7. | Channel estimation for OFDM systems based on comb-type pilot arrangement in frequency selective fading channels (1998) | 116 |
| 8. | Digital Sound Broadcasting To Mobile Receivers (1989) | 90 |
| 9. | Multidirectional Interpolation For Spatial Error Concealment (1993) | 87 |
| 10. | A modified remote user authentication scheme using smart cards (2003) | 86 |

Subsequently we can note the co-citation network of the key authors in the consumer electronics domain in Figure 25. Note that while we have discussed and analyzed similar results in the previous case study in considerably more details, here we only briefly discuss them because this case study is a demonstration of the generalized applicability of the methodology.

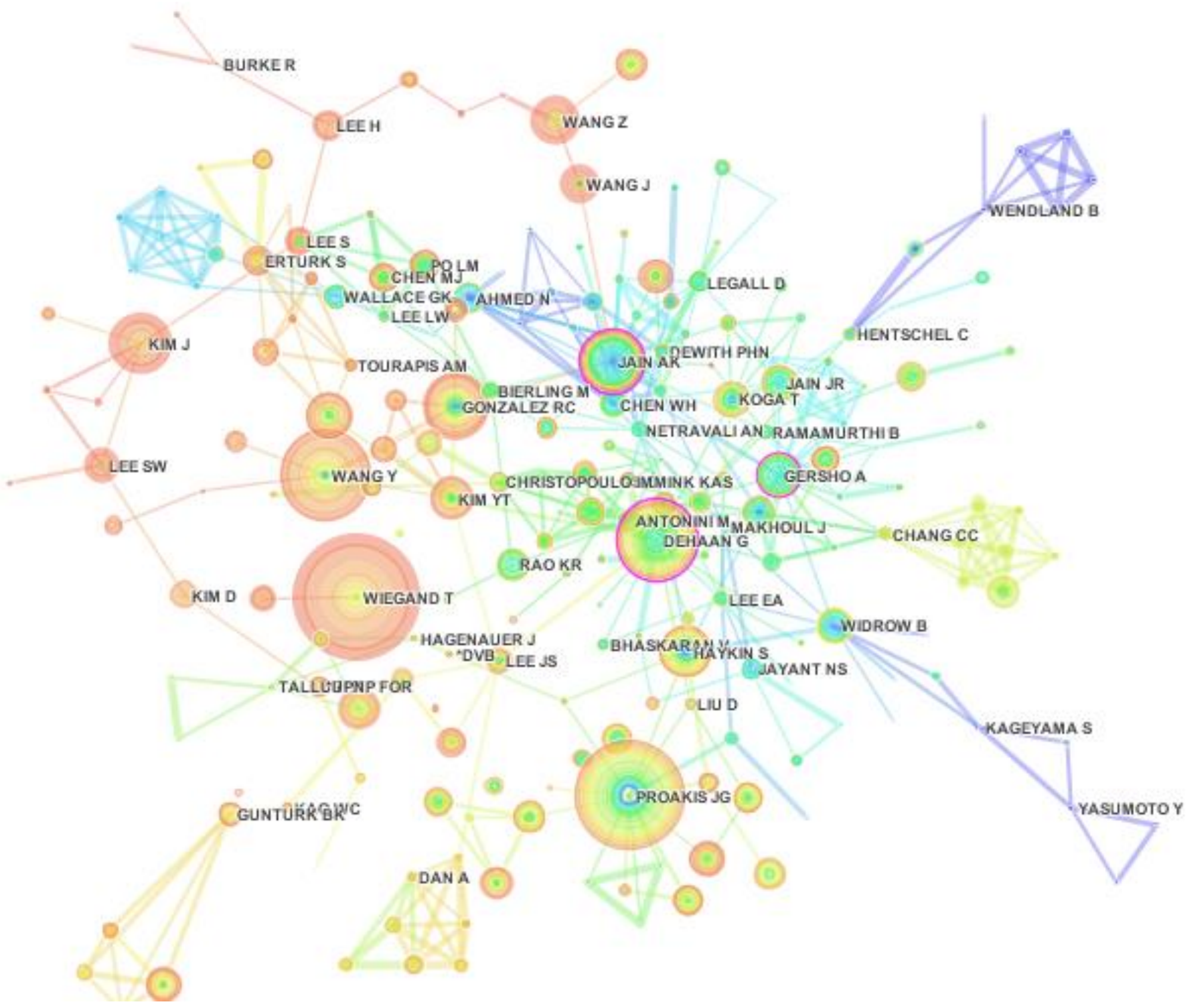

**Figure 25: Co-citation network in Consumer Electronics**

## 3.5 Summary of results

In this chapter, we presented the generalized ideas related to complex network modeling followed by two case studied of different types of complex interaction data. The goal of the case studies was to demonstrate how complex network modeling level of the proposed framework can be exploited to discover patterns of emergence in different entities constituting a cas. The two case studies were chosen based on the appropriate availability of interaction data from a well-structured data source i.e. the WOS. In this section, results of the two case studies are correlated with the general theme of complex network modeling level in the form of a set of summarized results.

### 3.5.1 Case Study I

In this case study, analysis was performed using different types of entities and their interactions in the agent-based computing domain.

Firstly, during the analysis conducted using Network Workbench, the top cited papers of the domain was identified from both the multiagent side of the domain (based in Computer Sciences) in addition to the agent-based modeling ideas (from ecology/social sciences). Next, using a clustering of index terms, the ideas were identified to originate from the agent-based modeling area. A major recent "turning point" signifying a trend was next identified based on some of Grimm's papers. An older identified turning point was based on Goldberg's 1989 paper on Genetic Algorithms, which tied in with the ecological cluster labeled as "Cannibalism". In addition, the prevalence of the "catchment" cluster was noted identifying the social aspects coupled with the "increasing returns" cluster identifying economics. Thus one single extracted network was able to demonstrate how these diverse areas are tied in together by means of identification of emergent patterns. The "persistence" cluster again identifies a "multiagent" related set of documents. Here, the oldest cluster is the "life history" cluster with roots in Biosciences. This strong intermingling of ideas can be noted to have autonomously guided the evolution of the agent-based computing domain. Whereas subsequent analyses only prove the point that there is further diversity in each of subject categories, authors, institutions and countries. However, while the domain has strong footing, being able to use visualization has allowed the examination of various complex interactions of the sub-domains, which would not have otherwise been possible to monitor without the use and application of network extraction, modeling and manipulation techniques.

### 3.5.2 Case Study II

In this case study, a complex network modeling and analysis of the consumer electronics domain has been performed. Using centrality measures has allowed a visual and quantitative identification of the top journals, the top authors as well as the top research papers in the consumer electronics scientific domain.

## 3.6 Broad applicability of the proposed framework level

In this section, an overview of how the proposed framework level can be used to conduct different complex network modeling based cas research studies is presented. It can be noted from the two case studies conducted in this chapter that each application domain in complex network modeling can have slightly different idiosyncrasies. However, complex network methods are suitable only in the case where specific target data is either readily available or else such data can be extracted by some means. This interaction data must primarily allow the creation of complex networks. In other words, the data columns must point out relationships between individual columns. Using these relationships, complex networks need to be designed and extracted. At times, the extracted networks contain considerably more information than needed. As such, it can be difficult to study the emergent patterns. As such, network nodes and links might need to be selectively eliminated or else new nodes and links might need to be developed inside the network. In addition, different network measures might need to be calculated for loading on to the different nodes and links. Based on these manipulations, the network nodes might also have to be selectively configured to display difference in terms of size, color or shape etc. In addition, different clustering algorithms might be applied to select clusters of nodes based on a common set of criteria. Subsequently network analysis might have to be repeatedly performed till a suitable network emerges allowing the studying of interactions of the various cas components.

While the cas methods in this chapter have been applied in the domain of citation data, the primary reason for this selection has been the stable and verified nature of this data. In other words, the methods used for CNA are equally applicable to different types of cas data ranging from social sciences (using survey instruments to develop relationship data etc.), biological sciences (such as using biological genomic databases) and telecommunication complex networks (using topological connectivity models of different nodes and peers in the communication network).

## 3.7 Conclusions

In this chapter, we have demonstrated the use of the proposed complex network modeling level of the framework using two different case studies. The use of Complex Network Analysis is shown to allow for the identification of emergent behavioral patterns in large amounts of citation data. Using two separate case studies in the domain of Agent-based computing and Consumer Electronics, the results demonstrate the effectiveness of the proposed approach. Results included identifying the key patterns and Scientometric indicators in research such as key authors, papers, Journals and institutions as well as the temporal aspects underlying the evolution of the entire agent-based computing scientific domain and sub-domains. In the next chapter, we present the exploratory agent-based modeling level of the proposed framework i.e. exploratory agent-based modeling for performing feasibility studies in multidisciplinary cas research.

# 4 Proposed Framework: Exploratory Agent-based Modeling Level

In the previous chapter, we gave an overview of the development of complex network models for cas based on the availability of real-world data by using two separate case studies. We discovered that by the use of complex networks, we could perform CNA of the data and discover interesting trends in the data such as finding emergent patterns. While in the ideal world, we would have all types of interaction data for cas, practically speaking most cas case studies suffer from a lack of suitable data for developing complex network models. As such, the rest of the framework modeling levels are concerned with the development of agent-based models of cas. While agent-based models of cas are simulation models, they do not fit in line with the traditional simulation models of engineered systems. As noted earlier in previous chapters, similar to the complex network modeling level of the proposed framework, exploratory ABM level is also an encompassing level. In other words, while on one hand, it is a part of the proposed framework, it allows other existing studies of legacy ABMs to be tied in with the proposed framework. In the next section, we develop a proposed methodology of performing exploratory agent-based modeling.

## 4.1 Research Methodology

Exploratory agent-based models are models which are developed by researchers to explore the concepts underlying the cas under study. The goals of developing these models are typically different from traditional computer simulation models such as involving queuing, Monte Carlo methods etc. Some of the possible goals of building exploratory agent-based models can be listed as follows:

1. The researchers want to explore if agent-based modeling is a suitable paradigm for modeling their particular problem.
2. The researchers want to develop a proof of concept that any further research in the chosen direction would be a viable possibility.
3. The researchers want to develop agent-based models as a means of demonstrating the applicability of the chosen model design of a cas system to prospective funding bodies.
4. The researchers want to develop agent-based models to explore exactly what kind of data would be needed for the sake of validation of these models without spending considerable amount of time on these models.

As we can note here, these models are at times not meant to be comprehensive or expected to give high quality research results. However, these models do allow the researchers to test the water and thus increase the possibility of proceeding on a larger-scale future project.

The most basic type of agent-based modeling can be considered as the exploratory agent-based modeling level as shown in Figure 26. This framework level is suitable for exploratory agent-based model design and development. While a number of researchers are known to follow exploratory agent-based modeling, previously the concept has not been formalized in a unified framework. The idea originated based on extensive observation of published cas models by researchers who like to first test the water and develop proof of concept models for further research. These models can be useful for assessing feasibility as well as for demonstrating how future research might be conducted. At this stage, these models do not have to be developed according to any detailed specification or even lead to any interesting results. Their goal is often to pave the way for future research by developing a proof of concept model. This frees the researcher to explore the cas domain and attempt to develop different models without any pressure. A large number of existing cas models can be considered as exploratory and thus come under exploratory agent-based modeling level of the proposed framework.

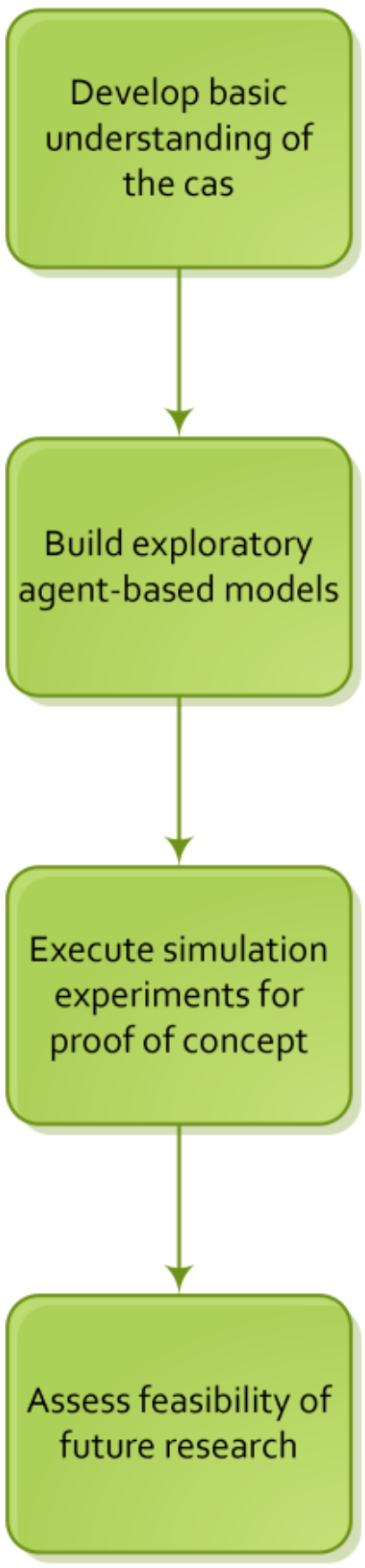

**Figure 26 Exploratory agent-based modeling level of the proposed framework (to assess research feasibility and proof-of-concept)**

One possible way of evaluating the success of exploratory ABMs and their associated problems is to examine the communication of various cas researchers. With the wide availability of open mailing lists for various agent-based modeling toolkits such as NetLogo, Repast-S, Mason and Swarm, such an observational approach is quite possible. While there are several mailing lists and tools, specifically we focus on NetLogo. There are sev-eral reasons for selecting NetLogo. Among other things, Railsback et al. notes NetLogo as the highest documented tool from the rest of the ABM tools[173]. They also highly recommend it for prototyping (exploratory) complex models as follows:

"NetLogo is the highest-level platform, providing a simple yet powerful programming language, built-in graphical interfaces, and comprehensive documentation. It is designed primarily for ABMs of mobile individuals with local interactions in a grid space, but not necessarily clumsy for others. NetLogo is highly recommended, even for prototyping complex models."

In this chapter, using a case study from the domain of complex communication networks, the application of exploratory agent-based modeling has been demonstrated. This proposed real-world application is based on coupling a set of P2P unstructured network search algorithms with household devices communicating with each other for intelligent discovery of user content in an "Internet of things".

The rest of the chapter is structured as follows:

We first give an overview of the selected case study along with a problem definition. Next, we present the design and implementation details of the exploratory agent-based models. Subsequently we discuss the results of the simulation experiments. Finally we conclude the chapter.

## 4.2 Case Study: Overview, Experimental design and Implementation

The chosen case study is the modeling and simulation of computing devices self-organizing themselves to develop a complex communication network infrastructure. Technically this is called the "Internet of things" [174] and the idea was initially introduced in 1999 by Kevin Ashton[34]. The basic idea is given as follows in Kevin's own words:

> "We need to empower computers with their own means of gathering information, so they can see, hear and smell the world for themselves, in all its random glory. RFID and sensor technology enable computers to observe, identify and understand the world—without the limitations of human-entered data."

The reason we are exploring this case study is that, to the best of our knowledge, there is no previous case study exploring the use of agent-based modeling in the domain of "Internet of things". Here, the goal is to firstly understand how agent-based modeling might be able to perform exploratory studies for the Internet of things. Secondly, this case study will also allow the use of extensive modeling and simulation demonstrating how exploratory case studies might be effective in the domain of studying new types of cas.

While internet of things is a very interesting concept in theory, till now, there have not been many practical applications. One possible reason of a lack of such applications is the inability of current modeling and simulation tools to model and combine different types of paradigms. As an example, as discussed in Chapter 2, while simulators are available to model peer-to-peer networks such as Oversim[155] and there are other simulators such as NS2 and OMNET++ suitable for the modeling and simulation of wireless sensor networks, practically it can tend to be very difficult, if not impossible, to combine two or more different paradigms in a single simulator without massive re-factoring of the simulator source code. In other words, to the best of our knowledge, existing simulators do not effectively allow the combination and application of protocols, ideas and entities from different complex communication networks domain. In addition, what is even harder to model is mobility because none of the previously well-known simulators handle mobility well. As an example, in traditional simulators, the locations of nodes are typically defined in the start of the simulation experiments. As such it is difficult to model the effects of Brownian random motion and other mobility models.

As such, here the goal is to explore the use of agent-based modeling and simulation in this domain and combine concepts from P2P networks along with the concepts of Wireless

Sensor Networks to develop a working simulation of the internet of things. The specific problem that we attempt to solve using agent-based modeling and simulation is defined as follows:

Existing computing media devices both have internet capabilities as well as file storage capabilities for various types of contents and media. However, a user who keeps a set of devices can have difficulty in locating e.g. a particular file from their various computing devices (such as Smart phones, Laptops, and other devices). In other words, we want to explore further how P2P search algorithms can be exploited for exploring these media devices for locating user files. Here are a few possible use case scenarios for such algorithms:

A. In an ambient assisted living initiative, consumer mobile phones can communicate with each other to discover internet access which then provides access to content from customized web services.

B. Julie is late for her meeting, gets in her car and realizes that she has forgotten her important files on one of her home computers. Using her mobile phone, she is able to perform a P2P search of her personal devices which connect with each other securely and she is able to access content from the computer in which she had left her files. This autonomous self-adapting hybrid network does not require prior configuration except any security measures that Julie has previously configured (such as the use of digital certificates for authentication and authorization of the devices).

C. Joe is in a meeting and Rick, his boss wants him to disseminate the meeting agenda to all the meeting participants on their local trusted phones. Joe's phone is able to connect to other local phones without using the internet, which Rick rightfully considers not secure enough for confidential company internal documents. As a company policy, Rick and all participants are using mobile phones which are not al-lowed internet access on company facilities. In other words, the devices form an internet of things to disseminate information.

D. In a media company, it has taken months to arrange a meeting for the busy ad designers with their clients in a nice comfy mountain resort. They are all sitting in various rooms discussing agendas. Mary has an important clip in her phone but the size of the clip is very large for uploading and downloading from the internet. She wants to share the clip with other designers. However, these other attendees are neither tech savvy enough, nor have high speed internet for fast sharing. She needs to disseminate the clip inside the room as well as in another room. Her phone recognizes other devices as she gives the phone the names of the trusted phones and she is able to send it. People in the same room receive the clip using faster transfer over local wireless network autonomously developed by the P2P applications while devices in the other room receive the clip both via the internet of things by using phones of meetings organizers in the hallways without requiring extensive configuration settings or using the internet.

Other use cases showing the utility of P2P algorithms coupled with wireless sensor networks in the domain of internet of things can be observed in Figure 27. As can be seen, applications can range from real-time weather monitoring, patient monitoring, business meetings, currency rates, condition monitoring of all of the family cars etc.

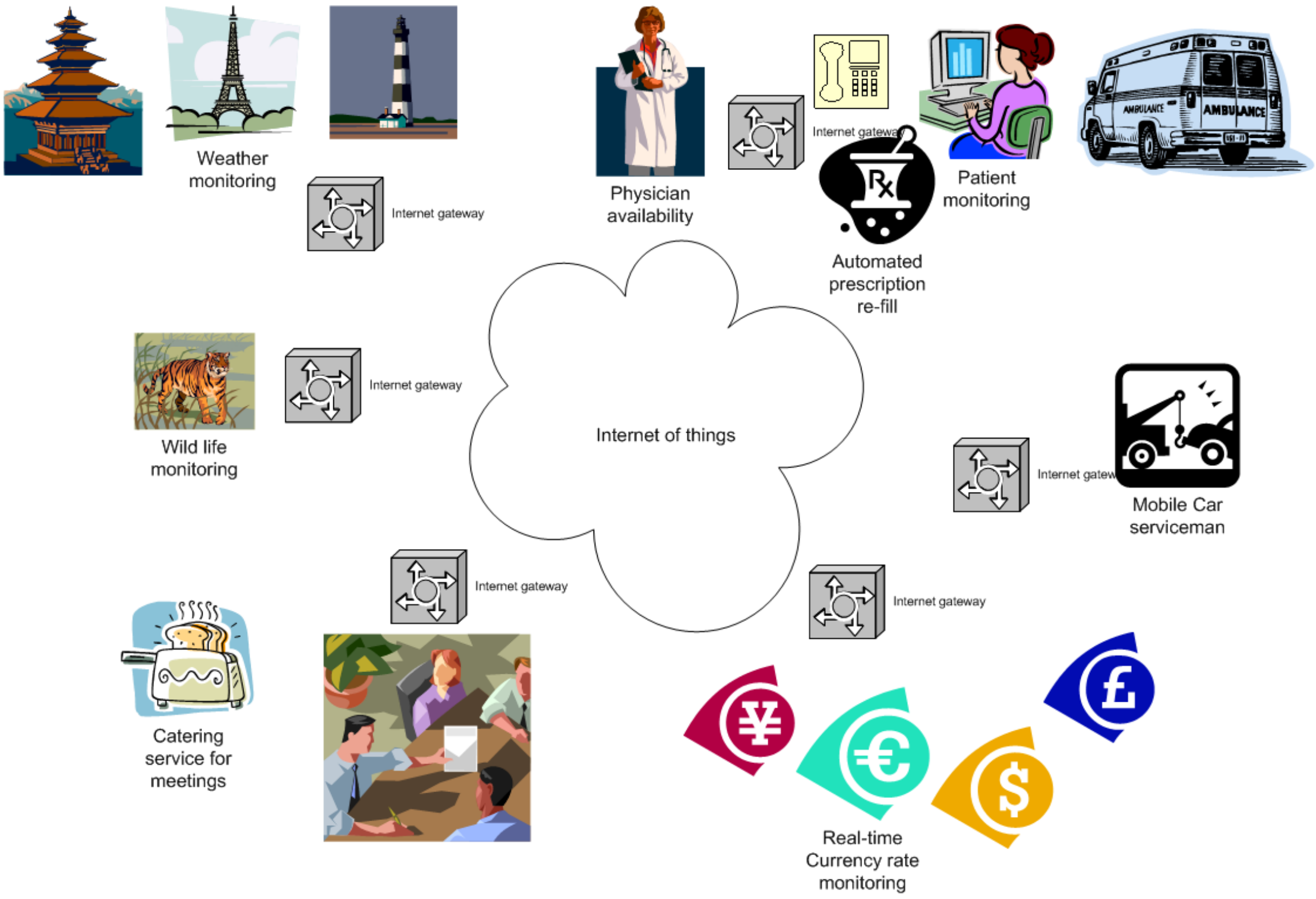

**Figure 27: Examples of Internet of Things**

In Figure 28, we can note some of the challenges in the design of hybrid networks which involve local as well as remote content access. Content can be accessed using two primary models. One is the Push model where content is automatically accessed on the devices by being "pushed" form other devices while in the "pull" model, the content is accessed by performing queries. P2P access is based on unstructured and structured overlay methods. In structured overlays, there is typically a Distributed Hash Table (DHT) which allows for indexing of content to a certain degree whereas unstructured overlay networks involve the use of algorithms which can access content without a structured network.

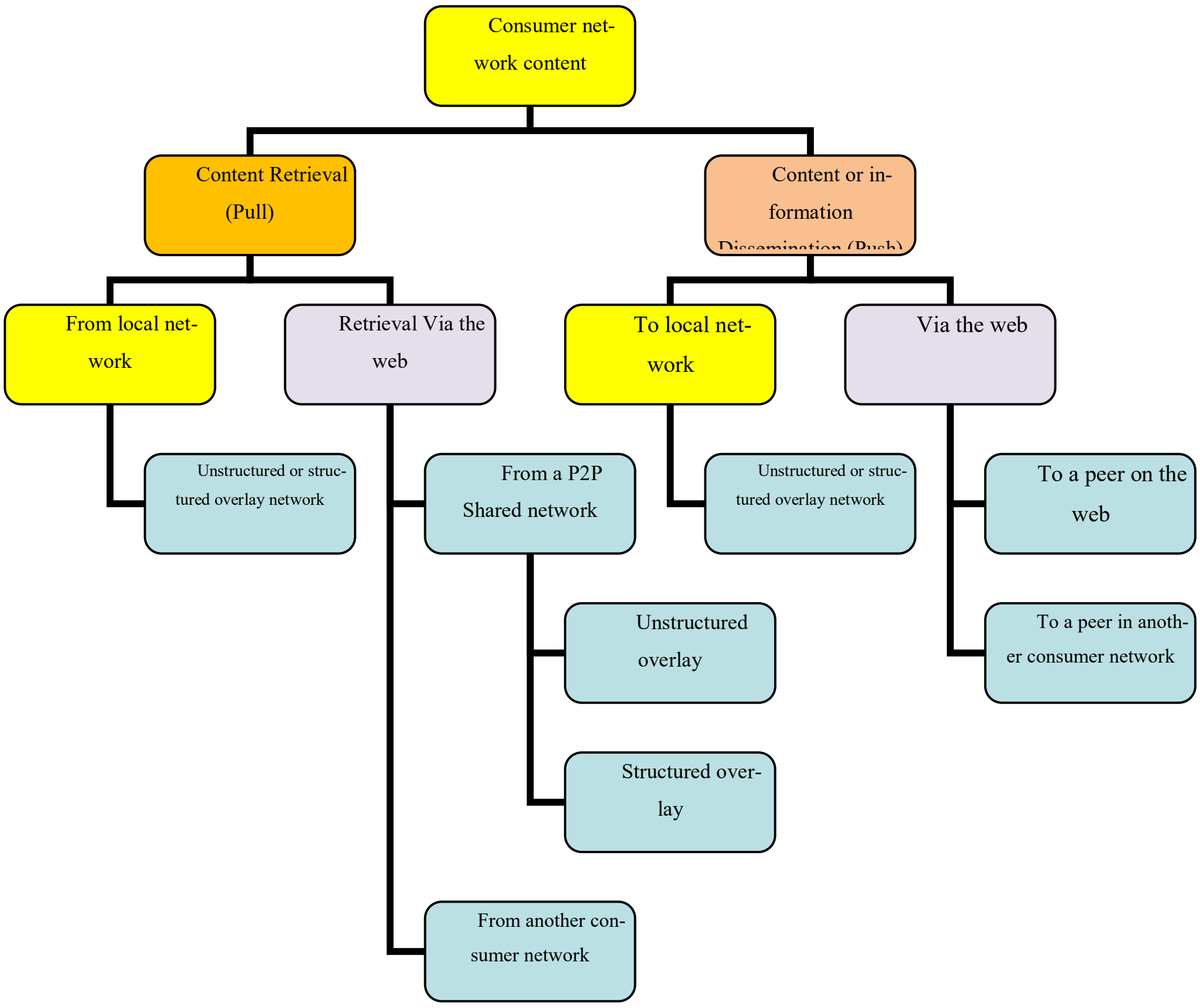

**Figure 28: Challenges in hybrid network design for Internet of things by using unstructured P2P algorithms for searching content in personal computing devices**

### 4.2.1 Goals definition

The key goal in this exploratory agent-based modeling case study is to develop an exploratory model for studying the feasibility of the proposed design of hybrid complex communication networks. By means of proposing an algorithm Self-Advertising Content Sources (SACS) which is similar to dropping bread crumbs around a house for locating content sources, each content source sends a message with a certain numeric value in its immediate vicinity. This gradient establishment process is demonstrated in Figure 29. Here S represents the content source. This could be a personal computer, a mobile phone or some other device. The other square tiles represent other computing devices located nearby. Now, how it can be physically implemented is by means of having S first locate other wireless communication devices within its immediate access based on its communication radius. These devices are then sent the gradient establishment message. Once the message reaches these devices, being the first in the vicinity, these devices keep the counter of the message and are shown by a 1 in the figure. These devices next search for other devices and send the messages but with a lower SACS value. This process continues till the SACS distance limit expires for the message in which case, it is not re-sent or else there is no other communication device within the communication radius.

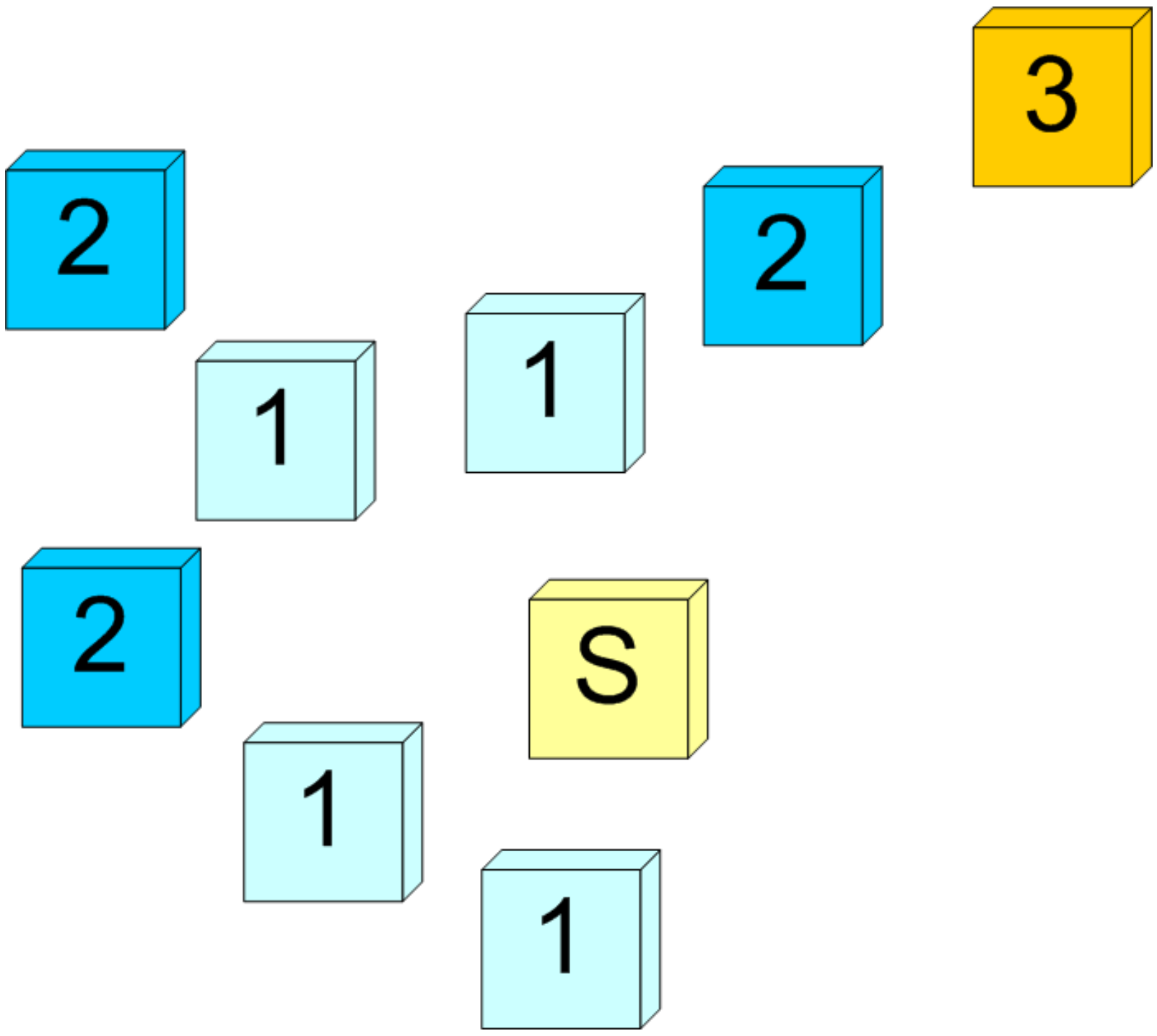

**Figure 29 Establishment of a gradient by SACS sources**

Subsequently after the establishment of a gradient, devices can query other devices in the P2P Internet of thing network. This is possible with the help of a TTL based algorithm as depicted in Figure 30. As can be seen, to avoid unnecessary flooding of the surroundings, the queries keep moving forward based on the gradient value. So, essentially the queries are able to locate content sources because they always move to a computing device with a lower gradient value.

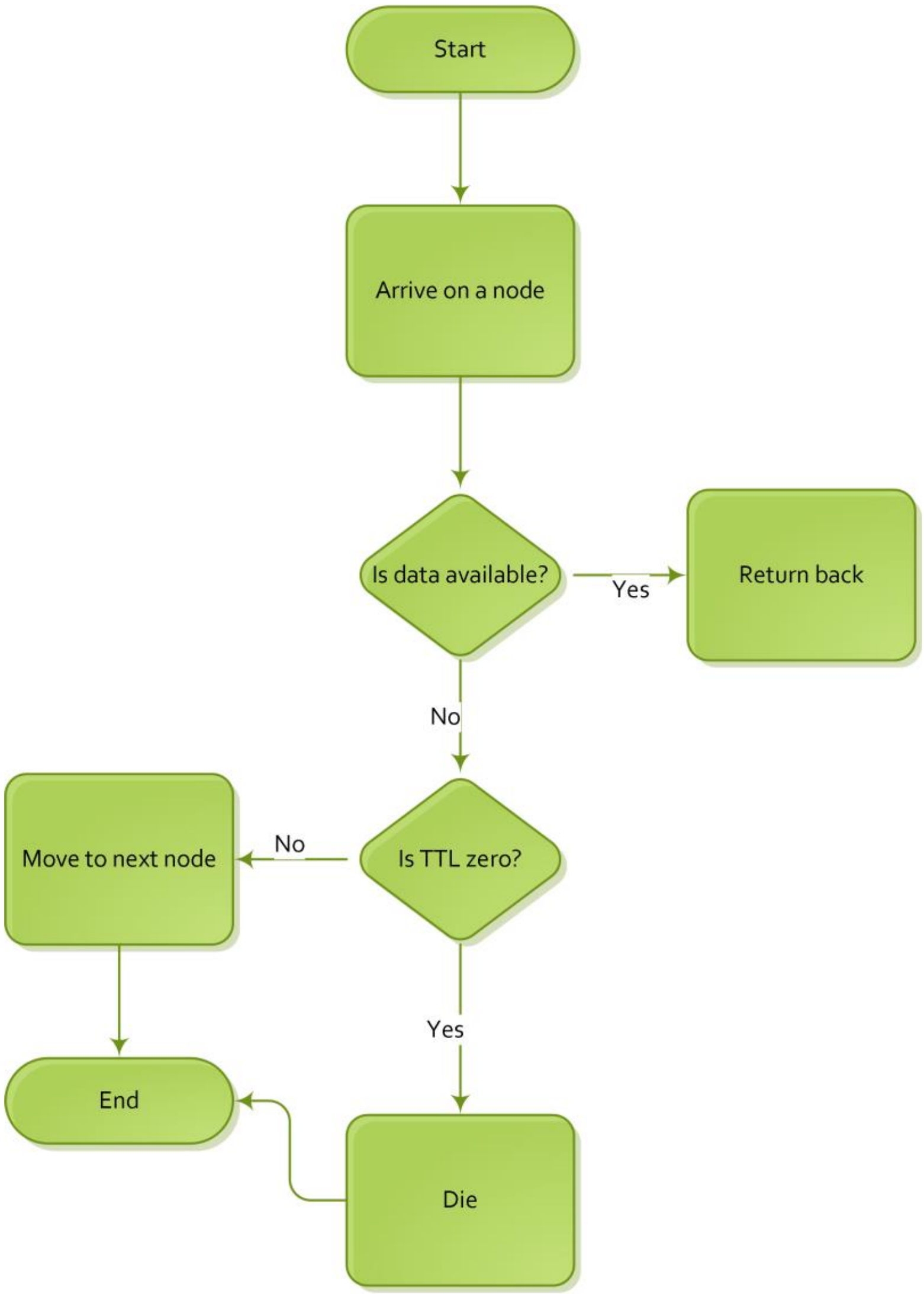

**Figure 30 Query algorithm flowchart**

### 4.2.2 Agent design

As discussed earlier in the previous chapters, Agent-based models have been used in a large variety of application domains. While some application domains involving social simulations, such Agents in Computational Economics (ACE) depict reproducibility, economic benefits and social interests, these descriptions are not inherent to agent-based design per se and are domain specific. As such, here we give detailed descriptions of agent-based models using various detailed design diagrams and implementation details in the complex adaptive communication networks domain using a case study of the Internet of things.

The agents in this simulation exploration are of two primary types:

1. Computing devices.
2. Message agents which are of two types
    a. Query Message agents.
    b. SACS setup message agents.

The detailed design description of these agents is given below:

#### 4.2.2.1 Computing device agents

These agents are designed to simulate various types of computing devices. The devices can range from mobile phones, handheld games as well as laptop computers to stationary computers. As such, there are the basic design requirements for these agents that can be listed as follows:

A. The agents must have a certain amount of memory to hold certain information.
B. In addition, these agents must be able to communicate with other agents in a certain communication radius.
C. The agents must be able to perform basic computations (such as retrieval of data from storage or using data structures).

The state chart diagram describes the Computing device agents. It is pertinent to note here that some of the computing device agents need to be considered as gateway agents. The idea of gateway agents is that in case the internet of things based algorithm (SACS) is unable to locate certain content or information locally, then it should be possible for message agents to find specific devices which are configured as gateway devices. These computing devices are essentially useful for allowing the local P2P network to extend its reach to the larger set of devices. As an example a person can have multiple devices at home as well as office. If the person requires a certain file, the search query can execute first locally in the vicinity of the house and if the query cannot locate the file, the search messages need to be routed on through a gateway node to the office gateway machine which will look for the required content/file in the office devices.

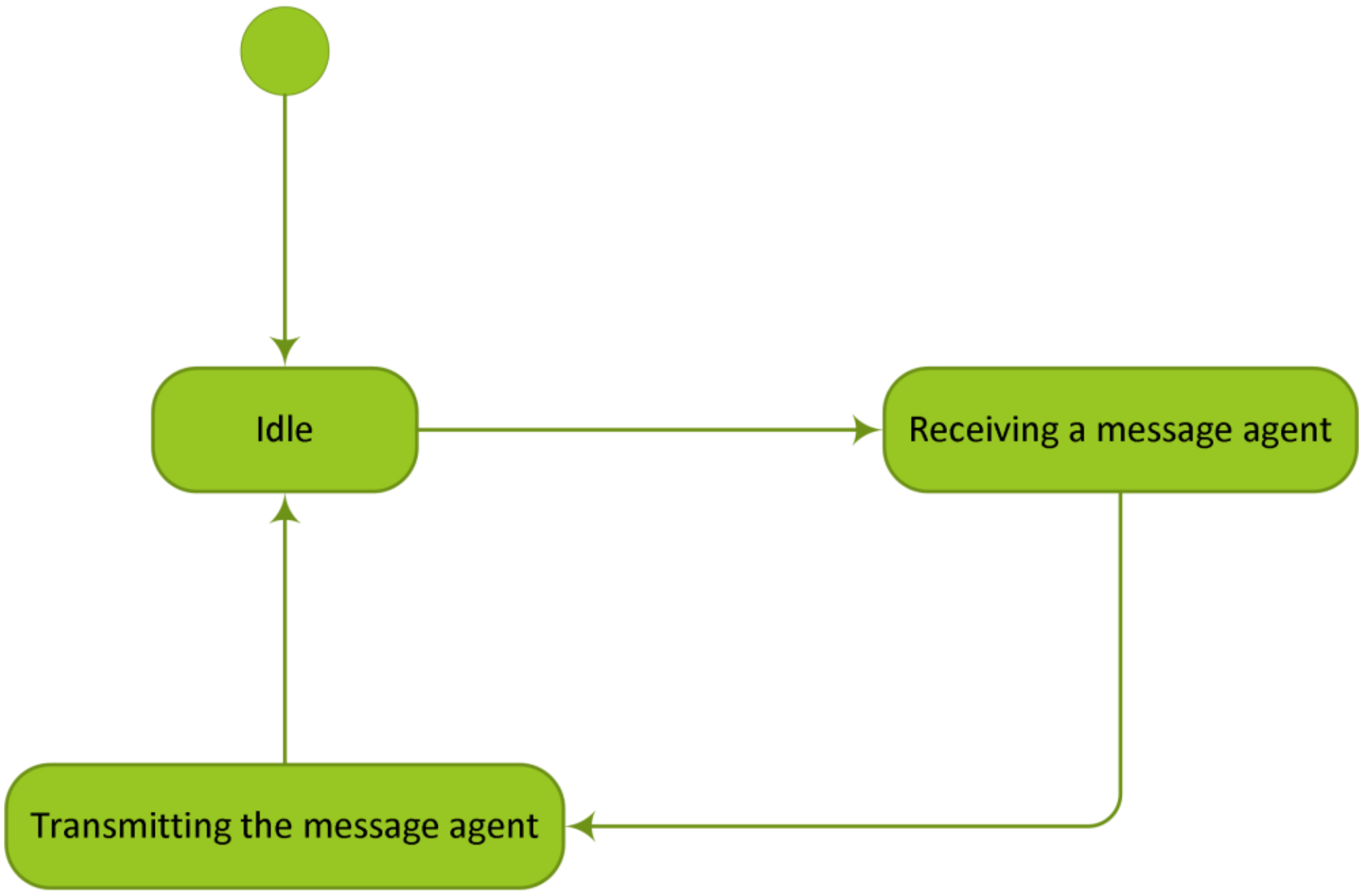

**Figure 31 State chart for Computing device agent**

The second type of computing device agents is the SACS source agents. The key idea is that each content source self-organizes its surroundings by dropping bread-crumbs for enticing query agents. The way it actually would be implemented in physical devices can be described as follows:

1. First the device sends a message within its communication radius. This message is basically a Hello message for finding out what is the list of active neighbor computing devices within communication range.
2. Next, these devices respond back to the sender giving their IDs.
3. The IDs of the neighbor devices are stored by the sender computing device.
4. Next, the sender device broadcasts the SACS setup message agent after assigning a TTL value to it. The reason for first finding the neighbor devices instead of an ini-tial broadcast is that the device must be sure that there are any active listening computing devices before it repeatedly sends these messages.

5. On receiving the message agent, the neighbor devices can subsequently forward the message based on the policy of the message agent, which will be described below.

Next, we discuss the design of the Message agents.

#### 4.2.2.2 Message Agent

Here we discuss the design of the message agents. Message agents are of two types. The first type is the SACS setup message agents and the second type is the query agents. Unlike the computing device agent, it has been primarily designed for forwarding the message agents, message agents themselves need to act more intelligently. Next, a description of both these types of message agents is given as follows:

1. SACS setup message agents are responsible for setting up a gradient in the vicinity of the SACS source nodes. As described above, these agents help establish a gradient starting from the initial communication radius of the SACS source node agents. SACS setup agents have a simple task to perform. Once they are on a specific device, they use the algorithm described in Figure 30 to hop from one device to the next. This is done by means of counting the distance (radius) by means of a hop count. As a result when the hop value reaches the end of the SACS radius, it implies the end of the line for the SACS setup process. However, during each hop, the agents also communicate their respective hop count value to the computing device agent, on which they are currently located. As such, the computing device agent stores this information locally for future queries. It is important to note here that SACS gradient can be established from a smaller radius to a large radius, where the radius here would be used in terms of communication hop counts. In case of smaller radii, there are chances that subsequently when queries are looking for a particular content source, they might spend a long time hopping randomly before they can locate a particular SACS gradient "scent" i.e. the hop value bread-crumbs left over by the SACS setup message agent in the initial phase.

2. The second type of message agents is the Query agent type. The goals of the query message agents are different from the SACS setup message agents. The query agents are initiated by a user logged on to one of the computing devices. As such, the goal of the query agent is to look for information. Now, the interesting thing to note here is that according to the small world complex network theory, a large number of real-world networks have nodes which can locate most other nodes using a small number of hops. As such, when an owner of a set of devices is looking for specific information, chances are that the query would be able to locate this content on one of the computers locally or within a few hops away from the person. Query agents are thus responsible for content location discovery based on the "smell" of the content discovered by means of the bread-crumb values dropped by the previous SACS setup messages during the initial setup phase.

### 4.2.3 Learning and adaptation in the system

As discussed in detail in the last two chapters, unlike traditional agents involved in multiagent systems, the agents in the simulation design here do not individually have intelligence and adaptation capabilities. However, the system as a whole has significant adaptation capabilities. Thus the nonlinear interactions of the different agents allow the entire system to emanate emergent behavior. The emergent behavior in this case is the discovery of content using the bread-crumbs in content vicinity spread earlier during the course of the execution of the distributed SACS setup algorithm. Here it can be noted that the bread crumbs are established here in a completely self-organized manner. Thus every particular assembly of computing devices will establish a completely new gradient and each case of randomly walking queries will behave differently. However irrespective of any variation in the configuration SACS algorithms need to be to be able to locate the required content sources.

### 4.2.4 Implementation description

Here, we describe the implementation details in the NetLogo tool. The model has been implemented in a modular fashion and is described as follows:

#### 4.2.4.1 Global variables

Firstly, the global variables depict some of the metrics which need to be collected over the due course for the simulation experiments. There are three key desired global output variables here:

1. The number of successful queries.
2. The total number of queries.
3. The total communication cost

In addition, there are several other input variables from the user interface which can be adjusted by the user either manually or else by using behavior space experiments as shown in Figure 32.

#### 4.2.4.2 Breeds

The "aggregation" basic of cas, as discussed earlier in chapter 2, can be noted here in this model. With the help of using two different types of agent breeds developed in the netlogo model, aggregation can be noted. These breeds can be described as follows:

1. Nodes breed

   This breed depicts the computing devices in the simulation. It has its own set of internal variables. These variables are of two types. The first ones are Boolean variables which determine the type of the node. These include whether or not, it is a gateway node, a query node, a goal node (i.e. a content source) and whether or not it has been previously explored till now. The other type of internal variable for the node is the SACS distance. This is the variable which is updated based on the hop distance i.e. gradient from the SACS nodes. SACS nodes are the nodes with the value of the Boolean variable *goal?* being true.

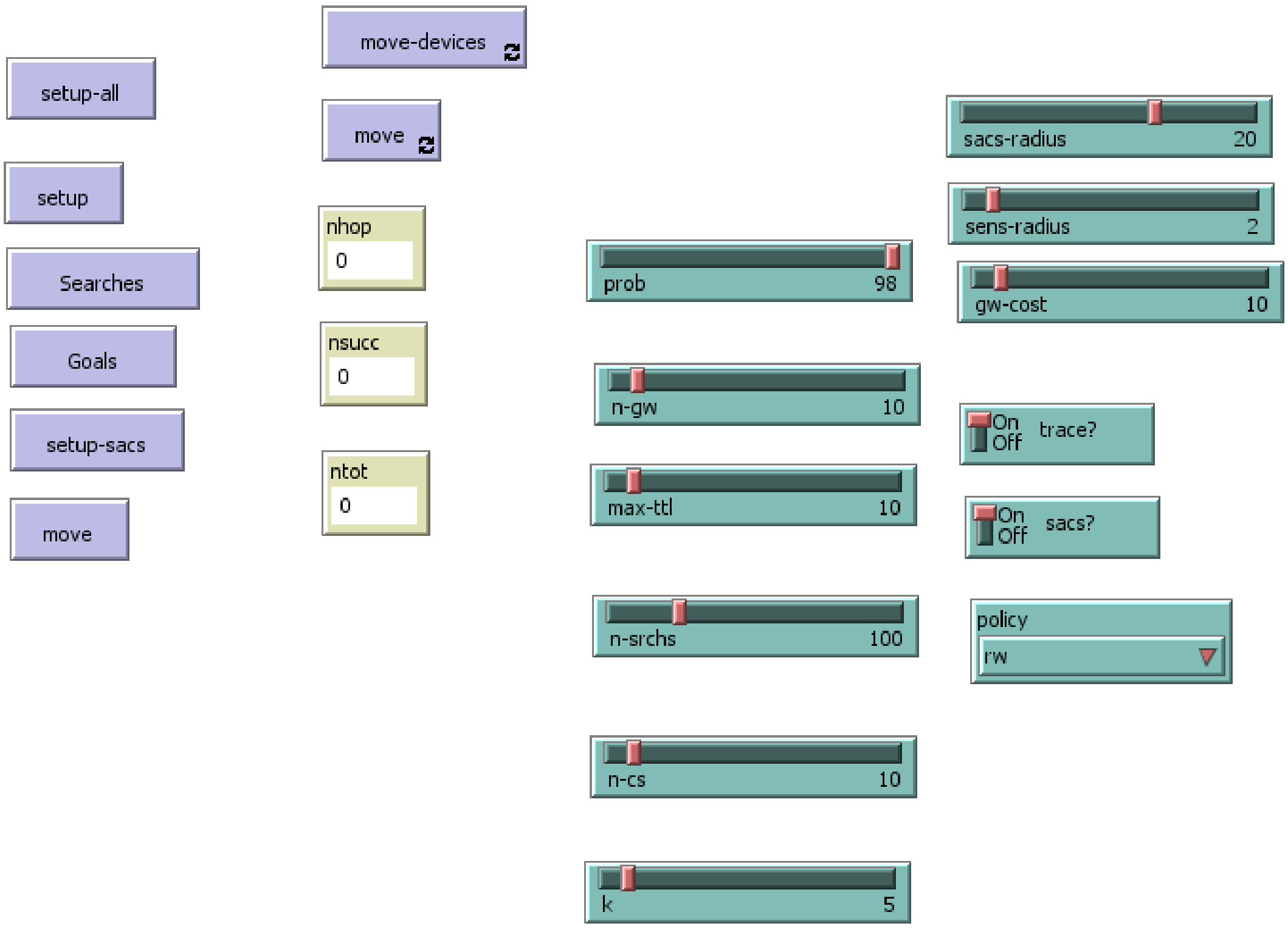

**Figure 32: User interface showing various configurable values for the simulation**

2. The second breed is for the messaging agents. These have been implemented as the "Query" breed. These agents have three internal variables for storing their state. Firstly, they have a variable which aggregates the current node agent, on which they are located. The third variable inside the query agents gives the concept of selective lookup by noting the particular type of content which this agent is searching for. As can be noted, both of these concepts represent the cas internal model basic concept as discussed earlier in Chapter 2.

After a discussion of the breeds developed in the agents, we next move to the implementation of the various procedures:

#### 4.2.4.3 Setup

The first function is the "setup" and allows the creation of the entire scenario for further simulation of the model. The overall setup function is shown in following Figure 33.

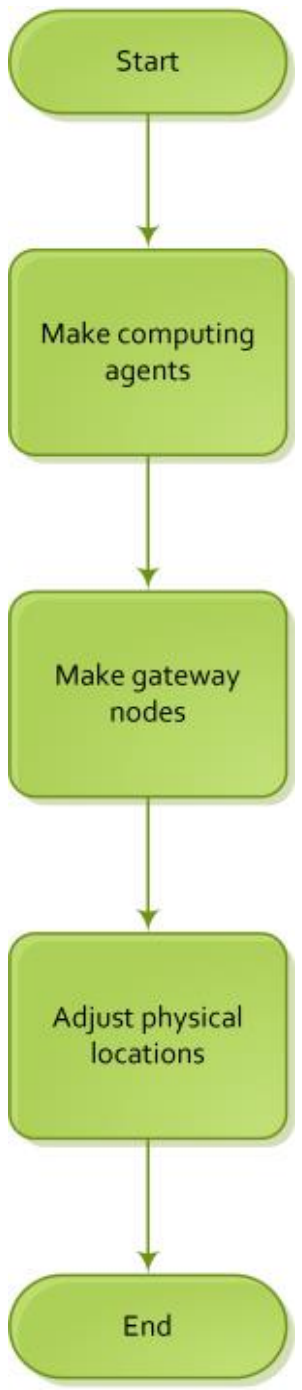

**Figure 33: Description of the setup function**

Here, we can note that the first part of the setup function is to clear all variables, agents and the environment of the simulation. This is important especially when multiple simulations are to be executed with the passage of time. The second part of the setup is related to making the computing agents. This function is related to constructing the computing device agents in the simulation. After the creation of the computing devices, the gateway nodes are next selected based on the input parameters. Finally, these gateway nodes adjust in terms of their locations in the Simulation world. The detailed description of each of these functions follows along with other functions.

#### 4.2.4.4 Make Computing Agents

This function is essentially the base for developing the computing agents according to the input variables. This procedure is based on invoking the patches in the NetLogo world. First of all, each patch creates a random number between 0 and 100. Next, this number is compared with the probability assigned via a global input variable. Based on a comparison of these two numbers, the patch might sprout an agent at this location. Next, the agent is initialized with certain values. Initially, the agent is given an unexplored status. The sacs distance is assigned equal to the sacs radius input global variable. Initially all nodes are given "false" as the Boolean value for both the start as well as the goal and the gateway variables. The shapes of the computing devices are next adjusted to be squares and they are slightly randomly moved to ensure that most devices will not cover up over previous devices.

#### 4.2.4.5 Make Gateway nodes

This function is concerned with creating the gateway nodes from the previously created computing device node agents. The working is based on a random selection of agents from these devices. The number of agents which are to be created as the gateway nodes is based on a global input variable. After making the node as gateway, its color is changed to yellow so that it is visible in the overall landscape.

#### 4.2.4.6 Adjust locations

This function asks each gateway node agent to perform certain tasks. First of all, the agent looks in its immediate location and notes if there are any other agents there. If it finds other agents located at the exact spot, the gateway node agent starts adjusting its loca-tion slightly. It executes a conditional loop and keeps taking a random turn and moving very slight till it is not immediately over another computing node. This exercise is important for both visualization as well as verification of the correct behavior in the case of execution of large scale simulations. The effects of the setup give a screen view as shown in Figure 34.

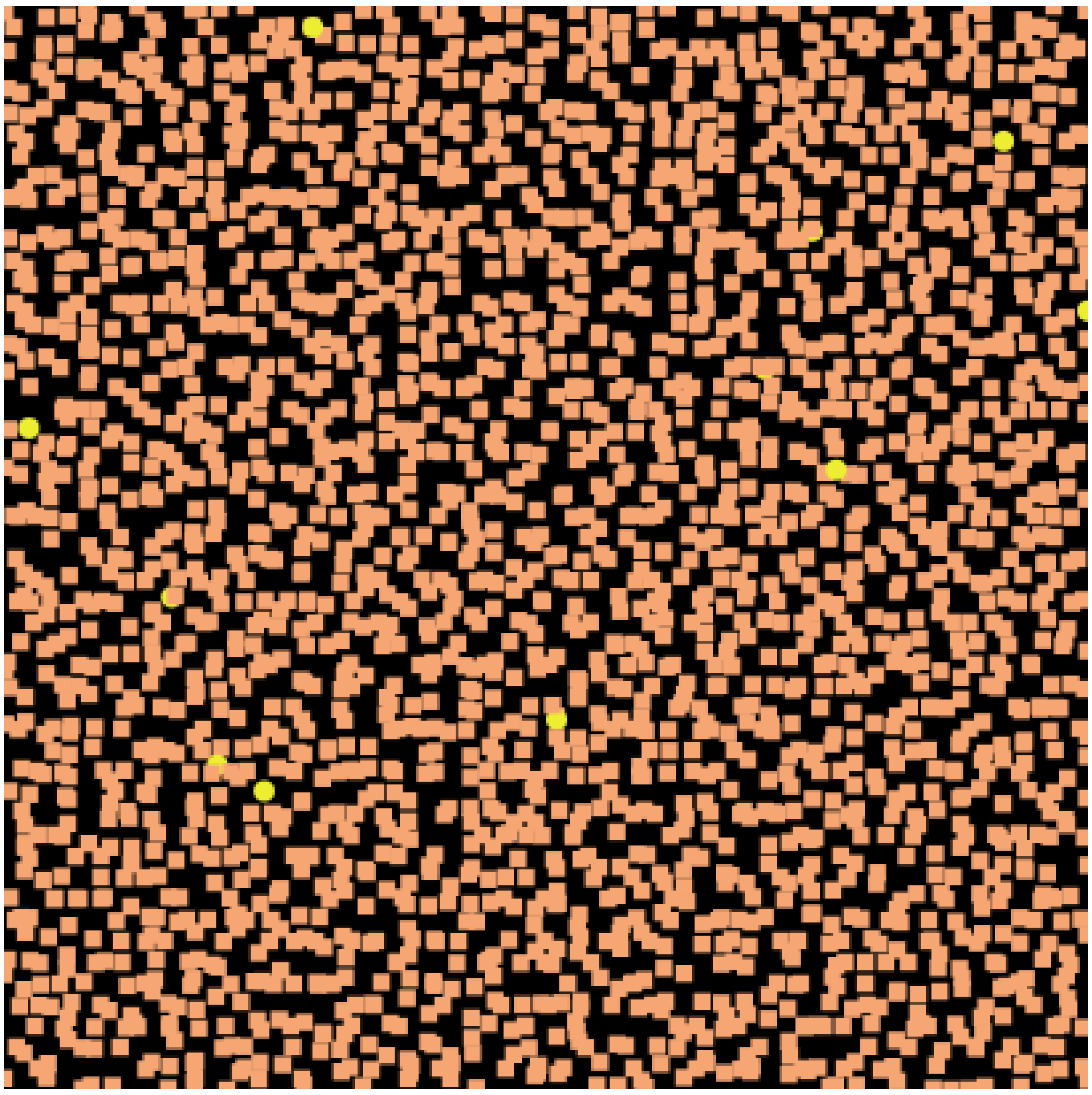

**Figure 34: Effects of the execution of the setup function with 2500 computing devices**

Note here that the square computing device agents have self-organized themselves based on simple rules resulting in a more realistic pattern. In addition, all gateway nodes are now completely visible on the screen.

#### 4.2.4.7 Searches

The “Searches” function has primarily got the task to repeatedly call make-search function. The number of times this function is called is again configurable from the User interface. This is a global variable slide which takes input from the user in terms of manual execution for the number of searches to be created randomly.

#### 4.2.4.8 Make Search

"Make search" function has two basic tasks. Firstly it randomly selects a computing device node agent. However, this selection is based on the criteria that the selected random node agent must not previously be a gateway node. It then calls the make-s-node function.

#### 4.2.4.9 Make-s-Node

This function is the actual function which creates the search node. The first thing that this function does is to change the "start?" Boolean variable to true. It thus tags the node as the location of the starting point of the queries. It also changes the color of the node to blue and its shape to a circle so that these nodes are visually recognizable. The location of the computing node is then stored in a temporary value. Afterwards, it executes the hatching of k query agents. These query agents are needed for the "k-random walker" algorithm from the domain of unstructured P2P networks. For each of the query agents, it calls a setup-query function.

#### 4.2.4.10 Setup-query

This function is called from the Make-s-node function. The goal of this function is to properly assign values to the query agents. After making the query agents a circle and assigning them green color, the function increments the total query count global variable. The query variable "loc" is next assigned an agentset made up of the agent where it is currently located.

#### 4.2.4.11 Goals

The “Goals” function repeatedly executes the Make-goal function. Here the n-cs global variable essentially reflects the total number of content sources which is another global input variable.

#### 4.2.4.12 Make-Goal

This function has got two main tasks. First, it selects nodes which are neither gateways, nor start nodes for queries. Secondly, it asks each of these nodes to execute the make-g-node function.

#### 4.2.4.13 Make-g-Node

This function is called only to have a centralized location for changing the attributes to each goal node i.e. a content source. Unlike most other functions in the simulation, this function is quite simple and only sets the goal Boolean variable to true in addition to changing the color of the node to red.

After calling setup, Searches and Goals functions, the screen now looks as shown in Figure 35.

#### 4.2.4.14 Setup-sacs

“Setup-sacs” is the key function in the SACS setup algorithm. It is used to firstly locate each of the goal nodes. Once all the nodes have been located, they are all asked to execute setup-sacs-d function in the turtle context.

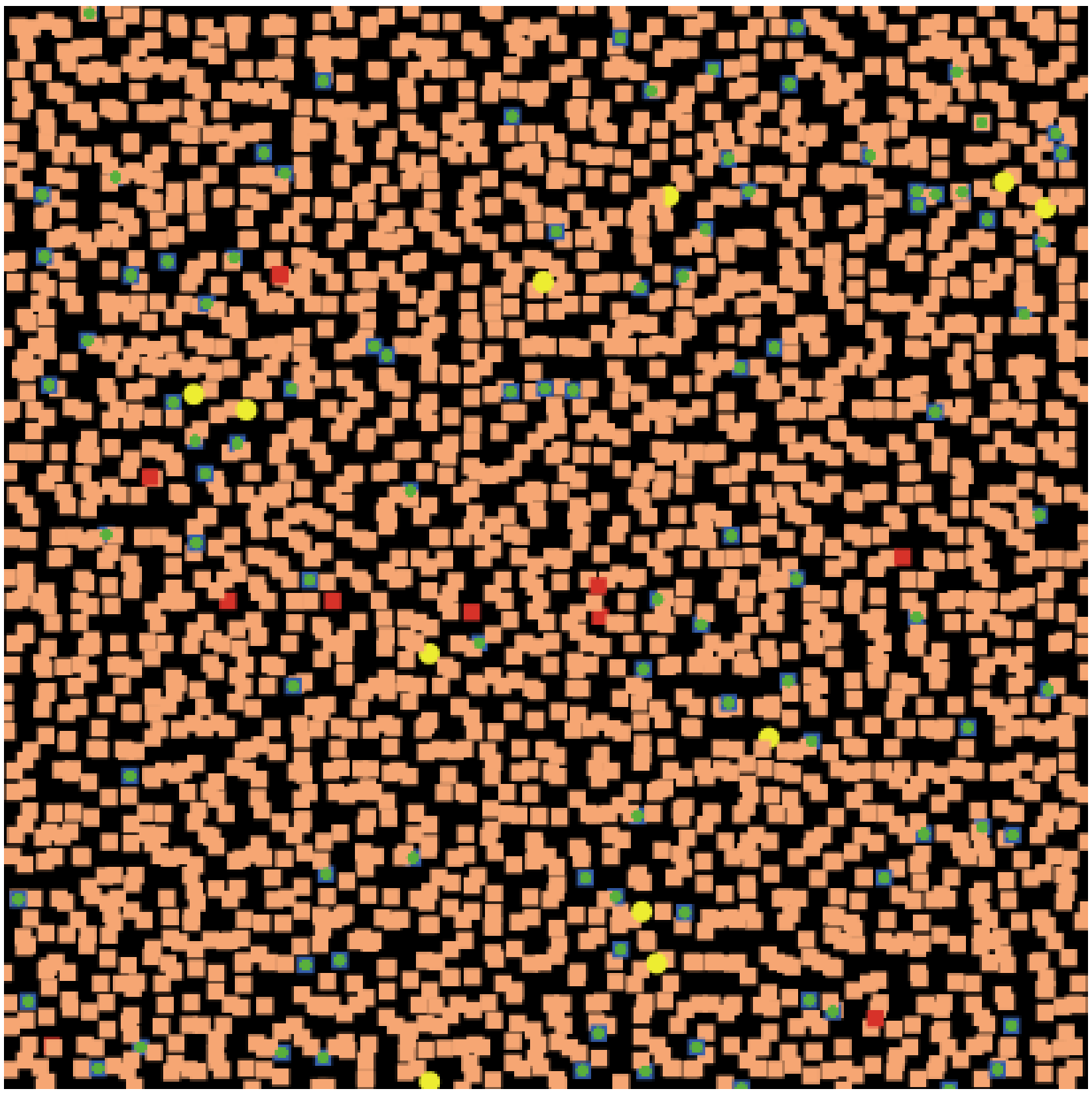

**Figure 35: View of the screen after the execution of the setup, Searches and the Goals functions**

4.2.4.15 Setup-sacs-d

"Setup-sacs-d" simulates the entire SACS setup algorithm. It is also a recursive algorithm but unlike traditional recursive algorithms, it is an agent-based distributed recursive algorithm. In other words, it executes on other agents. So, it essentially calls itself repeatedly till the SACS gradient is properly setup in the communication radius of each content source. Like all recursive algorithms, it needs to have a stopping condition. The stopping condition here is a complex stopping condition. It can be considered as a disjunction of two separate conditions. Firstly the algorithm will stop if there are no other "unexplored" computing nodes in its vicinity. Secondly, it will also stop if the calling argument "d" is greater than the global variable "sacs-radius". The "sacs-radius" global input variable is again configurable by means of a slider. The way this algorithm works is that first of all it equates the sacs-distance value to the minimum one as compared to the previous value. It then changes the label of the current computing node agent to reflect this sacs-distance value. It also makes the explored? Boolean variable true. If the sacs-distance is non-zero, it changes the color of the node to grey. This way, the content sources are the only ones which will visually stand out from the other nodes. After setting these basic attributes, it compares the current argument “d” with the sacs-radius global variable. If d is less than or equal to this value, it creates a new agentset. This agentset is formed of all other nodes in a certain communication radius but based on a condition of being unexplored till now. The communication radius is again a configurable value “sens-radius”. Now, this agentset is not guaranteed to be non-empty so it is tested for emptiness. If non-empty, each of these agents is asked to execute the same function again but with an incremented “d” value. Thus this process can continue till the SACS gradient is properly setup around all SACS content sources reflecting how the content sources can be Self-advertising their contents.

Afterwards, the simulation screen can be observed to reflect this setup as shown in Figure 36.

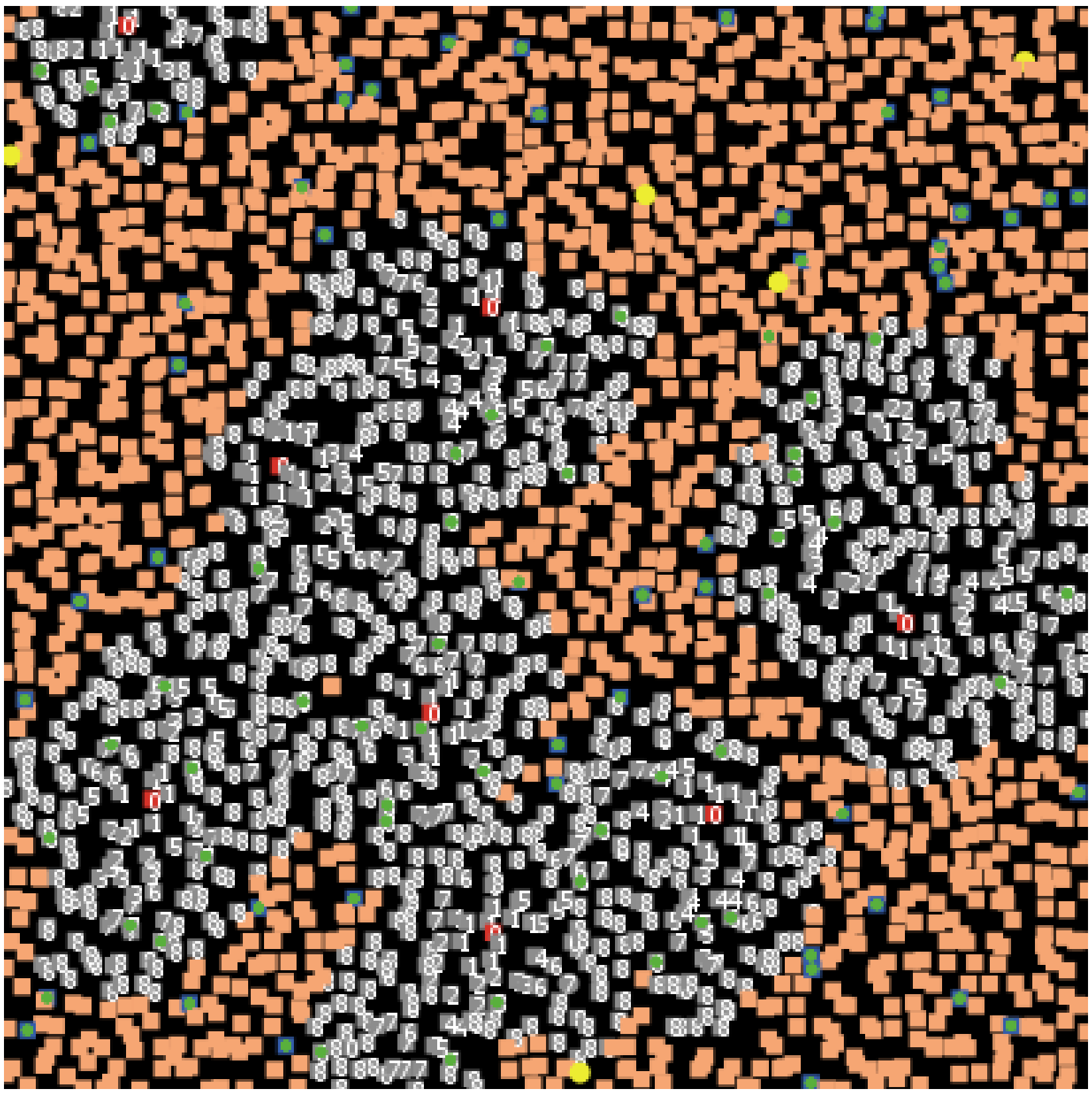

**Figure 36: Result of execution of SACS setup algorithm**

4.2.4.16 Move

This is the main function for the execution of the algorithm for queries as shown in Figure 37. The function itself is focused purely on the execution based on the existence of query agents. If any queries are there, it will continue by asking them to execute move-rw function otherwise it will terminate.

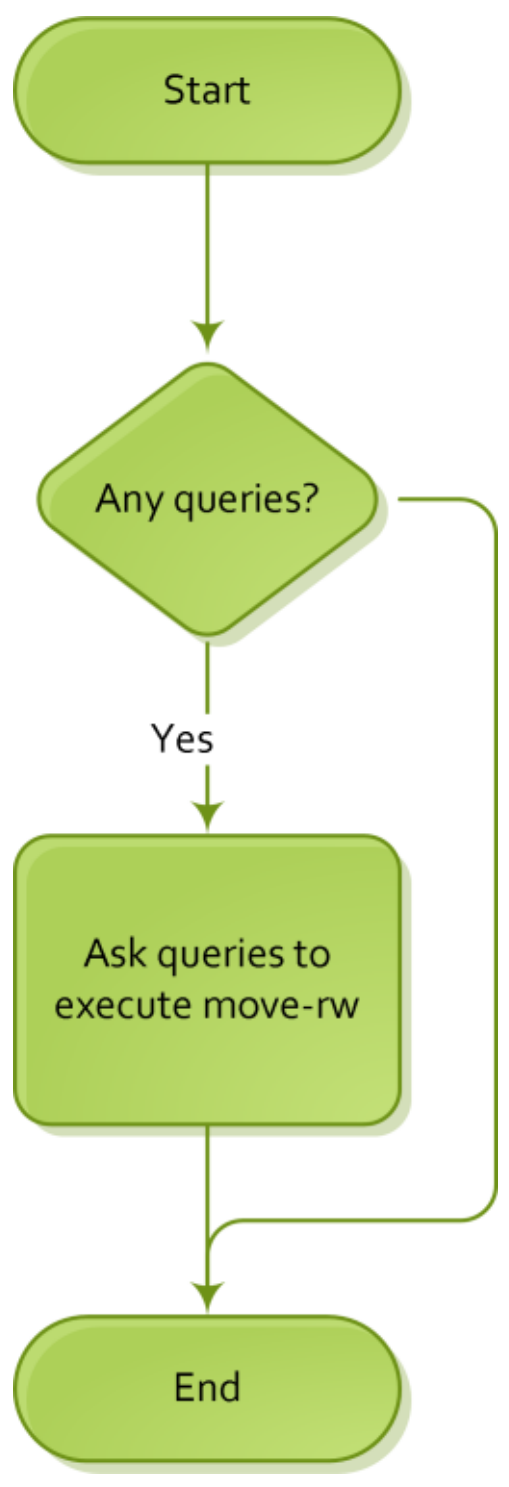

**Figure 37 Flowchart for move function.**

4.2.4.17 Move-rw

This is the main function for the execution of the algorithm for queries as shown in Figure 38. Here, we can note that the function executes by first creating a list of all nodes in the given sensing radius. This radius is basically equal in physical terms to the communication radius. As an example, in case of Bluetooth networks, it will be different as compared to Wi-Fi networks. In case there are no other devices in range, the query agent will simply die. If there are other devices, the next step is choosing one of the nodes as the next location. This is performed differently based on what is the selected mode of movement of queries from the user interface. Based on the Boolean variable "sacs?" the next node is either a random node in the case of a false value for this variable, or else one of the nodes with minimum of "sacs-dist" variable. Note that this represents the gradient previously established by the SACS nodes. As such, once this node is selected, the query agent can firstly move to the selected node. Next, it can update its internal variables such as TTL value by decrementing it and also storing the tagged location node inside the "loc" variable. After moving and updating these values, it needs to calculate the cost associated with this move. If the location is a gateway node, then the move will incur cost equal to "gw-cost", a user input configurable variable otherwise the cost will simply be incremented by 1. Finally before this function terminates, a call is made to the checkgoal function, which is explained next.

#### 4.2.4.18 Check-Goal

This function is primarily for goal verification for queries. In this function, the query checks whether it has reached either one of a gateway node or else a content source (goal node). If the current location agent satisfies either of these conditions, then the overall number of successful queries is incremented and then the query agent terminates itself. If it has not reached the goal nodes, then again it checks its TTL value. In case, the TTL value has reached zero, then again it dies.

4.2.4.19 Move-Devices

This function is to test the effects of mobility of devices on sacs. It randomly moves a percentage of the devices over time and when devices are out of range, it calls setup-sacs again to reset the sacs setup messages.

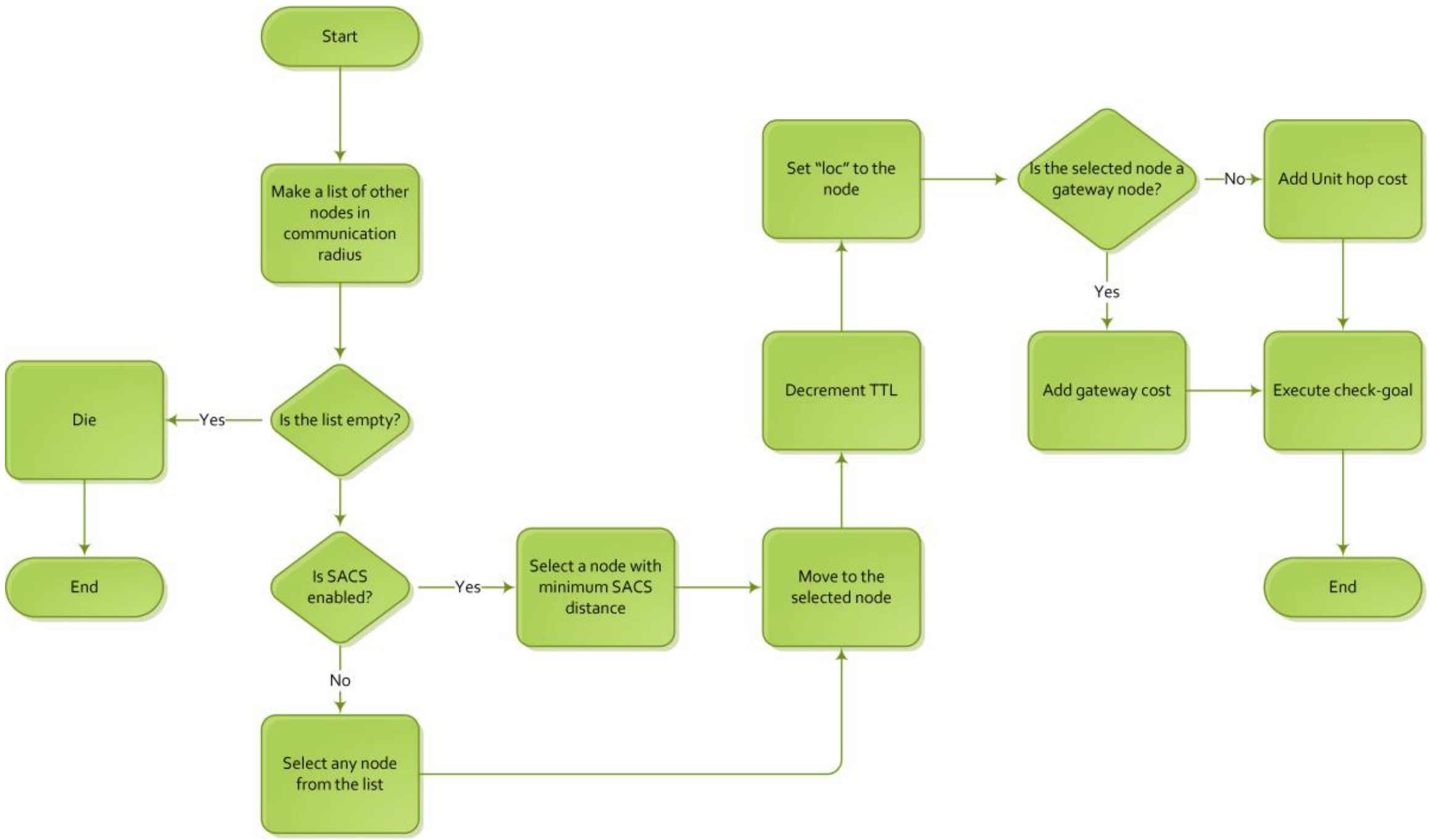

**Figure 38: Flow chart for the move-rw function**

## 4.3 Simulation Results and discussion

The goal here is to use agent-based modeling to develop working proof of concept models for modeling these concepts. Thus while small scale computing devices can be hand-simulated and might appear to be a trivial exercise which can be performed on a piece of paper, the specific goal of using agent-based modeling here is to be able to examine how this concept will work in the case of say hundreds of computing devices in say, a small community or a small town. The proof of concept models will allow us to examine the results based on hundreds of simulation runs. Thus it will allow us to discover the strengths and weaknesses, if any, of the SACS mechanism. By using extensive parameter sweeping, the simulation experiments should give a strong case for developing further models of SACS or else to abandon the concept if it does not scale for larger and more realistic settings of large-scale communication networks.

### 4.3.1 Metrics used

The input parameters and the outcomes of the simulation will be measured using the key metrics given as follows:

#### 4.3.1.1 Simulation parameters:

The input metrics to the simulation are as follows:

1. The number of computing devices used in the simulation
2. The number of content sources $n_{cs}$
3. The number of gateway nodes $n_{gw}$
4. The number of query source nodes $n_{src}$
5. Number of random walkers per source node $k_{rw}$
6. Cost in terms of hop counts for using a gateway node for locating a content source: $c_{gw}$

#### 4.3.1.2 Output metrics:

The key metrics in a P2P network are based on the number of lost queries as compared to the number of successful queries. The output parameters that are of interest in the evaluation of SACS are as follows:

1. Number of successful queries $S_{tot}$
2. Total cost of execution $C_{tot}$
3. Total number of queries $N_{tot}$

However, here the total number of queries will be constant and instead the cost of execution as well as the number of successful queries will be evaluated in depth.

### 4.3.2 Details of experimentation

The simulation execution is based on the execution of four separate Behavior Space experiments. The details of the experiment parameters and the reasoning are given below:

The Behavior space parameters for the experiments are given as follows in Table 12. We can note that some parameters are the same across all simulations while two key parameters are varied i.e. max-ttl and the sacs-radius parameters. Each of these variables in the simulations starts with a certain value. Subsequently these variables are incremented by the next value listed up until it reaches the third listed value as shown in the table. In addition, the last two experiments are conducted to evaluate the mobility of computing node devices.

**Table 12 Simulation parameters**

| Params | Exp I | Exp II | Exp III | Exp IV |
|---|---|---|---|---|
| gw-cost | 10 | 10 | 10 | 10 |
| n-srchs | 100 | 100 | 100 | 100 |
| sens-radius | 2 | 2 | 2 | 2 |
| n-gw | 10 | 10 | 10 | 10 |
| n-cs | 10 | 10 | 10 | 10 |
| max-ttl | 10 | [5 5 20] | 10 | [5 5 20] |
| sacs-radius | [0 5 20] | [0 5 20] | [0 5 20] | [0 5 20] |
| k | 5 | 5 | 5 | 5 |
| | Static | Static | Mobility | Mobility |

The descriptive statistics of the first experiment are given in Table 13. We can note here that the total simulations were 250 with each simulation repeated 50 times. The successful queries, total queries and the hop counts are also given.

**Table 13 SACS varying with constant TTL**

**Descriptive Statistics**

| | N | Minimum | Maximum | Mean | Std. Deviation |
|---|---|---|---|---|---|
| runnum | 250 | 1 | 250 | 125.50 | 72.313 |
| max-ttl | 250 | 10 | 10 | 10.00 | .000 |
| sacs-radius | 250 | 0 | 20 | 10.00 | 7.085 |
| nsucc | 250 | 16 | 399 | 219.51 | 108.612 |
| ntot | 250 | 500 | 500 | 500.00 | .000 |
| nhop | 250 | 3194 | 5055 | 3970.77 | 559.595 |
| Valid N (listwise) | 250 | | | | |

For the next experiment, the descriptive statistics are given in Table 14. Here the total set of simulation runs was 1000. The max-ttl as well as the sacs radius were both varied periodically in the course of the experiments.

**Table 14 SACS and TTL varying**

**Descriptive Statistics**

| | N | Minimum | Maximum | Mean | Std. Deviation |
|---|---|---|---|---|---|
| Runnum | 1000 | 1 | 1000 | 500.50 | 288.819 |
| max-ttl | 1000 | 5 | 20 | 12.50 | 5.593 |
| sacs-radius | 1000 | 0 | 20 | 10.00 | 7.075 |
| Nsucc | 1000 | 7 | 455 | 217.40 | 119.092 |
| Ntot | 1000 | 500 | 500 | 500.00 | .000 |
| Nhop | 1000 | 2125 | 9885 | 4541.87 | 1902.447 |
| Valid N (listwise) | 1000 | | | | |

The next two experiments were concerned with mobility of computing devices. The descriptive statistics are given in Table 15. The total simulations in this case were 250. The max-ttl was made constant and only the sacs-radius was varied.

**Table 15 SACS vary with constant TTL (mobility)**

**Descriptive Statistics**

| | N | Minimum | Maximum | Mean | Std. Deviation |
|---|---|---|---|---|---|
| runnum | 250 | 1 | 250 | 125.50 | 72.313 |
| max-ttl | 250 | 10 | 10 | 10.00 | .000 |
| sacs-radius | 250 | 0 | 20 | 10.00 | 7.085 |
| nsucc | 250 | 8 | 349 | 194.64 | 88.675 |
| Ntot | 250 | 500 | 500 | 500.00 | .000 |
| Nhop | 250 | 3369 | 5076 | 4113.39 | 479.169 |
| Valid N (listwise) | 250 | | | | |

The final set of experiments was conducted again to test mobility. This was similar to experiment 2 since it was based on varying both the max-ttl value as well as the sacs-radius values periodically. The descriptive statistics here are given in Table 16.

**Table 16 Experiment with TTL and SACS varying (mobility)**

**Descriptive Statistics**

| | N | Minimum | Maximum | Mean | Std. Deviation |
|---|---|---|---|---|---|
| runnum | 1000 | 1 | 1000 | 500.50 | 288.819 |
| max-ttl | 1000 | 5 | 20 | 12.50 | 5.593 |
| sacs-radius | 1000 | 0 | 20 | 10.00 | 7.075 |
| nsucc | 1000 | 6 | 430 | 208.20 | 111.068 |
| Ntot | 1000 | 500 | 500 | 500.00 | .000 |
| Nhop | 1000 | 1956 | 9883 | 4690.41 | 1916.672 |
| Valid N (listwise) | 1000 | | | | |

#### 4.3.2.1 Discussion of results

The first set of results is the details of variation of sacs-radius and its effects on the successful queries. Here we can note in Figure 39 that there are five values for the sacs-radius ranging from 0 to 20. Here the value 0 implies that the SACS is not used whereas on the y-axis, the number of successful queries are being listed. As can be noted here from the earlier descriptive statistics, each point represents a set of 50 repeated experiments and as such gives a pretty good realistic example of the outcome and cannot be considered as a mere chance. We note here that very few successful queries exist in the case of not using sacs (SACS=0). However, as we make a SACS radius value of 5, the number of successful queries almost increases 5 times. The increase however is not uniform with a further increase in SACS radius. Thus, the biggest jump in successful queries is for a small SACS radius. This is quite an interesting result because it implies that with a minimal use of SACS, there is a significant improvement in the success rate for searchers.

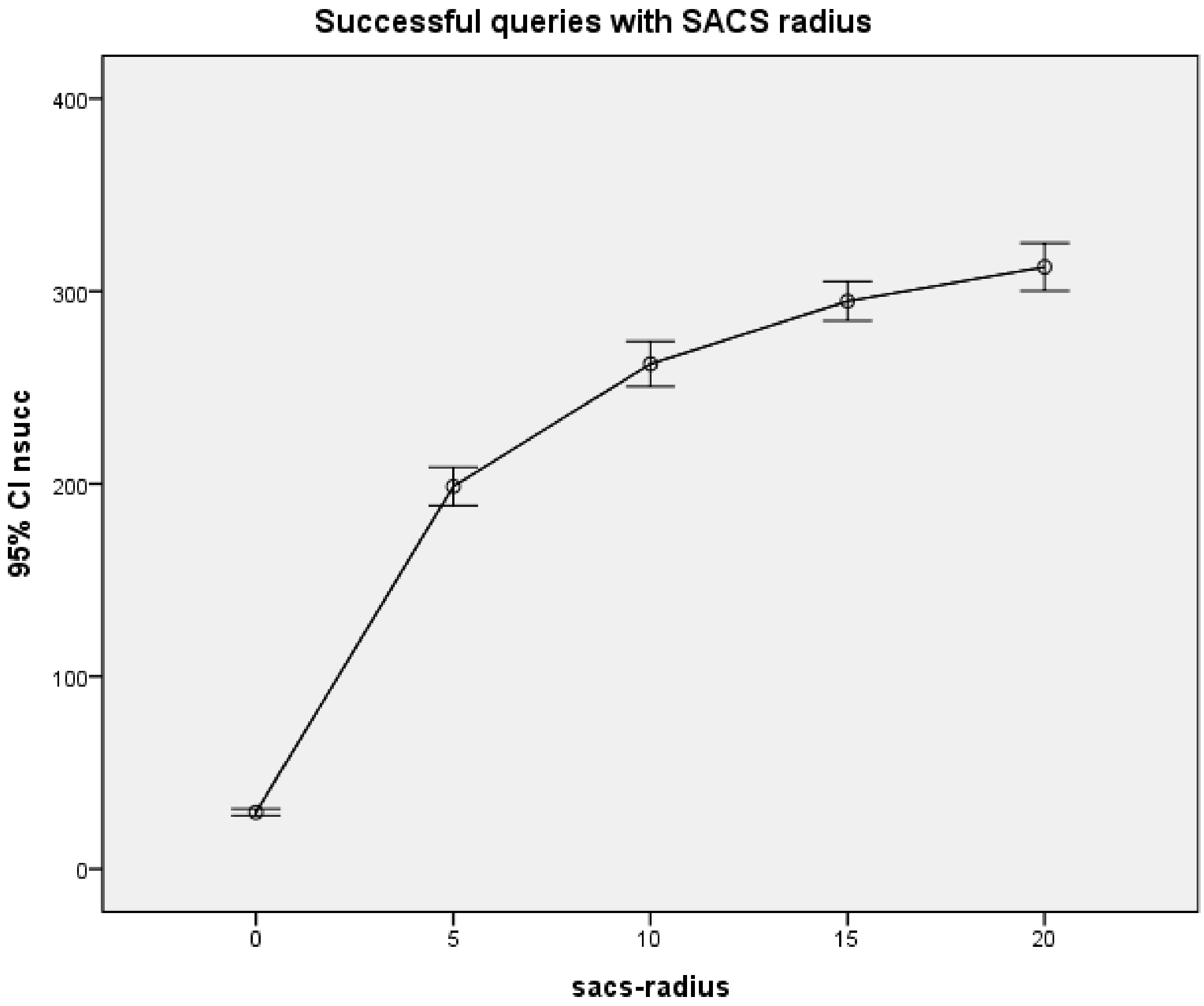

**Figure 39: 95% Confidence interval error plot for successful queries plotted against the increasing sacs-radius**

The next figure is based on the effects of increase of the SACS-radius on the overall cost of execution of the queries in terms of hop counts as shown in Figure 40. The results here are very similar to the successful queries in the previous figure. We notice that the cost is very large in the case of not using SACS close to 5000 hops. By merely turning on the SACS by giving sacs-radius of 5, the cost significantly decreases down to almost 3950 hops. There is a further decrease in cost when sacs-radius is made 10. However the jump is not that significant. This is even further noticeable when the sacs-radius is further increased. Thus it corroborates with the small-world phenomenon that just by using a small sacs-radius of 5, queries can easily locate the content sources.

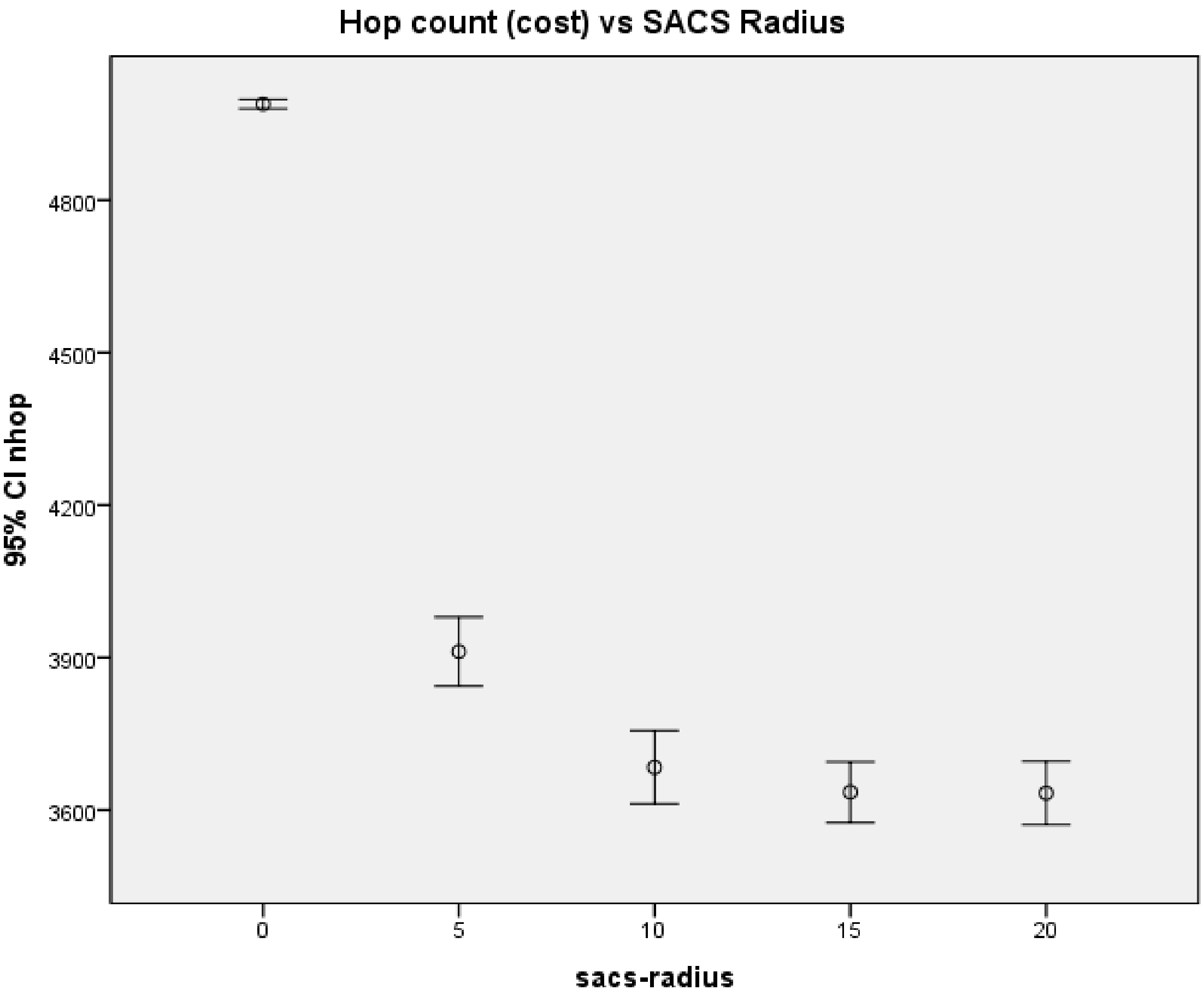

**Figure 40: 95% Confidence interval plot for hop count (cost) vs an increase of the SACS-radius**

In the previous figures, we note the first exploratory hypothesis about the effects of changing the sacs-radius on both the overall cost as well as the number of successful queries. The goal of the next plot as shown in Figure 41 is to test another hypothesis. This hypothesis was based on the TTL value of the queries. So, the big question to look for here was what if it was the TTL value of the queries that was actually resulting in the success of SACS. Also one more point to examine is how this effect would be coupled with the change in the sacs-radius. So, we can observer this effect using a scatter plot. In a scatter plot, different values of the simulation run can be noted. Here, the y-axis is labeled in terms of cost or hop count whereas the x-axis is giving details in terms of the TTL values of the queries. On the other hand, the data is color coded based on the sacs radius. Here, we can clearly see that the cost of queries without the use of SACS is always large. The difference for smaller TTL values is not as pronounced as larger TTL values. As a matter of fact, the higher the TTL values, the more pronounced is the effect of a better cost for all sacs-radius values. And from max-ttl value of 10 onwards, the difference in all experiments is very clear in terms of significant reduction of hop counts/cost.

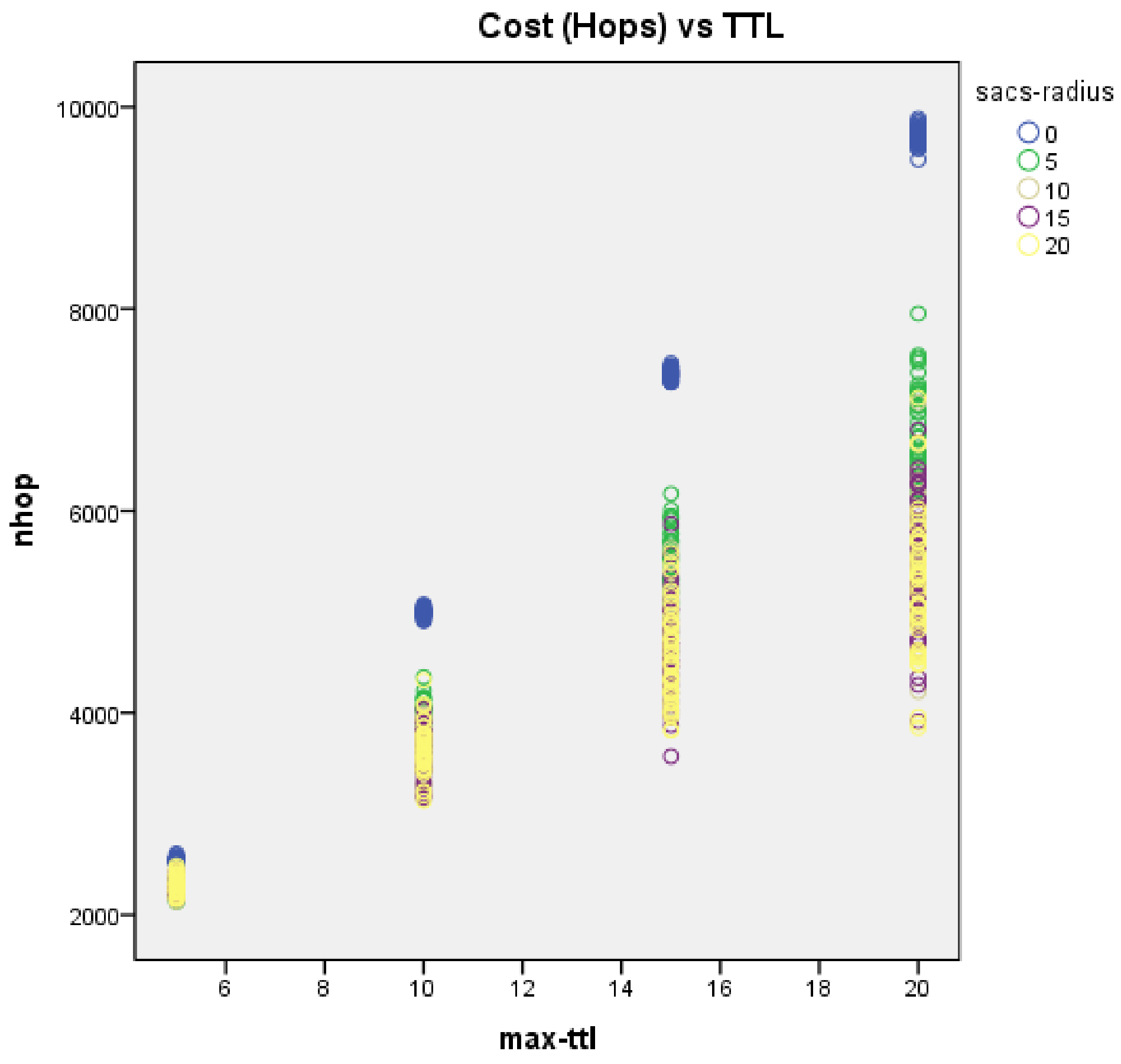

**Figure 41: Scatter graph showing the variation of the cost in terms of hop count with a change in max-ttl and sacs-radius**

For verification of this hypothesis in terms of successful queries, we plot the next graph in Figure 42 using mean successful queries on the y-axis color coded according to different sacs-radius values. The x-axis depicts the max-ttl values. Now, we can note here that for max-ttl = 5, sacs-15 appears to have the best possible successful query values. However, the numbers of successful queries for k-random walk without SACS remains constantly and significantly lower than any of the SACS radius simulation experiments. Furthermore, we note that sacs-radius = 5 reflects the most benefit as it is compared with the implementation model not executing SACS. However, the differences between all sacs values are initially very small for max-ttl of 5. Subsequently however, for an increase in sacs-radius values, there is a corresponding increase in the successful queries. This effect is however, not as pronounced as the difference between non-SACS and sacs-radius = 5 successful queries.

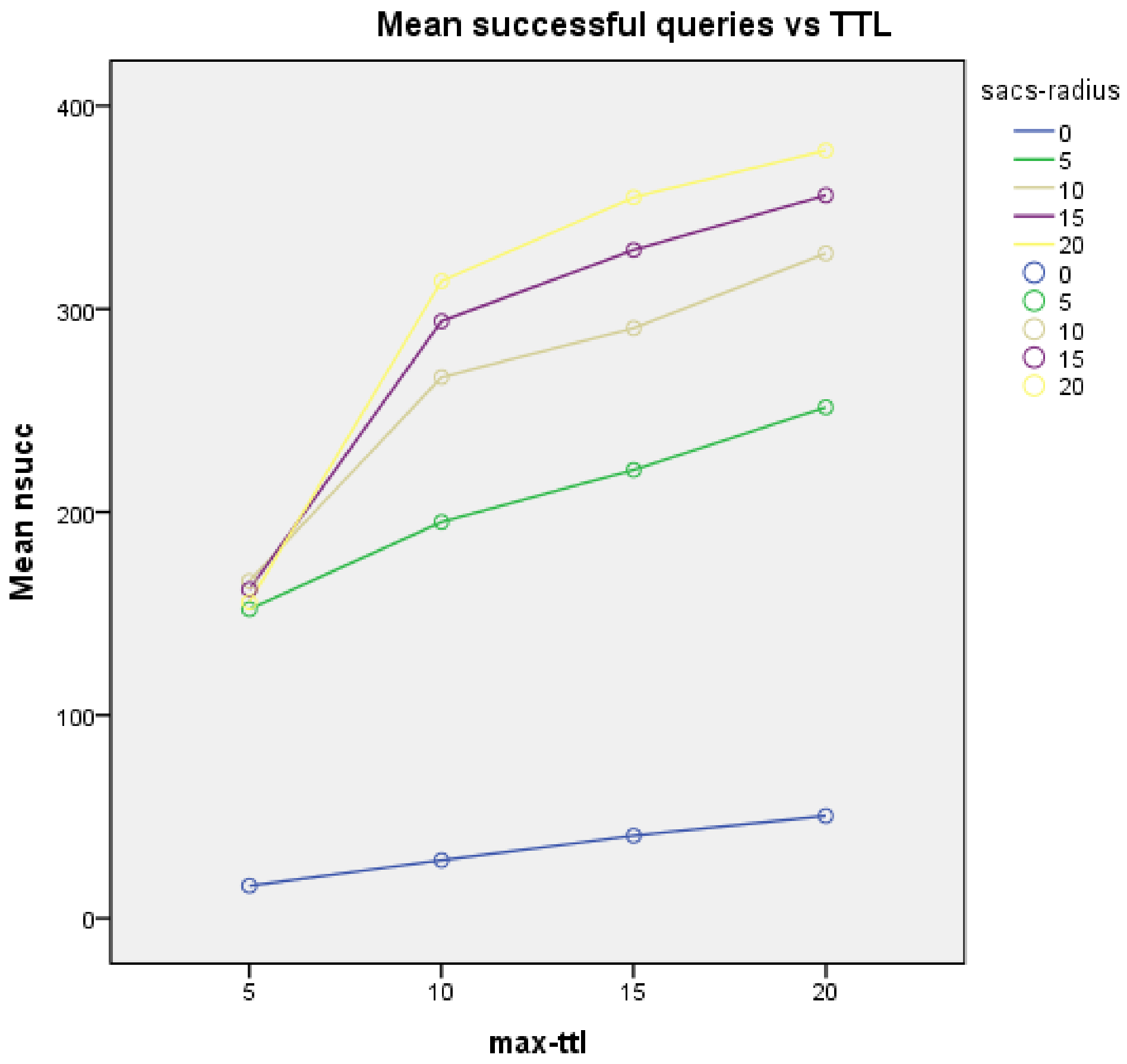

**Figure 42: Mean successful queries with a variation in max-ttl value and a variation in colors based on different sacs-radius values**

Up until this point, several different hypotheses have been tested and it has been discovered that SACS indeed can be a better alterative as compared to k-random walk algorithm from the domain of unstructured P2P networks. However, it is important to note here that till now, the scenarios are for static nodes. While this is good in theoretical terms, this is not a reflection of a realistic physical environment. In a realistic physical environment, devices are not stationary. Rather, they tend to move in a Brownian random motion as noted by Groenevelt et al. in [175]. Gonzalez et al. [176] note similar patterns in human mobility models. As such, the next set of experiments is concerned with mobility modeling of the computing device agents. The idea here is to evaluate the results of using SACS algorithm in the case of mobility modeled as Brownian random motion.

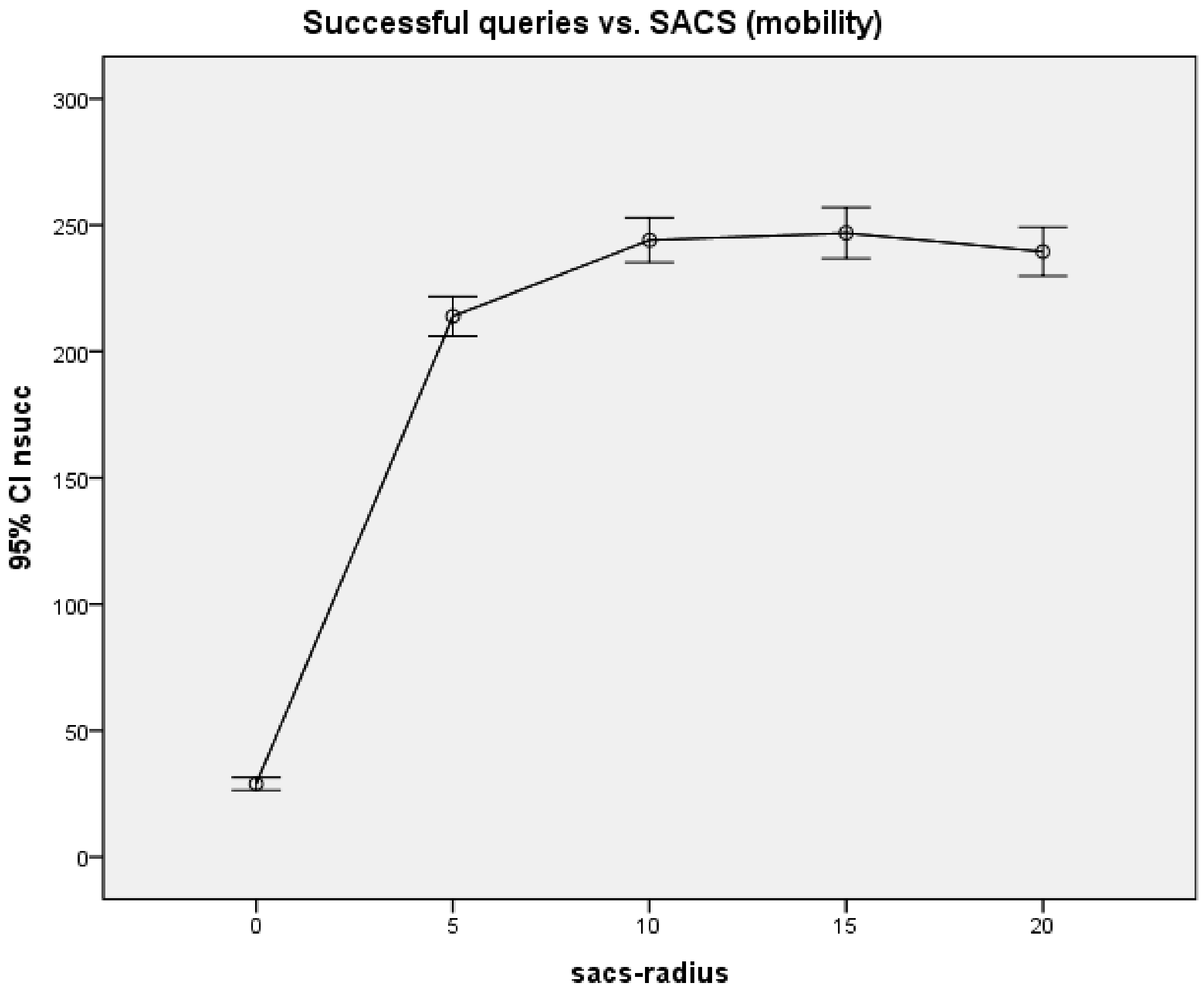

**Figure 43: 95% confidence interval plot showing the effects of increasing sacs-radius on the number of successful queries in the case of node mobility**

The first plot using mobility is for the evaluation of the effects of changing the sacs-radius by means of measuring the number of successful queries. Here, in Figure 43, we note that with different sacs-radius values, the results are still much better than the k-random walk reference algorithm. However, we also note an interesting observation that the value of successful queries increases from sacs-radius of 5 to sacs-radius of 10 but tapers off at sacs-radius of 15. And furthermore sacs-radius of 20 actually starts to reduce the number of successful queries. In other words, with the mobility problem, because of per-haps the motion of the intermediate nodes over time, some of the queries tend to get lost if there is a very large sacs-radius.

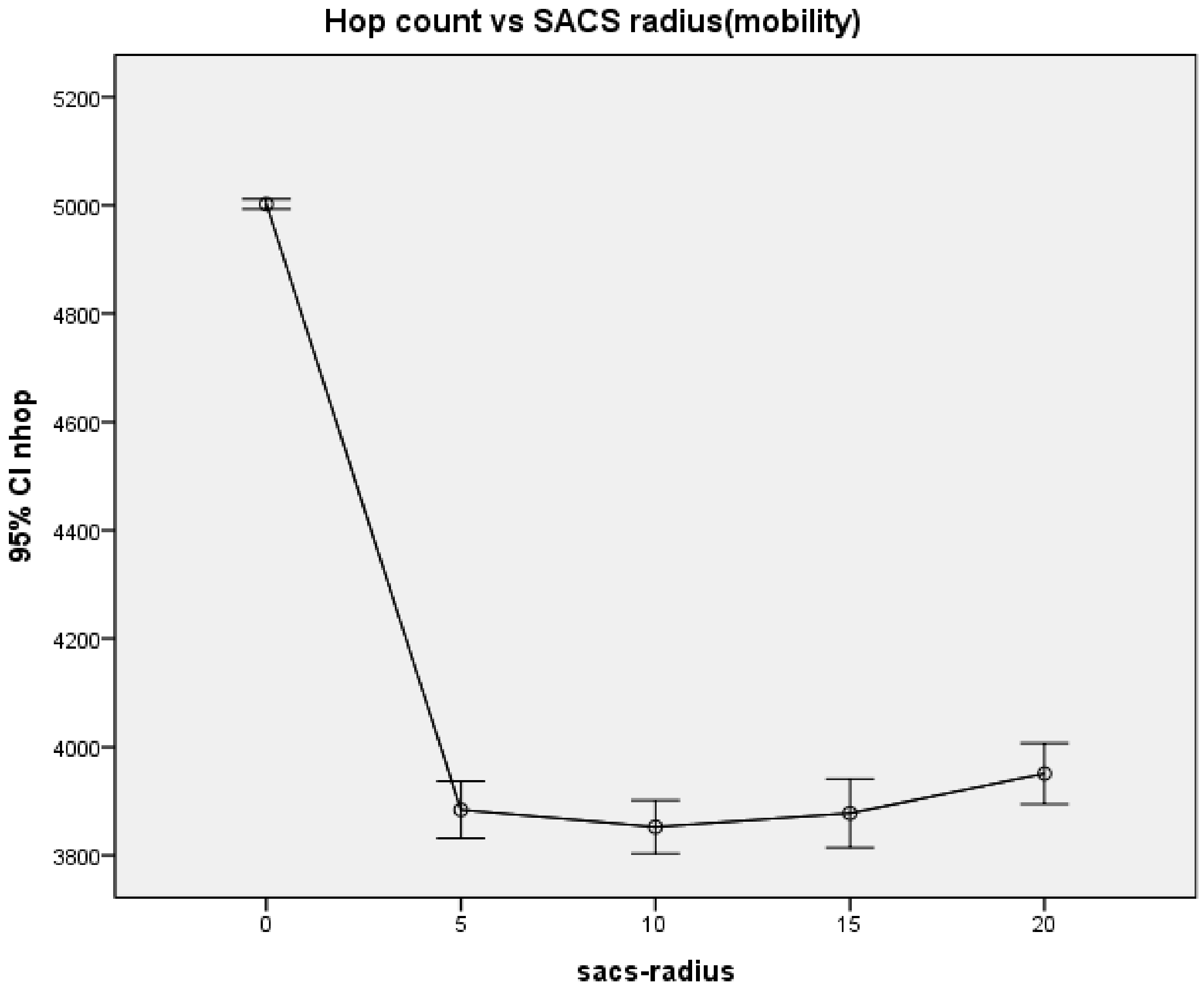

**Figure 44: 95% confidence interval plot for the cost in terms of hop count with a variation of the sacs-radius value in mobility of computing devices scenario**

While we have noted the effects of the successful queries, next we want to evaluate the effects of sacs-radius on the total cost of the search in terms of hop counts. Here, we can again note the significant jump in cost from no SACS algorithm to sacs-radius = 5. Now, interestingly unlike the previous plot, after sacs-radius is varied to 10, the value of cost tapers off a bit and actually increases from sacs-radius of 10 to sacs-radius of 15 and eventually sacs-radius of 20. Thus, we note here the negative effects of the mobility on the SACS algorithm resulting in a deterioration of results as nodes move about randomly.

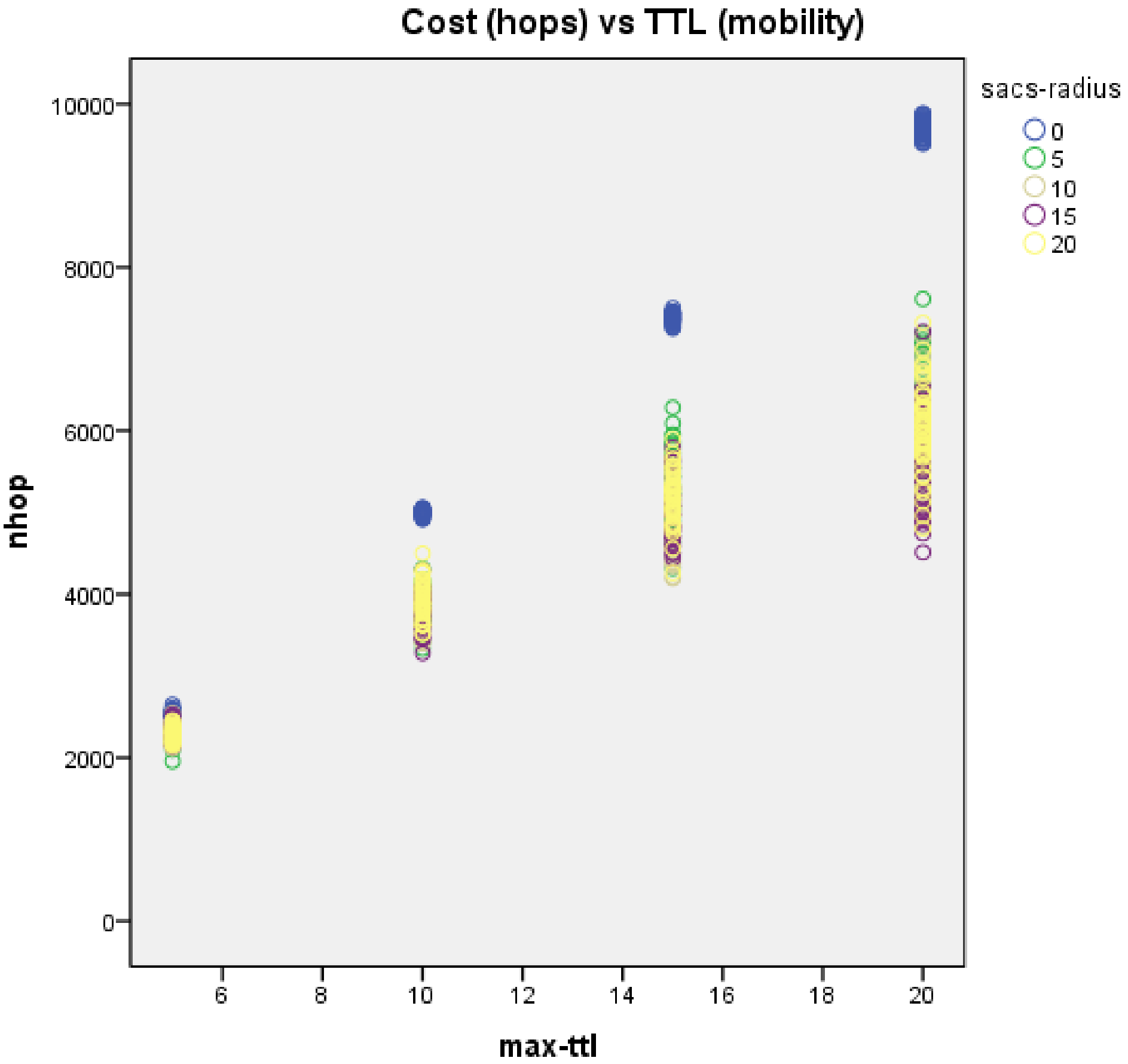

**Figure 45: Scatter plot of hop count cost with a variation in max-ttl value color coded according to sacs-radius in mobility modeling**

In the next set of plots, we want to note the bigger picture so we use scatter plot and note the cost as it varies with a change in the max-ttl value and the sacs-radius as shown in Figure 45. In this case, we note again that the cost associated with the algorithm in case of mobility is always better than the cases of not using SACS. However, the cost difference shows that increasing a max-ttl value does not have much impact in terms of differences between different sacs-radius values. One important point to note here is that initially for sacs-radius = 5, we note the lowest cost but as max-ttl increases, the cost of sacs-radius = 5 tends to become the highest and the lowest cost being that of sacs-radius = 15 at max-ttl value=20.

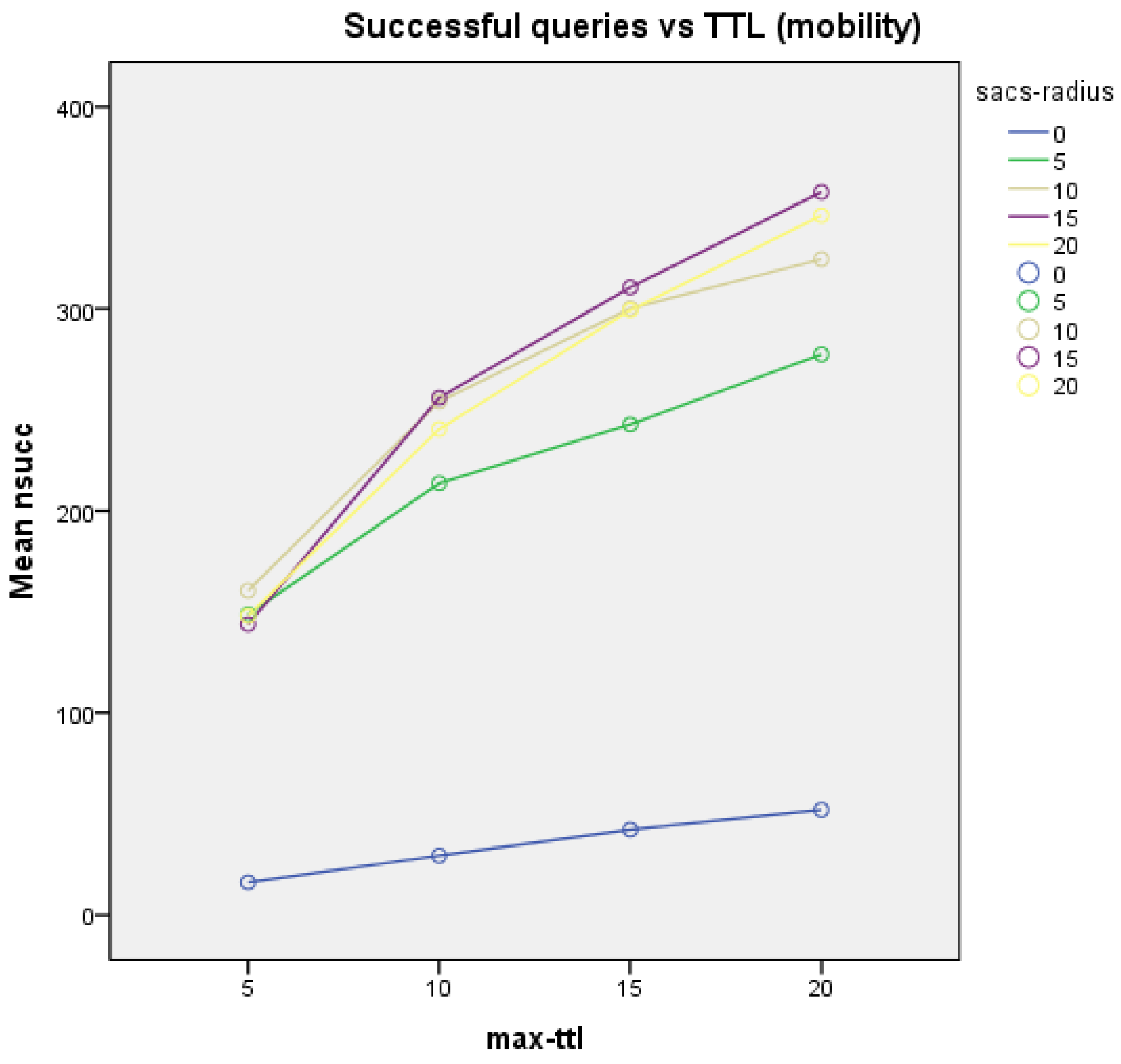

**Figure 46: Average values of successful queries with respect to a variation in max-ttl value color coded according to sacs-radius in the case of using mobility modeling**

In the next plot, we test this hypothesis in the case of successful queries. Here for validation, instead of using a scatter plot, we use a mean value plot. In this case, we note again the significant difference between non-SACS algorithm and different SACS runs. Now, in terms of successful queries, initially, sacs-radius of 10 has the largest number of successful queries. However sacs-radius of 15 wins when max-ttl value increases to 10 and keeps winning till the value of 20. As such, we can note here that mobility models have interesting effects on the SACS algorithm but SACS performs quite strongly even in this case.

### 4.3.3 Discussion of Intelligence in cas Agents

While certain types of heuristics and decision support are added inside "intelligent or rational" agents, the goal of a cas simulation is different. Intelligent engineered autonomous agents can use game theory to make choices, execute evolutionary algorithms to find best chromosomes with the highest fitness values and execute Monte Carlo simulations to find the best possible outcomes based on knowledge of input distributions. The idea of cas simulation is based on using simple but numerous agents in a system similar to life.

As discussed earlier in previous chapters, in living systems, individual cells behave in very simple ways. The total interaction of most cells involves simple operations such as sensing chemicals, emitting chemical molecules, using chemical molecules and cellular motion. However, based on these few simple rules and interactions, large and complex multi-cellular organisms are formed. In terms of learning and adaptation, unlike a typical MAS, where a single agent learns from either supervised, unsupervised or reinforcement learning schemes, agents in agent-based models have very simple models of internal adaptation. However, the real adaptation is collective. As such, while cells in the body have very simple states, the global cas in the form of e.g. the body of a mammal adapts collectively to different events in the environment. An example is that of blood clotting. While each particular cell always performs the same actions that it was designed for, the body as a whole adapts if there is say a wound or a bruise. Different cells are created or guided by chemicals to a particular spot by the body and result in clot formation. A detailed description of the differences between these two types of agents was given earlier in Chapter 3 in the Scientometric Analysis of Agent-based computing domain. In this chapter, we propose the concept of exploratory agent-based modeling level of the proposed unified framework for the modeling and simulation of cas.

## 4.4 Conclusions

In this chapter, we have given details of how a completely new cas domain application can be modeled using exploratory agent-based modeling. We have used an application which, to the best of our knowledge, cannot otherwise be modeled using any other simulator easily because this particular application is from the "Internet of Things" domain. Modeling Internet of things requires a flexible modeling paradigm since it uses concepts from the domain of P2P networks, ambient intelligence, pervasive computing and wireless sensor networks.

Our exploratory agent-based modeling of the SACS algorithm demonstrates clearly how different cas researchers can use agent-based modeling to perform extensive simulation experiments using parameter sweeping to come up with comprehensive hypotheses. Our results demonstrated the effective testing and validation of various exploratory hypotheses. We demonstrated how one hypothesis can lead to the next and how simulation experiments can be designed to test these hypotheses. In the next chapter, our goal will be to further exploit agent-based modeling by demonstrating how ABMs can be described by means of a pseudocode based specification coupled with a complex network model and Centrality "footprint" to allow for a comparison of ABMs across scientific disciplines.

# 5 Proposed Framework: Descriptive Agent-based Modeling Level

In the previous chapter we have demonstrated how agent-based modeling can be used to develop exploratory models of cas. The key idea in exploratory modeling was to use agent-based models to supplement research studies with simulation models with the goal of evaluating feasibility of further research.

An examination of the interchange of messages between researchers associated with the mailing lists of agent-based modelers can be considered to offer a cross-sectional view of the cognitive modeling problems faced by various cas researchers. It is pertinent to note here that the number of NetLogo-users mailing list[177] subscribers amount close to 3, 800 subscribers and the list contains almost 13, 000 archived messages to date. Here, we can develop a closer examination of the cognitive problems faced by researchers since it demonstrates that while helpful, exploratory modeling approach intrinsically lacks key features which are of importance to cas researchers. Some of these missing features result in problems as can be noted as follows:

A. Firstly, exploratory agent-based models tend to have a narrow focus. While multi-disciplinary cas researchers are interested in exploiting modeling constructs from one scientific research domain for use in another, there is no easy way to transfer knowledge related to model concepts in this manner. With models developed primarily for a single specific scientific application, researchers find it difficult to transfer knowledge from one exploratory ABM to another cas application domain even though apparently there are inherent similarities between the models. To the best of our knowledge, the only two means of transferring constructs is to either re-use source code snippets or else replicate models based on textual descriptions (Both approaches shall be discussed critically later in this chapter). So, a model built for say, tumor growth might have useful constructs which could be used in a completely different scientific domain such as data aggregation in Wireless Sensor Networks. However, even with this intuitive similarity, there is no easy translation of model constructs from one domain to the other using a non-code but high fidelity description of the model allowing for an almost 1-1 correspondence of model constructs between the specification and the ABM.

B. Another set of modeling problems can be observed in the case of model comparison of different case studies and application domains. Again, to the best of our knowledge, there is currently no mechanism for model comparison across cas domains and applications. As an example, Social Scientists have well-developed agent-based models and concepts for gossiping [20]. Intuitively, this could perhaps be very similar to how a forest fire spreads in a forest. In other words, the way fire moves from a single tree or a group of trees to its vicinity appears similar to gossip propagation models in communities and cultures. So, it is inherently interesting to be able to compare these models. However, once again, there is no well-known standard or formal means of comparison of these abstract concepts gleaned from Agent-based Models across domains. Currently, typical descriptions available for models are textual in nature. Textual descriptions offer a limited ability to perform comparisons. If such comparisons are performed, they are qualitative in nature and not quantitative. As such, having the ability to compare different agent-based models quantitatively could offer deeper and objective insights into different types of agent-based models without either requiring the model to be implemented or else executed for collecting simulation data.

C. A third setback of exploratory ABMs is the difficulties in teaching model constructs. While there is notable community interest in this domain as can be observed by the close to 800 educators enrolled in the NetLogo-Educators mailing list[178], the only current well-known option of teaching ABMs is via code tutorials. However, these tutorials are often incomplete, relate to a specific application domain only or else require learning a specific set of Language Libraries to develop simple ABMs.

As a means of exploring these problems further, we evaluated random snapshots of topics and message data in a typical fortnight of "netlogo-users" mailing list. Results of the experimentation shown in Table 17, demonstrate that almost 77% of topics and 85% of messages are on "how to model".

On a further semantic examination of the message contents, a key problem can be clearly identified that most cas researchers have difficulty extracting ideas from computer programs or code snippets. This problem includes the difficulty in both describing models as well as in relating "implicit mental models" to program code. Thus, while a large number of open source models are freely available online, a large number of mailing list users still tend to ask fundamental modeling questions; questions which apparently demonstrate the difficulty of most users in understanding ABMs using only code.

**Table 17 Categorical Sampling of Netlogo-users messages for a recent two weeks period (Aug 2011)**

| *Topic* | *n-Topics* | *%* | *Messages* | *%* |
|---|---|---|---|---|
| Errors | 2 | 6.45 | 3 | 3.49 |
| How to model? | 24 | 77.42 | 73 | 84.88 |
| Extending/Connecting tools | 5 | 16.13 | 10 | 11.63 |
| Total | 31 | | 86 | |

In this chapter, as a first step towards resolving these communication problems, we propose an extension to exploratory agent-based modeling. Our proposed DescRiptivE Agent-based Modeling (DREAM) approach has been developed with the following user-centric design goals:

A. DREAM should extend exploratory Agent-based Modeling by allowing for a quantitative comparison of different models without requiring coding or execution of simulation experiments.

B. Focus on better and more effective documentation has numerous benefits as noted by Parnas[179] . It can also assist in reverse engineering and model replication as noted by Li[180]. Thus, in addition to a textual description, the proposed approach should allow for examination of an ABM visually similar to that demonstrated earlier in Chapter 3. As discussed earlier, visual models allow the model to be examined and analyzed from an abstract perspective rather than requiring researchers to look at the source code. Such an approach would thus prove useful in model and cas comparison, description during teaching and in general, more effective knowledge transfer and communication between multidisciplinary cas researchers.

C. Thirdly the proposed methodology should allow for a translation from these different sub-models such as from a visual (Complex network-based) model to pseudocode specification model to an agent-based model. Apparently this is possible only if there is a high level of coherence between elements of the sub-models.

The proposed DREAM approach is based on merging complex network modeling approach with pseudocode-based specification model to specify Agent-based Models. It also minimizes learning curve because Complex Networks is an existing area of research interest for cas researchers as noted earlier in Chapter 2. As demonstrated earlier in Chapter 3, Complex Networks allow for exploring data visually as well as quantitatively. Whereas coupling a pseudocode based specification model allows for a one-to-one correspondence of the non-code description with the ABM.

The structure of the rest of the chapter is as follows:

First, a generalized description of the proposed DREAM methodology is given. Next as a validation of this modeling methodology, an experimental design of a complete case study example is developed. This is followed by a discussion of results of the case study as well as a critical review of the proposed methodology with existing methods of documenting and teaching Agent-based models such as “code templates” and “textual descriptions”. Finally we conclude the chapter.

## 5.1 Research Methodology

An examination of communication of cas researchers over the past several years in mailing lists related to Agent-based modeling (such as netlogo-users[177], Repast-interest[181] etc.) coupled with an examination of cas scientific literature as examined in detail earlier in chapter 3 reveals that typically cas research can start in several different ways. Some of the possible cases can be listed as follows:

1. Researchers can start with a concrete research question and need to develop a simulation model to explore it further.
2. Researchers can start with a generalized cas domain but without a specific research question in mind. In this case, they subsequently want to explore the intrinsic dynamics of the domain further using a simulation study.
3. Researchers can have interest in relating multidisciplinary cas studies to discover generalized patterns related to their particular application case study/scientific domain. As such, they are looking to relate concepts from other models to their area of interest either for developing models or else for comparison of these models with their models.

Based on these set of ideas, next we give an overview of the proposed DREAM methodology.

### 5.1.1 An overview of DREAM

In this section, we describe the proposed DREAM methodology in detail. Here the goal is to provide details in the context of how cas research studies are typically conducted and how DREAM can be applied to any agent-based modeling case study.

Development of descriptive agent-based models for inter-disciplinary communication, comparison and learning

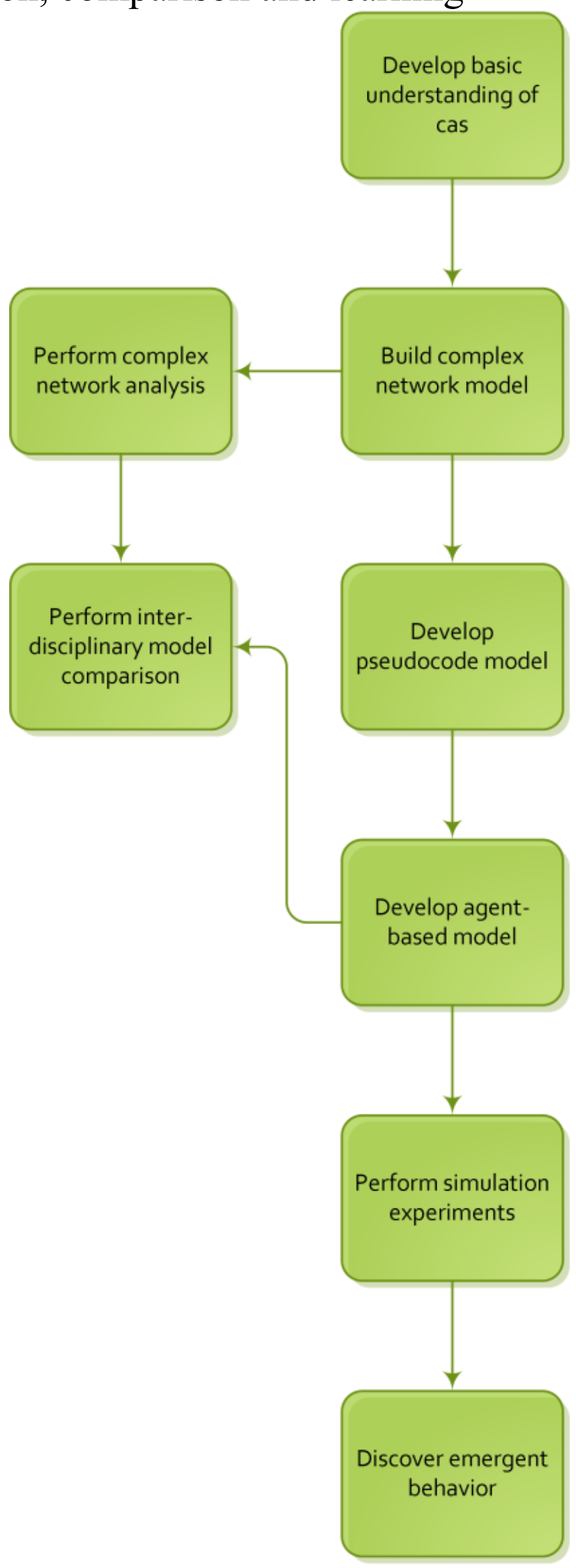

**Figure 47: Descriptive agent-based modeling level**

In contrast to exploratory agent-based modeling level, the proposed descriptive agent-based modeling level of the framework entails developing a Descriptive Agent-based Model (DREAM), which comprises of a complex network model, a quantitative centrality-based “footprint” based on the complex network model coupled with a pseudocode-based specification model. The goal of this framework level is to allow for inter-disciplinary comparative studies and a free exchange of information and ideas without regards to scientific disciplinary boundaries. So, if a model has been developed in say, Biological Sciences, the model should be comparable to models developed in Social Sciences both visually as well as quantitatively. We can observe further details of this framework level in Figure 47.

To start with, as a means of generalization, we start by presenting a definition of DREAMS specification framework. The presentation of this definition is inspired by the Discrete Event System Specification (DEVS) formalism proposed by Concepcion and Zeigler[182].

A DREAM D is a structure

$$D = < N, S, A >$$

Where

N is a Complex Network model

S is a specification model

A is an Agent-based model

Where A can be further expanded as following

$$A = < G, AG, PT, LN, P, E >$$

Where

G represents the global variables including both the input as well as the output variables

AG represents the agents

PT represents the patches

LN represents the links

P represents the procedures

E represents the simulation experiments

One way of understanding how DREAM can be used is by an examination of how an ABM is structured in NetLogo as shown in Figure 48. Thus N, the complex network sub-model can be translated into a specification model S by handling each of these sections as shown in the figure. Furthermore, the specification model S can be translated to the ABM A and vice versa informally.

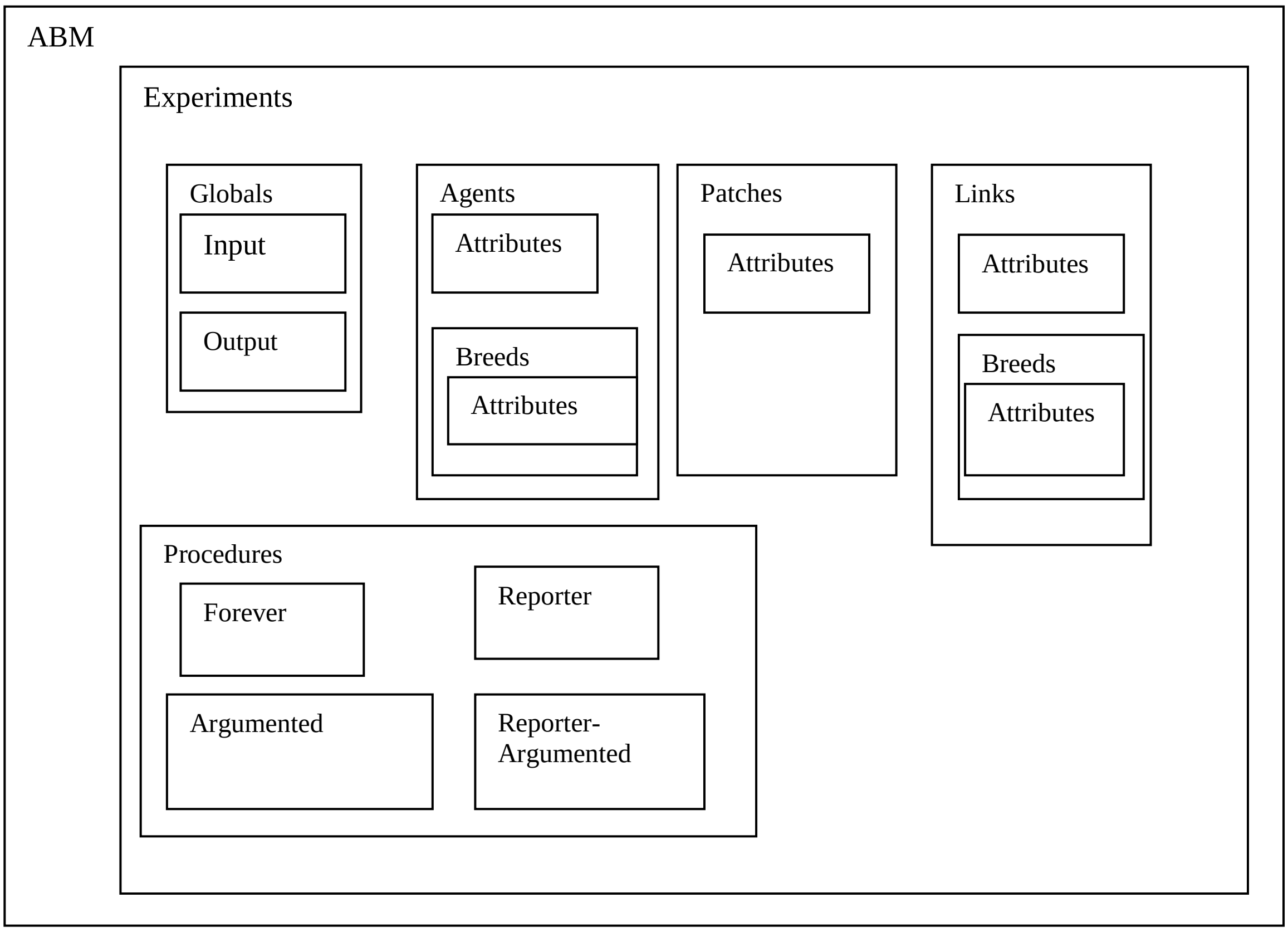

**Figure 48: An overview of ABM constructs in relation with the specification model constructs**

Here, we can note from the diagram that a complete ABM encompasses various constructs. The first of these is the “globals” construct. As discussed earlier in the description of the baseline network model, this construct corresponds directly to the one in the network structure. Thus, globals can be of two types i.e. the input globals which constitute the model configuration parameters allowing the user to vary the behavior of the model either manually or else programmatically and the output globals which typically represent metrics associated with the simulation results. So, while input globals modify the behavior, output globals represent the quantitative output metrics.

Other than the globals, the specification model can have particular attributes for the general breed of agent (turtle in the case of NetLogo). If however, there is need to use specific types of agents, breeds can be defined which can be designed with further attributes.

Similar to agents, we have links which can themselves have attributes as well as be classified using new breeds. If new link breeds are defined, occasionally, they themselves are assigned new attributes. Patches, on the other hand, are general purpose constructs and while they can also be assigned attributes, it is not possible to sub-class them into making breeds.

While different Agent-based Modeling tools have different ways of specifying procedures (such as using methods associated with a class in the case of using an Object-Oriented programming paradigm such as Java), here we give a generic definition without requiring procedures to be tied with agents (similar to use of procedures in Logo-based tools). If procedures are of general purpose or utility type, they can come directly under the umbrella of “procedures”. However, if the procedures represent specifically one simulation cycle (in other words, they need to be called repeatedly), then they would come under the classification of “forever” procedures. In addition, if they return one or more values, then they are classified as “reporter” procedures. If they take in arguments, they are classified under “argumented” procedures. Finally, procedures which both take in values as argu-ments and also return some values are classified as “reporter-argumented” procedures. Eventually, in a well-structured ABM, the entire set of constructs might be useful for one or more simulation experiments. Thus, it can be noted here that this division of constructs has been designed so as to specifically allow the specification model to have a direct one-to-one correspondence with the ABM constructs as noted earlier in the network model.

The proposed DREAM methodology has been designed to allow transition from the various sub-models such as the complex network sub-model to the pseudocode-based specification model and the agent-based model by allowing for a high degree of coherence between these sub-models. An overview of how translation of various models is possible can be seen in Figure 49. In this figure, we can note that the modeling research ideas related to a cas are translated firstly on the basis of an initial implicit cognitive mental model. The first step towards developing the DREAM model is to develop a complex network model. Subsequently for interdisciplinary comparison, the research can proceed by means of either performing a complex network analysis or else by an informal expansion of the network model into a pseudocode-based specification. In the case of a complex network analysis, a result digital quantitative footprint emerges which can prove to be useful for comparing different models. On the other hand, the network model can subsequently be informally expanded in the form of a pseudocode-based specification model with the idea of improving communication between researchers. Another possibility is that researchers can use an existing agent-based model and subsequently, reverse engineer it into a pseudocode-based specification model. Validation to ensure the correct translation of these models could be performed by means of informal methods such as testing, face validation and code walkthroughs etc.

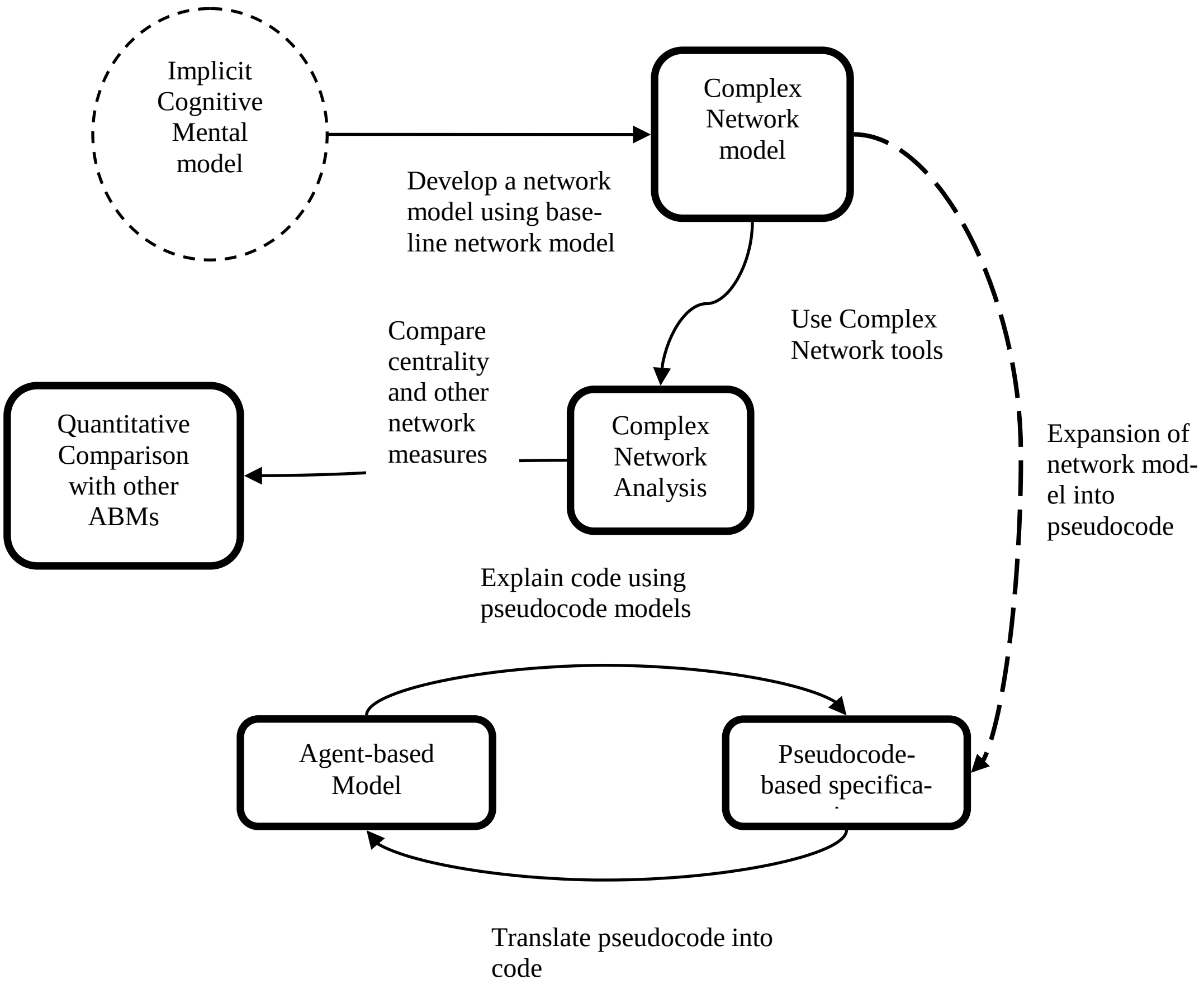

**Figure 49: Pictorial representation of the DREAM Methodology**

As noted by Zeigler [183] real-world complex systems need to be modeled using a multifaceted approach. As such, we can note here that the DREAM methodology can also be used in several different ways. One way of starting modeling is when a researcher is able to come up with a tentative implicit mental model of the proposed agent-based simulation model, this mental model can be converted to a network sub-model using a proposed baseline agent-based network model, which can serve as a template for developing agent-based models[6]. This baseline network sub-model shown in Figure 50 can be used to develop a complex network sub-model of new agent-based models by expanding each of the leaves with application case study related constructs. In the next section, we give details about the development of this complex network model.

[6] For the purpose of this discussion, we shall be focusing on building agent-based models using the NetLogo tool.

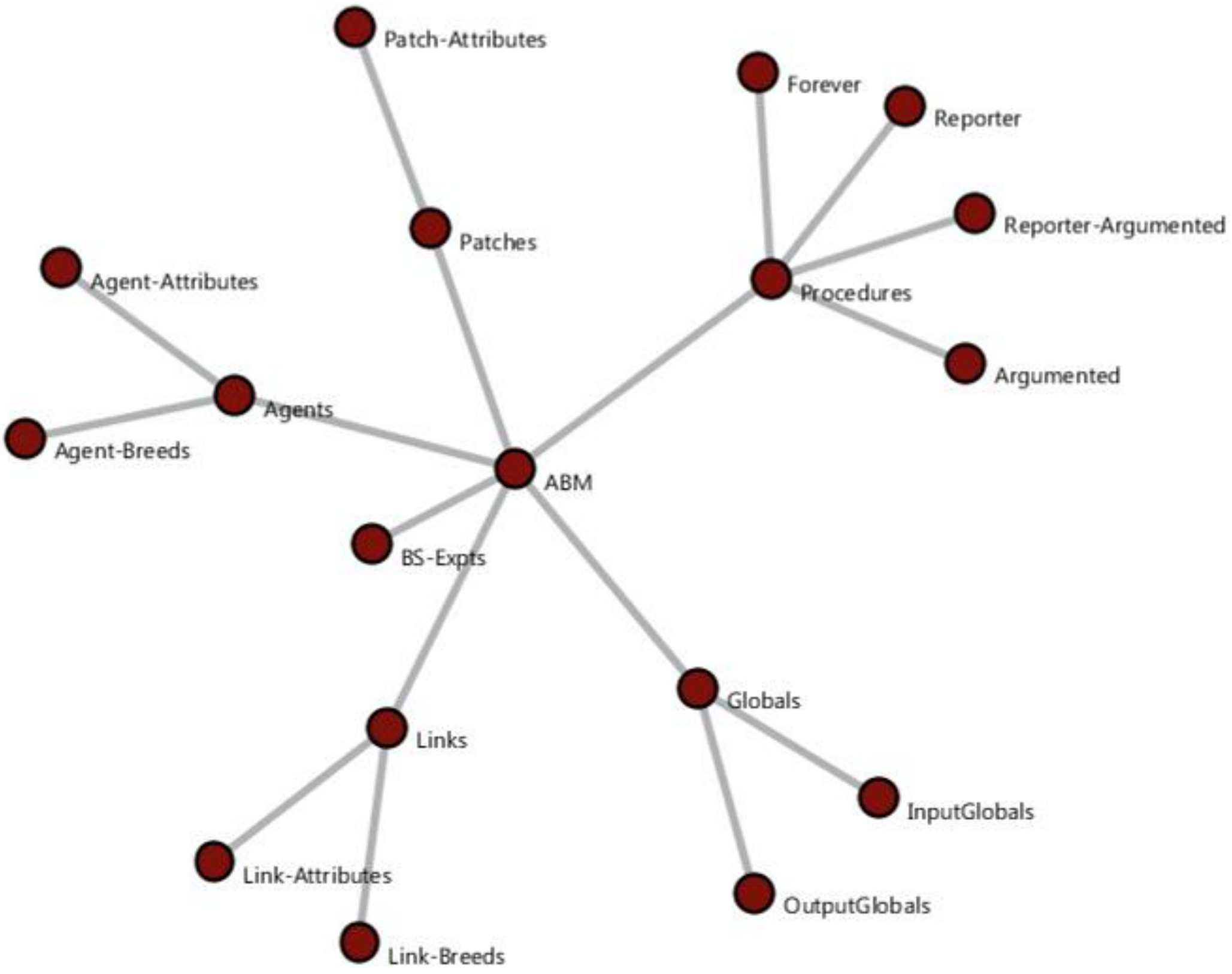

**Figure 50: Baseline Network template Model of a complete ABM**

### 5.1.2 Design of the Complex Network Model

Initially, to facilitate interactive study, researchers can start by developing a network sub-model on paper. This baseline network sub-model has been designed to serve as a template model. Leaf nodes from this model can first be expanded by connecting them to new nodes in the domain of the application case study. Therefore, at this stage, researchers can mind map[184] the network sub-model to describe the agent-based model at an abstract level.

The ABM is at the center of the network and it is connected to various nodes. If we first start with the global variables, we note that they are of two specific types. One of these is for giving inputs to the model, typically using User Interface (UI) elements (such as slider, switches etc.) while the other are the output global variables, which are expected to help in retrieving the simulation metrics from the model. These variables cannot be generalized and would vary on a case-to-case basis. So, cas researchers can add new input and output global variables by extending the network from these two types of global variables. Using the network approach, this step can be performed without worrying about actually going into the details of how the model is going to be practically implemented.

After deciding the global variables, the next step could be to start looking at adding various agent types or attributes, as needed. Now, there are two possibilities of using agents; either the generic agent breed (i.e. Turtle) can be used or else one or more new breeds can be created. If the generic breed is used then it might still be assigned suitable attributes. In that case, child nodes can be added to the "agent-attributes" node. If however, new agent-breeds are to be created, they would be attached to the "agent-breeds" node. Subsequently any attributes that the particular breed is assigned may be next attached to these agent breeds.

After deciding the breeds, a possible next step can be to decide upon using links and patches. Patches cannot have further breeds however new breeds for links can be defined. Not every agent-based model will have use for specific features related to patches or links so it would depend on the specific model requirements. Similar to the agent breeds, link breeds can also be defined, if needed. If no new link breed is defined, then the standard links can be given new attributes as suited to the case study.

After these object types have been defined for the model, the designer can next focus on the procedures. If the new procedures are general purpose or helper/utility procedures, nodes with their names can be attached to the "procedures" node directly. If not, then in the case of "forever" procedures, which represent one simulation execution cycle, procedure nodes can be connected to the "forever" node. Other options for parent nodes include the "reporter" procedure node for procedures which return values and "argumented" procedure node, for procedures which expect arguments for completing their processing. Finally procedures can also be defined such that they can both take in arguments as well as report values. Such procedures can be added as children to the "reporter-argumented" parent node. These names of different nodes are however given primarily as a template and not meant to be strict in the sense that if in a particular case study, some specific type of procedures are not required, then the related parent node can be removed from the network sub-model. In other cases, new types of nodes may be added if it allows for a better representation of the concepts related to the model.

After deciding upon the names and types of the procedures, the last part of the network that needs to be finalized is the different experiments that are to be executed once the simulation has been completed. Each of these is developed as a single child node attached to the "BS-Expts" parent node.

### 5.1.3 Analysis of the network sub-model

After the network has been designed on paper, it can be modeled on the computer. To start with, a suitably formatted text file (e.g. using a comma separated or tab-separated format) can be generated with columns depicting the nodes. An example table representing the baseline model is given in Table 18.

**Table 18 Table showing the format for creation of complex network using Cytoscape**

| *Node from* | *Node to* |
|---|---|
| ABM | BS-Expts |
| ABM | Globals |
| Globals | InputGlobals |
| Globals | OutputGlobals |
| Procedures | Reporter-Argumented |
| ABM | Agents |
| Agents | Agent-Breeds |
| Agents | Agent-Attributes |
| ABM | Procedures |
| Procedures | Forever |
| Procedures | Reporter |
| Procedures | Argumented |
| ABM | Patches |
| Patches | Patch-Attributes |
| ABM | Links |
| Links | Link-Breeds |
| Links | Link-Attributes |

There are several ways in which networks can be extracted from a text file and imported into Network tools such as the Network Workbench[162]. One possible method is to manually create a plain text file compatible with one of the standard file formats. Subsequently, the file can be imported in a complex network modeling tool and visualized using a suitable layout algorithm such as the GEM Layout in GUESS[167].

In addition to model conversion, the network sub-model can now be analyzed using CNA as discussed in detail earlier in Chapter 2 and demonstrated in Chapter 3. Such an analysis would thus allow for several benefits such as:

1. The different centrality measures would serve as a "digital footprint" of the Agent-based Model.
2. The ability to discover relationships between ABM-based complex network sub-models across domains or applications quantitatively.

After having developed and analyzed the network sub-model, specification templates can be used for developing the pseudocode-based descriptive specifications for the ABM by expanding relevant nodes from the ABM complex network sub-model. These templates are described in the next sections.

### 5.1.4 Goals of a pseudocode-based specification model

After the construction of the network sub-model, it can be next expanded into a pseudo-code-based specification model. For developing this specification, the design goals of this part of the DREAM methodology can be listed as follows:

1. Robin et al.[185] review facts about “learning curve theory” in the domain of programming. They note the general acceptance of the fact that “novice/amateur” programmers can take up to 10 years of programming practice and experience to turn into “expert” programmers. As such, the proposed specification syntax should have a short learning curve by being easy to use for multidisciplinary cas researchers to minimize transition difficulties without adding further to this learning curve.
2. It should be based on using a flexible set of terms in either natural language or very basic formal mathematical terms (such as use of sets or predicates).
3. It should conform to the constructs used in an agent-based model. In other words, the specification needs to have a strong coherence with the source code for the agent-based model.

In other words, the key design constraint in the template syntax design is to make it comfortable for learning and use by multidisciplinary researchers such as social scientists, life scientists or linguists to use comfortably. The pseudocode part of this syntax can thus be a comfortable mix and match of any syntax or form which is comprehensible and preferred by the specific cas researchers

To summarize, based on the specific user-centric design requirements of DREAM specifications, the guidelines for writing the pseudocode are relaxed and not based on any type of stringent rules as the entire goal of this modeling approach is to improve communication without causing hurdles or requiring an extensive learning curve for multidisciplinary researchers. Pseudocode used can be either “pidgin” code, a technical term for pseudocode using formal mathematical symbols or else plain English language pseudocode. The original idea of pidgin code is from pidgin ALGOL (short for ALGOrithmic Language)[186].

### 5.1.5 Templates for the specification model

Here in this section, generalized translation templates are presented. These templates allow the expansion of the complex network model to agent-based model and vice versa.

The first template is for defining a new breed. As noted below, it starts with a name and description of the breed. Type can be either an “agent” or a “link”. Finally, the breed can be given any number of “own” or “internal variables”. Writing this detailed description thus allows the researcher to expand and refine the ideas in their implicit mental models by writing down the specification models.

| |
|---|
| Breed **BreedName**: Single Sentence description of breed<br>Type: *Agent* \| *Link* |
| *Internal Variables*: <variable $_1$, variable $_2$, . . , variable $_n$.><br>**variable$_1$**: Conceptual role and description of the variable $_1$<br>**variable$_2$**: Conceptual role and description of the variable $_2$<br>. . .<br>**variable$_n$**: Conceptual role and description of the variable $_n$ |

After defining the breeds, a possible next step could be to define the different types of input and output global variables which will be used in the simulation. By starting with these variables, the designer can think about the research questions, in general or the problems to be explored, in particular. Input variables can be described in terms of the user interface elements whereas output variables can be defined without any relation to these elements.

$\{$**Input | Output**$\}$ **Globals:** <variable $_1$, variable $_2$, . . , variable $_n$.>

---

*Sliders*:
**variable$_1$**: Description of the variable
**variable$_2$**: Description of the variable
. . .
**variable$_n$**: Description of the variable $_n$

*Switch*:
**sw$_1$?**: Description of the switch
**sw$_2$?**: Description of the switch
. . .
*s***w$_n$?**: Description of the switch

*Input*:
**ip$_1$?**: Description of the input
**ip$_2$?**: Description of the input
. . .
**ip$_n$?**: Description of the input

*Output*:
**op$_1$?**: Description of the output metric
**op$_2$?**: Description of the output metric
. . .
**op$_n$?**: Description of the output metric

---

Next, we present the template for different types of procedures. Procedures can be of numerous types as discussed earlier. By using this particular template, the researchers can give details of the procedures they intend to use in the model. The template firstly describes the input (pre-conditions or the arguments taken in by the procedure). Output is the resultant or expected outcome of the procedure. Execution can be used to note how the procedure is called and during what phase of the simulation. Next, the context of the procedure is defined; contexts can be agent, patch or observer types. Observer context is useful for general purpose execution or if multiple sub-contexts are going to be used in the procedure. Finally, the procedure is defined in terms of pseudocode.

Procedure **$proc_1$**: Description of procedure

*Input*: Description of inputs and pre-conditions expected by the procedure
*Output*: Resulting outcome in the simulation
*Execution*: Called by which procedure and during which process
*Context*: Observer | Agent | Patch
**begin**

1. Appropriately indented Pseudocode step
2. Appropriately indented Pseudocode step
3. Appropriately indented Pseudocode step
4. . . . .
5. Appropriately indented Pseudocode step

**End**

Next, the specification template for experiments is presented.

Experiment **$exp_1$**: Description and goal of the experiment
*Inputs*: < $ip_1$ , $ip_2$ , . . .$ip_n$ >
*Setup procedures*: < $spr_1$ , $spr_2$ , . . .$spr_n$>
*Go procedures*: < $gpr_1$ , $gpr_2$ , . . .$gpr_n$>

*Inputs*:
**$ip_1$**: {constant value} | {[$value_1$ , $value_2$ , . . . $value_n$ } | {[$value_1$→ increment→Finalvalue]}
**$ip_2$**: {constant value} | {[$value_1$ , $value_2$ , . . . $value_n$ } | {[$value_1$→ increment→Finalvalue]}
. . .
**$ip_n$**: {constant value} | {[$value_1$ , $value_2$ , . . . $value_n$ } | {[$value_1$→ increment→Finalvalue]}
*Stop condition*: Pseudocode
*Final commands*: Pseudocode

In this template, we note that experiments are first described in the top of the schema. Next, their "setup" and "go" procedures can be defined. "Setup" procedures are those procedures which will only be called at the start of a simulation experiment. "Go" procedures, on the other hand, will be executed with each simulation step. Stop condition represents when a particular simulation run will stop execution. Final commands are the commands which might be executed once the simulation is terminating. Input values can be specified either as a single value, or else a group of values or in terms of a three value list (with start, increment and final value as its members).

## 5.2 Case study: Overview, Experimental design and Implementation

In the previous section, we have given details of the DREAM methodology by providing details of how a proposed case study in agent-based modeling may proceed. As a validation of the proposed methodology, in this section, we give an overview of the case study, experimental design and the implementation. This includes the development of the DREAM model consisting of a complex network model, pseudocode-based specification model, a quantitative centrality-based “footprint”, description of the implementation as well as the design of the simulation experiments.

### 5.2.1 Overview

The case study follows the complex adaptive communication networks concept introduced in the previous chapter. In this case study, the goal is to explore a well-known artificial cas concept of “flocking” and use it to evaluate how Wireless Sensor Network nodes behave in the vicinity of complex behavior.

It is essentially based on the “Boids” model of flocking behavior introduced earlier in chapter 2. Boids is an example of an artificial cas. In other words, it is an in-silico representation of a set of real-world agents (e.g. birds or fish or insects etc.). The initial idea was propagated based on an exploration of how flocks of animals and birds group together and form complex structures based on seemingly simple and local rules. Thus, using rules such as cognitive perception of nearby “boids”, every Boid agent is able to adjust the direction such that it neither collides with its neighbor boids nor follows any general direction. The way this model technically works is based on a set of calculations. In a real living cas, the calculations are deemed to be performed automatically by means of perception of motion, however in the artificial life scenario (simulated cas), the calculations have to be actually performed. This interesting simulation model[187], commonly available with agent-based simulation tools such as NetLogo exhibits intelligent collective behavior based on local rules followed by the components ("boids" in this particular case).

5.2.1.1 Problem statement

While the behavior of boids and flocking has been well-studied in literature[100], to the best of our knowledge, no well-known study has been conducted previously in the domain of examination of how physical wireless sensor nodes would react in the presence of a group of flocking "boids". As discussed earlier, Wireless sensor networks are made up of sensor nodes and most, if not all, simulators of WSNs are based on extensions of existing communication network simulators. As such, these simulators have difficulty in modeling any advanced concept such as "mobility modeling" or else "environment modeling". This is actually a problem because WSNs are not primarily designed for communication rather network communication is simply a means by which the WSN extracts sensing information from individual sensor nodes since this environmental sensing is the primary goal of developing a WSN application[188]. However, it has taken considerable research efforts in the domain of communication networks to come up with custom algorithms which suit the specialized nature and limited battery problems[189] related to the WSN nodes[190]. As such, a key focus area of these networks has previously been on getting the prototypes and the communication of simple messages across the network working (routing). More recent advanced algorithms include focus on data aggregation and fusion[191]. However, previously, to the best of our knowledge, the problem of developing a cas model for WSNs has not been studied. WSNs themselves constitute a cas because of how they need to self-organize for performing common goals such as monitoring and sensing[73]. In addition, if we couple a cas environment of flocking boids, it poses a set of interesting questions as to how to study the interactions of two cas with each other in the same simulation model. The goals of this simulation study can thus be noted as follows:

1. Evaluate the use of simple sensors in sensing complex cas behavior.
2. Evaluate the cas effects in these sensors and note the effects on sensing this complex behavior when different sensor nodes start to die (due to battery failure). In other words, the nodes themselves are eliminated from the simulation. So, assuming the results of the first set of experiments demonstrates if sensing can somehow be performed, what would be the effects on sensing as WSN nodes die down due to a battery failure at different time instances during the simulation.

There are numerous interesting complexity problems which pose a unique set of modeling challenges in this case study as it reflects a merger of an artificial cas along with the physical notion of a set of sensor nodes randomly deployed in the same terrain. Some of these problems can be noted as follows:

1. Complexity due to the special nature of the environment (constituted by the boids as well as the other sensors)
2. Complexity due to a random deployment model[192].
3. Complexity due to the gradual possibility of reduction in the total number of sensors available to sense the proximity of the boids.

### 5.2.2 Agent design

In this section, we describe the design of the agents used in the simulation model. Primarily there are two types of agents which will be used in the simulation which are described next.

5.2.2.1 Boid agents

Boid agents or Boids are the mobile agents in the model. These agents exhibit complex behaviors. The complex behaviors are collective however as in different real-world cas, similar to the discussion in the previous chapter, these agents are able to have particular attributes and behaviors described as follows:

1. Visibility and ability to view

   Boids are able to view other Boid agents and can focus on using a certain set of individuals as their visible neighbors. In other words, this particular group of individual boids inadvertently decides the direction chosen by the Boid agent in the near future.

2. Mobility

   Boids are able to move about the screen. This motion is in general dictated by the neighbors. In other words, each Boid agent decides upon a certain direction to follow based on its neighbors in a certain "visibility" radius.

3. Change of direction

   The direction followed by each Boid is what results in the flocking behavior. Thus this change of direction is based on certain calculations which we shall describe next.

Boid agents change their directions based on the following three functions:

### *5.2.2.1.1 Separate function*

This function is primarily for ensuring collision avoidance. The way it works is that each agent first locates its nearest Boid neighbor. Next, it perceives the agent's heading and slightly changes its own direction.

#### *5.2.2.1.2 Align function*

This function has a different goal. It allows the agent to change its direction towards a certain heading. This heading is calculated by first enlisting other neighbor Boid agents, which this agent considers as its flockmate neighbors. Next, their headings are calculated and averaged. Based on this heading, the agent slightly turns towards this heading. The net effect is to align in the general direction of the nearby agents.

#### *5.2.2.1.3 Cohere function*

This function is called alongside the align function. While the previous function discovered the average flockmate heading, this particular function is concerned with the angles of the other flockmate agents towards this agent. Thus, first the function requires the discovery of the angles of these agents to itself but adds 180 degrees to these angles. The reason for that is because that way, it will give the direction from this agent's perspective instead of the other agents' perspectives. Then these angles are averaged. Finally the agent adjusts its own heading towards this general direction slightly.

### 5.2.2.2 Node agents

In the same ABM as the boids model, the case study requires the placement of the sensor node agents. These agents are randomly deployed. Such kind of deployment follows the concept of "smart dust" devices[193] which can be deployed either in space[194] or on land. In the case study, these agents have communication capabilities. Thus, these node agents can communicate with a base station which is located within their transmission radius (single-hop). As such, they do not require multi-hop communication paradigms. The reason for choosing a single-hop network sub-model is to allow focus on the sensing aspect because if multi-hop networking was used, a considerable energy can be lost in developing the routing algorithms. However, in this scenario, the goal of these sensors is to detect

"boids". The physical implication of detection can possibly involve the use of "infra-red" or "image-processing" as a means of proximity detection of moving boids. In addition, binary sensors with acoustic detectors have previously been discussed by Kim et al. [195].

##### *5.2.2.2.1 States of Node agents*

Node agents have several different states as shown in Figure 51. Firstly, the agents are active (in other words functional). If they are functional, next they can go into sensing state if they detect a Boid in their proximity or else non-sensing state otherwise. However, if battery power goes down, these agents would then go to inactive state.

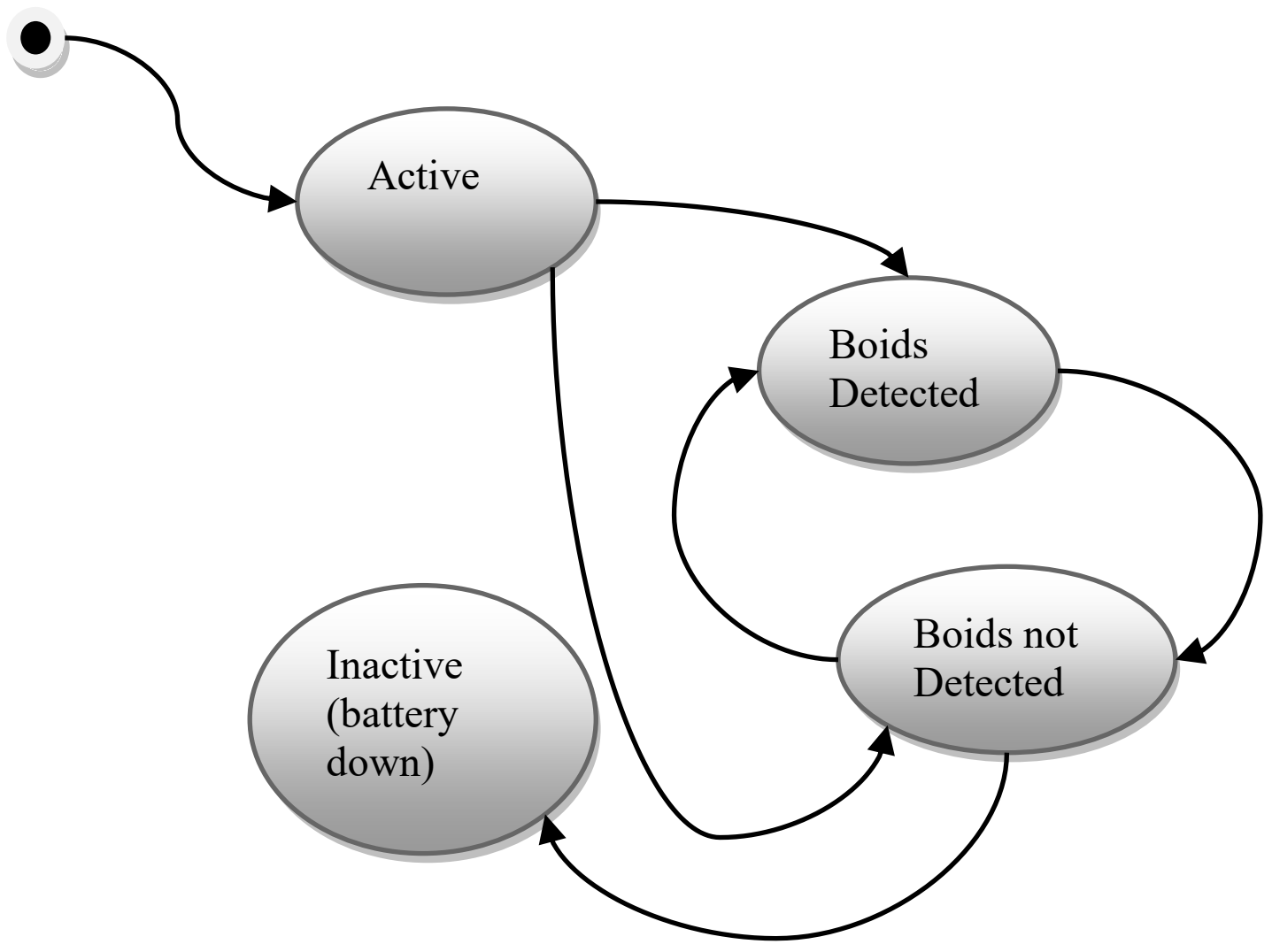

**Figure 51: State chart diagram for sensor node agents**

### 5.2.3 Network Model

As noted earlier in the chapter, DREAM can be initiated here by developing a paper model of network design by extending the baseline network model. The network sub-model can be imported into a network modeling tool. The result can be seen as shown in Figure 52.

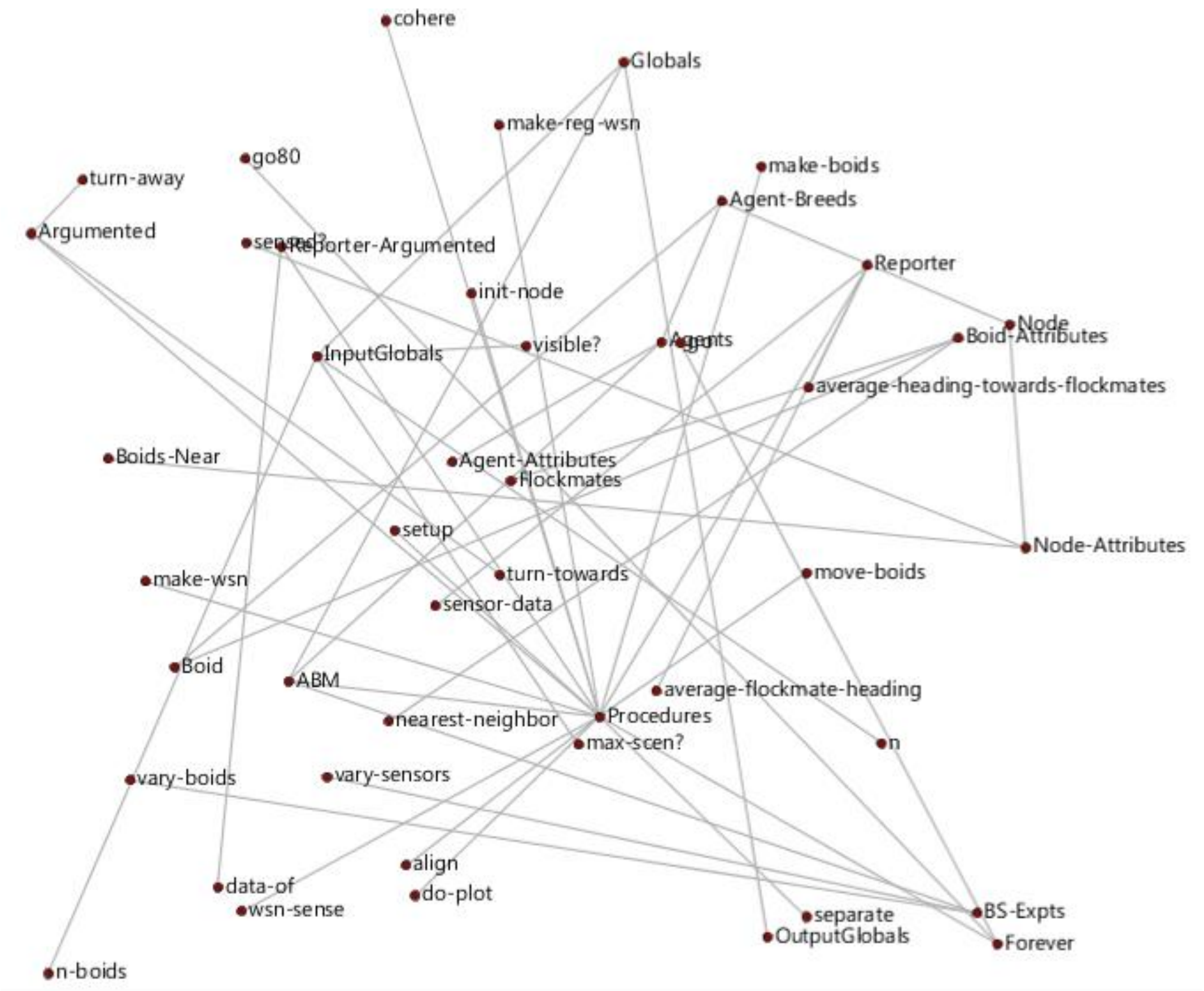

**Figure 52: Initial view of imported network**

In its current state, the network is difficult to comprehend. As such, there is need for network manipulation. After the network has been further manipulated using various layout algorithms, the resulting network can be viewed in Figure 53.

This network model will be described in detail and a CNA will be performed next. The results will be discussed in the results section of the chapter.

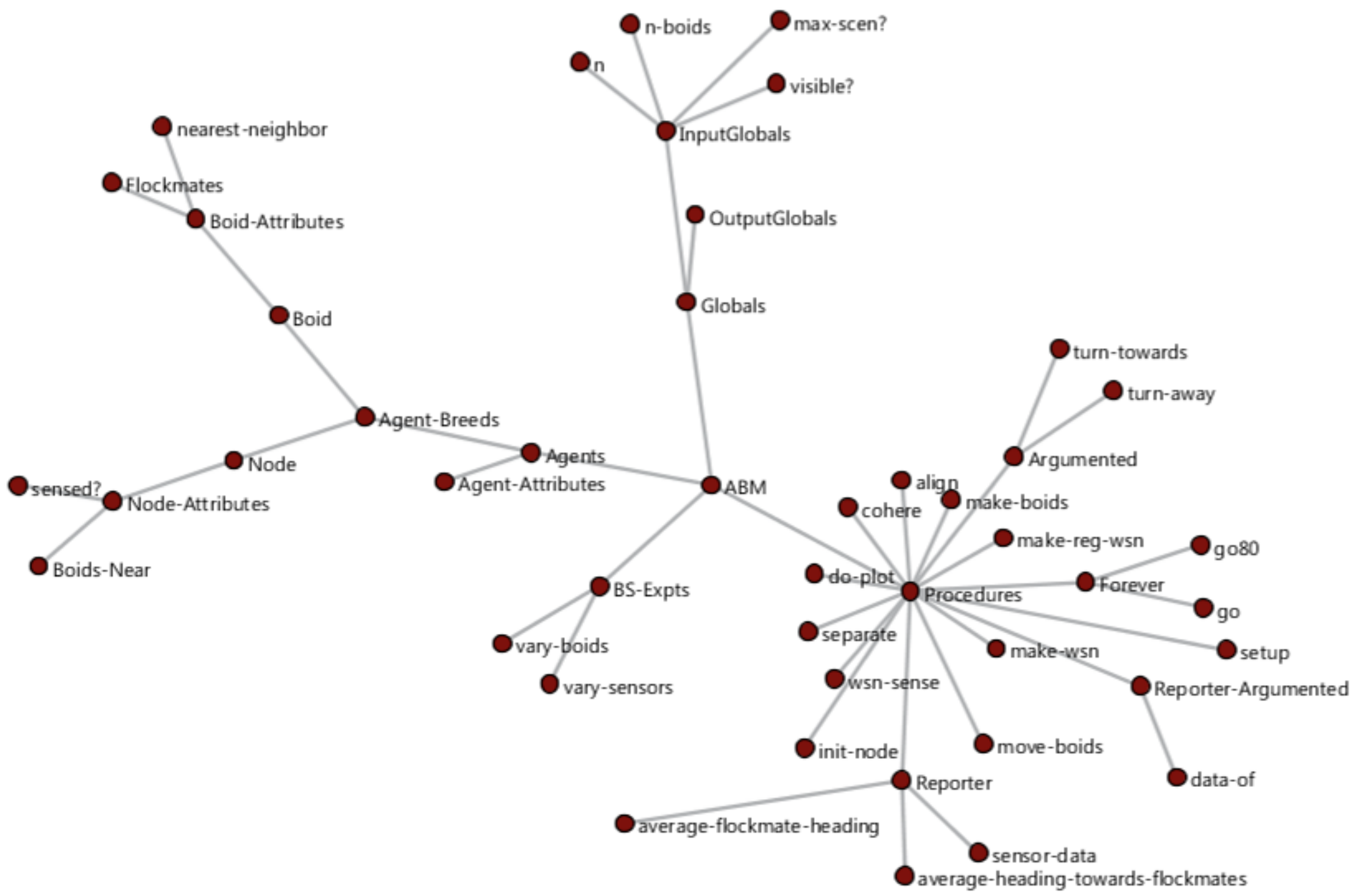

**Figure 53: Network view after manipulation algorithms**

### 5.2.4 Pseudocode-based specification

Next we develop a detailed translated specification model expanded from this network sub-model as a means of describing the model in detail. We first start by describing the agents and breeds.

#### 5.2.4.1 Agents and Breeds

The model has two breeds and their definitions are given next in the form of a specification model as below:

| Breed **Node**: This breed represents the Wireless Sensor Nodes |
|---|
| *Internal Variables*: <boids-near, sensed?><br>**boids-near**: All agents of breed boid which come closer to the given sensor node<br>**sensed?**: A Boolean variable which represents the state of the sensor in response to sensing |

Now, the “Node” breed here is first described in the specification model. Afterwards we note the internal variables that will be used in the simulation. There are two specific inter-nal variables here. One is the “boids-near” variable. This variable is used to keep an “agentset” of all nearby boids. The variable is key to the discovery of flocking which is one of the basic simulation requirements for the design of the sensor nodes.

On the other hand, “sensed?” variable is a Boolean variable. It represents the state of the sensor. In other words, when the sensor is active and is detecting boids in its vicinity, this Boolean variable will turn true. However, if the sensor is active and not detecting any boids in its sensing radius, the value will turn false.

The next breed is the “boid” breed, which is described below in the specification model as follows:

| Breed **Boid**: This breed represents the Boids (birds) |
|---|
| *Internal Variables*: <nearest-neighbor, flockmates><br>**nearest-neighbor**: A boid agent nearest to the agent itself<br>**flockmates:** An internal model agentset of other boids which this boid considers as flockmates |

The breed “Boid” represents the Boids agent type in general. Now, this particular agent breed/type has two internal variables. The first internal variable is the “nearest-neighbor” variable. Nearest neighbor is a variable which will be holding an agent. This single agent will be reflecting the “internal model” of the agent regarding which other environmental agent is physically near to this agent. This modeling concept reflects the cas concepts discussed earlier in Chapter 2. This internal model of the Boid agent keeps updating over time. We would like to note here that this concept of internal model matches with the concept of “Belief” in the well-known “Belief-Desire-Intention framework by Rao and Georgoff[196] used to model rational agents in multiagent systems.

Next internal variable of the Boid agent reflects the flockmates “agentset”. Flockmates are again related to the belief of the current agent. In other words, the agent has this “internal model” noting that this set of agents is its flockmates. Now, since every Boid agent might be moving in the simulation as well as continuously updating this variable, the fact that one agent has other agents listed as flockmates does not guarantee that the other agents will also keep this agent as flockmate.

After developing a specification model of the agent breeds, we next develop a model for the various global variables. These variables are important for the configuration of the simulation model.

#### 5.2.4.2 Globals

Here we describe the global variables for the simulation model. Now, in this model, the key input global variables here are four in number. These are described next in the specification model given below.

---

**Input Globals:** <n, n-boids, visible?, max-scen?>

*Sliders*:

**n**: Used for giving input for the number of sensors in the simulation
**n-boids:** Used for giving input for the number of boids

*Switch*:

**visible?**: Switch used to remove boids from the screen for only observing patterns in sensors
**max-scen?**: Switch used to create maximum sensors with one sensor per patch

---

In this specification, we can note clearly now that two of the variables viz. "n" and "n-boids" both represent inputs which are provided by the user via a "slider" UI element in the NetLogo simulation model. Whereas "visible?" and "max-scen?" are switches or in other words, represent Boolean input variables. "n" is used to configure the number of initial sensors in the simulation. These sensors will be randomly deployed when the initial simulation screen is setup. Having "n" as a slider variable allows for a gradual change in the input configuration as needed in different experiments. It also assists in minimizing the effects of random sensor node deployment in the model. So, while one particular simulation experiment might result in specific results due to certain placement of sensor nodes, repeating the simulation several times (e.g. 50 times in this particular case study) will ensure that the specific deployment does not skew the eventual results.

"n-boids" slider variable is used to enter the number of boids used in the simulation. The reason for having this variable configurable is to test the validity of hypotheses with a variation in the number of Boid agents. So, essentially this helps answering questions such as:

- If a hypothesis appears to be true with a certain number of Boid agents, would the same be true with lesser agents or else more agents?

The same can be noted about the slider described earlier i.e. depicting the initial number of sensor nodes in the simulation.

The switch "visible?" is especially useful in case we do not want to observe the simulation based on the boids. In other words, if we only want to examine the patterns being made in the randomly deployed sensors, then we can use this switch by turning it to false.

The other switch "max-scen?" is for the maximum scenario of sensors in the sense that in this case, sensor nodes will be located in the form of a lattice over the entire simulation world. Thus each "patch" of the simulation will behave as a single agent. This could be useful when instead of a randomly deployed sensor network, we want to use a set of sensors which are placed at exact locations. This would again allow us to examine the validity of the hypotheses that are to be tested in the simulations.

#### 5.2.4.3 Procedures

After describing the breeds and the global variables, we next start to describe the various procedures which will be part of the model. These procedures can be of different types and we shall use the templates described earlier in the methodology section as a means of describing and expanding each procedure in detail.

The first procedure is the setup procedure. The specification model for the setup procedure is given below as follows:

Procedure **setup**: Setting up the simulation

*Input*: Uses global parameters from the User Interface
*Output*: All agents and patches are setup for simulation
*Execution*: Called at the start
*Context*: Observer
**begin**

1. Clear the screen.
2. Clear all nodes
3. Clear all patches
4. Create WSN nodes
5. Create Boid nodes

**End**

The "setup" procedure is the key procedure to every agent-based simulation model. While the actual name of the procedure might be different, similar procedures are needed to setup the scenarios in a simulation. Here, it can be noted that in this particular "setup", it needs to be ensured that the remnants of the previous simulation experiment do not disturb the next simulation. As such, the previous results are also cleared before the execution of the next simulation. This would allow the simulation experiments to eventually run in an automated fashion. Next, each of the patches is sent a message. This message is to change their particular color to "white". The reason for doing that is to be able to better observe the simulation visually during the animations. After this, three different functions are called next. Each of these functions sets up part of the scenario. Dividing the setup into multiple functions allows for code localization. Thus, if needed, the particular functions/procedures can be modified without changing the global setup procedure. Next we describe each of these procedures. The first of these procedures is the "make-wsn" procedure described as follows:

Procedure **make-wsn**: Creating the nodes

*Input*: Uses global parameters from the User Interface
*Output*: All nodes are created after this function call
*Execution*: Called by setup procedure
*Context*: Observer
**begin**

1. **If** maximum sensors are needed
2. Create maximum node agents
3. **Else**
4. Create agents based on required numbers

**End**

"make-wsn" is a procedure which will setup the sensor nodes in the simulation experiments. It is called without any particular context (in other words, labeled as the observer context). Next, we describe the "init-node" procedure as follows:

Procedure **init-node**: To Initialize the nodes

*Input*: Uses internal variables of node
*Output*: The node is configured according to the norms of the simulation
*Execution*: Called by make-wsn
*Context*: Agent
**begin**

1. Initialize active sensor nodes displaying their state

**End**

Note here that this procedure has an "agent" context. In other words, this implies that the procedure can be used by changing specific attributes of the agent. Here the goal is to initialize a single Node agent for the simulation. This procedure is called by the "make-wsn" procedure. It essentially sets the color, shape, size and the initially active state of the sensor before the start of the simulation.

The next procedure is the "make-reg-wsn" procedure:

Procedure **make-reg-wsn**: For Creating nodes according to inputs

*Input*: Uses global variables from User Interface
*Output*: The entire wsn is configured according to the norms inputs
*Execution*: Called by make-wsn
*Context*: Observer
**begin**

1. Create nodes
2. Place nodes at random locations in the simulation world

**End**

This procedure is used to create the randomly deployed sensor network. It uses global variables from the User interface elements and is executed by the "make-wsn" procedure. It creates "n" nodes and then firstly initializes each node by calling "init-node" and secondly sets their location randomly in the simulation world. Next, it demonstrates the use of individual intelligence of the agents by allowing "node" agents to relocate based on the proximity of other nodes. This will allow the simulation to be more realistic since in the real physical world, the chance of two sensors at the same location is not very probable. It is important to note here that while this mobility is part of the procedure, it is not part of the capability of the sensor nodes in this case study. Simply mobility of nodes is used here to have a better and more realistic placement for simulation execution.

We next write the specification of the "make-boids" procedure given as follows:

Procedure **make-boids:** For creating boids according to inputs

*Input*: Uses global variables from User Interface
*Output*: Boids are created and initialized
*Execution*: Called by setup
*Context*: Observer
**begin**

1. Create boids
2. Initialize boids
3. Place boids at suitable random locations in the simulation environment

**End**

The "make-boids" procedure is a basic setup procedure. It is used to create and setup agent attributes such as the shape, size, color and location of the Boid agents.

We next write the specification of the "go" procedure. The "go" procedure is responsible for executing a single step in the simulation. As such, it needs to be called repeatedly. It takes in global parameters from the user interface. Firstly, it is expected to increment a global environment specific tick count. While simulation runs can execute without incrementing the tick count, it generally makes concrete steps in the simulation, which can be later measured (as we shall want to do in "go80" procedure). Next this procedure calls "move-boids" and "wsn-sense" procedures. These procedures are responsible for code execution pertaining to the Boid and the Node agent respectively. Finally plotting code is executed by calling "do-plot" which will be useful for observing the simulation evolution temporally as shall be explained subsequently.

---

Procedure **go**: For executing one step of the simulation

---

*Input*: Uses global parameters from the User Interface
*Output*: Equates to a single step of each agent in the simulation
*Execution*: Called repeatedly for execution of the simulation
*Context*: Observer
**begin**

1. Move boids in the environment
2. Allow sensors to sense their vicinity for boids
3. Update simulation plots showing active sensing boids

**End**

---

Next, we write the specification model of the "move-boids" procedure. As discussed earlier, this procedure is designed to be the primary procedure for Boid agents. The general functionality here is similar to standard published "Boids" models and extends NetLogo model library code. The basic idea in boids simulation is for boids to be close to their perceived flockmates but not too close. So, here the Boid agent looks at its nearest-neighbor and if it notes that it is too close, it calls "separate" otherwise it calls "align" and "cohere". After adjusting the heading using these functions, the agent moves forward by a small amount.

Procedure **move-boids**: For moving the boids for a single simulation step

*Input*: Uses global parameters from the User Interface
*Output*: Moves each simulated boid by a single simulation step
*Execution*: Called by the go function
*Context*: Observer
**begin**

1. Let boids maintain a list of flockmates
2. Let boids adjust their headings by means of separation, alignment and coherence

**End**

The first of the direction adjusting procedures that we write the specification for next is the "separate" procedure and given as follows:

Procedure **separate**: For moving the boid away from the direction nearest neighbor is moving

*Input*: Uses no specific input parameters
*Output*: Moves the boid agent away from the nearest-neighbor boid
*Execution*: Called by move-boids procedure
*Context*: Agent
**begin**

1. Update which of the nearby flockmates are nearby
2. Find the heading of the nearest neighbor
3. Slightly change direction so as to avoid future collision

**End**

The logic in this procedure is based on messaging. The Boid agent requests its "nearest-neighbor" for its heading. In a more advanced simulation, the neighbor can decide whether or not to give the heading back. However, in this case, since the goal here is to examine the effects of the sensing in sensor nodes, the "Boid" agent gets the heading from the "nearest-neighbor" and then passes this as an argument to the "turn-away" procedure.

Next, we write the specification of another boid-related procedure "align" as follows:

Procedure **align**: For moving the boid toward the general direction of flockmates

*Input*: Uses no specific input parameters
*Output*: Moves the boid agent towards the average flockmate heading
*Execution*: Called by move-boids procedure
*Context*: Agent
**begin**

1. Update a list of current flockmates
2. Based on the flockmates' directions, find a suitable average heading
3. Change the direction to this average heading

**End**

We move next to describe the procedure "cohere" using the templates as follows:

Procedure **cohere**: For moving the boid for cohering with the flockmates

*Input*: Uses no specific input parameters
*Output*: Moves the boid agent towards an average value of heading for cohering
*Execution*: Called by move-boids procedure
*Context*: Agent
**begin**

1. Find the average heading towards flockmates
2. Slightly adjust direction towards this average heading

**End**

The next procedure "wsn-sense" is related to the sensor Node agents as well as the Boid agents. Here in "wsn-sense" procedure, the Node agents are used to check their proximity for Boid agents. If they are discovered then the state of the Node agents is changed both visually as well as internally.

Procedure **wsn-sense**: Allows Node agents to change state based on sensing of boids

*Input*: No specific input values
*Output*: Changes the state of the sensor nodes based on the proximity of Boid agents to the Node agent
*Execution*: Forever procedure called from the user interface
*Context*: Observer
**begin**

1. Make a list of boids nearby
2. Update state of sensor node based on if there is any boid nearby

**End**

The final procedure here is the "do-plot" procedure which performs the plotting.

Procedure **do-plot**: For plotting the current status of Node agents

*Input*: None
*Output*: Updates the plot with the current set of Node agents who are sensing Boids nearby
*Execution*: Called from the go procedure
*Context*: Observer
**begin**

1. Plot the number of active sensing sensor nodes

**End**

### 5.2.5 Experiments

There are two experiments that we shall be using in this simulation. The first one demonstrates the effect of the variation of number of sensor Node agents while the second one shows the effects of the variation in the number of Boid agents.

Experiment **vary-sensors**: Experiment with the effects of changing the number of Node agents
*Inputs*: < n-boids, n, visible?, max-scen? >
*Setup procedures*: < setup>
*Go procedures*: <go>
*Repetition*: 50

*Inputs*:
**n-boids**: 50
**n**: [100→ 100→1000]
**visible?**: true
**max-scen?**:false
*Stop condition*: Ticks=1000
*Final commands*: None

Experiment **vary-boids**: Experiment for noting the effects of changing the number of Boid agents
*Inputs*: < n-boids, n, visible?, max-scen? >
*Setup procedures*: < setup>
*Go procedures*: <go>
*Repetition*: 50

*Inputs*:
**n-boids**: [100→ 100→1000]
**n**: 1000
**visible?**: true
**max-scen?**:false
*Stop condition*: Ticks=1000
*Final commands*: None

## 5.3 Results and Discussion

In this section, we present the detailed results and discussion based on the complex network model design as well as the simulation experiments proposed earlier in the case study specification model.

### 5.3.1 Complex Network Analysis of the Network sub-model

We start by performing manipulation of the network shown earlier in Figure 52. The first analysis is by calculation of the degree centrality measure discussed in detail earlier in Chapter 2. Once the degree measure has been calculated for each node, it is subsequently loaded/added to the corresponding node as an attribute. Subsequently this centrality measure is used to first resize and then colorize each node. The final visualized network is shown in Figure 54.

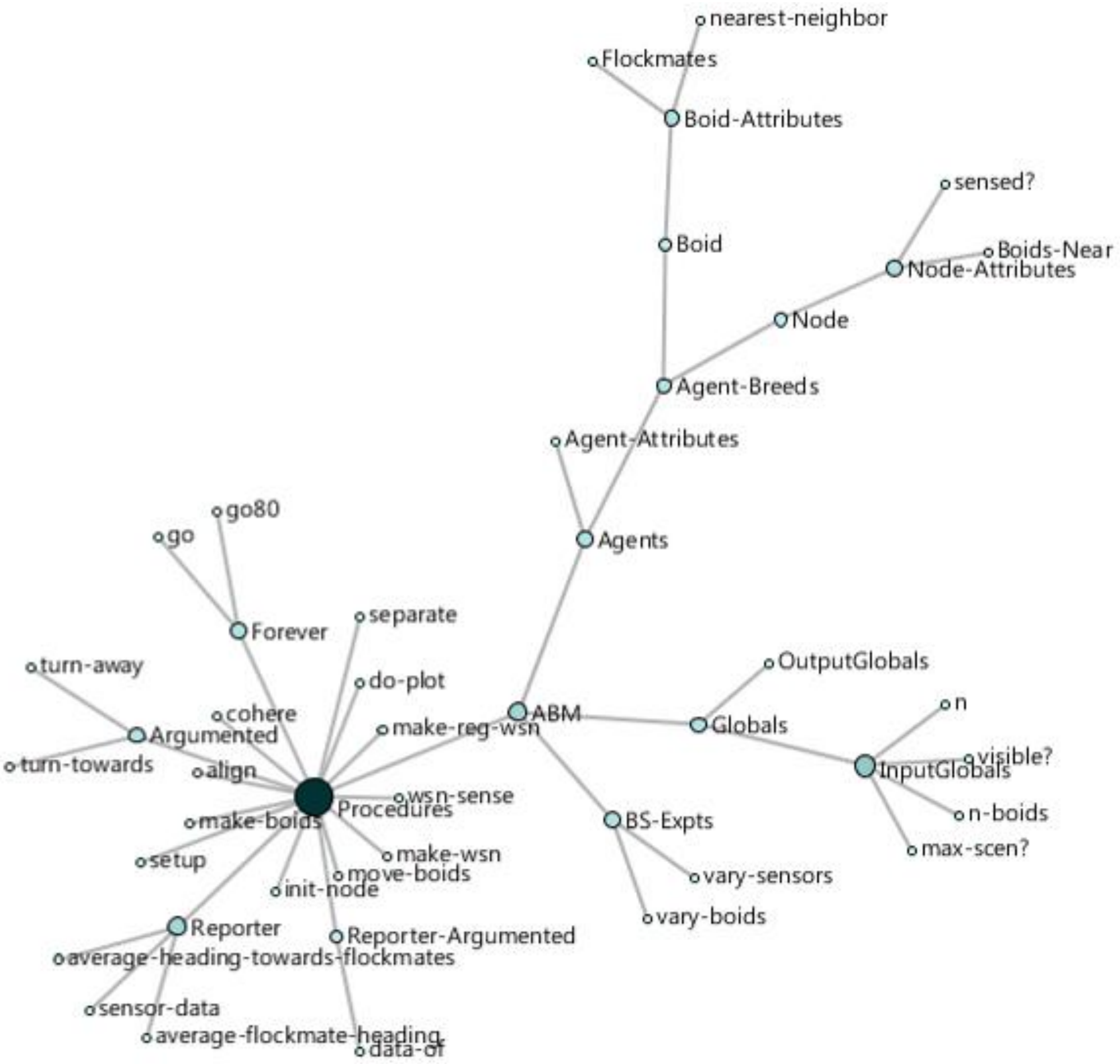

**Figure 54: Complex Network Model of Boids and WSN simulation nodes resized and colorized according to degree centrality**

Here, the first point to note is that we can clearly note peculiar characteristics of this ABM visually without needing to look at the source code specifically. If another complex network of a different ABM were constructed, it might have significantly different characteristics. Now, firstly, we note the size of the "procedures" node, which is larger and darker than other nodes because of the large number of procedures in this particular model. There are also several "forever" and "Reporter" procedures which have resulted in larger node sizes for these as well. We can also note how the network sub-model allows an examination of the agent-breeds and their respective attributes. Finally, the global variables as well as the two behavior space experiments are visible in the network sub-model.

While the degree centrality gives an interesting overview of the network, there are other quantitative measures which give further topological details as shall be examined next in terms of detailed CNA. Next, we calculate various quantitative network measures of the complex network presented in Table 19. One way of looking at this table is considering this as a digital network footprint of this particular ABM. Thus, another ABM could possibly have a completely different network footprint thus allowing for a quantitative similarity analysis and comparison similar to the use of minutia and other features extracted from human fingerprints for use in person identification and fingerprint indexing systems[197].

**Table 19 Table of Eccentricity, Betweenness and Degree Centrality measures of the complex network**

| *Id* | *Eccentricity (%)* | *Betweenness (%)* | *Degree (%)* |
|---|---|---|---|
| ABM | 15.36 | 0.00 | 4.44 |
| Agent-Attributes | 0.00 | 0.00 | 1.11 |
| Agent-Breeds | 3.20 | 8.85 | 3.33 |
| Agents | 3.36 | 6.19 | 3.33 |
| Align | 0.00 | 0.00 | 1.11 |
| Argumented | 3.84 | 5.31 | 3.33 |
| average-flockmate-heading | 0.00 | 0.00 | 1.11 |
| average-heading-towards-flockmates | 0.00 | 0.00 | 1.11 |
| Boid | 0.00 | 0.00 | 2.22 |
| Boid-Attributes | 5.76 | 0.00 | 3.33 |
| Boids-Near | 0.00 | 0.00 | 1.11 |
| BS-Expts | 3.84 | 1.77 | 3.33 |
| Cohere | 0.00 | 0.00 | 1.11 |
| data-of | 0.00 | 0.00 | 1.11 |
| do-plot | 0.00 | 0.00 | 1.11 |
| Flockmates | 0.00 | 0.00 | 1.11 |
| Forever | 3.84 | 5.31 | 3.33 |
| Globals | 5.76 | 5.31 | 3.33 |
| Go | 0.00 | 0.00 | 1.11 |
| go80 | 0.00 | 0.00 | 1.11 |
| init-node | 0.00 | 0.00 | 1.11 |
| InputGlobals | 7.68 | 7.08 | 5.56 |
| make-boids | 0.00 | 0.00 | 1.11 |
| make-reg-wsn | 0.00 | 0.00 | 1.11 |
| make-wsn | 0.00 | 0.00 | 1.11 |
| max-scen? | 0.00 | 0.00 | 1.11 |
| move-boids | 0.00 | 0.00 | 1.11 |
| N | 0.00 | 0.00 | 1.11 |
| n-boids | 0.00 | 0.00 | 1.11 |
| nearest-neighbor | 0.00 | 0.00 | 1.11 |
| Node | 2.88 | 7.96 | 2.22 |
| Node-Attributes | 3.84 | 7.08 | 3.33 |
| OutputGlobals | 0.00 | 0.00 | 1.11 |
| Procedures | 20.16 | 37.17 | 17.78 |
| Reporter | 5.76 | 7.96 | 4.44 |
| Reporter-Argumented | 14.72 | 0.00 | 2.22 |
| sensed? | 0.00 | 0.00 | 1.11 |
| sensor-data | 0.00 | 0.00 | 1.11 |
| Separate | 0.00 | 0.00 | 1.11 |
| Setup | 0.00 | 0.00 | 1.11 |
| turn-away | 0.00 | 0.00 | 1.11 |
| turn-towards | 0.00 | 0.00 | 1.11 |
| vary-boids | 0.00 | 0.00 | 1.11 |
| vary-sensors | 0.00 | 0.00 | 1.11 |
| visible? | 0.00 | 0.00 | 1.11 |
| wsn-sense | 0.00 | 0.00 | 1.11 |

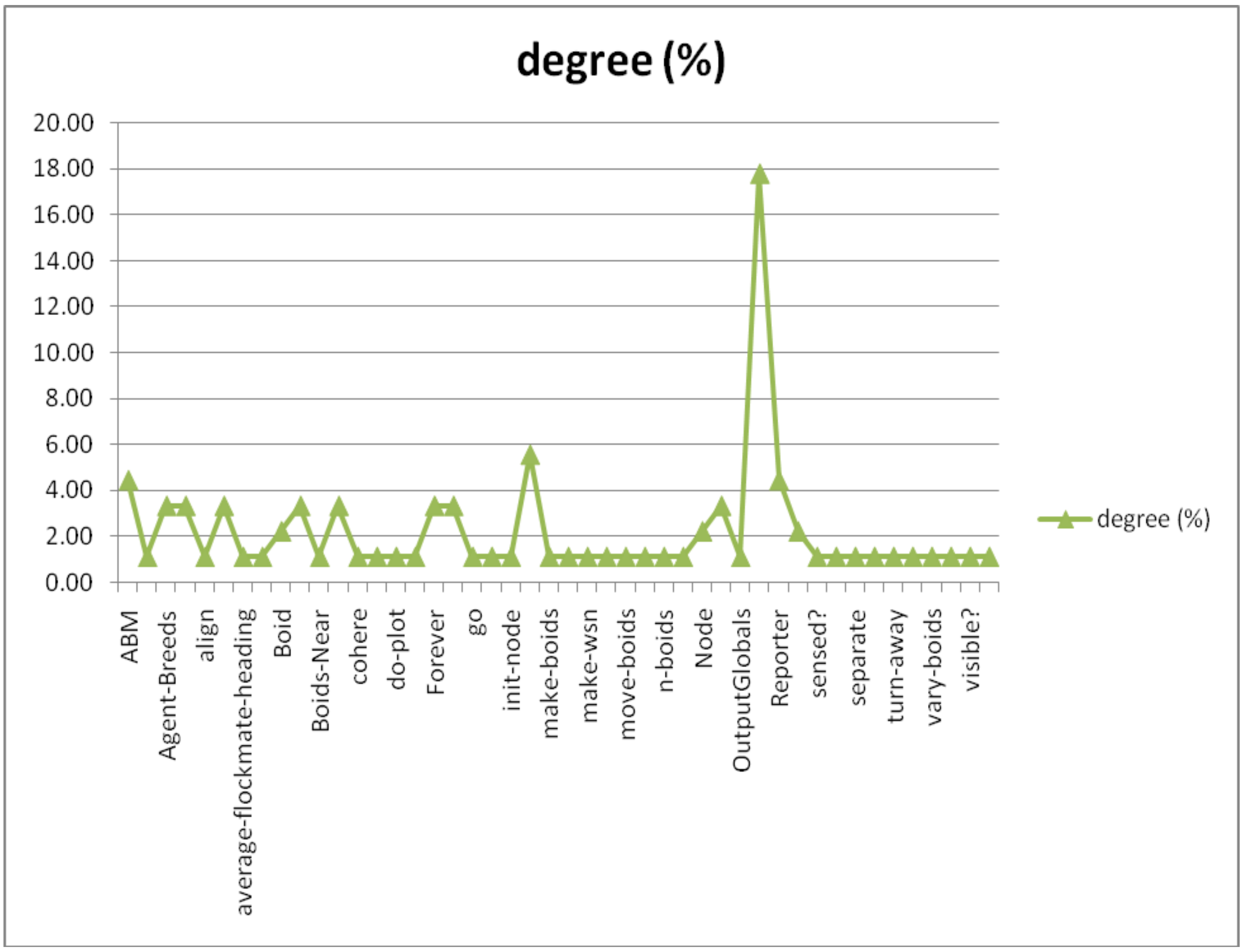

**Figure 55: Plot showing the Degree Centrality**

The degree centrality analysis is plotted in Figure 55. The first point to note from this analysis is that the degree centrality of the "procedures" node is the highest. The second highest degree centrality is the "input-globals". These represent the inputs which are entered from the NetLogo user interface model.

Next, the plot showing the eccentricity centrality is shown in Figure 56. Here, we note that the highest eccentricity centrality value is again for "Procedures" node. However, the second highest centrality node is that of "ABM" itself followed closely by "Reporter-Argumented" which depicts such procedures which have both reporters as well as arguments. This is followed by "InputGlobals".

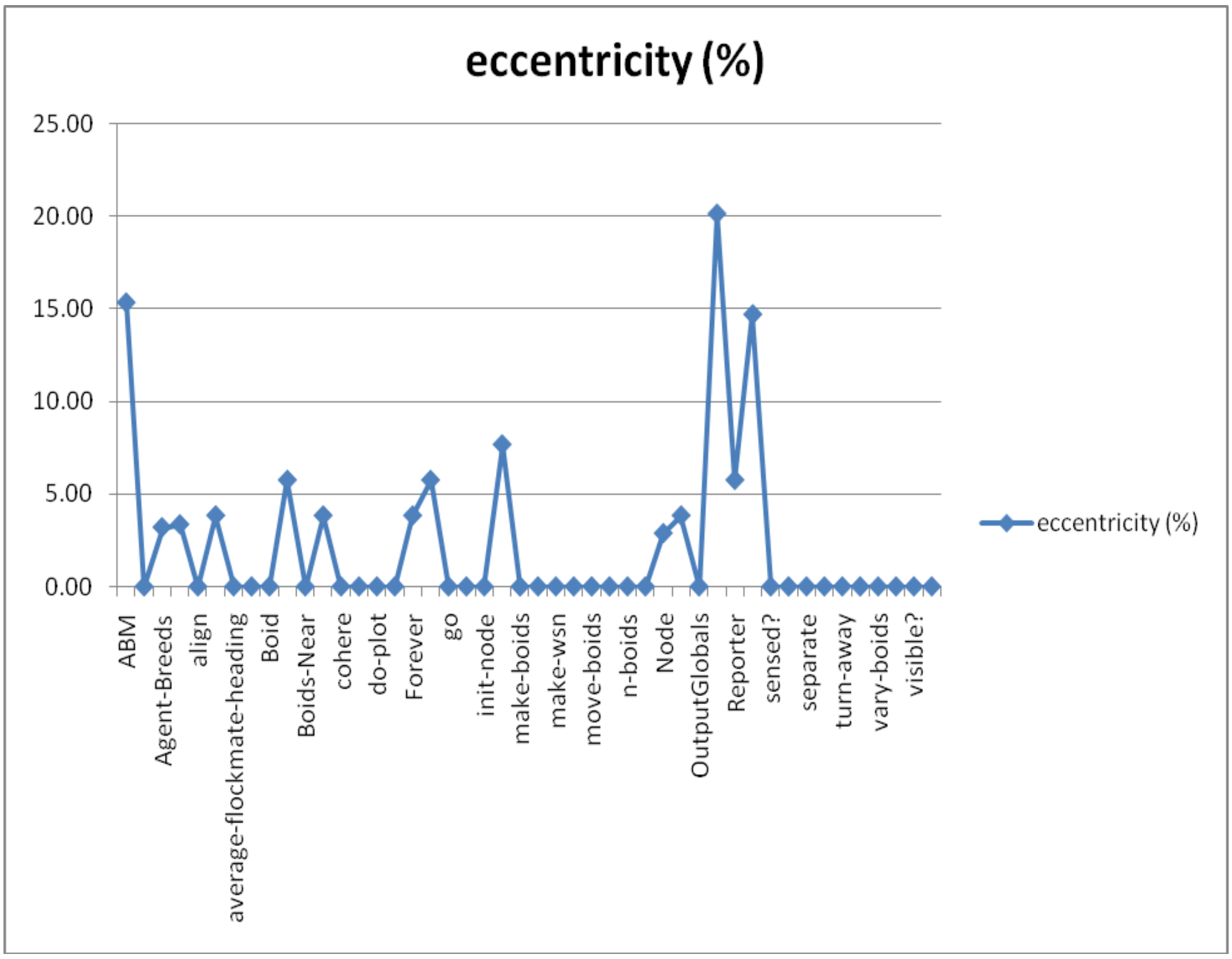

**Figure 56: Plot showing the eccentricity centrality**

Finally, the plot showing the betweenness centrality[198] is shown in Figure 57. In the case of betweenness centrality, following "Procedures" which has the highest centrality value, we find "Agent-Breeds" to have the second highest value followed by "Reporter", "Node", "InputGlobals" and "Node-Attributes".

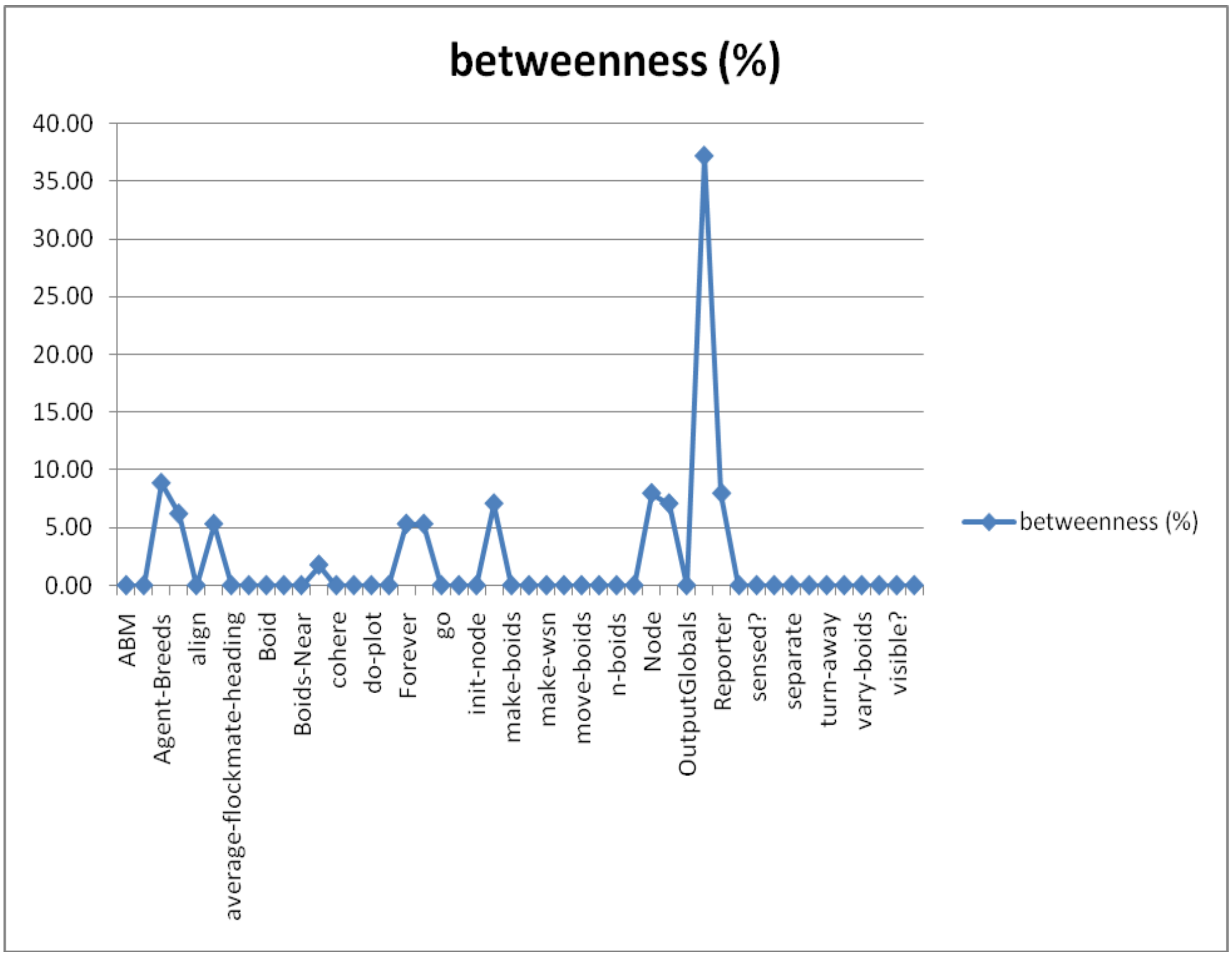

**Figure 57: Plot showing the betweenness centrality measure**

The implications of these network topological measures are numerous. Firstly, as discussed earlier, having a network allows for a visualization-based analysis and comparison of various agent-based models without resorting to the source code. Secondly, quantitative measures can be used to create a network footprint database of models for a repository of different ABMs corresponding to numerous multidisciplinary studies.

### 5.3.2 Simulation Measurements

While the previous section discussed the results of the CNA of the network sub-model of the case study, starting from this sub-section, we focus on the simulation experiments and their results. We note that the key measurement in this set of simulations is *"Sensed"* (represented below by $S(t)$). *Sensed* is the mathematical aggregation (sum) of measurements of all sensors at any given time $t$. So, being a binary sensor, every time a sensor turns on, it results in an increase in the *Sensed* value. Thus, *Sensed* would always be less than or equal to the total number of available sensors in the world at that instance ( $N_s(t)$ ) as follows in equation(2.6).

$$S(t) \leq N_s(t) \qquad (2.6)$$

If $n_d(t)$ represents the state of a single sensor which is detecting "boids" in its proximity, from a total of $i$ active sensors at a given time, $S(t)$ would be translated as the sum of values of all currently sensing and active proximity sensors formally defined as follows in equation (2.7):

$$S(t) = \sum_{d=0}^{d=i} n_d(t) \qquad (2.7)$$

#### 5.3.3 Results from Simulation Visualization

In this section, we give a detailed overview and discussion of the simulation results from the agent-based model constructed using the pseudocode-based specification model. We start by giving the details of how flocking boid agents are visualized in the simulation model. In the following Figure 58a, we can note the boids are in the simulation shown here as "bees". Grey sensor node agents are deployed randomly in the simulation. However, those sensor nodes which are close to the Boid agents detect them and change the color and size to black.

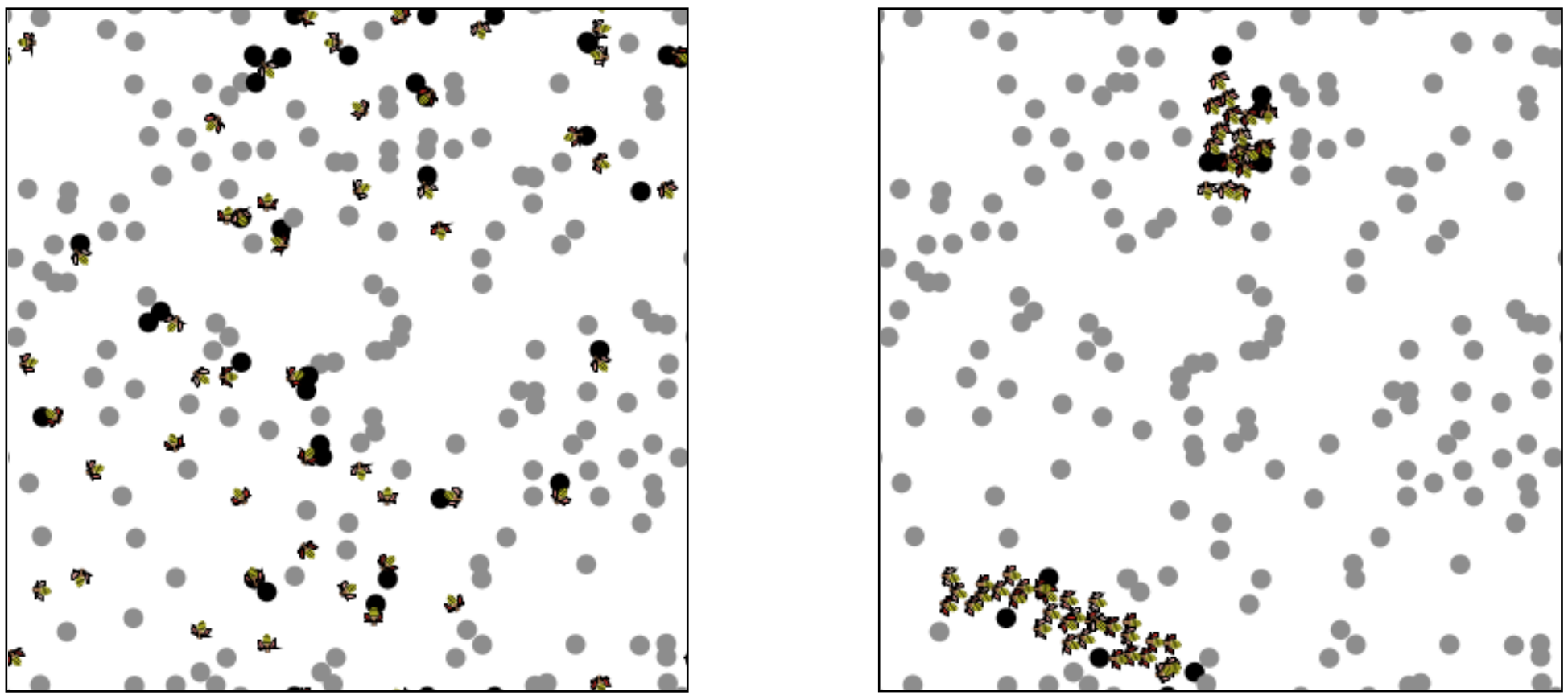

**Figure 58:(a) shows unflocked “boids” while (b) shows flocked “boids” and the sensor nodes sensing the boids, show up as black on the screen.**

With the passage of time, we note in Figure 58b that the boids flock together considerably and the number of sensors detecting the boids has greatly reduced. In other words, from the simulation model, we can directly correlate how the *Sensed* measured value should correspond and vary with flocking.

While the above figures demonstrated the concept of detection and sensing of the binary sensors, we need to validate the simulation parameters by checking to see if the simulation actually works. This is performed visually in Figure 59a and b. Here, the simulation execution has been performed by turning on the tracks for the mobile Boid agents and removing the visible sensors and all other agents from the simulation screen. We can note here that at the start of the simulation, the Boid agent tracks are more or less random and not coherent. However, in the later case, the Boids have flocked together and while the overall movement of the flocks might be considered as random, the agents have flocked in particular ways towards directions followed by the nearest Boids giving the example of an intelligent emergent collective behavior of flocking.

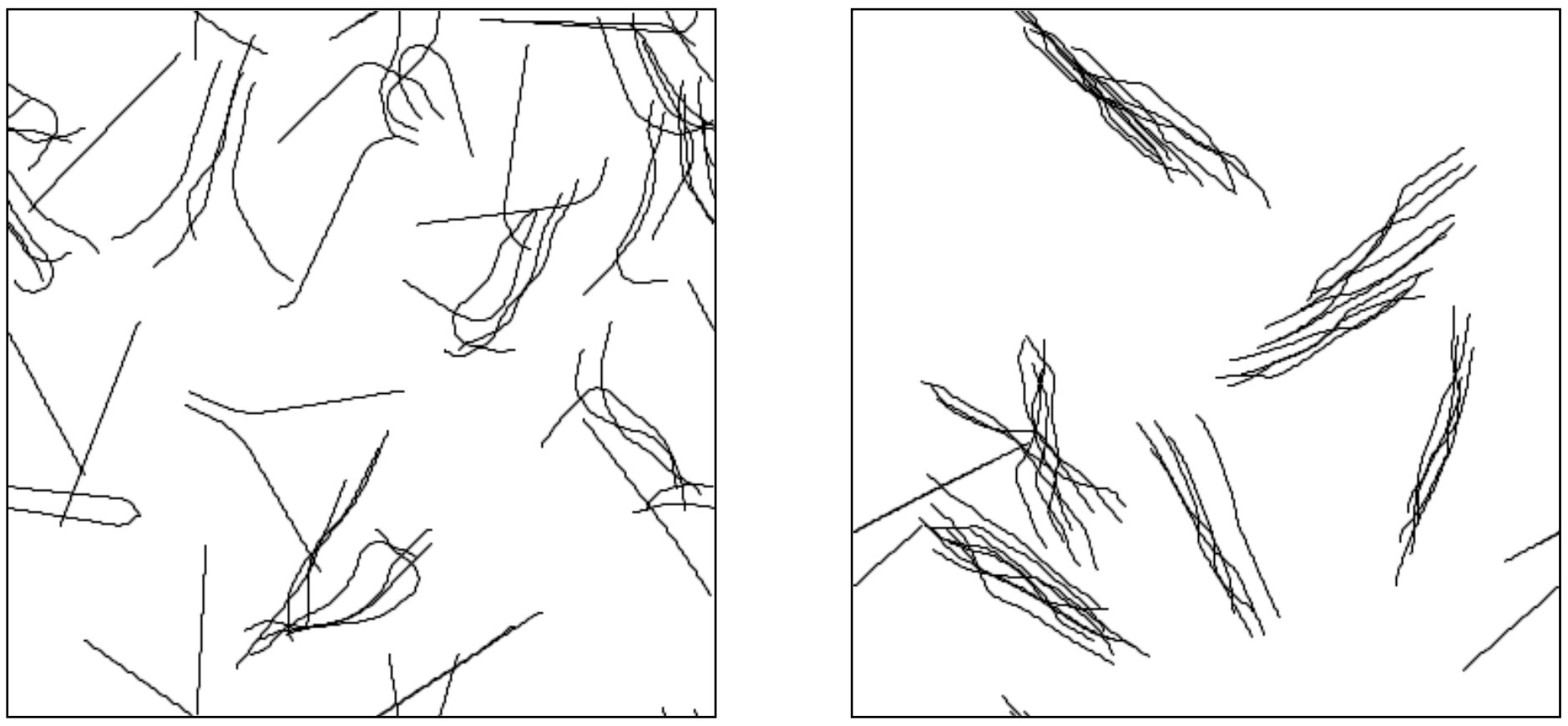

**Figure 59: Sensed Motion tracks of "boids" showing emergence of flocking behavior (a) Unflocked (b) Flocked**

### 5.3.4 Quantitative results

As discussed earlier, there were two key experiments in this case study. The first was to test hypotheses related to the sensing of Boid agents with respect to a variation in the total number of Boids whereas the second experiment was related to the examination of measurements in relation to variation to the total number of sensors. In this section, we give detailed analysis of the simulation results.

To start with, the descriptive statistics of "vary-boids" experiment are given in Table 20. We can note here that the total number of simulation steps were close to 0.5 million. The total simulation runs "Runnum" were 500 where each run was repeated 50 times to ensure that the results would minimize stochastic effects of the simulation experiments. Each experiment was executed for 1000 steps. The *Sensed* measurement varied from a value of 118 to 999 with a mean value of around 670 and a standard deviation of around 244. Here we can also note that the number of sensor Node agents was set to a constant value of 1000 while the number of Boid agents was varied from 100 to 1000 with increment of 100.

**Table 20 Descriptive Statistics of vary-boids experiment**

**Descriptive Statistics**

| | N | Minimum | Maximum | Mean | Std. Deviation |
|---|---|---|---|---|---|
| Runnum | 500500 | 1 | 500 | 250.50 | 144.337 |
| n-boids | 500500 | 100 | 1000 | 550.00 | 287.228 |
| N | 500500 | 1000 | 1000 | 1000.00 | .000 |
| Step | 500500 | 0 | 1000 | 500.00 | 288.964 |
| Sensed | 500500 | 118 | 999 | 670.48 | 244.277 |
| Valid N (listwise) | 500500 | | | | |

We can note the descriptive statistics of experiment "vary-sensors" given in Table 21. In contrast to the first experiment set, we can observe here that firstly the variation in the Sensed value is considerably different from the previous set of experiments varying from 1 to 267. The total number of experiments is still 1000 but the number of Boid agents is set to a constant value of 50. In contrast, the number of sensor Node agents varies from 100 to 1000. Each simulation was executed for a duration of 1000 simulation steps (or simulation seconds).

**Table 21 Descriptive statistics of experiment vary-sensors**

**Descriptive Statistics**

| | N | Minimum | Maximum | Mean | Std. Deviation |
|---|---|---|---|---|---|
| Runnum | 500500 | 1 | 500 | 250.50 | 144.337 |
| n-boids | 500500 | 50 | 50 | 50.00 | .000 |
| N | 500500 | 100 | 1000 | 550.00 | 287.228 |
| Step | 500500 | 0 | 1000 | 500.00 | 288.964 |
| Sensed | 500500 | 1 | 267 | 63.46 | 37.980 |
| Valid N (listwise) | 500500 | | | | |

### 5.3.5 Graphs

The previous sub-section gave details of the simulation experiments. In this section, we use graphs as a means of evaluation of results obtained from the above two set of simulation experiments.

To start with we first plot a 95% Confidence Interval plot of the Sensed value with a variation in the number of Boid agents as shown in Figure 60. Looking at this graph, we can see the bigger picture since we can note here that there is actually a variation in the sensed value with the number of Boids. As the number of Boids increase, the Sensed value also increases verifying that the model for sensing is working correctly. However, it should also be noted that the variation in the Sensed value is not linear. This nonlinearity is essentially a validation of the hypothesis discussed earlier in this chapter regarding the case study noting there is a high level of intrinsic complexity in the mixing of two different cas systems to form a heterogeneous cas model. The graph used is an error graph. The fact that each point is actually having a very small set of high and low values shows the fact that this is a consistent phenomenon from this particular case study.

In this graph, we noted how the *Sensed* value varied with a variation in the number of Boid agents. Next, we want to evaluate this phenomenon further. So, we plot the mean value of Sensed vs. simulation time.

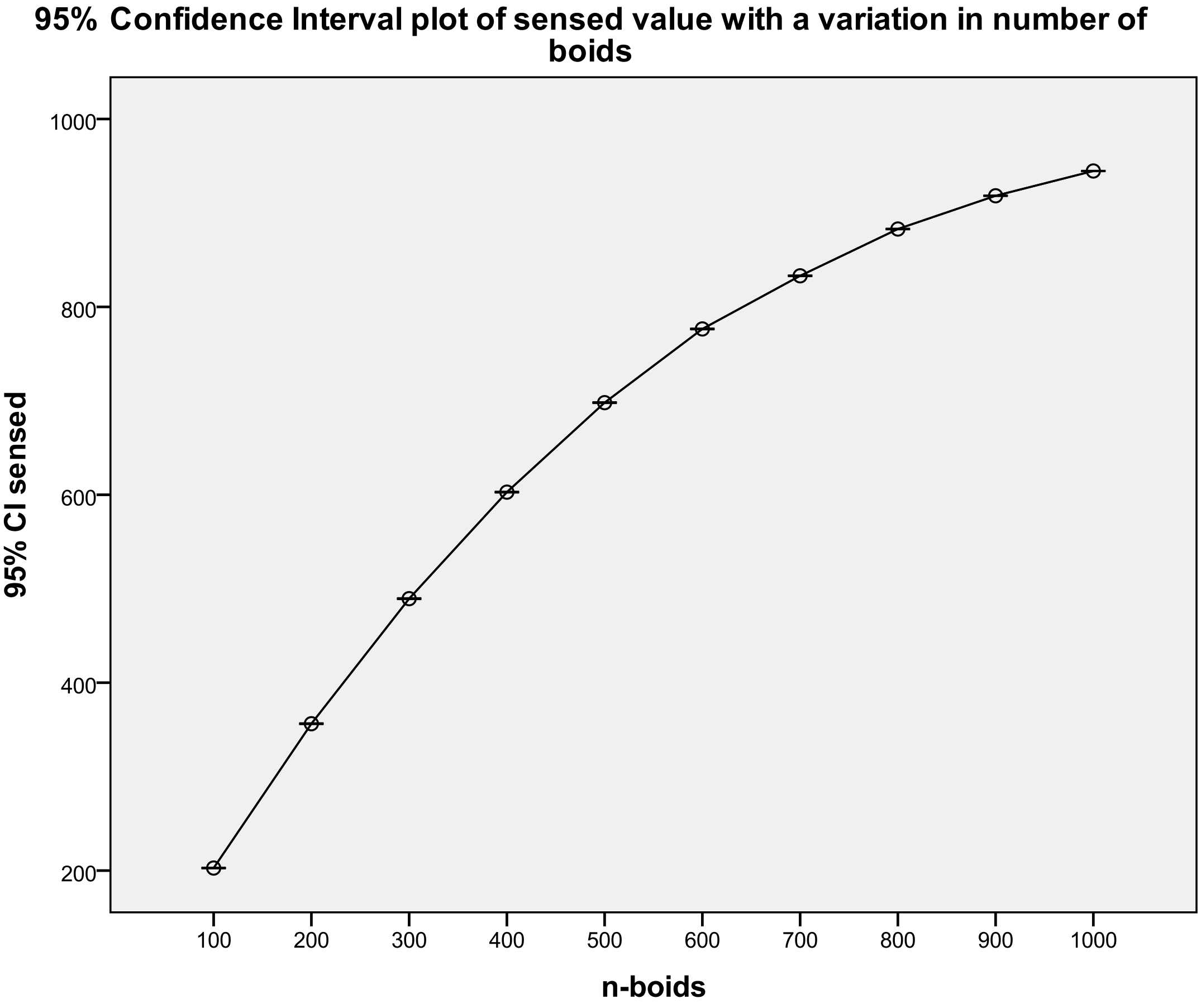

**Figure 60: 95% Confidence Interval plot of sensed vs. number of boids**

However, to ensure that we are not looking at a particular set of instances of the simulation and that the hypothesis is valid for a large variation of the simulation experiments, we plot mean values for different number of Boid agents as shown in Figure 61.

Here, the first thing we can note from this chart is that in general, with the passage of time, as the Boids tend to flock over time, the sensed value keeps on decreasing. In other words, the hypothesis that we developed earlier while examination of the visualization of the simulation discussed earlier about sensing is thus proven valid here. The second observation we can note here is that the variation in the *Sensed* value over time is more considerable when the number of Boid agents is smaller. With an increase in the number of Boid agents, due to a constraint of the simulation space limitation, the variation in the Sensed value appears to be small but noticeable. So, we can infer another hypothesis that even with a large number of Boids, flocking can still be detected with a relatively high degree of sensitivity.

Next, we note the interesting perturbations in the simulation experiments with the number of Boid agents visible more clearly at 500, 600 and 700 at time steps from 350 to 700. These interesting nonlinear effects are again a demonstration of the internal nonlinearities of this heterogeneous cas system. Thus, while emergent behavior of flocking is observable and quantifiable at the global scale from this graph, there are significant small nonlinearities which might not be observable if only that section of the graph was to be used as a test case.

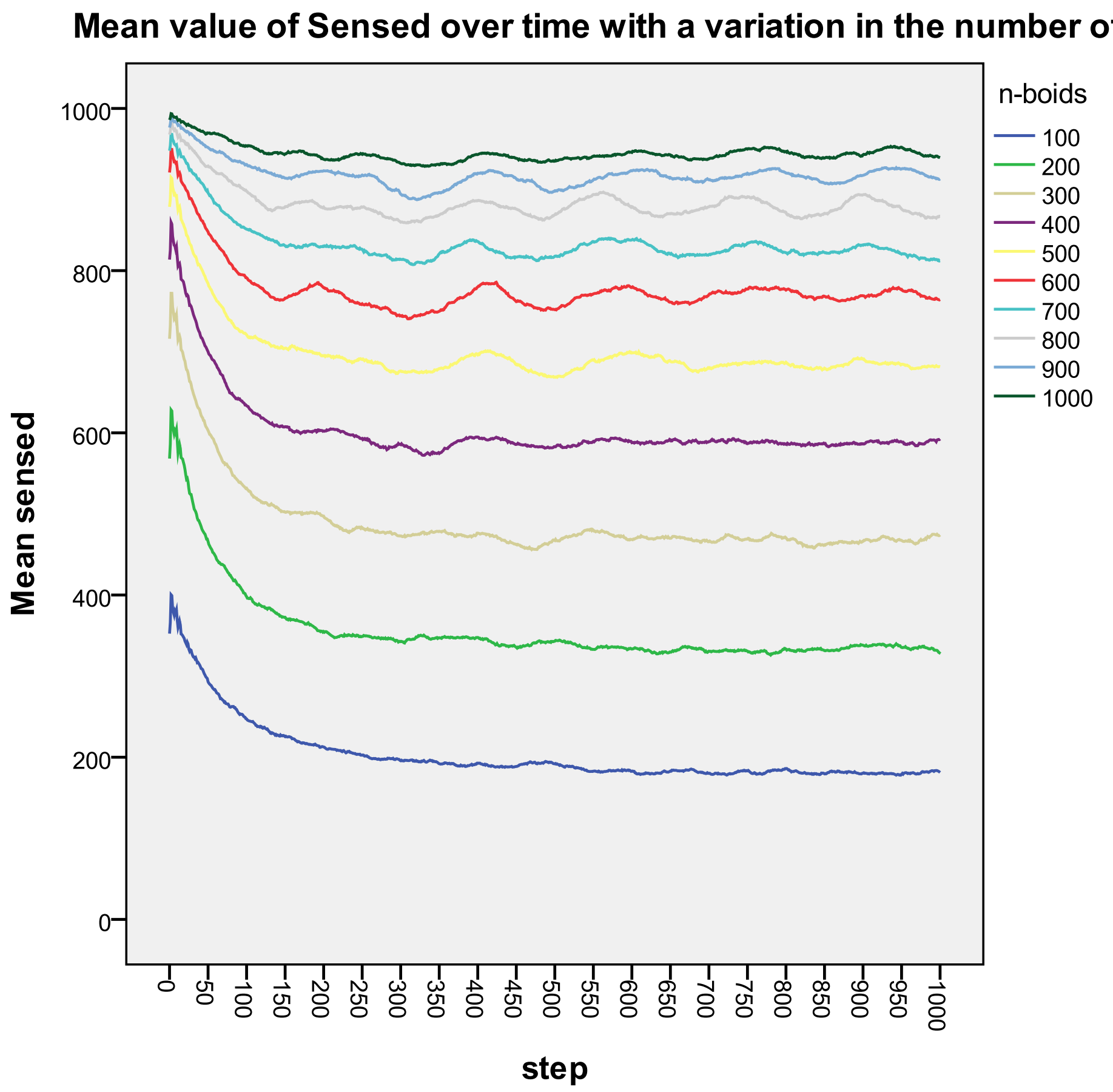

**Figure 61: Mean value of sensed over time with different number of boids**

This also demonstrates the effectiveness of the use of agent-based modeling in general and the particular DREAM methodology in particular, showing how it allowed for a detec-tion of complex global emergent behavior based on nonlinear interactions of a large number of components.

In the previous two graphs, we evaluated the effects of a variation of the number of Boid agents in the simulation. Next, we focus on experiments where the number of Boids is constant but the number of sensor Node agents varies. While the previous set of experiments gave consistent results testifying to the testing and validation of several hypotheses relating to sensing, we next want to verify and infer new hypotheses regarding the number of sensors. In other words, we want to validate that the previous results were valid in the case of a large variation in the number of sensor Node agents and were not a manifestation of the particular setting of the number of sensor Node agents.

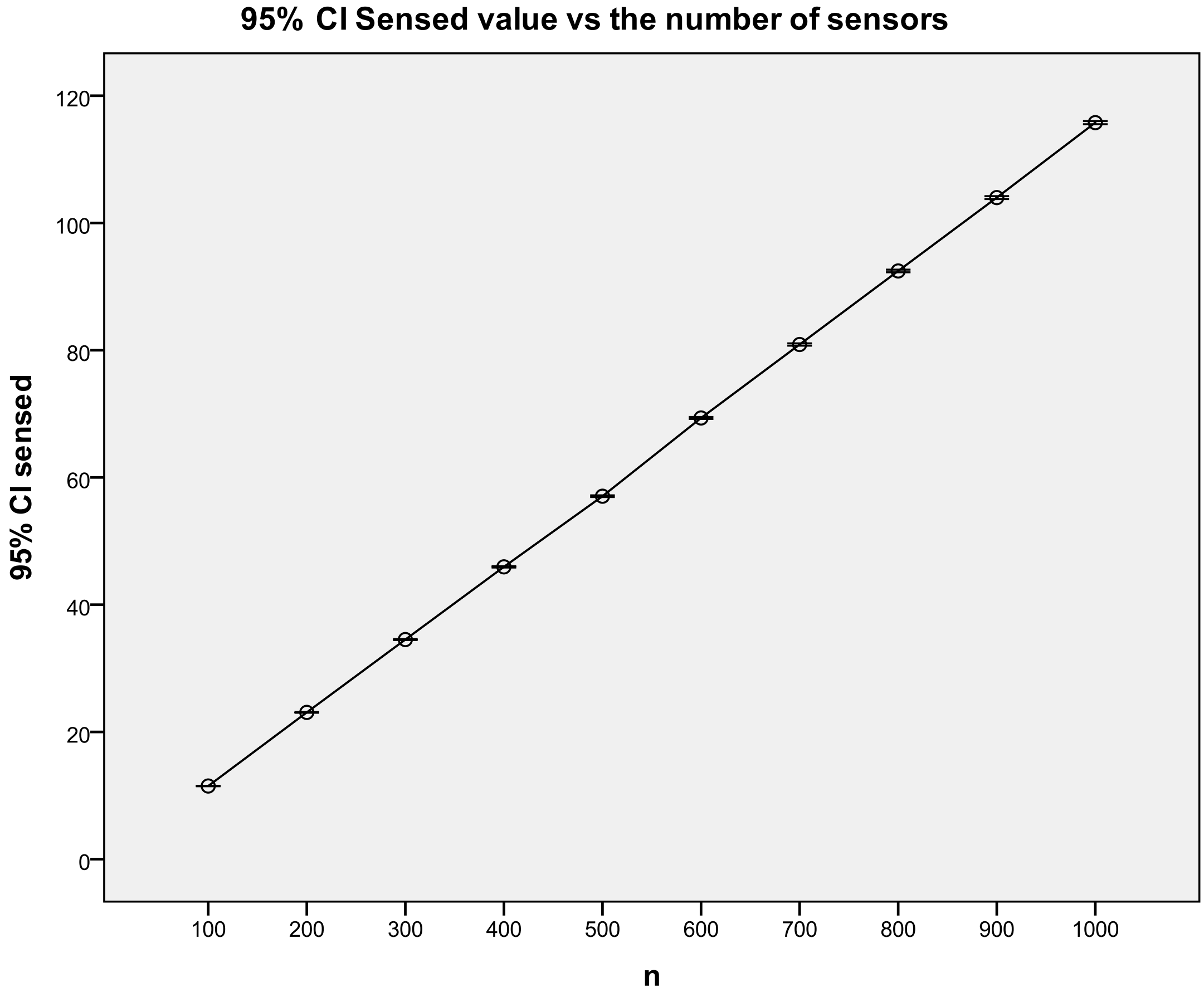

**Figure 62: 95% Confidence Interval graph of Sensed with a variation in the number of sensors**

The first graph that we evaluate here is a 95% confidence interval error graph shown in Figure 62. This graph shows a straight line. In this case, the straight line is a very good indication because it shows that the results of the *Sensed* value from the previous simulations are all valid. The change in the number of sensors does not affect the results and thus the results were all consistent and valid.

The next graph is useful as an extension of the previous graph. In this case, we plot the mean values of *Sensed* against simulation time. We also plot multiple number of sensors in the same graph so that we can see the net effects of sensor Node agents in the simulation. The important hypothesis here to test is whether or not the hypothesis regarding sensing nodes holds as sensor nodes are losing their battery power. Thus we note here that the variation of *Sensed* value still holds over time.

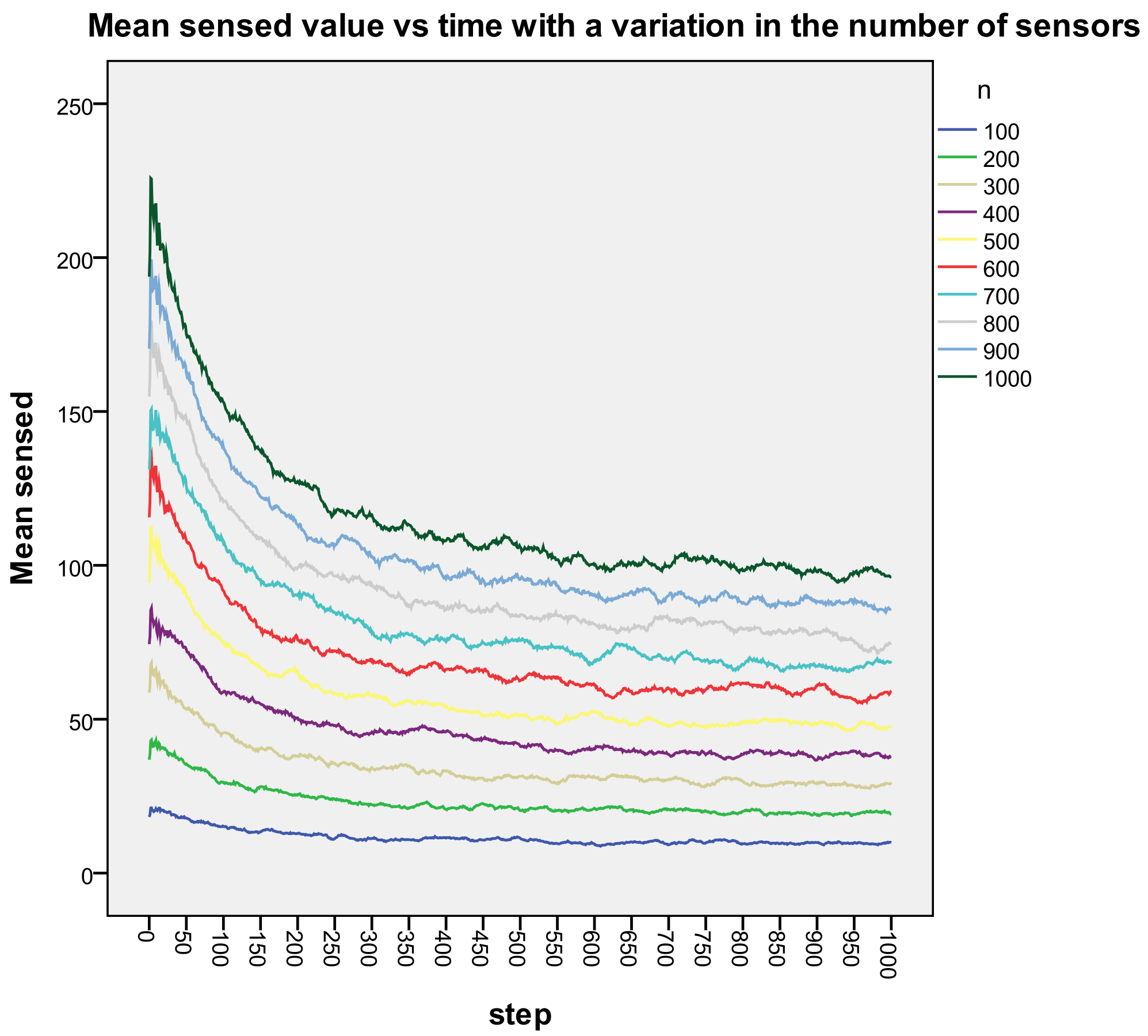

**Figure 63: Mean Sensed value vs. simulation time with a variation in the number of sensors**

What is really interesting here is the fact that even with a relatively small number of sensor Node agents (100), the flocking can still be detected or "sensed" thus further validating the hypothesis that the binary sensors can actually behave intelligent collectively and detect flocking behavior of the Boid agents. However, we can also note that if the number of sensor Node agents increases, the variation in the initial *Sensed* value to the final value is more pronounced. In other words, this helps verify the simulation further that it is working correctly and performing the expected behavior.

### 5.3.6 Significance of findings

The significance of the results of these simulation experiments can be noted by the number of possible applications as a result of developing a better understanding of this hybrid cas system. In short, if randomly deployed sensors can be used to detect apparently distributed mobile flocking, then some possible applications can be given as follows:

1. Sensing can be used for the identification of collective behavior of people using an image processing or infrared approach. In other words, this technique can be used to detect any group of people such as a group of "shoppers" flocking to a particular set of shops in a mall even though there is a large crowd of thousands of other people moving around.
2. This same approach can be used to detect a group of malicious attackers moving through a crowd thus allowing for the detection of the group "flocking" without cohering too much as that close grouping might have raised suspicions of the security personnel.
3. Another application of this approach is in the domain of detecting a group of stealth aircraft flocking towards a common mission. Stealth aircrafts are known to be invisible to radars in certain conditions. However these aircrafts are physically visible to the naked eye or in the visible light. As such, while it is difficult to infer a coordinated operation based on only the detection of a single aircraft, if there were a randomly deployed group of sensors with cameras or other sensing equipment pointing to the sky, these wirelessly connected sensors could essentially detect a group of stealth aircraft flocking and moving towards a common goal. Using simple shape detection image processing algorithms, each sensor can simply note if a "large" object passes over them. And this detection of "proximity" could be used to quantify flocking at a monitoring station collecting the data from these sensors.

### 5.3.7 Critical Comparison of DREAM with other approaches

The previous sections were related to discussion of the simulation experiments and results of the case study. Here, we relate the DREAM approach in the bigger context of describing agent-based models.

As discussed earlier in this chapter, DREAM approach is based on an informal methodology which allows for conversion between a complex network model, a digital footprint based on centralities and a pseudocode-based specification model. This conversion is essentially tied in with an informal validation process which can be followed by means of face validation, code walkthroughs and testing.

While, to the best of our knowledge, there is no other well-known approach in literature which describes agent-based models descriptively and quantitatively similar to our proposed DREAM approach, here we perform a comparison of the approach to two somewhat related approaches of knowledge transfer and description of agent-based models:

#### 5.3.7.1 Template Software approach

A common approach to sharing models, as noted by Santa Fe Institute's Agent-based modeling software Swarm's wiki page, is the template software approach[130]. This is ba-sically a code-based approach for comparison of models. The way this works is that template software for NetLogo, Swarm, Repast and MASON etc. are publicized on the website. Interested researchers download and adapt the source code for their particular usage. The website defines template software as code which has been designed with the following features in mind:

- For use in new models by means of copying and modification.
- As a means of tutorial-based teaching/learning.
- As examples of programming tasks.
- In some cases, for comparison of various ABM tools/software.

As such, it is clear here that templates are useful but, each of them is firstly based on code and thus specific to single platforms/languages. They also are not suitable for non-code or quantitative comparison of models, as proposed via the DREAM approach. The benefits of moving away from pure code to other models which do not require detailed knowledge of programming is clearly obvious here since at times, the designer of the simulation (Simulation Specialist) might be a different person than the cas researcher.

#### 5.3.7.2 ODD approach

Grimm et al. have proposed the Overview, Design concepts, and Details (ODD) protocol which is a text-based descriptive approach for documenting agent-based models. This approach has been demonstrated in ecological and social sciences [199]. The initial ODD proposed protocol[101] was subsequently updated and is now known as the "Updated ODD protocol". As compared with the proposed modeling approach, there are certain key differences given as follows:

Firstly, the description in ODD is textual. In other words, it can be considered as a way of describing agent-based models using a checklist of proposed headings. As such, it does not allow a quantitative comparison such as the use of Complex Network methods as demonstrated in the proposed modeling methodology.

1. ODD does not offer a visualization-based approach where researchers can visualize an entire ABM to compare it with other ABMs.
2. ODD also does not offer a one-to-one translation of models from the actual model to the source code. The proposed DREAM methodology allows this by means of an informal translation process.
3. ODD also does not allow for detailed descriptions of algorithms.

Being textual and not using any pseudocode, flow charts, state charts or other algorithm description models, it is difficult to see how a model described using ODD can be reproduced exactly. As Edmond and Hales [200] have previously noted that ". . .the description of the simulation should be sufficient for others to be able to replicate (i.e. re-implement) the simulation". Other problems with purely textual descriptions of agent-based models have been noted by Wilensky and Rand[145] who describe the difficulties in the replication of results based on textual description of Axelrod. It is pertinent to mention here that Axelrod is one of the pioneers of agent-based modeling in Social Sciences [25] with publications in the domain dating back to the 1980's[201]. Wilensky and Rand note that "In the replication experiment detailed herein, differences were discovered between the Wilensky-Rand model and the Axelrod-Hammond model, and the replicated model had to be modified to produce the original results. Moreover, the process that was required to determine that the replication was successful was complicated and involved unforeseen problems." Thus to summarize, while textual descriptions are effective in giving basic ideas about a simulation, it is difficult to directly implement these ideas without the use of pseudocode as proposed by the descriptive ABM methodology in this chapter.

## 5.4 Conclusions

In this chapter, we have proposed DREAM, an extension of exploratory agent-based modeling. DREAM allows for quantitative, visual and pseudocode based comparison and description of agent-based models. We have described a detailed case study as a means of demonstrating the effectiveness of using DREAM for modeling and simulation of various cas problems and inferring hypotheses. We have also critically reviewed DREAM with other prevalent approaches in the domain of agent-based modeling.

In the next chapter, we complete the last part of the proposed unified framework by developing a methodology of Verification and Validation of agent-based models using in-simulation validation. The proposed Virtual Overlay Multiagent System (VOMAS) approach is demonstrated by application on three different and separate case studies demonstrating its effectiveness in a number of multidisciplinary application case studies.

# 6 Proposed Framework: Validated Agent-based Modeling Level

In the previous chapter, we explored Descriptive Agent-based Modeling. DREAM allowed the developing of a non-code description of the ABM using a combination of a complex network model, a quantitative "footprint" of the ABM as well as a pseudocode-based specification model. However, while it solved some of the key methodological problems of ABM design, the proposed framework would still be amiss, if it did not allow for a methodology of developing "trustable" models. Trust in any simulation model is arbitrary because the fact of the matter is that a simulation model, however well-designed, is only as good as its correlation with the real-world or "imagined"[7] phenomenon, which it is expected to map. So, quite simply, a simulation model needs to be first verified so that it is working as expected and then validated to ensure it models the simulated phenomena close enough at the required level of abstraction as dictated by the particular requirements of simulation. However, as discussed earlier in Chapter 2, this particular type of verification and validation in the case of agent-based modeling cannot be considered in the same light as other more traditional simulation models. Next, we give a detailed description of the problem statement of verification and validation of agent-based modeling in the next section.

[7] Imagined such as in computer games and entertainment software, where the simulation is specifically designed not to reflect the real world.

## 6.1 Problem statement

As discussed earlier in Chapter 2, while the problem of verification and validation is applicable to any modeling and simulation approach, in the case of agent-based modeling of complex adaptive systems, there are certain peculiarities which make their validation different from that of traditional simulation models:

- The simulation design might not be required to develop a quantitative validation model because the phenomena might be observable and not easily quantifiable. In other words techniques such as Statistical validation or Empirical validation[140] used alone, might not always be the most appropriate choice for validation [144].
- Fagiolo et al.[202] note that ABMs have characteristics such as bottom-up modeling, heterogeneity, bounded rationality and nonlinear networked interactions which make them especially difficult to validate in the same way as other simulation models. In addition, they note that two of the key problems in ABMs are "the lack of standard technologies for constructing and analyzing agent-based models" and "the problematic relationship between agent-based models and empirical data".

In this chapter, we propose a solution to these problems by proposing a generalized methodology for in-simulation verification and validation of agent-based models using a Virtual Overlay Multiagent System (VOMAS). Our approach is based on existing ideas of verification and validation from different Agent-based Computing disciplines such as Social Sciences, Software Engineering and Artificial Intelligence to develop a validation approach which can be customized for application to multidisciplinary cas case studies. As a means of demonstration of the generalized applicability of our proposed approach to various cas domains, we present three separate example case studies from ecological modeling, WSNs and social sciences.

The rest of the chapter is structured as follows: First we give a description of the generalized methodology of in-simulation validation. Next, we present experimental design for the development of VOMAS for three example case studies. Next, in the results and discussion section, we present the simulation results followed by a detail discussion of benefits of using the VOMAS methodology in the validation of the particular case study model. We also critically compare the VOMAS methodology with other validation approaches of ABMs. Finally we conclude the chapter.

## 6.2 Proposed validate agent-based modeling methodology

In this section, we develop the generalized VOMAS methodology. To start with, we first examine how the problem of verification and validation of computer simulations is considered by simulation experts.

A generalized view of the verification and validation process of computational simulations has been presented by Sargent[203] as shown in Figure 64. This diagram encompasses the conceptual model as well as the computerized model. As we can note here, the validity of the conceptual model is performed by means of analysis and modeling. Whereas from the conceptual model to the computerized simulation model, programming and implementation can be evaluated by performing computerized model verification. However, to ensure that the computerized model is a true representation of the actual problem, a set of thorough experimentation is needed since it allows for a confirmation of the operational validity. Thus in the case of ABMs, the proposed methodology has to allow for a flexible means of performing computerized model verification as well as operational and conceptual validation.

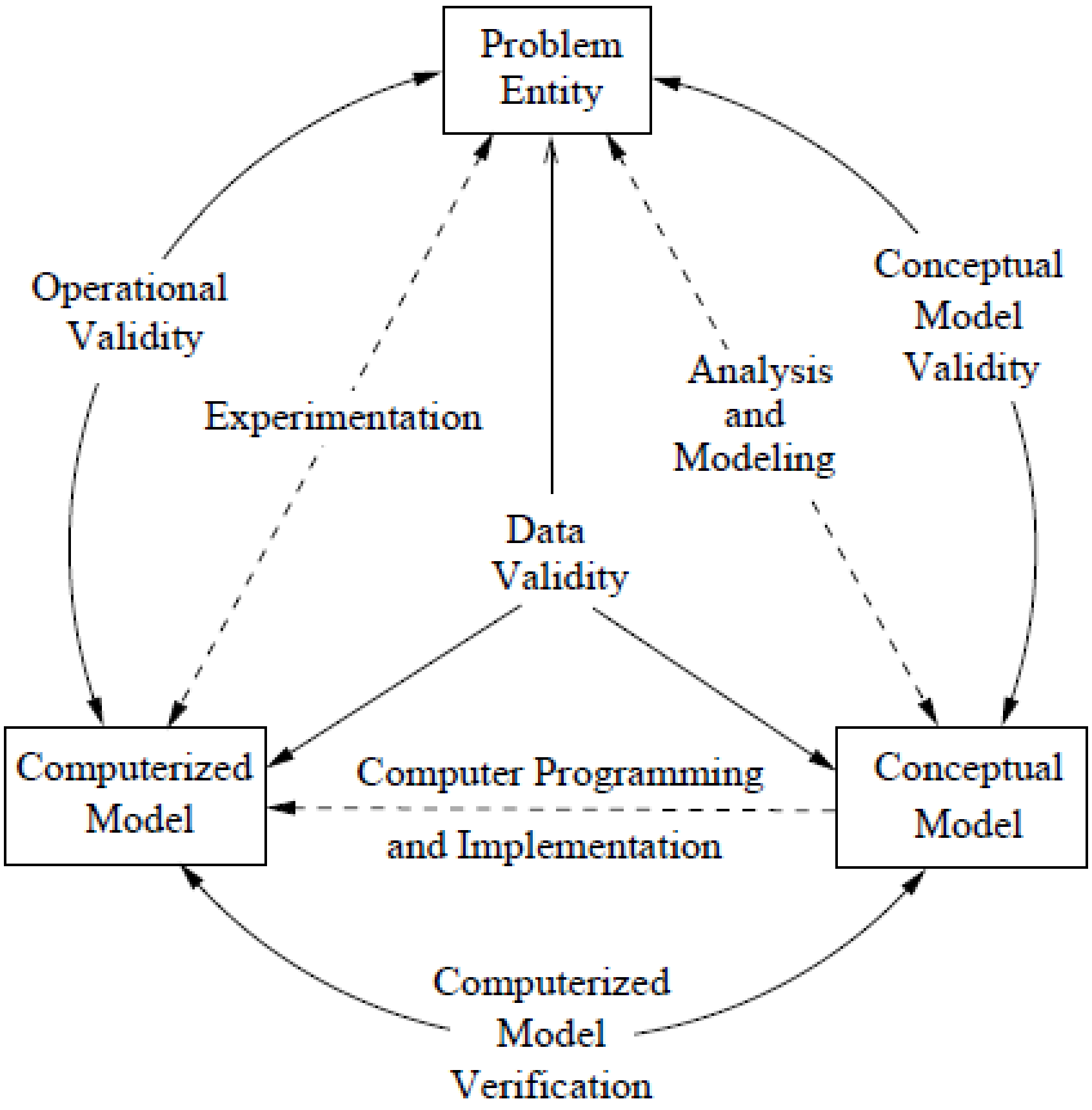

**Figure 64: A view of the modeling process, figure adapted from Sargent [203]**

The validated agent-based modeling level of the proposed framework can be noted in Figure 65. In this level, the development of the agent-based model is followed similar to a Software Engineering approach. As such, the SME and the SS work together in teams to develop requirements for the validation of the proposed model by determining invariant contracts to be added to the simulation. These invariant contracts allow the model to be customized for any validation paradigm and thus can be used based on the particular case studies. Based on these meetings, the SS develops the cas ABM by adding a Virtual Overlay Multiagent System (VOMAS) which allow in-simulation validation of the agent-based model using the invariant contracts.

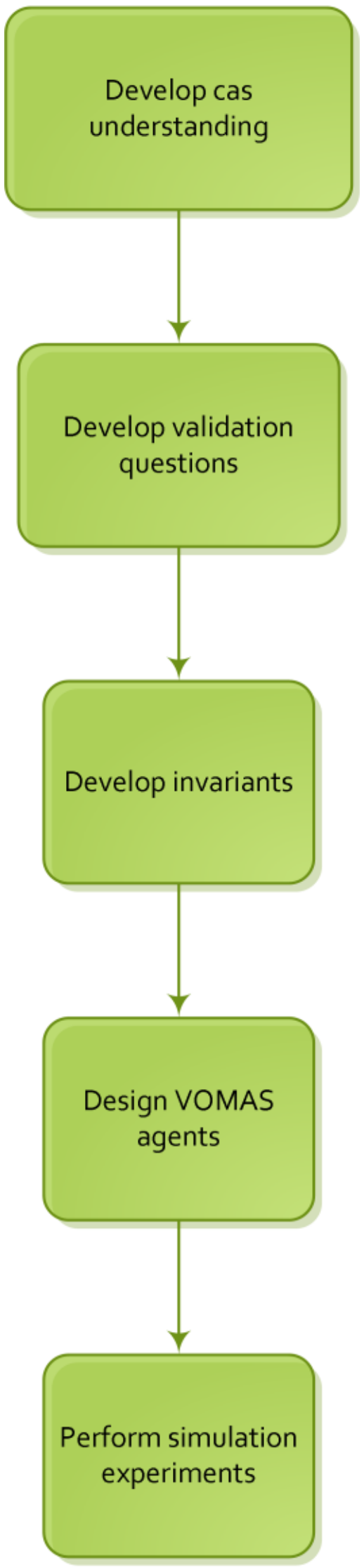

**Figure 65 Validated agent-based modeling framework level**

### 6.2.1 Detailed overview of origins of VOMAS concepts

The proposed VOMAS-based methodology has basis in three different set of basic ideas. These ideas originate from cooperative multiagent systems, software engineering and social sciences and are noted below as follows:

#### 6.2.1.1 VOMAS basis in Software Engineering Team concepts

The temporal and interactive structure of building a VOMAS ties in closely with how Software Engineering projects are constructed because the Software Engineering discipline is entirely focused on developing software based on teams of various sizes. The idea stems from the fact that simulations are also pieces of software however, to the best of our knowledge; there is no standard methodology which uses Software Engineering practices in the development of agent-based simulations. The VOMAS approach is thus structured on the basis of software engineering and requires the Simulation Specialist (SS) to work with the domain exert end user researcher or Subject Matter Expert (SME) in large scale research simulation projects. This leads to an iterative approach where the SS develops different models and diagrams (e.g. using DREAM approach) and gets approval from the client SME to proceed in each step.

A typical software development team can consist of various roles such as Manager, Software Engineer, Analysts and Quality Assurance Engineer. While one or more team members might have the same hat or role, or might have different roles depending upon the stage in a particular project, all work together with the end-users to develop a piece of software in line with the user requirements[204].

In the proposed methodology, the end users are nontechnical but multidisciplinary researchers who are experts in their own domains. Thus while they are not assumed to have advanced knowledge of modeling and simulation domains, they can intelligently work with the SS to develop and test the simulation models. While they might not correctly identify exactly what features are to be instilled in the software, they might be able to give constructive feedback on exactly how the simulation model can be proven correct in terms of their domain-specific terminology. Thus an ecologist would be able to look at a particular fire model and affirm that this is exactly how fires propagate in a particular area and also note as to what would be needed to make a model be realistic. In addition, a Biologist researcher investigating precancerous tumors might be able to observe at a tumor growth model and affirm if the model is looking close to the implicit mental model well-known to biologists and exactly what should the SS be looking for in an ABM tumor model for validation. Therefore, by means of using different models such as those earlier proposed in DREAM modeling, the end user researcher SME can gradually becomes comfortable with the software concepts helping the SS in developing a simulation model which qualifies the user requirements.

#### 6.2.1.2 VOMAS basis in software engineering concepts

Traon et al. [205]note that Meyer's "Design by contract" concept[206] allows for easily including formal specification elements (referred here specifically to software-based invariants, pre-conditions and post-conditions) into a software design. Using these built-in and programmed contracts, software engineers use these software constructs as "embedded, online oracles". The proposed methodology uses the design by contract concept. It extends the idea to the domain of agent-based simulation models of cas. It allows for building invariants based on pre- and post-conditions inside agent-based models by having VOMAS agents inside the simulation model of cas which dynamically observe the simulation and provide visual and log-based feedback to the SS. An invariant *I* is technically defined as follows:

"If a set of pre-conditions $C_{pre}$ holds in an ABM *M*, then *M* is guaranteed to undergo a state change which will result in the post-conditions $C_{post}$."

Thus periodically extracted information from the simulation model can then be used to evaluate different invariants, as and when required by the particular case study under guidance from the SME. Thus, unlike the design by contract model of software design, where contract becomes an integral part of the software model, these agents need to be designed to use minimal system resources. VOMAS agents can thus be designed to act on their own to periodically evaluate the simulation state and results.

As a consequence of invariants, we can also define an exception *E* as follows:

"If pre-conditions $C_{pre}$ do not hold in an ABM *M*, then *M* is not guaranteed to undergo any particular state change of interest to the SME"

Thus, such an exception behavior can be noted by means of execution of the simulation model. The reduction in the number of exceptions generated from the simulation experiments is thus an indicator of the improvement in the validation of the model. In addition, exceptions can be used to validate the Invariant itself. This can be performed by means of limiting all the conditions in a model to a specific set thus allowing for a focus on only removing just one of the pre-conditions. Now, with different variations of other inputs, exceptions can be noted. Next the pre-condition can be inserted and this would allow for the validation of the Invariant by a complete removal of the exceptions. This is shown graphically in the following Figure 66. We can note here the process of designing and testing Invariants by means of changing pre-conditions in the proposed methodology.

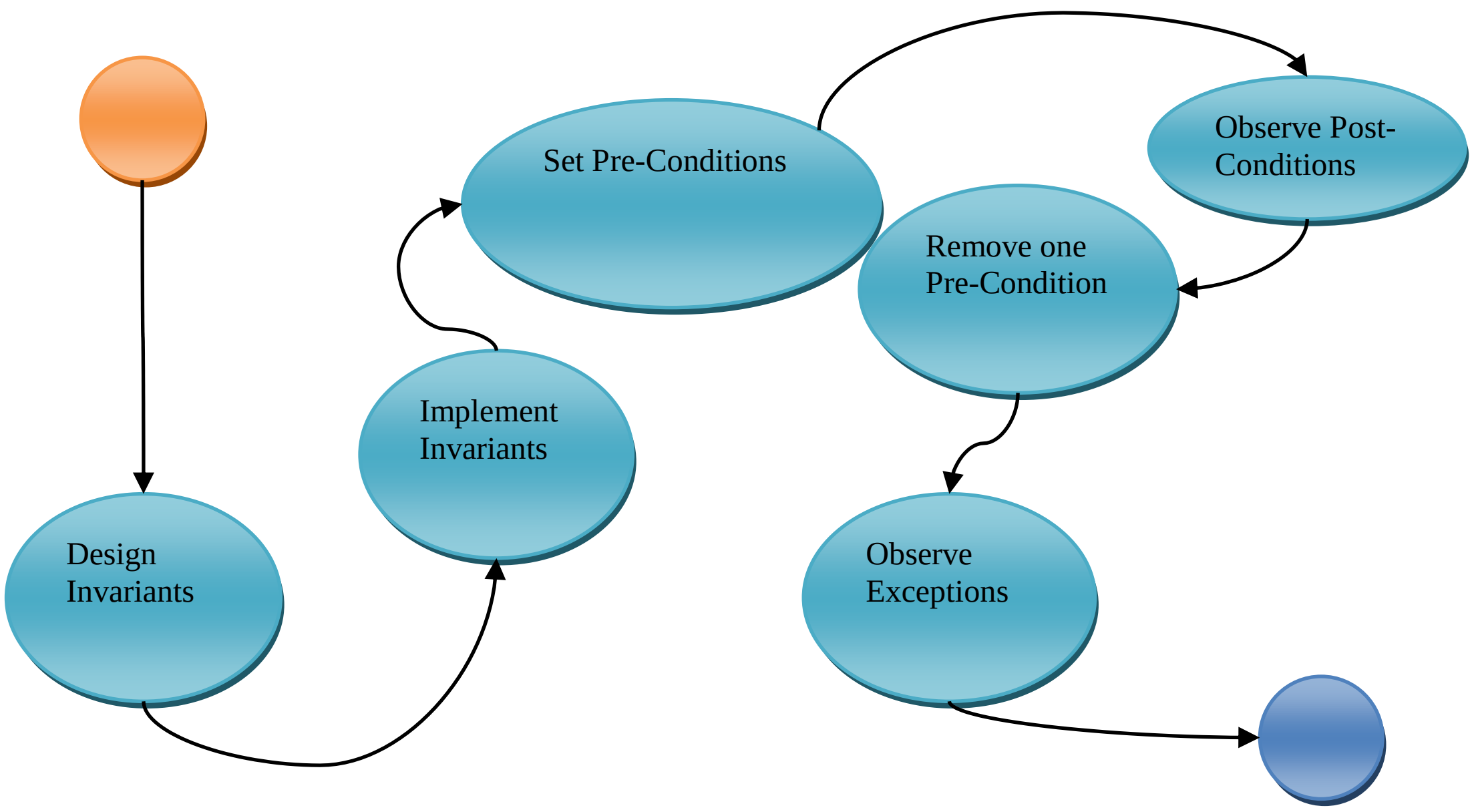

**Figure 66: Design and testing of VOMAS Invariants by means of manipulation of the pre-conditions**

6.2.1.3 VOMAS basis in multiagent system concepts

Panait and Luke [95] note that cooperative multiagent systems are a common paradigm in multiagent models where different agents work together to a common goal. Our proposed methodology has its third basis in the concept of cooperative agents. Thus, SME can use any suitable combination of agents inside a ABM which satisfy the requirements of validation. There are two possible extremes of this development. Either the developed agents are part of the ABM themselves or else the agents are developed separately and constitute agents observing the simulation from within simulation. The actual idea is to be able to facilitate the SME in developing the validation model and would thus vary on a case-to-case basis.

In other words, a VOMAS model built for one application domain or even one aspect of an application domain could vary considerably from another VOMAS model built for a different purpose. However, one primary design goal of VOMAS agents has to be so as not to disturb the results of the simulation model since simulation results can be are very sensi-tive to computation. If the computation performed by the VOMAS agents is too extensive, then it might end up disturbing the entire simulation. Thus care needs to be taken in the development of the in-simulation validation model so that the agents periodically report results without affecting the actual simulation. In addition, VOMAS agents must also use a minimal of system resources to extract information. The diagram in Figure 67 gives details of how the overall proposed in-simulation methodology is structured.

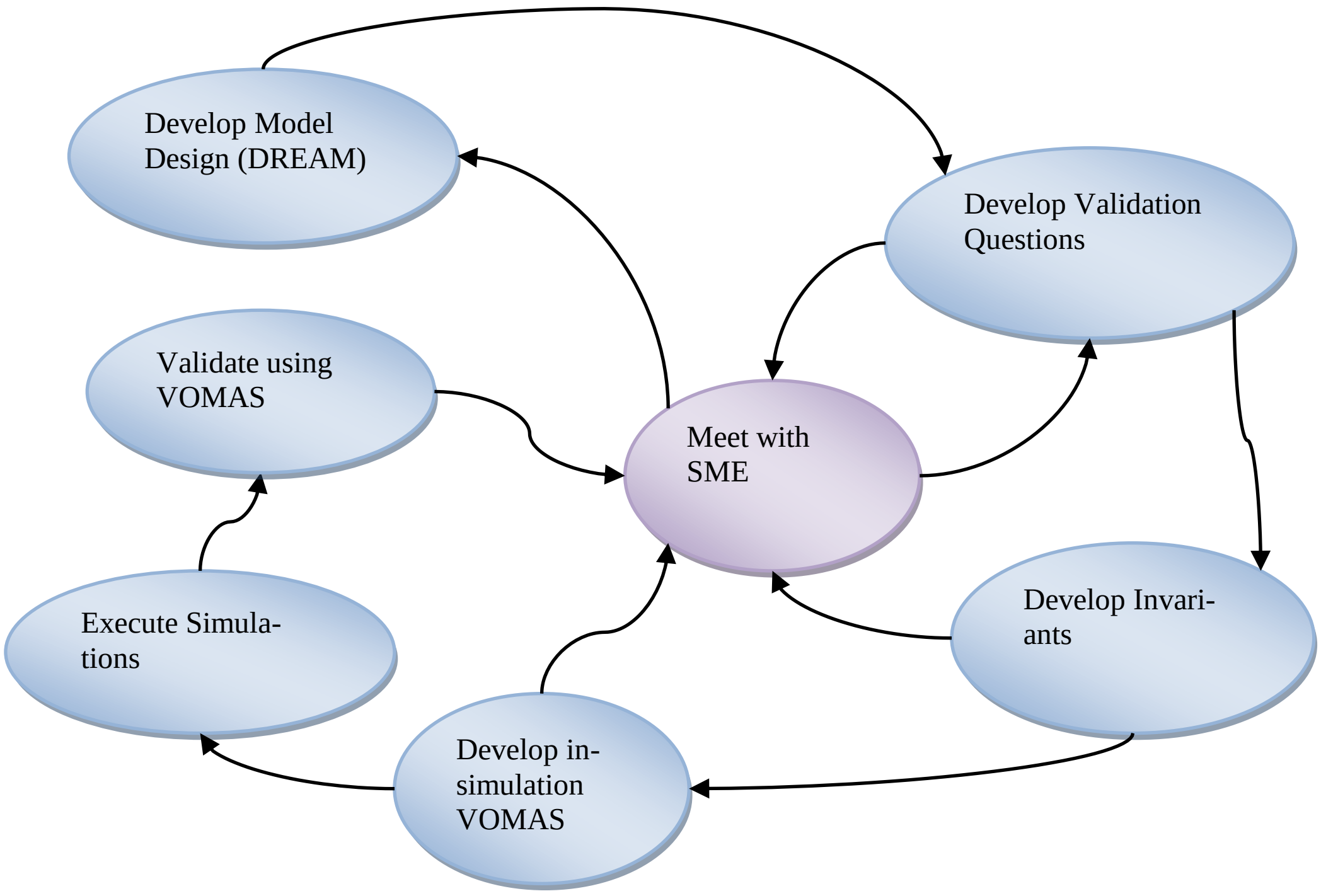

**Figure 67: Methodology of building In-Simulation Validated ABMs**

We can note here that the central activity of the proposed methodology is communication and collaboration with the SME. The methodology here is given from the perspective of the SS. As such, every step of this methodology is linked with meeting and communication with the SME. How it can actually be made possible is by ensuring that the methodology starts when the SME and SS meet and start to plan the simulation-based research project. Collaboratively, the two teams of multidisciplinary researchers i.e. the SMEs and the designers of the ABMs i.e. SS formally meet and develop the design of the ABM e.g. using DREAM. Next, they discuss what would be good validation questions. The validation questions should be succinct assisting the team to develop invariants along with any needed pre- and post-conditions.

The next step in the methodology is to develop invariants along with the suitable pre- and post- conditions. Finally the decision is to be made as to what are the specific agents which will be part of the simulation. The questions at this stage depend on the particular case study and would vary in different cases depending upon the level of computation required by the agents. So, in some cases the agents can be embedded in the actual simulation while in other cases, the agents can be observing the simulation and validating the invariants. Subsequently, the model is developed and gradually the invariants are added to the ABM.

After adding each invariant, the model can be verified using the method for testing invariants using VOMAS presented earlier in Figure 66. After the basic testing, the invariants can all be considered working and next the simulation experiments can be executed. The results can subsequently be used by the team to evaluate the effectiveness of the ABM in representation of the real-world phenomena

### 6.2.2 A Taxonomy of Agent-Based Validation techniques using VOMAS

In this sub-section, we examine some of the possible methods by means of which agent-based models can be validated by means of using VOMAS as shown in Figure 68. Since agent based models traditionally have one or more agents, what the physical implication in the real-world is entirely up to the designer i.e. the SS. Agents act as elements or pieces in a chess board and thus can be, at times, placed spatially in the simulation, where distance between agents in the simulation is important either for reduction of clutter or else for a concept of "distance" modeled by the designer. In other cases, non-spatial concepts could be used in the model since spatial concepts might not make sense in that particular application case study or discipline. In case of spatial models, it is also entirely possible that the exact distance may not be important, but the links between agents could be important. A detailed description of each of these follows.

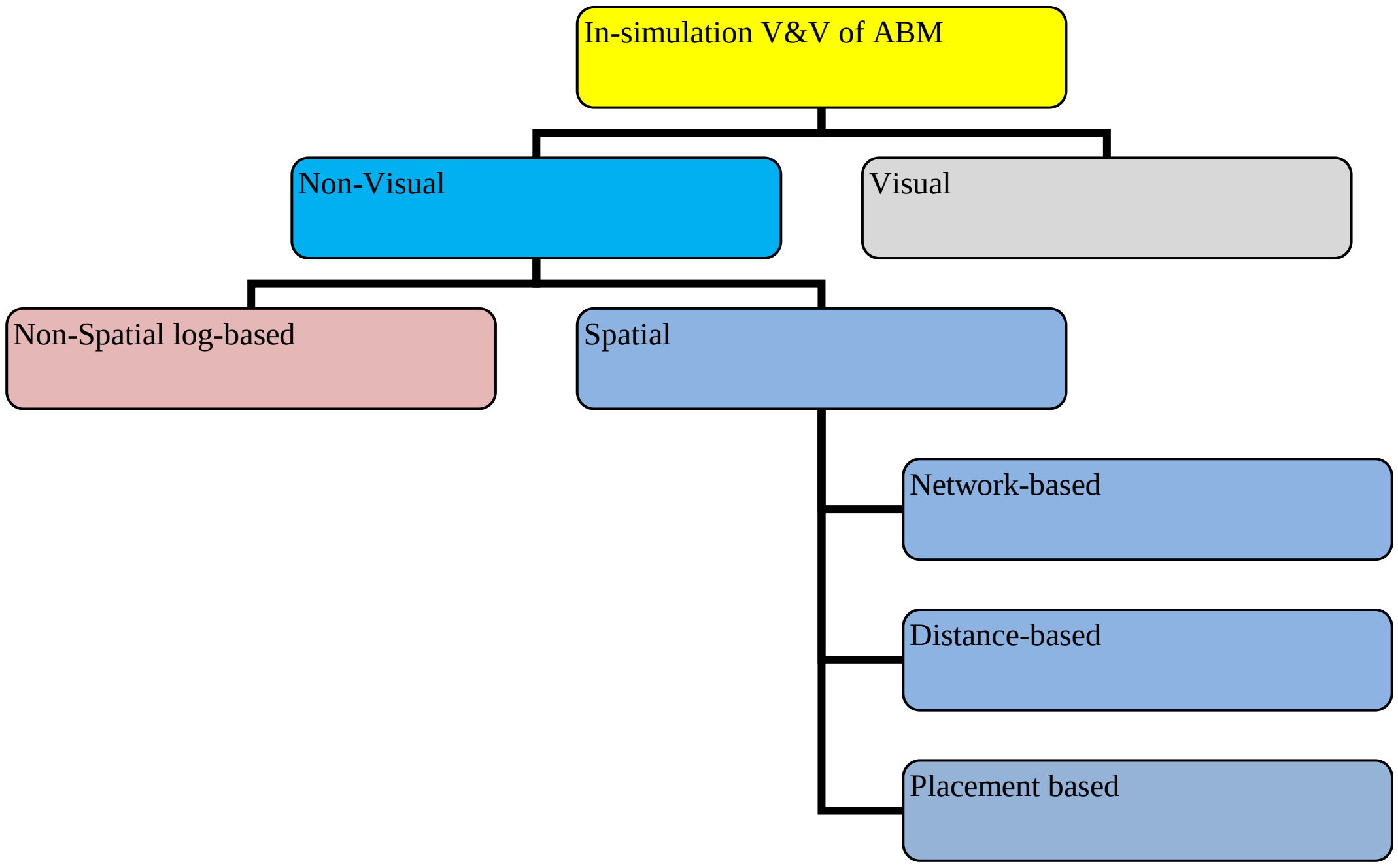

**Figure 68: A Taxonomy of Agent Based Validation techniques**

1. Visual Validation:

   Visual validation is related to the face validation technique[203] where the model animation or results can be validated by the SME to see if the behavior is similar to that expected in the actual system.

2. Non-visual Validation:

   Non-visual validation is a generalized methodology of validation which would be anything that is unrelated to purely face validation concepts.

   a. Spatial Validation:

In spatial validation, the placement of agents in the simulation is important. Thus either the agents are connected based on a network or else it is that the agents are place on a 2-D or 3-D plane and their distance or placement is important in terms of usage in the particular case study.

b. Non-Spatial Validation:

In non-spatial validation, the validation can be performed by means of developing extensive logs. These can ensure all data is stored for future usage. While this kind of validation allows for data storage, it also can result in some problems. Firstly, the size of the logs might grow considerably. And with size, the time to store the log might interfere with the simulation time. In addition, data might be too difficult to analyze due to significant size of logs.

## 6.3 Case studies: Experimental Design and Implementation

In the previous section, we developed the general methodology of in-simulation verification and validation of ABMs. In this section, we present the overview, experimental design and implementation details for three separate case studies from different disciplines to demonstrate the generalized applicability of the proposed validation methodology and framework. The first discipline is in the domain of ecological modeling where we develop a forest fire model for evaluation of the effects of forest fires on the overall forest structure as well. The second area is in the domain of WSNs, where we develop a specialized complex network model of multi-hop WSNs for observing the forest fire effects demonstrating the applicability of the proposed methodology in different ways. The third model is from the domain of quantitative social sciences in the domain of developing a complex network-based model of researcher evolution over time by means.

### 6.3.1 Case Study I: Ecological modeling of forest fire spread and forest re-growth

In the first case study, we explore the use of VOMAS in a Cellular Automata (CA) model of a forest fire simulation. To start with, we first evolve a basic and subsequently an advanced CA model of the forest fire spread. Next, validation questions and subsequently a VOMAS model for cross-model validation is developed using the Fire Weather Index (FWI), a standard mathematical model for forest fire prediction. In the next section, we describe in detail the model and experimental design.

#### 6.3.1.1 Basic simulation Model

We start by developing a forest fire model. In the real world, a forest fire spread can be based on different parameters such as:

1. The cause of the fire
2. The wind speed and direction
3. The amount of rain and snow in the past
4. The humidity level
5. The current rate of rain or snow
6. The particular type of Trees
7. The structural aspects of the forest

To develop the basic forest fire model, we thus need to develop several modules incrementally as given in the next sub-sections.

##### *6.3.1.1.1 Forest Creation Module*

In this module, our goal is to develop the forest scenario. Our basic model extends upon the Forest Fire model given in NetLogo model library[207] which is a simple model with no advanced parameters for configuration or any means of realistic forest fire simulation. As such, we first develop a forest made up of trees covering a random set of locations. However, to ensure we can control the tree coverage, we give the tree coverage model a probability ($P_{cov}$). Based on the value of $P_{cov}$, which varies from 0 to 100, we can build an environment model with trees. With a $P_{cov}$ value of 68%, we can see the forest model as follows in Figure 69. We can also note here that the wind direction is made visible by means of a dial at the top.

#### *6.3.1.1.2 Basic Fire module*

After the forest has been created, we can develop a baseline fire model. The way the fire module works is that at a given random probability of fire ($P_{fire}$), a fire can start at a random location. Now, the fire has three distinct states:

- Started
- Spreading
- Dying

In the "started" state, fire has just been initiated. After the fire has been created, it is given the direction of spread based on the "Wind direction" parameter. After creation, the fire next moves to the "Spreading" stage. In the spreading stage, we use a parameter "tic". This parameter is initially zero but as fire spreads, this parameter gradually increases. Thus depending upon the parameter "intensity", the fire keeps growing up until it reaches the "intensity" which serves as a threshold to stopping the fire. The "Dying stage" gradually serves as a means of stopping the fires and is reflected by a gradual change of color.

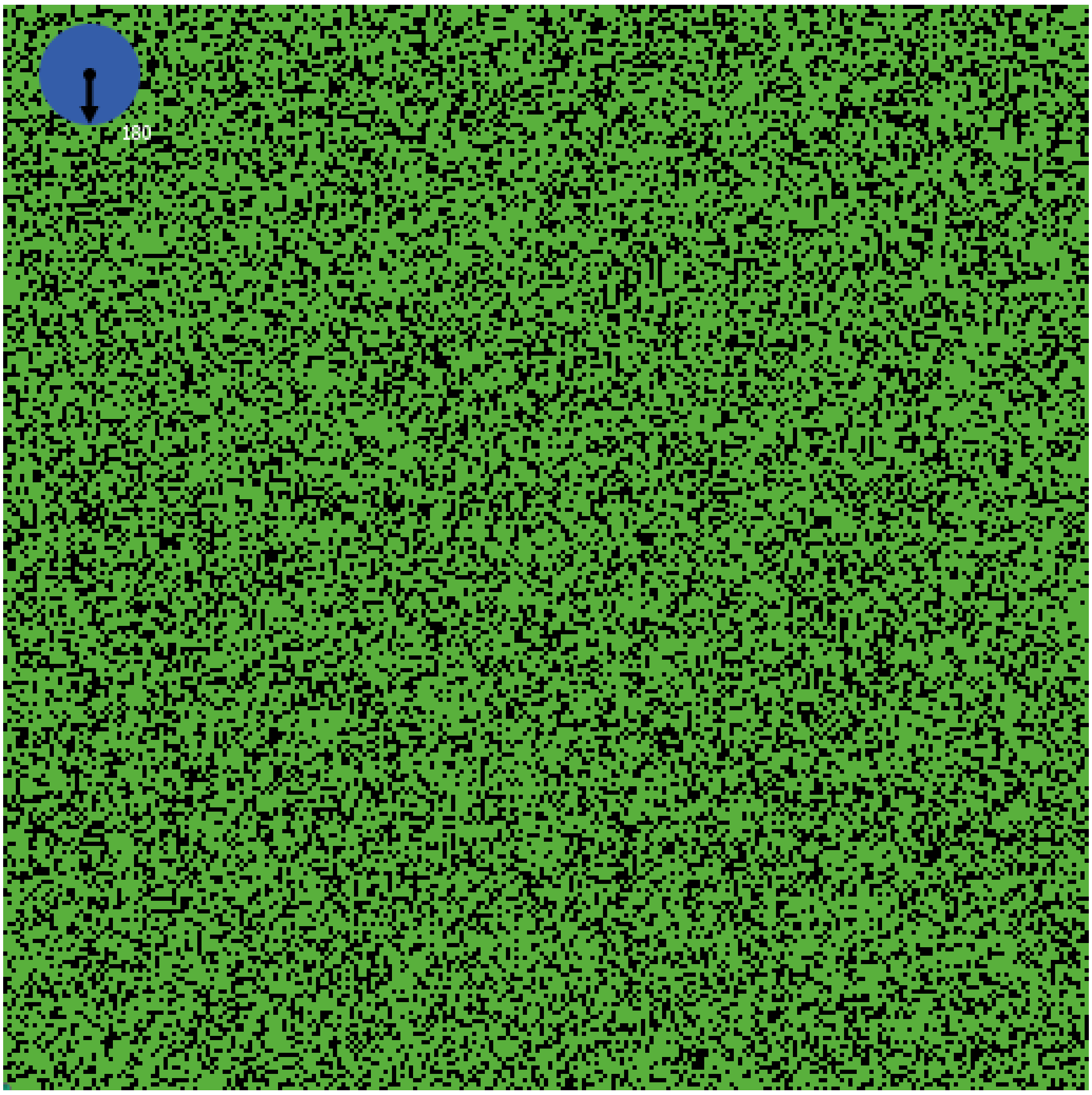

**Figure 69: View of the forest before the fire**

The effects of a fire can be seen in Figure 70. We can note here that the edges of the fire are a different color than the center of the fire because it represents that the trees have been destroyed in the center by now.

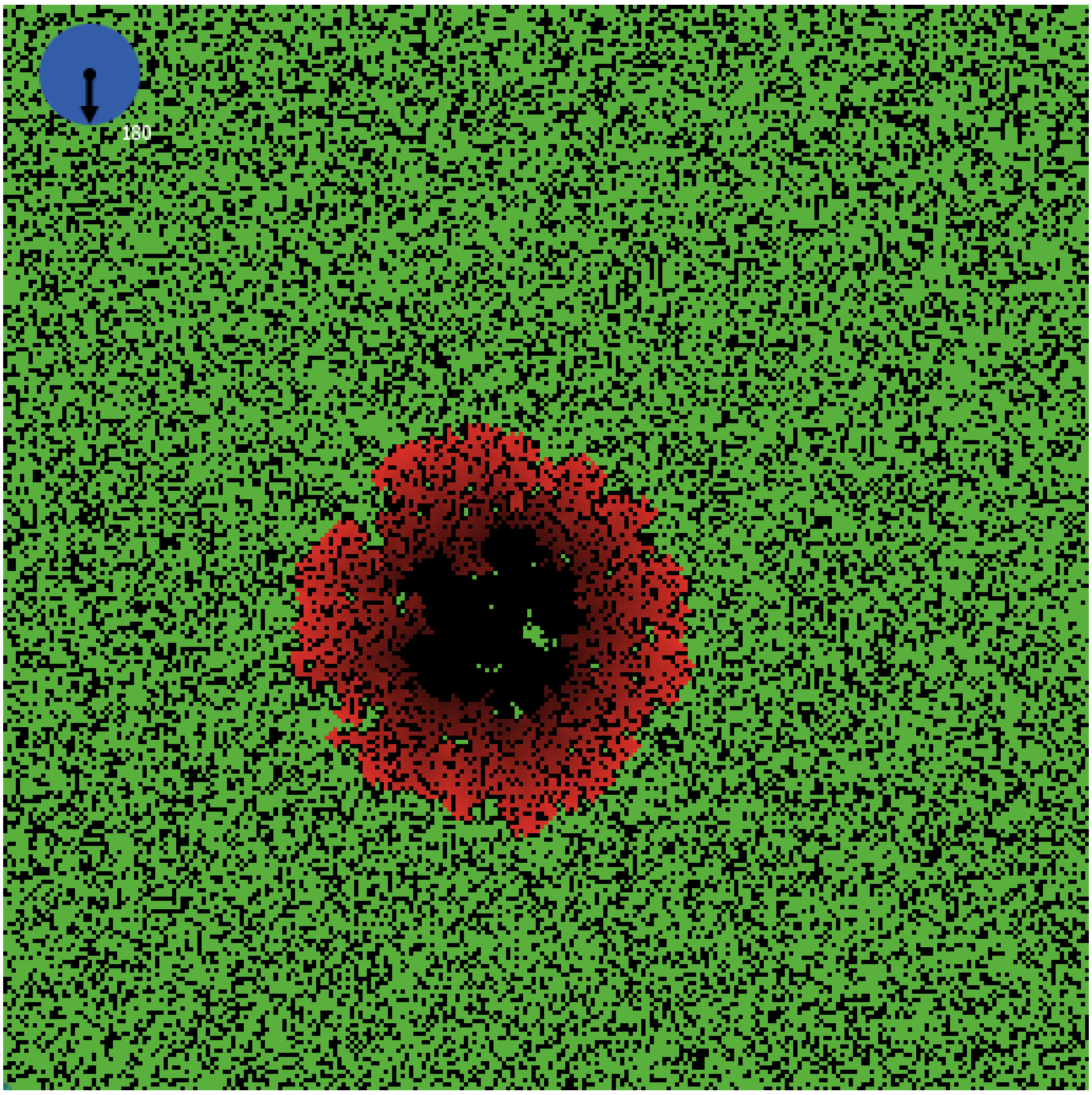

**Figure 70: Forest fire spread**

In addition to being able to simulate single fires, the baseline model can support more than one fire as shown in Figure 71.

#### 6.3.1.2 Advanced Forest Fire model version 1

While the basic or baseline forest fire model gives an interesting simulation, it does not give any realistic fire results except the use of wind direction. We next develop an advanced forest fire model. In the advanced forest module, we want to simulate long term fires so we want to model the effects of tree re-growth other time. In other words, while forest fires come and destroy the forest, periodically the tree cover grows as well. So, to be able to evaluate the advanced patterns formed due to the effects of various fires as well as the tree re-growth, we develop tree re-growth capability.

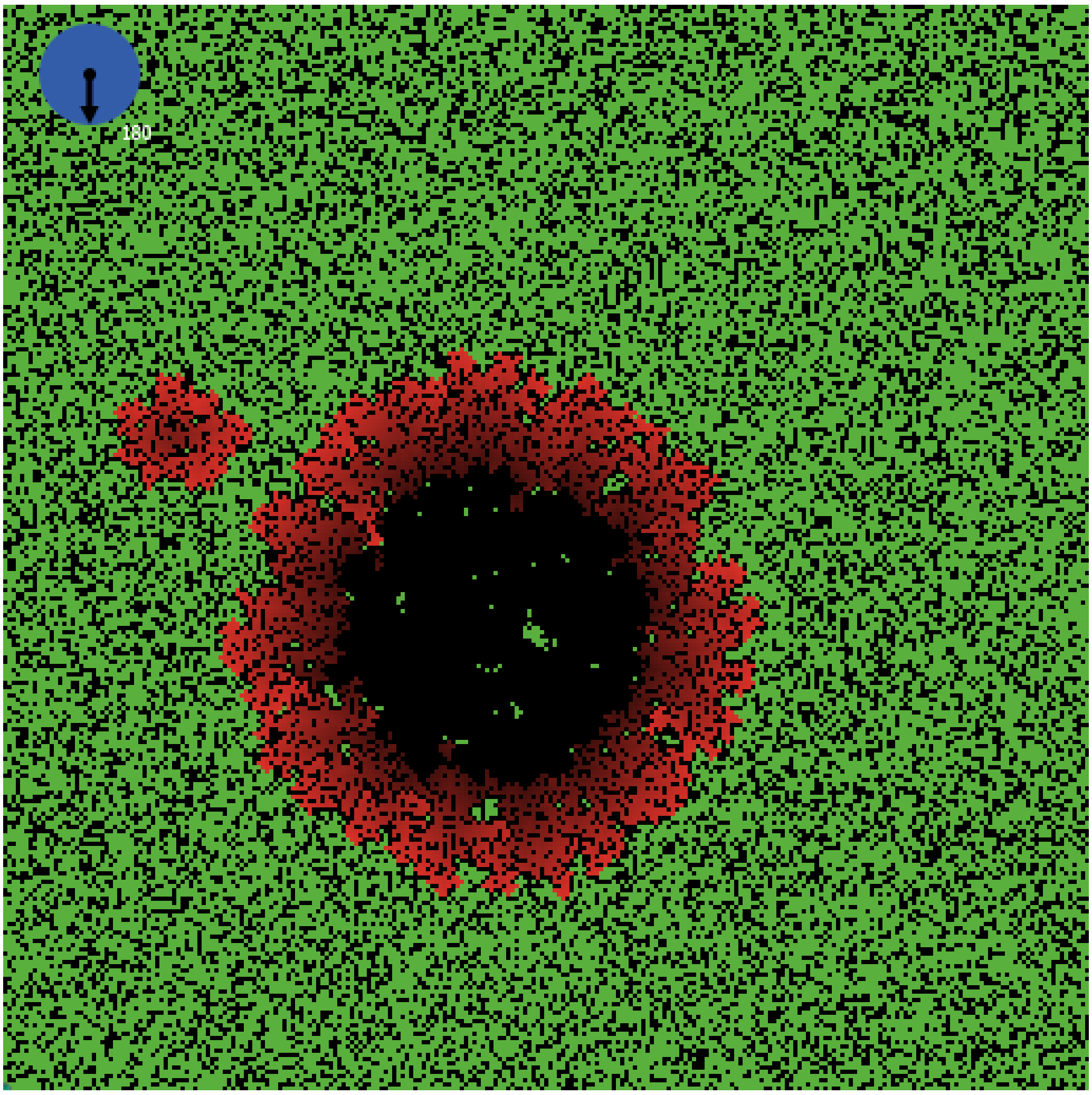

**Figure 71 Baseline forest fire model with two consecutive fires**

Based on a re-growth percentage and a re-growth period, our simulation can be developed which can give rise to more advanced effects in forest ecological simulation. As an example of tree re-growth based patterns with three different random forest fires, we see the pattern as shown in Figure 72.

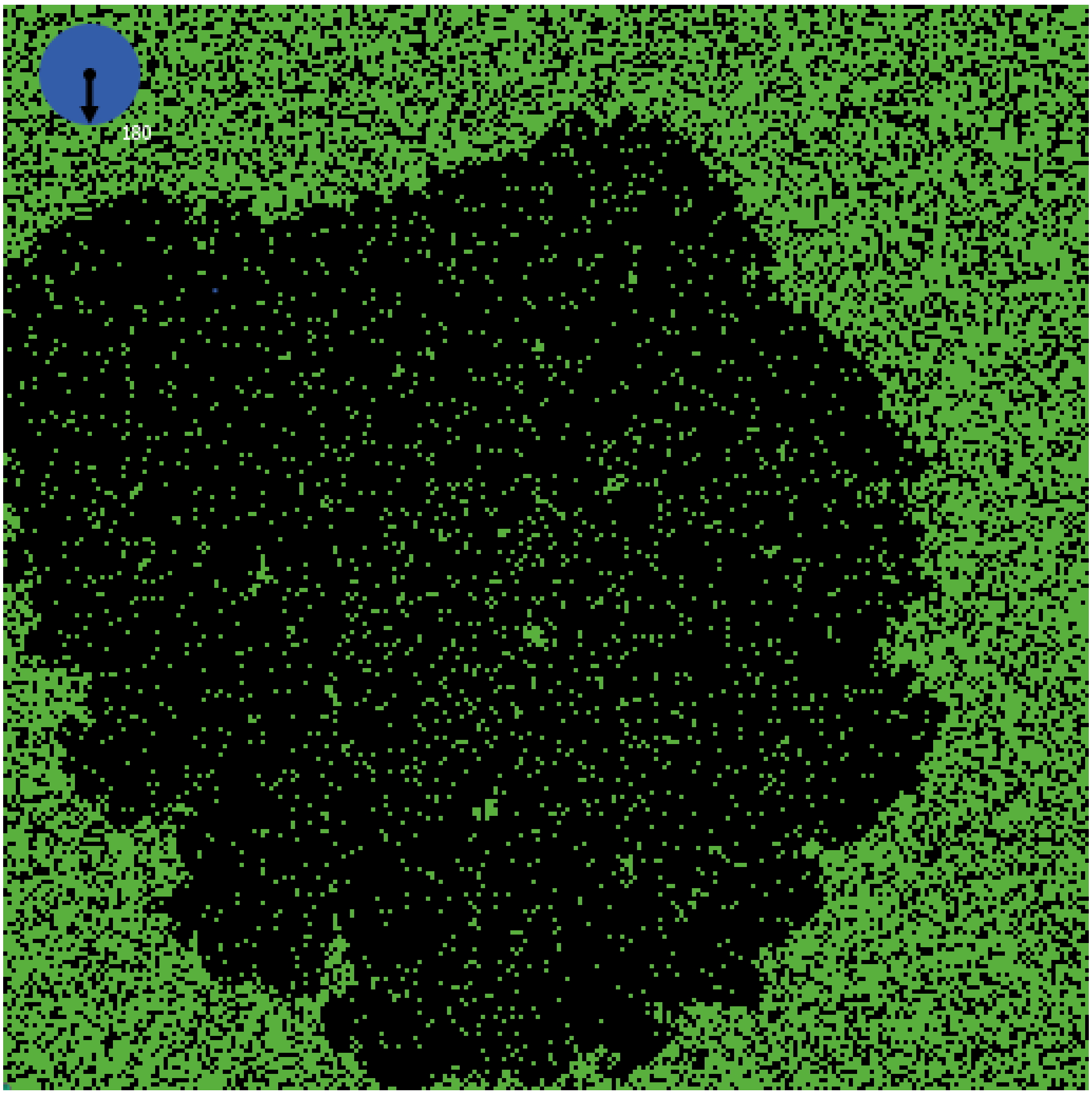

**Figure 72 Advanced forest fire model version 1 with tree re-growth**

6.3.1.3 Advanced Forest Fire model version 2

To make the forest fire scenario more realistic, we add several different scenarios and parameters. The first of these is the effect of rainfall as shown in Figure 73. The effects of rainfall are that the humidity of the location of the forest where rain is occurring is increased by a random value over time.

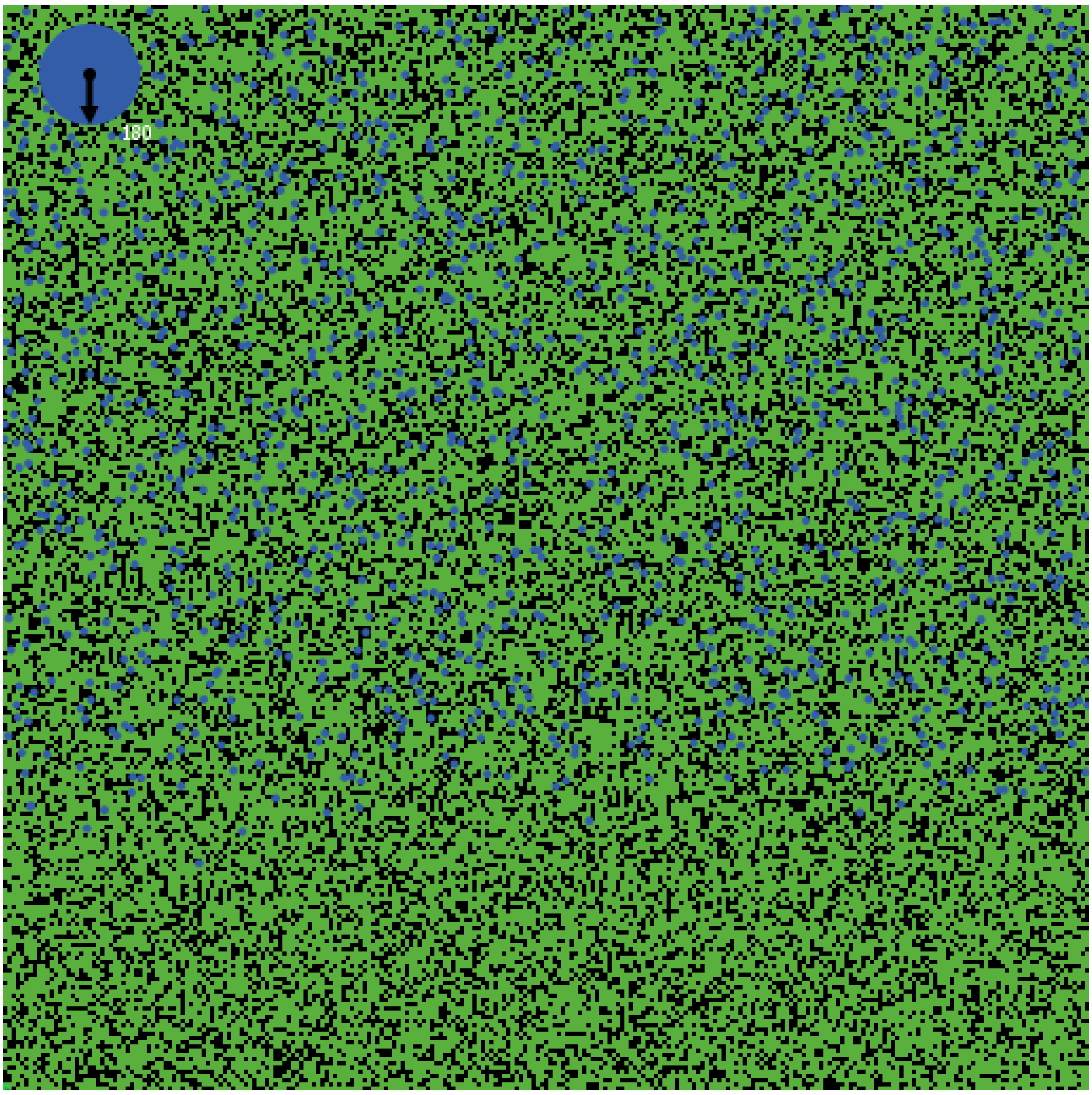

**Figure 73 Advanced forest fire model version 2 with rainfall in part of the forest**

In addition, we allow for coupling the effects of forest fires along with snowfall as shown in Figure 74. The physical impact of the snowfall is that the humidity as well as the temperature of the location is affected over time e.g. with an increase in humidity level and a decrease in temperature.

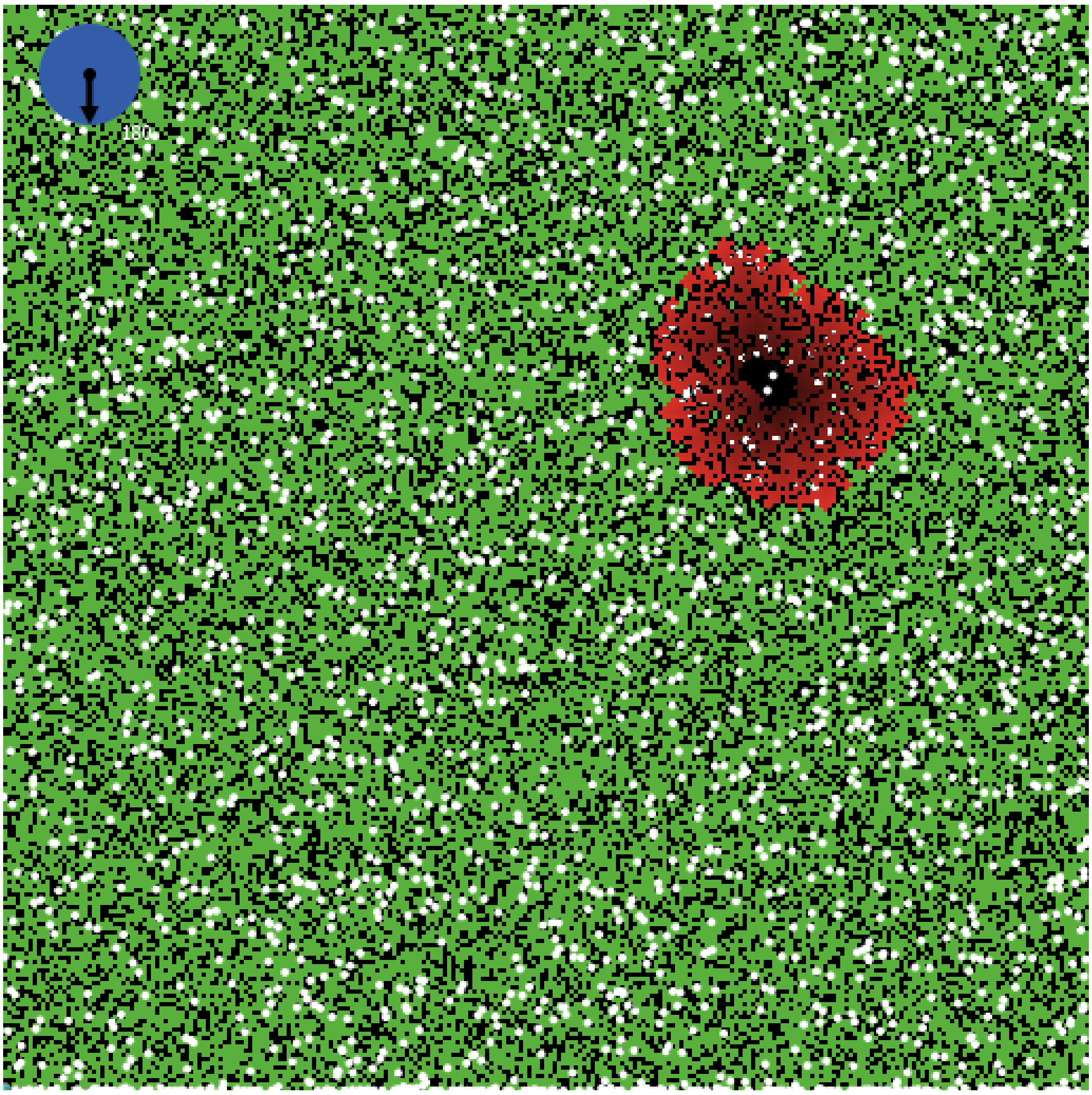

**Figure 74 Advanced forest fire model version 2 with snowfall effects**

#### 6.3.1.4 Validation Question

Up until now, we have developed the advanced forest fire model which contains forest creation, forest re-growth, weather effects simulation and forest fire simulation modules. The goal of this section is to demonstrate how this model can subsequently be validated by means of building an associated VOMAS model. Thus, while the entire model can be described in a real case study in the form of a DREAM model described earlier in the last chapter, here we only focus on the validation part. As described earlier in the VOMAS Methodology, we shall start by first writing our validation question in the model. Our key validation question can thus be stated as follows:

**Question**: How can we validate that the developed forest fire model is representative of actual forest fires?

Now, the way this could be practically answered is by looking for either a repository of data for forest fires or else a standard forest fire model. By performing an examination of previous literature such as [208] by Alexandridis et al., it becomes clear that while there is data such as either the temperatures of forests or else data in terms of the final results of fires, it is very difficult to get any data of the exact structural aspects of forest fire spread over time. In other words, the data availability is limited to Geographical Information Systems (GIS) based data visible via satellites. This data does not demonstrate the micro-dynamics of the forest fires. As such, instead of this approach, which would not have guaranteed model validity, we have to use a different approach. One possible approach in the absence of real data is thus to perform cross-model validation or "docking" of the models as noted by Axtell et al. in [209]. As such, here we use this approach but in contrast to Axtell's approach, which was to develop two different agent-based models, our ABM which will be validated against a formal mathematical model of the FWI. FWI is a standard model in use by several countries including France and Canada. In France, the index has previously been calculated and gradually updated for the last 40 years and similar practices are prevalent in Canada as noted by Lawson et al. in [210]. Being such a stable Mathematical model thus allows FWI to be a suitable candidate for use in model validation.

The FWI is calculated using weather values input in a complex set of equations as noted in [211] allowing prediction of forest fires[212]. In other words, FWI can be used to perform cross-validation of our simulation model.

#### 6.3.1.5 Software Contract

The next step in our proposed methodology is to develop software contracts for the simulation model. As such, based on the validation question that we have discussed in the previous section, we can focus on the following contract:

**Invariant Contract:** *If the pre-condition of a "tree is on fire" is true, then a post-condition of "FWI value in the range of Fire Danger index" is true.*

Thus, we need to be able to validate the FWI value in the simulation. Next we design the VOMAS agents to cater for this validation.

#### 6.3.1.6 VOMAS agent design

A naïve way of designing a VOMAS could be to have each cell calculate the FWI value over time. However, practically speaking there is a problem in this implementation because the FWI calculation requires calculation of almost 40 Mathematical equations (as we shall examine next), a number of which require complex mathematical calculations such as exponential etc. So, if we were to execute a simulation that has an environment of only $300 \times 300$ and every cell of the simulation calculates the FWI in each simulation run, then it would imply that for each simulation second, these equations would require $90000 \times 40 = 3600000$ calculations per simulation second. In other words, such an implementation would itself result in an overhead that might invalidate the entire basic simulation. As such, this would defeat the whole purpose of the forest fire simulation. As an alternative to this design, we thus propose using a random number of trees as VOMAS agents which calculate the value of the FWI and validate the value against fires. In other words, any trees, which have a FWI value in the range of fire danger index, might have fire. In addition, any trees having fire should have an FWI value in the danger range.

The actual calculation of FWI is based on items called the “Blocks”. The blocks are given as follows in Figure 75. A brief description follows of the different components is given as follows summarized from [1], [213].

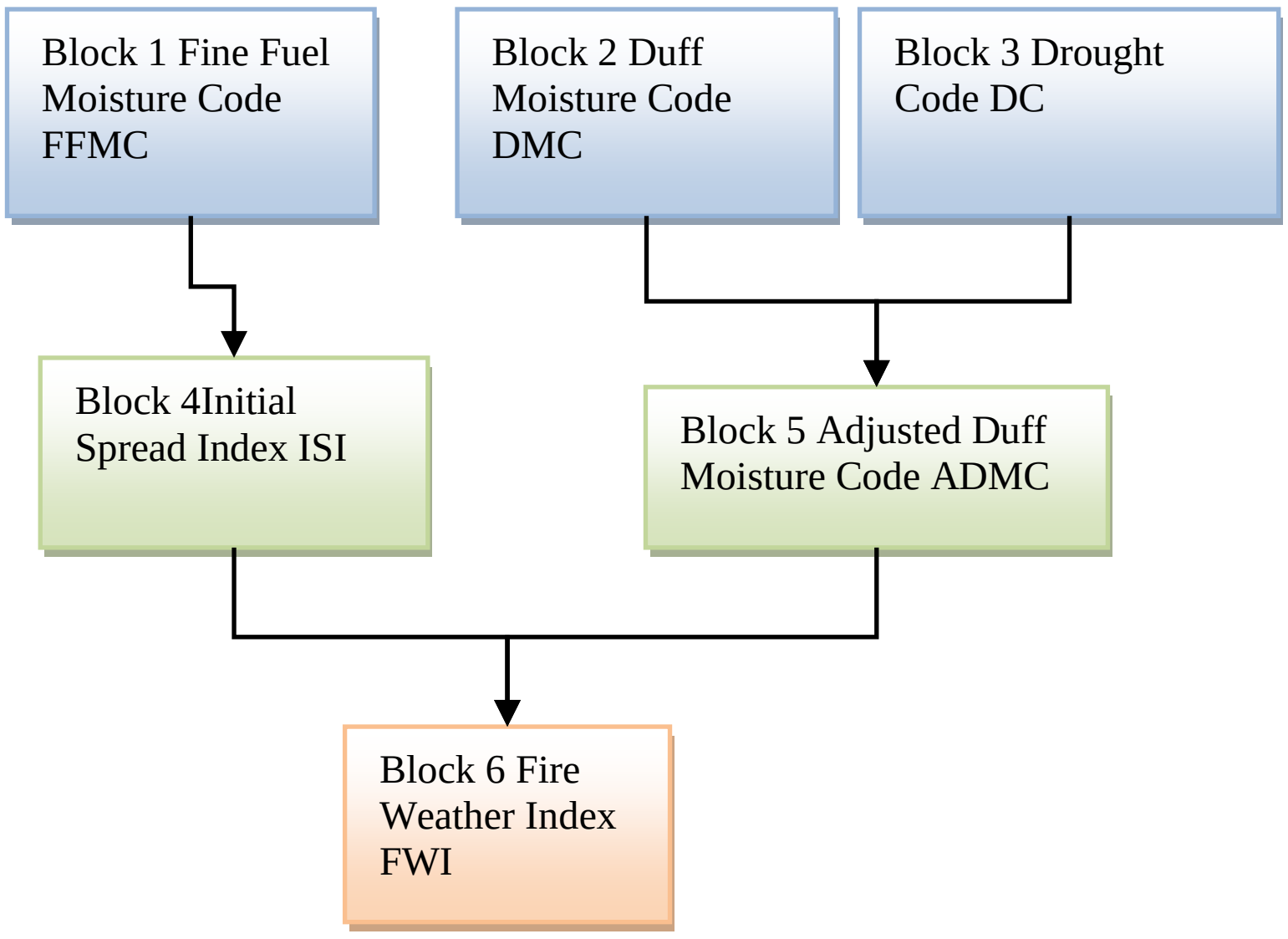

**Figure 75: FWI calculation, figure adapted from Wagner[1]**

### *6.3.1.6.1 Fire Weather Index (FWI)*

FWI [1] provides the assessment of the relative fire potential based on weather conditions. It depends on several sub-components. It is primarily calculated from the Initial Spread Index (ISI) and Build Up Index (BUI) to provide an estimate of the intensity of fire spread. In general, FWI can be considered to indicate the fire intensity based on a combination of an expected rate of fire spread along with the total amount of fire fuel consumed. Next, we discuss the sub-components needed to calculate FWI.

#### *6.3.1.6.2 Fine Fuel Moisture Code (FFMC)*

FFMC represents the moisture contents of litter and fine fuels, 1-2cm deep, with a typical fuel loading of about 5 tons per hectare. FFMC indicates the ease of ignition or in other words, it formalizes the ignition probability.

#### *6.3.1.6.3 Duff Moisture Code (DMC)*

DMC is an indicator of the moisture content of the duff layer, which is essentially a layer of compacted and organic matter in decomposing state. It predicts the probability of fire ignition as a result of lightening. It is also representative of the rate of fuel consumption.

#### *6.3.1.6.4 Drought Code (DC)*

Drought Code is an indicator of average moisture content of the deeper layers of compacted organic matter. DC is indicative of long-term moisture conditions and determines fire resistance to extinguishing. It is also an indicator of fuel consumption.

#### *6.3.1.6.5 Initial Spread Index (ISI)*

ISI indicates the rate of fire spread immediately after ignition. In combines the FFMC and wind speed to predict the expected rate of fire spread.

#### *6.3.1.6.6 Build Up Index (BUI)*

BUI index is a combination of the DMC and DC codes and it indicates the total amount of fuel available for combustion. DMC has the bigger influence on the BUI.

#### *6.3.1.6.7 Calculation of FWI*

The details of the calculations pertinent to our simulation implementation are discussed below however further details of these parameters are out of scope of and interest of this thesis in general although they are publicly available in [211].

Block 1 is the Fine Fuel Moisture Code (FFMC) and is calculated based on the rainfall, relative humidity, wind speed and the temperature as follows (F):

$$
\begin{aligned}
&FFMC\\
&m_o = 147.2(101 - F_o)/(59.5 + F_o)\\
&r_t = r_o - 0.5\\
&m_r = m_o + 42.5rf(e^{-100/(251-mo)})(1 - e^{-6.93/r_f})\\
&mr = m_o + 42.5rf(e^{-100/(251-mo)})(1 - e^{-6.93/rf}) + 0.0015(mo - 150)^2 r_f 0.5\\
&E_d = 0.942H^{0.679} + 11e^{(H-100)/10} + 0.18(21.1 - T)(1 - e^{-0.115H})\\
&E_w = 0.618H^{0.753} + 10e^{(H-100)/10} + 0.18(21.1 - T)(1 - e^{-0.115H})\\
&K_o = 0.424[1 - (H/100)^{1.7}] + 0.0694W^{0.5}[1 - (H/100)^8]\\
&K_d = ko * 0.581e0.0365T\\
&k_l = 0.424[1 - (100 - H/100)^{1.7}] + 0.0694W^{0.5}[1 - (100 - H/100)^8]\\
&Kw = k1 * 0.581e^{0.0365T}\\
&m = E_d + (m_o - E_d) * 10^{-Kd}\\
&m = E_w - (E_w - m_o) * 10^{-Kw}\\
&F = 59.5(250 - m)/(147.2 + m)
\end{aligned}
$$

Block 2 is the Duff Moisture Code (DMC). It is based on the rainfall, relative humidity and the temperature. DMC (P) is calculated as follows:

$$
\begin{aligned}
&DMC \\
&r_e = 0.92r_o - 1.27 \\
&M_o = 20 + e^{(5.6348 - Po/43.43)} \\
&b = 100 / 0.5 + 0.3P_o \\
&b = 14 - 13\ln P_o \\
&b = 6.2\ln P_o - 17.2 \\
&M_r = M_o + 1000re / (48.77 + br_e) \\
&P_r = 244.72 - 43.43\ln(Mr - 20) \\
&k = 1.894(T + 1.1)(100 - H)Le*10^{-6} \\
&P = P_o(or\,\mathrm{P_r}) + 100K
\end{aligned}
$$

Block 3 is the Drought code (DC) and is based on rainfall value and the temperature. DC (D) is calculated below as follows:

$$
\begin{aligned}
&DC \\
&r_d = 0.83ro - 1.27 \\
&Q_o = 800e^{-Do/400} \\
&Q_r = Qo + 3.937rd \\
&D_r = 400\ln(800 / Qr) \\
&V = 0.36(T + 2.8) + Lf \\
&D = Do(orDr) + 0.5V
\end{aligned}
$$

Next, we note the Block 4, which represents the Initial Spread Index (ISI). ISI (R) is calculated as follows based on the FFMC calculated earlier:

$$
\begin{aligned}
&ISI \\
&f(W) = e^{0.0811W} \\
&f(F) = 91.9e^{-0.1386m}[1 + m^{5.31} / (4.93*10^7)] \\
&R = 0.208f(W)f(F)
\end{aligned}
$$

Next, we calculated the Build-up index (BUI) or the Adjusted Duff Moisture Code (ADMC) represented as U is calculated next as follows:

$$
\begin{aligned}
&BUI \\
&U = 0.8PD / (P + 0.4D) \\
&U = P - [1 - 0.8D / (P + 0.4D)][0.92 + (0.0114P)^{1.7}]
\end{aligned}
$$

Finally these values are used to calculate the FWI as follows:

$$
\begin{aligned}
&FWI \\
&f(D) = 0.626U^{0.809} + 2 \\
&f(D) = 1000/(25 + 108.64e^{-0.023U}) \\
&b = 0.1R_f(D) \\
&\ln S = 2.72(0.434\ln B)^{0.647}
\end{aligned}
$$

As can be seen, these calculations are necessary for the accurate calculation of the FWI value in our case study. Thus, VOMAS agents will be performing these calculations as required. The results of the VOMAS experiments of the Forest Fire Simulation case study will be further discussed in the results section.

### 6.3.2 Case Study II: Environmental Monitoring using Wireless Sensor Networks

In the previous case study, we developed a Forest Fire Simulation case study. In this second case study, our goal is to extend the previous case study and use a set of connected wireless sensor network nodes to monitor the forest fires. As a means of getting to this stage, we first need to develop a formal model for a randomly deployed WSN by choosing one of the standard models.

Technically, a connected WSN is represented as a Graph or Network $G = (V, E)$. Where $V$ is the set of vertices and represents the sensor nodes while $E$ (edges) represents the connectivity of the nodes. Thus for any two nodes $u, v \in V$, $(u, v) \in E$ if the sensor node $u$ is adjacent to $v$. The classical connectivity model for WSNs is the Unit Disk Graph (UDG)[214] where if any two nodes are considered adjacent if and only if their Euclidean distance is at most 1. The UDG [215] has been considered as unrealistic connectivity models. It also does not cater for the facts such as that real world sensors might not have Omni directional antennas and also that any small obstacles might result in disconnection[188]. An advanced version of the UDG is the General Graph (GG) model which is a general un-directed graph. As such, every node can be considered as an adjacent to every other node. However, this model is again unrealistic because it can be too generous and thus it might result in nodes which cannot be reached from a node having connections. As such, both these models are considered as two extremes for realistic models. However, our proposed simulation model for WSNs follows a flexible realistic version of network models which has been considered as a reasonable model between the two extreme modeling paradigms called the Quasi Unit Disk Graph (QUDG) proposed by Kuhn et al.[216]. Thus our WSN model would be formally defined as follows:

Our proposed WSN model consists of nodes in a two dimensional Euclidean plane in $\mathbb{R}^2$. All pairs of nodes with Euclidean distance $\rho$ for some given $\rho \in (0,1]$ are considered adjacent. In addition, pairs of nodes between distances $\rho$ and 1 may or may not be neighbors. It is a well-known fact that for $\rho = 1$, QUDG becomes a UDG[188]. As such, building a QUDG using the NetLogo interface, we can see the simulation output shown as follows in Figure 76.

Next, the models for forest creation, forest fire and the WSN simulation are integrated resulted in a single model. This sensor model is able to extract useful information such as temperature and other parameters from the location of the sensor inside the simulation environment. With a description of the basic connected WSN model, we next move to the validation question part of the proposed methodology.

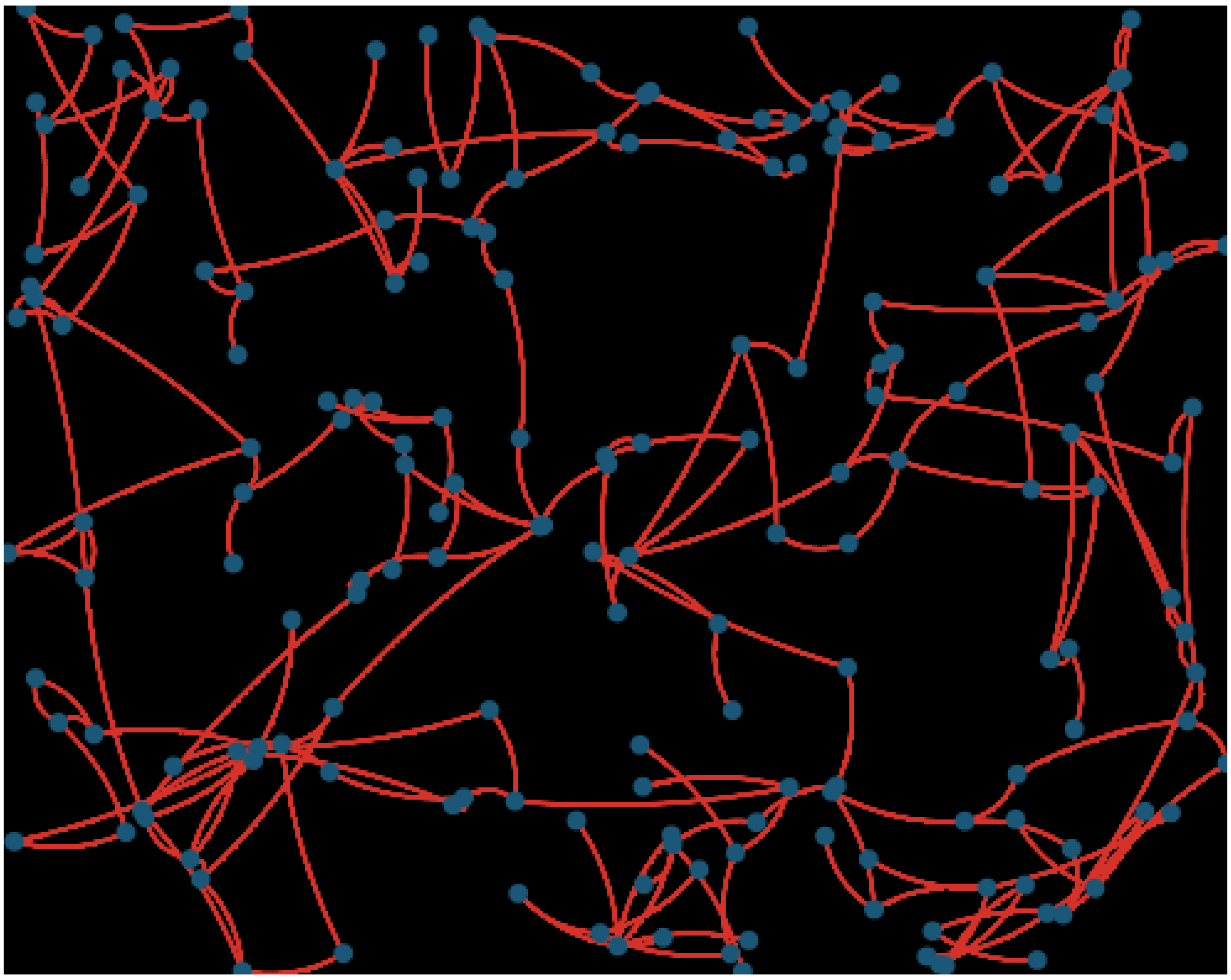

**Figure 76: WSN network simulation as QUDG with n=200 sensor nodes**

6.3.2.1 Validation Question

The simulation in this case study is based on a set of sensor nodes connected together. In the previous chapter, we have simulated WSNs with a unit distance from the sink with the goal of the WSN being solved using single hop communication. In that case, the nodes did not require connectivity between each other for communications unlike the multi-hop scenario in the current case study. Here, each node is expected to confirm the temperature, humidity and other parameters from the local area and then report to a remote base station. So, in this case, our validation question can thus be defined as follows:

**Question:** How can we validate that the WSN is accurately monitoring the forest fire simulation?

The solution to this question lies in following the validation of the previous ecological case study however we shall demonstrate here as to how VOMAS development can vary from one case study to another even in closely related ABMs. We can note that in the previous scenario, we used the FWI to validate if the forest fire generated was "valid" by comparing the FWI value with the status of the fire location. In this case, we need to redesign the VOMAS to validate the sensing capability of the sensor node.

#### 6.3.2.2 Software Contract

In this particular scenario, we can use the following Invariant Contract:

**Invariant Contract**: *If the pre-condition that "sensors are correctly reading the values from their neighborhood" is true, then the outcome of the FWI calculation by the sensors would result in a post-condition of "correct prediction of fire in the vicinity of the sensor node".*

#### 6.3.2.3 VOMAS agent design

As noted in the generalized methodology, our next step is to decide upon the design of the agents which will be used for performing the validation. In this case, since we want to validate the effects of the sensor nodes, it makes sense to use some of the sensor nodes as VOMAS agents and then use them to compare the validation criteria (FWI) with that of their neighboring trees. Thus, if there is a fire in the neighborhood of a sensor node agent, the sensor should predict fire based on the FWI value. The simulation results of this case study will again be evaluated in the results section of the chapter.

### 6.3.3 Case Study III: Researcher Growth Simulation

While our first case study was from the domain of ecological Sciences and the second was from telecommunications in particular, our third case study has been chosen from the domain of Social Sciences. The goal of choosing different case studies is to be able to show the generality of the proposed in-simulation validation methods by the use of VOMAS. In this particular case study, our goal is to simulate the evolution of researchers over time by means of an integration of ABM and Complex Network concepts.

To be specific, we start by an examination of the research process. The research process can be considered in the light of two clear aspects. The first aspect is related to how researchers perform research and get it published. The second aspect is the eventual complex adaptive emergent effects of the popularity of specific researchers, publication Journals or institutions over time. This particular aspect of research has previously been discussed and evaluated empirically using case studies earlier in Chapter 3. Here we first examine the first aspect of the publication and research-related processes.

#### 6.3.3.1 Research Process I: Publication

A first examination of how researchers perform research can be observed in Figure 77. Here the research workflow starts when a researcher or a group of researchers start working individually or collectively on one or more innovative ideas. The ideas are correlated with existing literature to understand a feel of the state of the art. The researcher acquires data, performs analysis and then subsequently attempts modeling/simulation and/or experimentation in an iterative fashion striving to achieve successful results based on a research success criteria. After apparent success, the researchers embark on the task of writing the results in the form of a report. Once this article has been submitted for review, it is evaluated by impartial and anonymous referees. The reviewers examine the report for originality in terms of advancing the state of the art and subsequently submit their reviews to the Editor. Eventually the referee report can result in either an acceptance for publication or else a rejection by means of provision of a set of reviews offering guidance to the authors. In the case of a rejection, the researcher is at least armed with these reviewer's comments, so s/he can start to work again on the same project and perform more experimentation, generate better results, and the process starts over again.

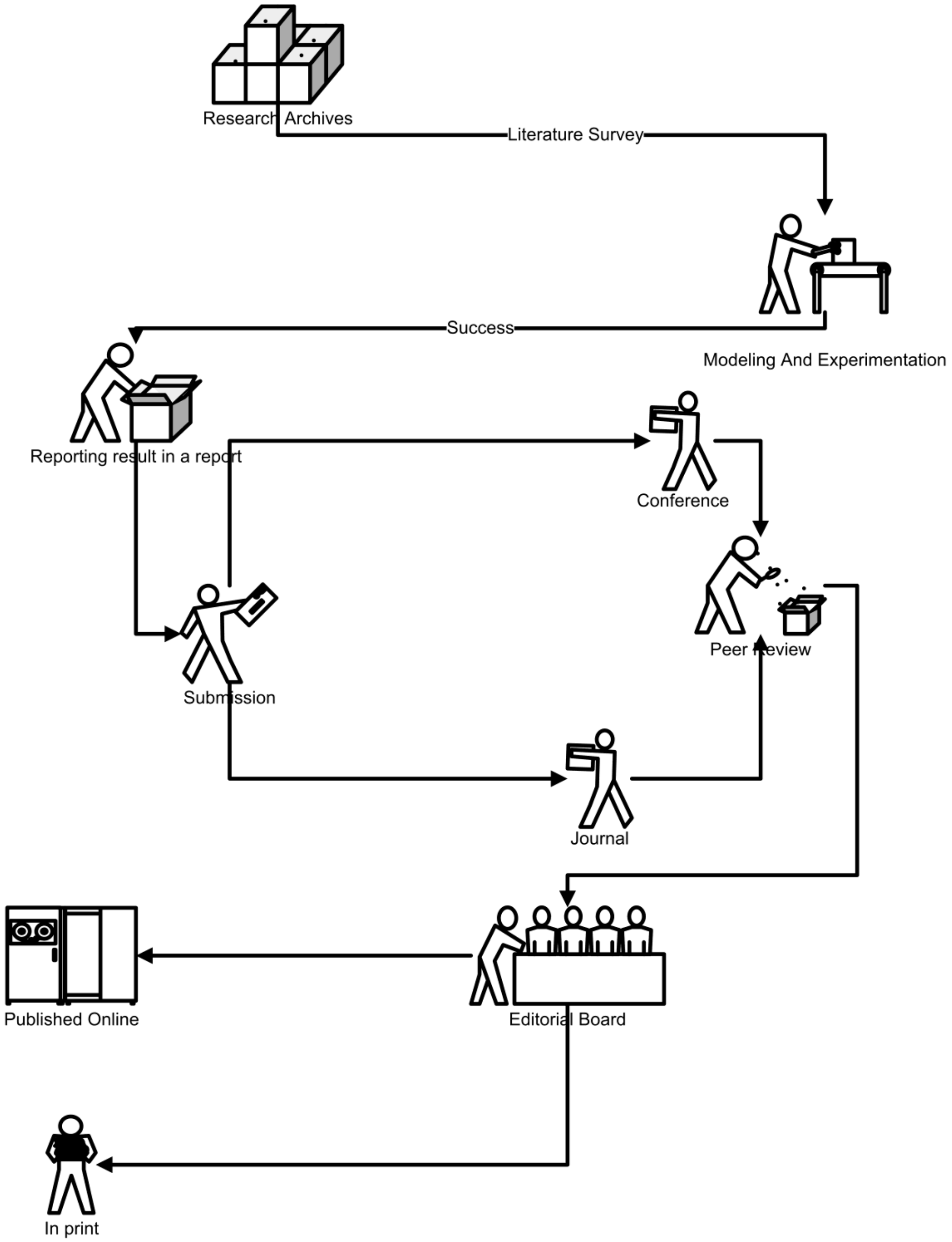

**Figure 77: Research Process I: Generalized research publication Workflow**

#### 6.3.3.2 Research Process II: Citations

After a paper has been published and is available for citation by other researchers by being listed in standard indices (such as the Google Scholar, Scopus or Thomson Reu-ters), the next step in the publication life cycle is the process of citation. Citations are the means by which the importance of a particular scientific article can be measured quantitatively. In other words, when other researchers read and consider a paper to be important, they cite the paper in their own papers and the process continues. Interestingly this process proceeds completely autonomously and thus cannot be strictly influenced by any single entity. Eventually the citation process can, at times, result in a high prestige of a researcher. This can be quantitatively measured by means of indices such as the h-Index. This measurement was first proposed in 2005 by Hirsch [217]. In general, the impact of a scientist can be measured by being highly cited by other authors. Egghe [218] presented a detailed overview of the inter-relation of Hirsch with related indices. These indices have been considered as effective measures of a scientist's impact or research productivity.

Hirsch index is formally defined as:

"A scientist has index $h$ if $h$ of his/her $N_p$ papers have at least h citations each, and the other $(N_p - h)$ papers have no more than h citations each."

So, as an example, suppose an author has 6 papers: 5 of them have 4 citations and the sixth has 1 citation. Then $h = 4$. It is important to note here that the calculation of h-Index is not simple. It involves several tasks which are performed behind the scenes. These include intelligent retrieval of information from some standard indexing source (such as Google Scholar or SCI, SSCI or Scopus etc.) and then calculation of the index by measuring the citations of each published paper. A well-known tool for the calculation of Hirsch index used extensively by researchers is Ann Harzing's "Publish or Perish" [219].

Earlier in Chapter 3, we examined the process of performing Scientometric analysis of citation data in depth by means of case studies. We noted earlier that the process of performing CNA involves pruning of nodes. However, while a standard network manipulation practice, automatic pruning can result in the loss of important and, at times, central nodes from the network inadvertently. Here the goal of this case study is to propose an agent-based model for simulation of an alternative Complex Network Modeling approach. This approach is the Temporal Citation Networks (TCN), which involves a two step process from bi-graph based author-paper Complex Networks such as those studied earlier in Chapter 3. The formal methodology of developing these TCNs as an experiment of the VOMAS methodology is as follows:

- First step is the removal of paper to paper links in the network resulting in a great reduction in the total number of links in the network.
- The second step is to visually align the network nodes in line with the h-index of the researcher.

After application of the two steps, the resultant network representation is capable of representing different features which were otherwise impossible to note earlier using only author-paper complex networks. These can be clearly noted in Figure 78. Here blue nodes represent the papers depicting the total number of citations inside the node. The black colored nodes represent the researchers, which are aligned in the 2-D Euclidean plane with a height equal to the h-Index of the researcher, scaled to the particular vertical dimension of the viewing plane. Thus, the goal of the case study is to evaluate the use of TCN in representing researcher evolution over time (i.e. temporally).

#### 6.3.3.3 Agent Description

The model contains two different types of agents i.e. the researcher agents and the research paper agents. The researcher agents are developed to model the researchers. As such, they contain two variables for state maintenance. The first of these is “num-papers” representing the total papers for the researcher. The other variable “my-papers” is representative of all the paper agents. In other words, this variable allows for an aggregation of these agents.

On the other hand, the research paper agent has three different variables. The first variable is the "tend-to-be-cited" variable. This variable ranges from 0 to 1 and is useful for simulation of random papers. The second variable is the "num-cites" which represents the current number of citations of the paper. The third "my-res" stores a pointer to the researcher, to which this paper belongs.

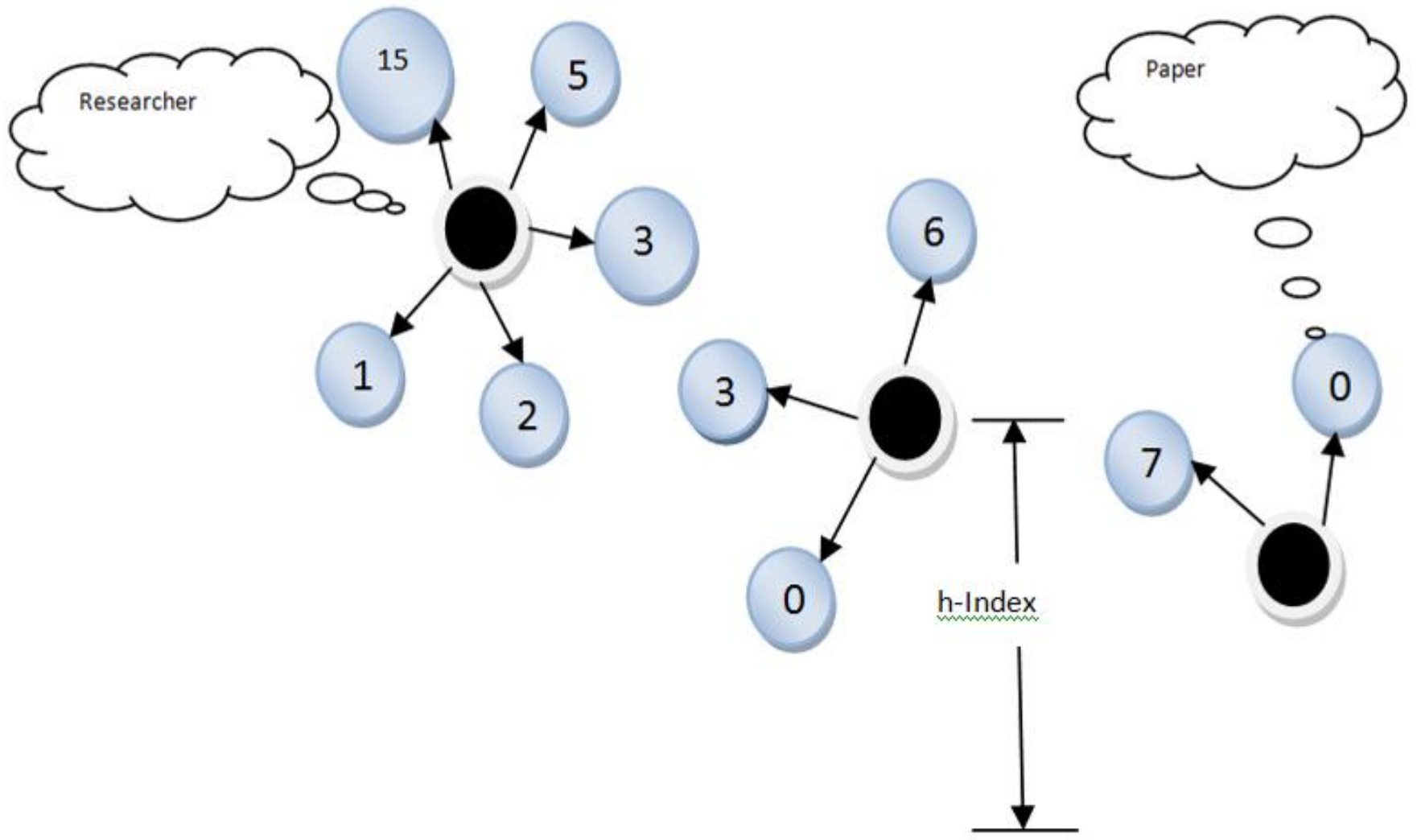

**Figure 78: TCN with black nodes depicting researchers placed according to their Hirsch index and blue nodes depicting individual research papers and citations**

In TCN, each paper will be replicated for each author. In other words if a paper has been co-authored by n authors, there will be n instances of the paper in the TCN. This allows for a reduction in clutter as well as helps in a succinct visualization to allow focusing on the evolution of the researchers.

#### 6.3.3.4 Validation Question

In this case study, we shall examine two different validation questions:

**Validation Question 1**: If the ABM of the TCN is able to represent any number of researchers, would the number of nodes in the TCN be similar to the value predicted earlier empirically?

**Validation Question 2**: Is the Hirsch index calculation of the ABM correct in terms of the actual data?

#### 6.3.3.5 VOMAS agent design

Here, for the first validation question, we shall design VOMAS agents who create a set of random researcher and paper agents to validate the representational capabilities of the ABM. However, for the second research question, we need VOMAS agents which can take in real researcher data and simulate agents based on the evolution of this particular researcher. Thus we have the following software contracts:

**Invariant I**: *If for any TCN, the pre-condition that "any random number of researchers are correctly represented" is true, then the post-condition that "the total number of links should be significantly lesser than the number of links in a author-paper network with the same data" is also true.*

**Invariant II**: *If the pre-condition that "researcher evolution data is correctly represented in a TCN" is true, then the post-condition that "the height of each node should correspond correctly to the Hirsch Index (h) of the researcher over time" should also be true.*

## 6.4 Results and Discussion

In the previous section, we developed three different case studies from multiple disciplines. Here our goal is to present the results of the experimentation involved in the validation of the simulations. It is pertinent to note here that while we can perform consid-erably more experimentation here similar to previously done in earlier chapters in each of these case studies, we shall however only focus on verification and validation of the effects of the case study to demonstrate the effectiveness of the proposed generalized methodology. Our focus here will be on the different effects of the simulation execution as well as the resultant effects on the VOMAS validation. Here, we shall demonstrate how VOMAS based validation builds on empirical validation methods while adding value addition in the form of being a customizable software engineering modeling scheme.

### 6.4.1 Case Study I

In our first case study, our goal was to simulate the forest and forest fire dynamics in considerable details. As such, we developed a set of advanced ABMs which allowed for evaluation of different aspects of the case study.

To perform the simulation experiments, we notice that for validation, we have to evaluate the following invariant which was described in the previous section:

"*If the pre-condition of a "tree is on fire" is true, then a post-condition of "FWI value in the range of Fire Danger index" is true.*"

Next we discuss first the simulation parameters followed by the results of simulation experiments for proving the validity of this invariant.

#### 6.4.1.1 Simulation Parameters

The experiments were conducted to empirically validate the increase of temperature associated with the forest fire spread. The simulation parameters are given below:

Number of VOMAS agents $n_v \in \{50, 100, 2000\}$

Forest density $d_{tree} \in \{60\%, 65\%, 70\%\}$

Fire delay in terms of simulation days $\Delta t_{fire} \in \{10, 30, 180\}$

Normalized average fire growth rate (area per simulation hour):

$$\frac{df_{spread}}{dt} = \{0.001, 0.002, ..., 0.9\}$$

Tree re-growth rate (new trees per simulation day in forest) $\frac{dn_{new}}{dt} \in \{0.001, 0.005, 0.01\}$

Initial average temperature of forest $t_{ave}(^{o}C) \in \{5, 10, 20, 30, 40\}$

Average relative humidity (%) $h_{ave} \in \{10\%, 20\%, ..., 50\%\}$

#### 6.4.1.2 Validation discussion

The results of the first set of experiments can be seen in Figure 79a. Here, we note how detected forest fires intensity increases with an increase in detected temperature. The second result in the Figure 79b depicts the effect of the probability of fire ignition with an increase in the FFMC.

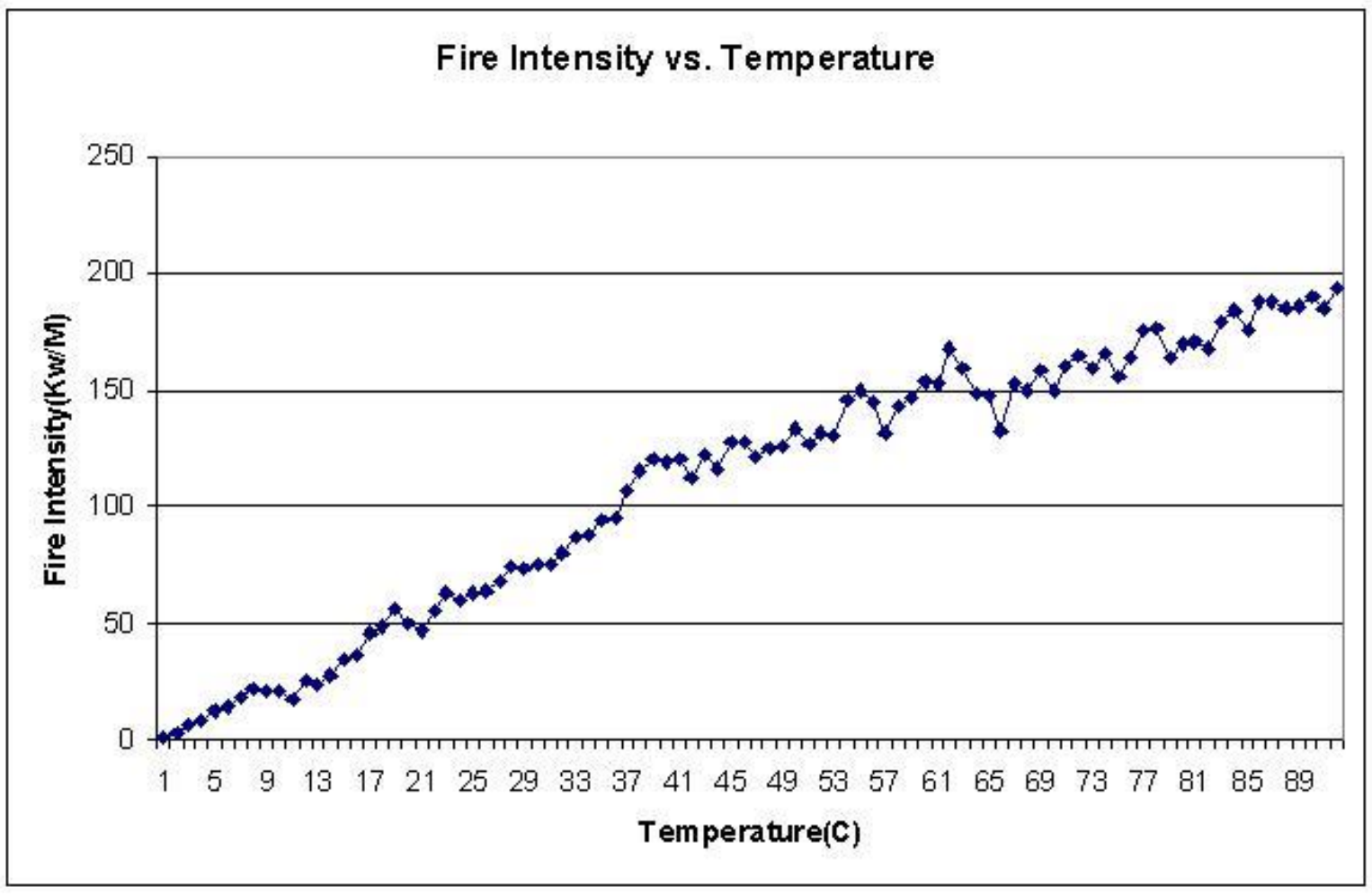

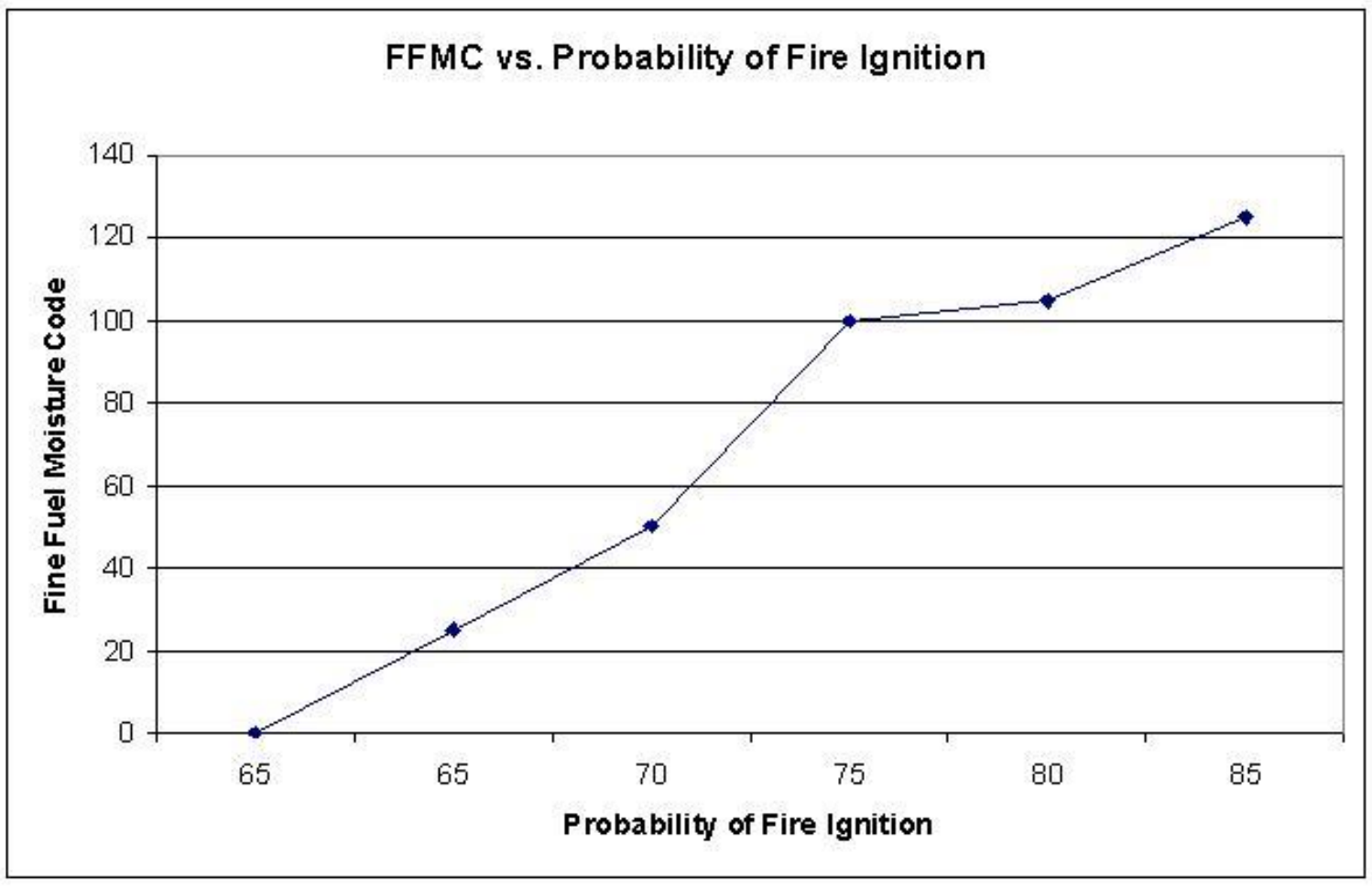

**Figure 79: Relationship between (a) Fire Intensity and temperature, (b) Probability of fire ignition and fire fuel moisture code**

We also note the relationship between simulated FWI vs. the FFMC value as a means of verification of the correct working of the model as given in Figure 80. By means of a realistic reflection of the relation of the different values, we can thus deduce that the FWI calculations are realistic and working.

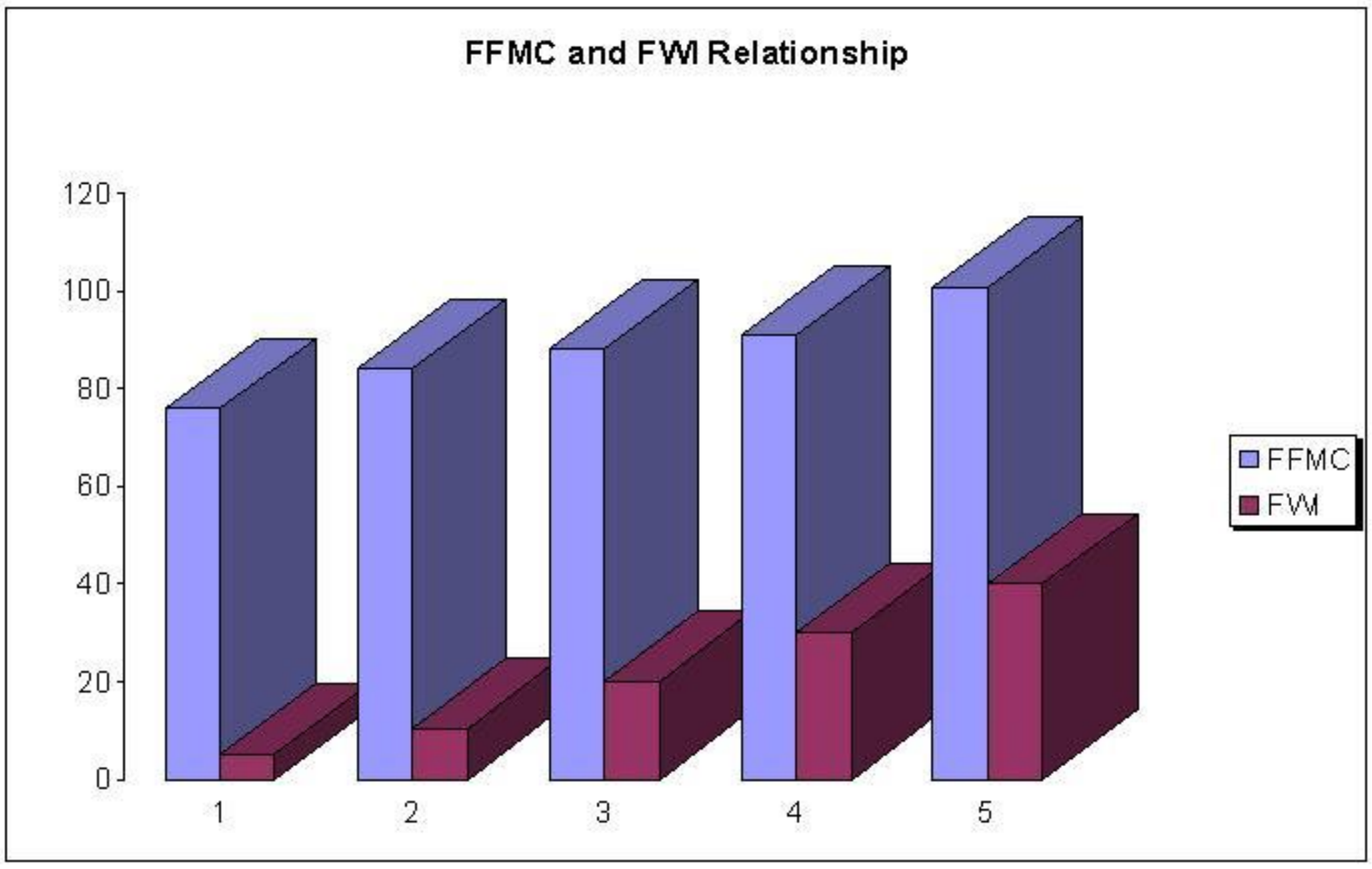

**Figure 80: Relation of FFMC with an increase in FWI**

The table for testing the invariant is given in Table 22.

**Table 22 Table of Fire Danger Index values**

| FWI value | Fire risk |
|---|---|
| 0-5 | Low |
| 5-10 | Moderate |
| 10-20 | High |
| 20-30 | Very high |
| 30+ | Extreme |

For validation of the invariant, the results were plotted as shown in Figure 81. As can be noted, the randomly placed tree agents who are acting as VOMAS agents are successfully able to detect the fire.

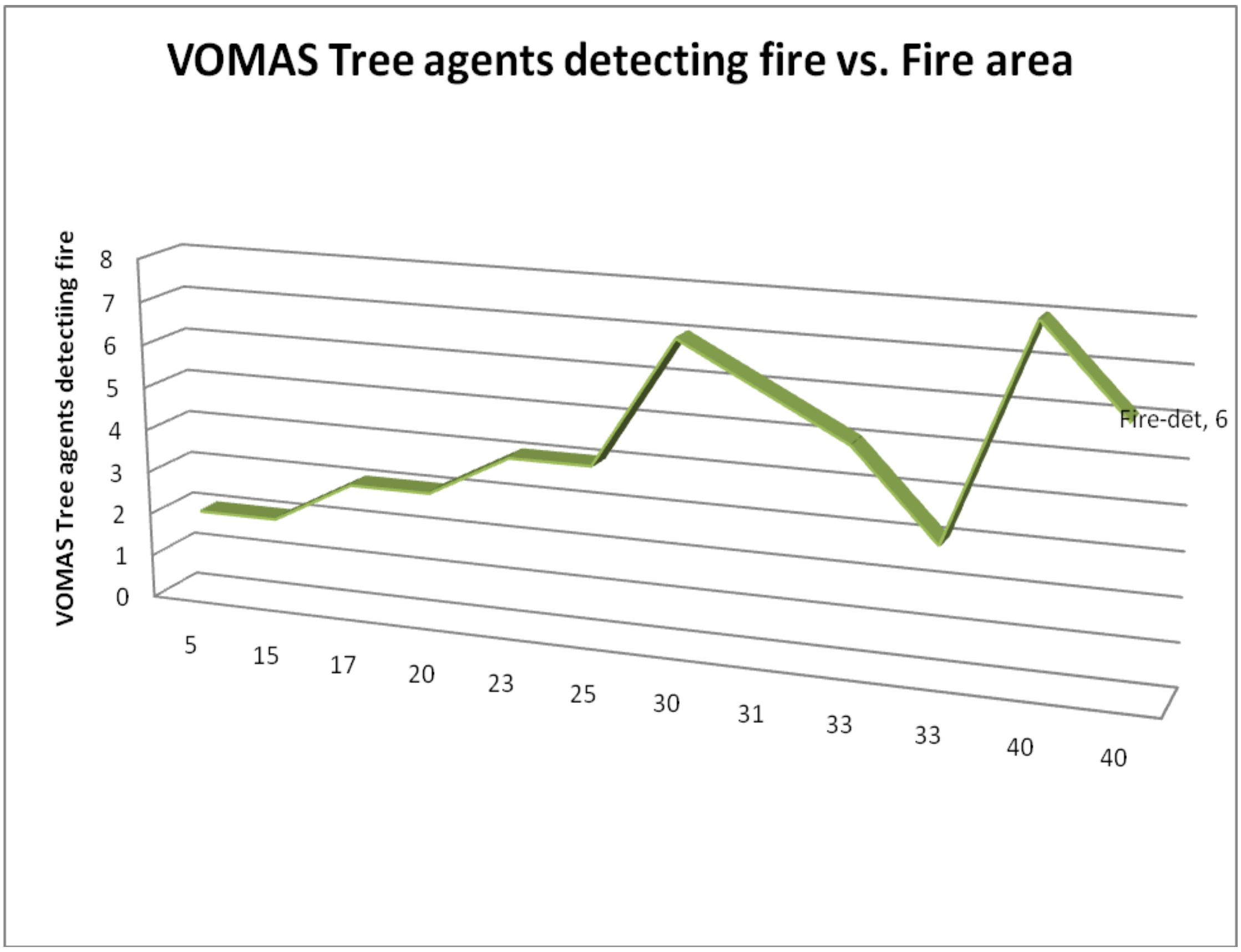

**Figure 81: Tree VOMAS agents detecting fire vs. Fire area**

In terms of the detected FWI in the case of fires, we can observe the detected behavior in Figure 82. We can note here that the FWI value detected is with the fire danger as shown earlier in the fire danger index table. Here, while the VOMAS agents are scattered throughout the simulation environment, they are still able to detect the effects of the forest fires of reasonable sizes.

Thus, we have noticed that by means of VOMAS, we have been able to perform validation of the case study model in addition to using empirical validation methods.

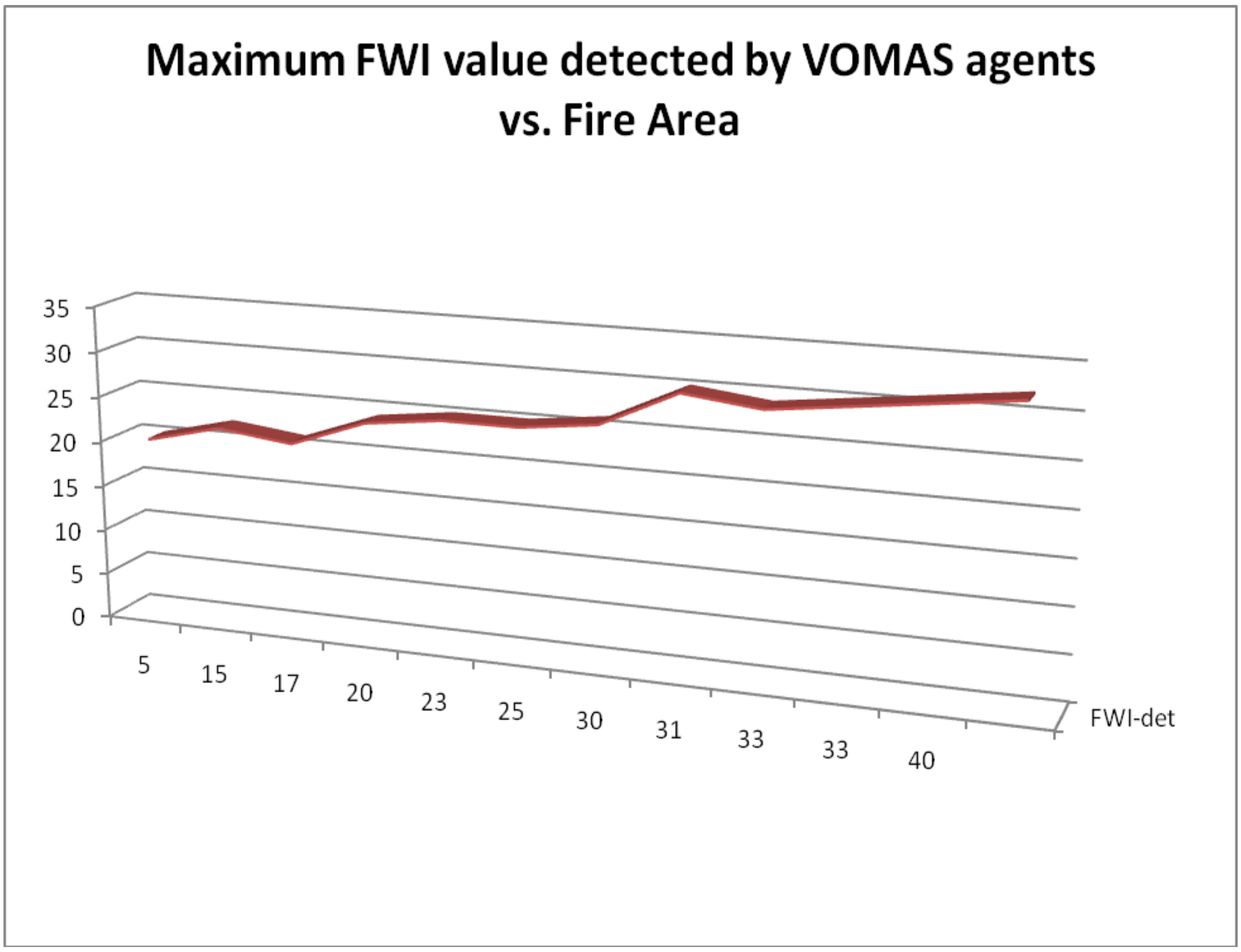

**Figure 82: Maximum FWI detected by VOMAS tree agents plotted vs. the area of fires**

### 6.4.2 Case Study II

In the previous case study, we validated the ecological simulation of forest fires. In the second case study, our goal was to combine the earlier case study of fire simulation with a multi-hop environmental monitoring WSN simulation developed as a QUDG as explained earlier in the chapter.

We can recall here, the invariant which was designed for use in the validation of this case study ABM. The invariant can be re-stated as follows:

*If the pre-condition that "sensors are correctly reading the values from their neighborhood" is true, then the outcome of the FWI calculation by the sensors would result in a post-condition of "correct prediction of fire in the vicinity of the sensor node".*

#### 6.4.2.1 Simulation Parameters

Since this ABM is based on the previous ABM which was on ecological modeling of the structural evolution of a forest, some of the parameters of the experiments are similar to the previous case study. While others are new, being related to the sensor nodes as given below:

Number of VOMAS agents $n_v \in \{50, 100, 2000\}$

Forest density $d_{tree} \in \{60\%, 65\%, 70\%\}$

Fire delay in terms of simulation days $\Delta t_{fire} \in \{10, 30, 180\}$

Normalized average fire growth rate (area per simulation hour):

$$\frac{df_{spread}}{dt} = \{0.001, 0.002, ..., 0.9\}$$

Tree re-growth rate (new trees per simulation day in forest) $\frac{dn_{new}}{dt} \in \{0.001, 0.005, 0.01\}$

Initial average temperature of forest $t_{ave}({}^{o}C) \in \{5, 10, 20, 30, 40\}$

Average relative humidity (%) $h_{ave} \in \{10\%, 20\%, ..., 50\%\}$

Number of sensor nodes $n_{sensors} \in \{100, 200, ..., 500\}$

Normalized distribution of sensors (in terms of sensors per square simulation area)

$$\frac{n_{sensor}}{A_{sim}} \in \{0.0013, 0.026, 0.013\}$$

Maximum sensor communication radius $\max(R_{comm}) \in \{20, 30, 40, 50\}$

Maximum sensor sensing radius in terms of simulation measure $\max(R_{sens}) \in \{2, 5, 7, 10\}$

6.4.2.2 Validation discussion

Here, our basic simulation model can be viewed in Figure 83. As can be noted from the simulation output, the sensors are all connected with each other however not every sensor will be directly accessible from each other sensor using a single hop. Thus, the information in this WSN needs to be aggregated and confirmed with local sensors before it is forwarded to the remote sink.

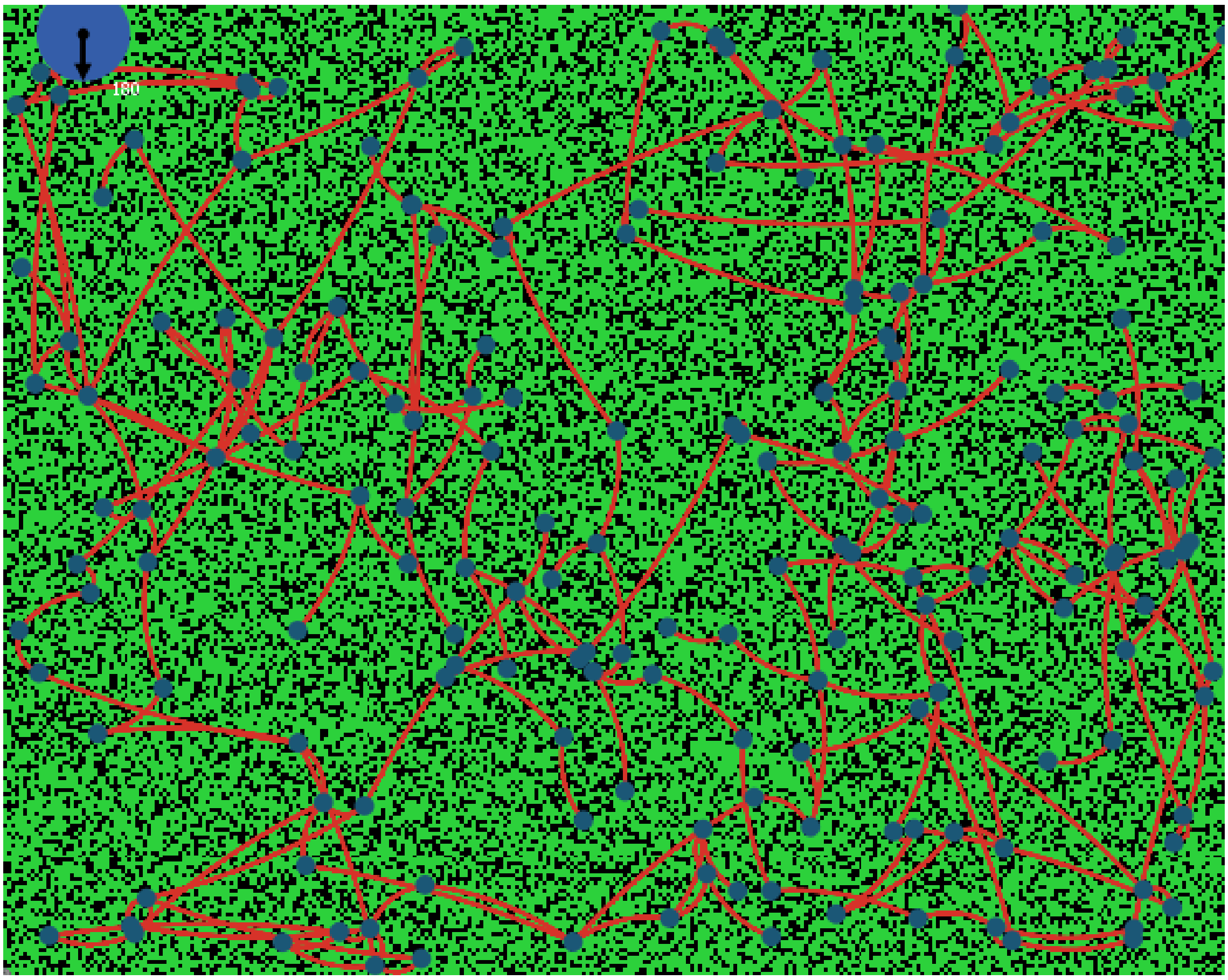

**Figure 83: Basic simulation model of multi-hop WSN modeled as a QUDG monitoring forest conditions**

When the simulation is actually executed, each of the WSN nodes measure local parameters and display them as labels as can be noted in Figure 84. In this particular design, as noted earlier, the number of sensors is limited as compared with the number of tree nodes in the case of environmental modeling; as such the calculation of FWI in these nodes periodically (e.g. every 1000$^{th}$ tick) would not be a problem for simulation execution.

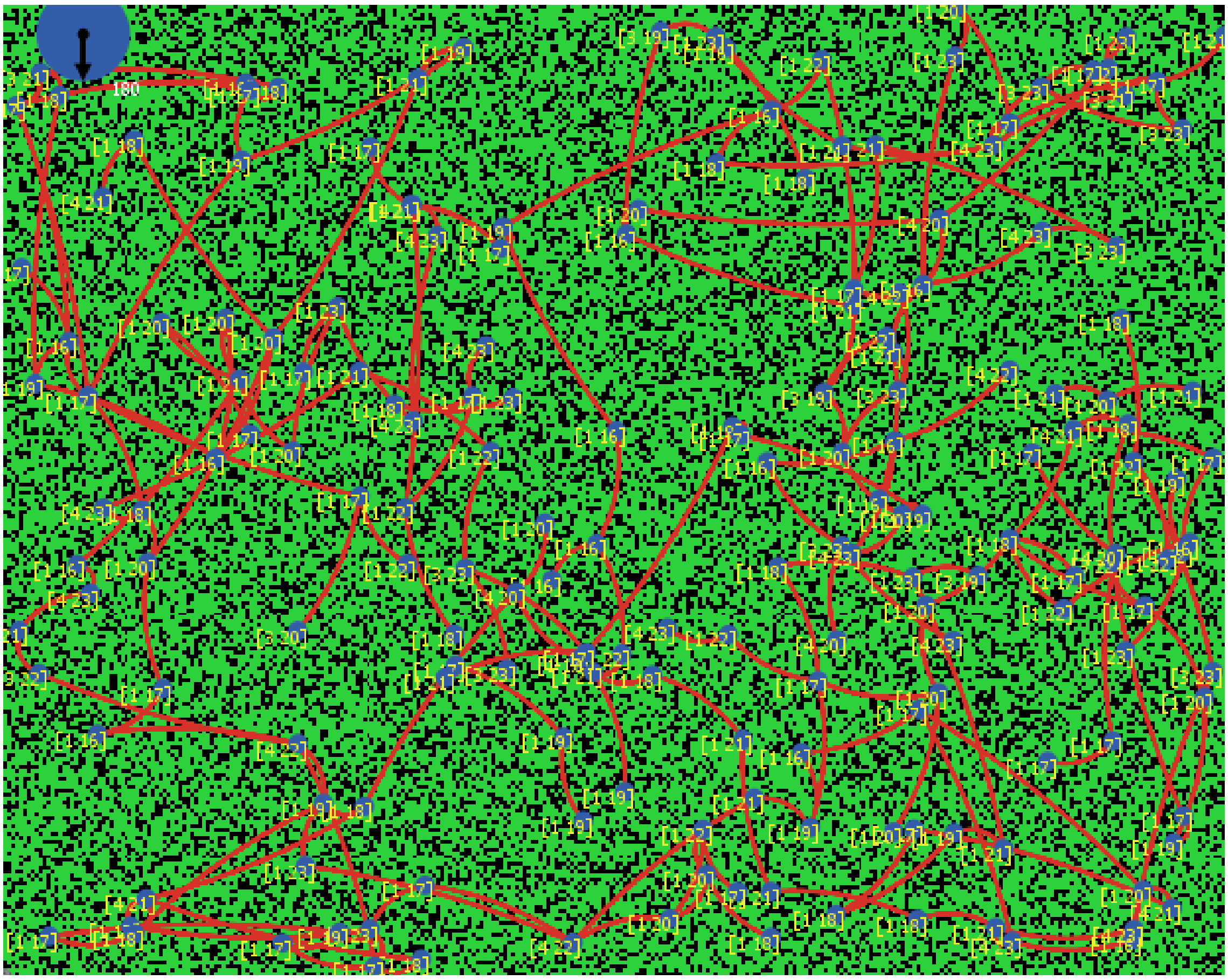

**Figure 84: WSN monitoring the local environment displaying measured parmeters**

The structure of the forest as well as the weather evolves over time with fires, snowfall and rainfall occurring at random instances. The result of a simulation where the fire has caused an impact on a major section of the forest monitored using the WSN can be observed next in Figure 85. Here, we can also note that one of the center nodes demonstrates the fire was very recent. This phenomenon can be noted visually by means of observing the changing of color of the appropriate WSN node to yellow.

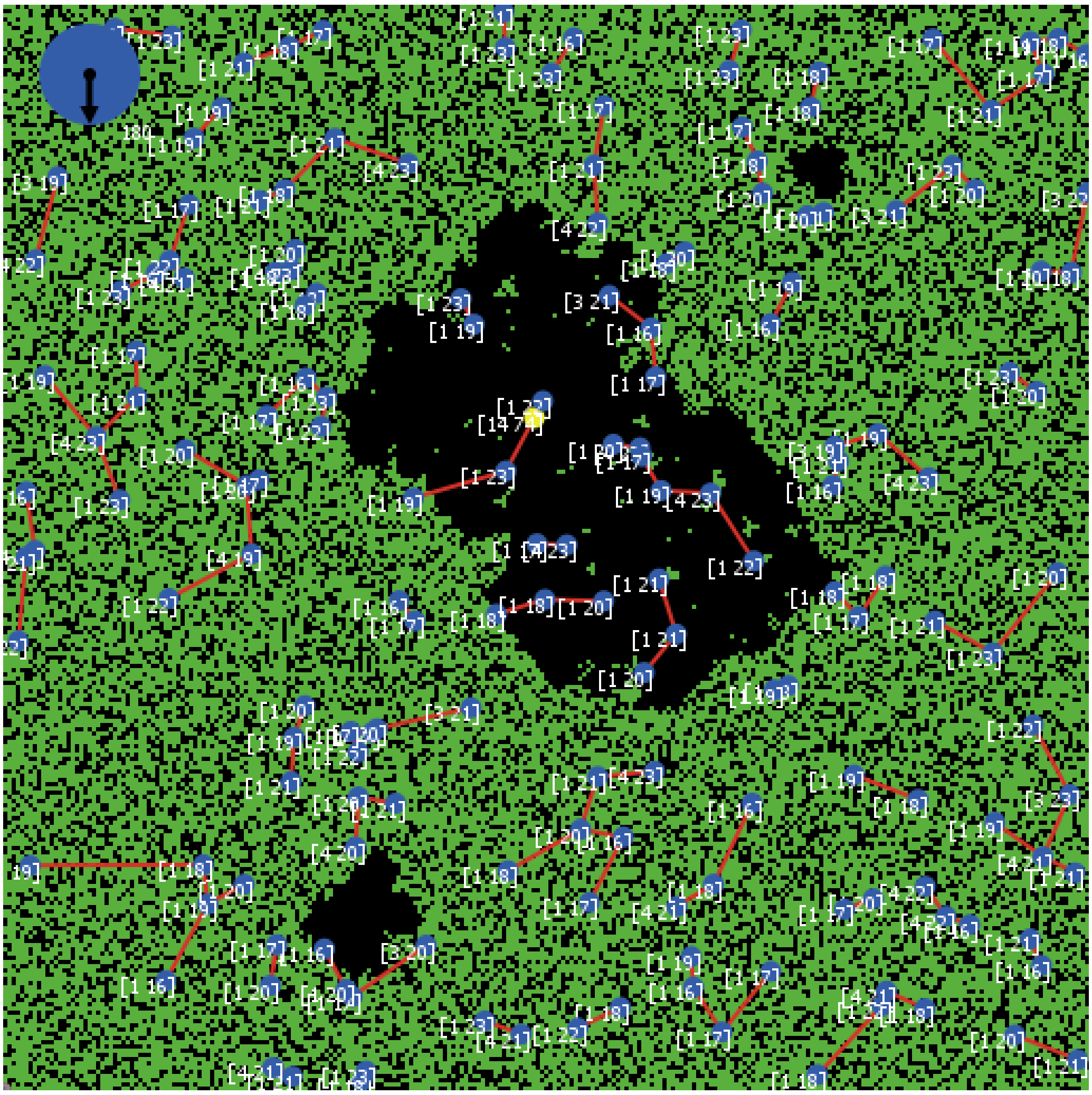

**Figure 85: Wireless Sensors recording forest temperature and calculating FWI**

Different simulation parameters added to the simulation model can now be seen in the configuration panel shown in Figure 86.

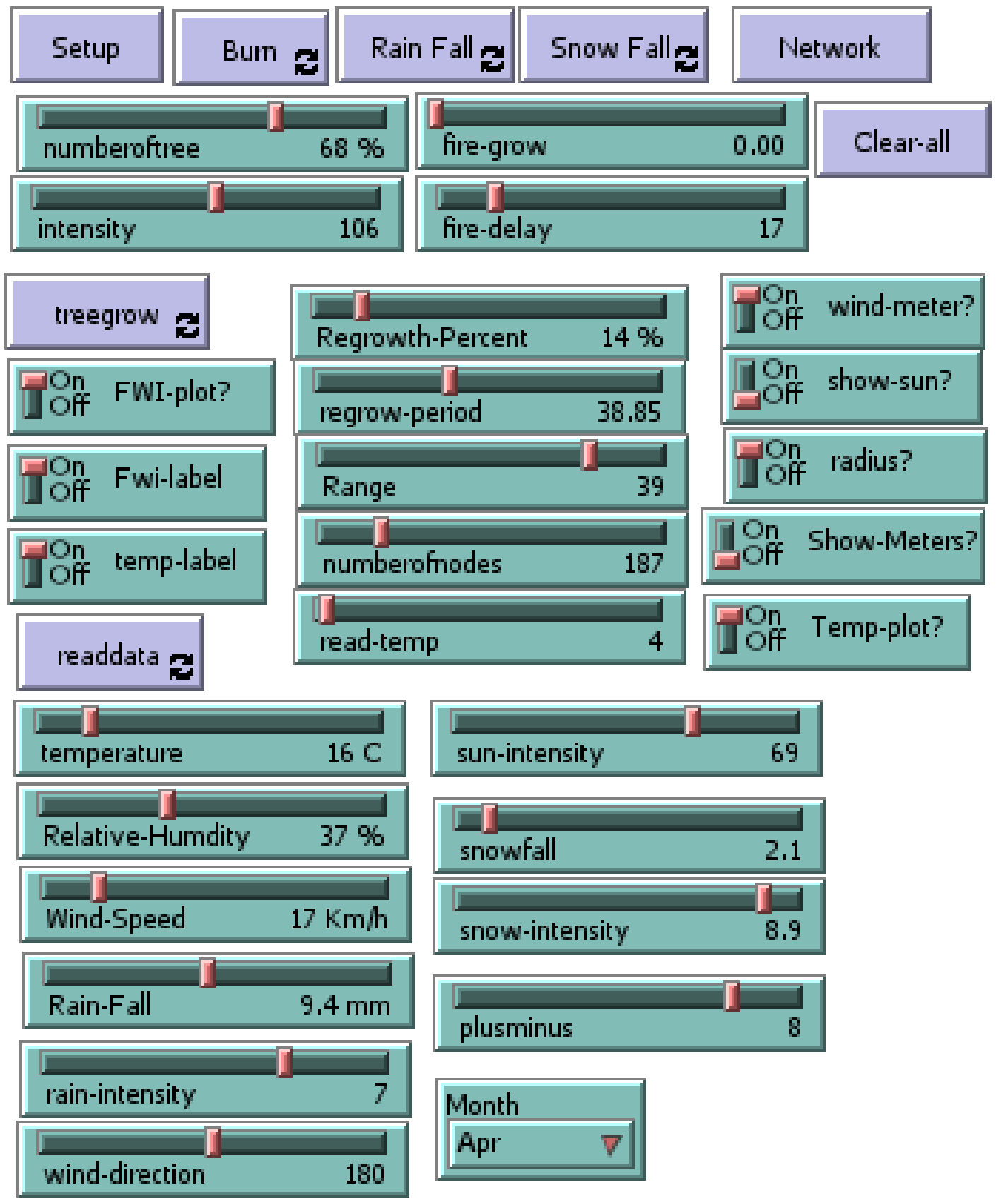

**Figure 86: Configuration Options of simulation models of combining Forest Fire simulations with monitoring simulation by a WSN**

A detailed view of the simulation of several fires in a forest with observed FWI values is given.

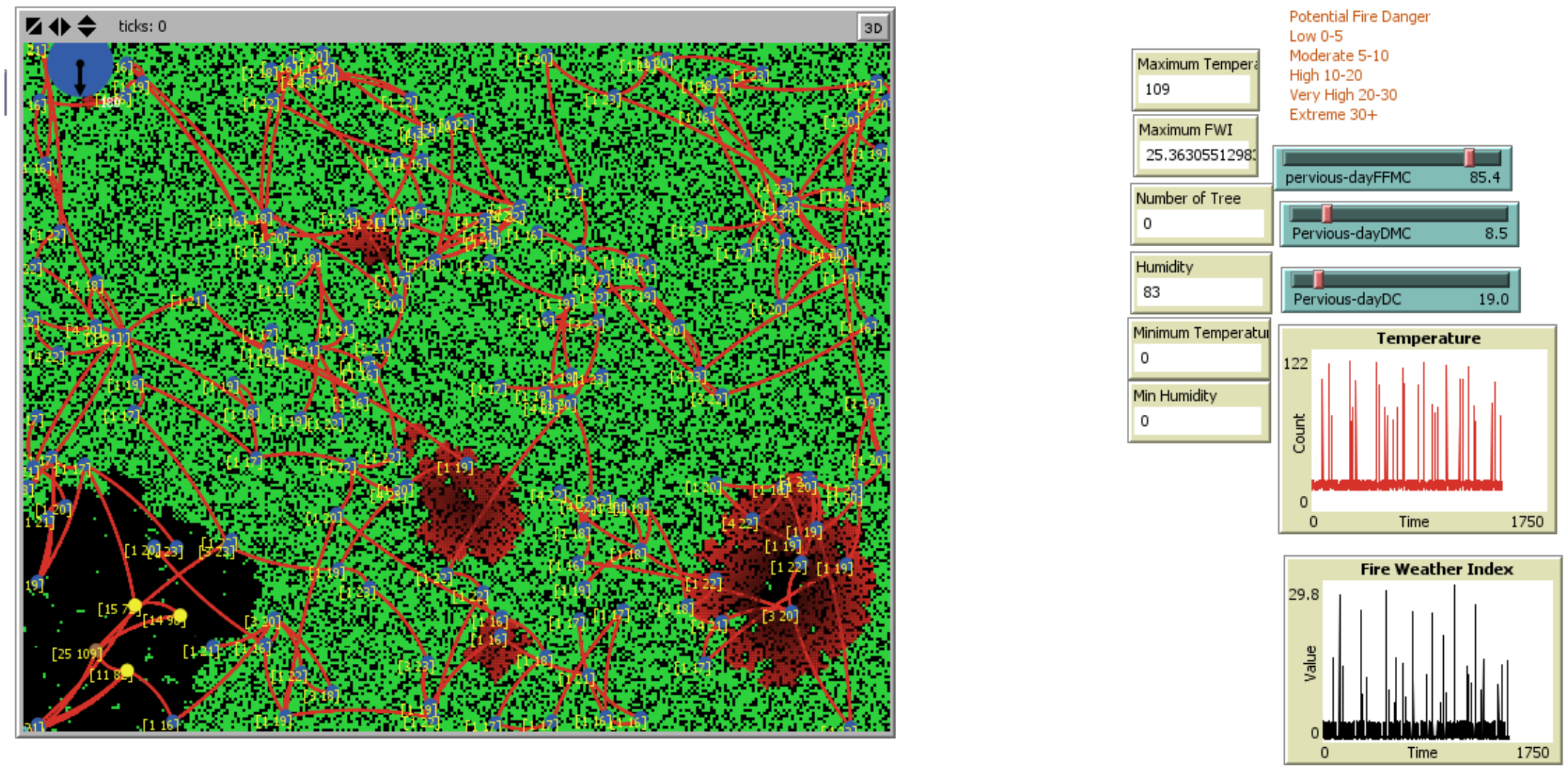

**Figure 87: Effects of monitoring large set of fires and FWI value**

As we can see in Figure 87, the temperature as well as the observed FWI values exhibit similar variations. In addition, we note that the observed value of FWI greater than 20 is demonstrating the effects of the forest fires correctly validating the ability to monitor forest fires using the sensors.

For a formal validation of the fire detection invariant, we plot the earliest time of fire detection vs. the average distance of forest fires as shown in Figure 88.

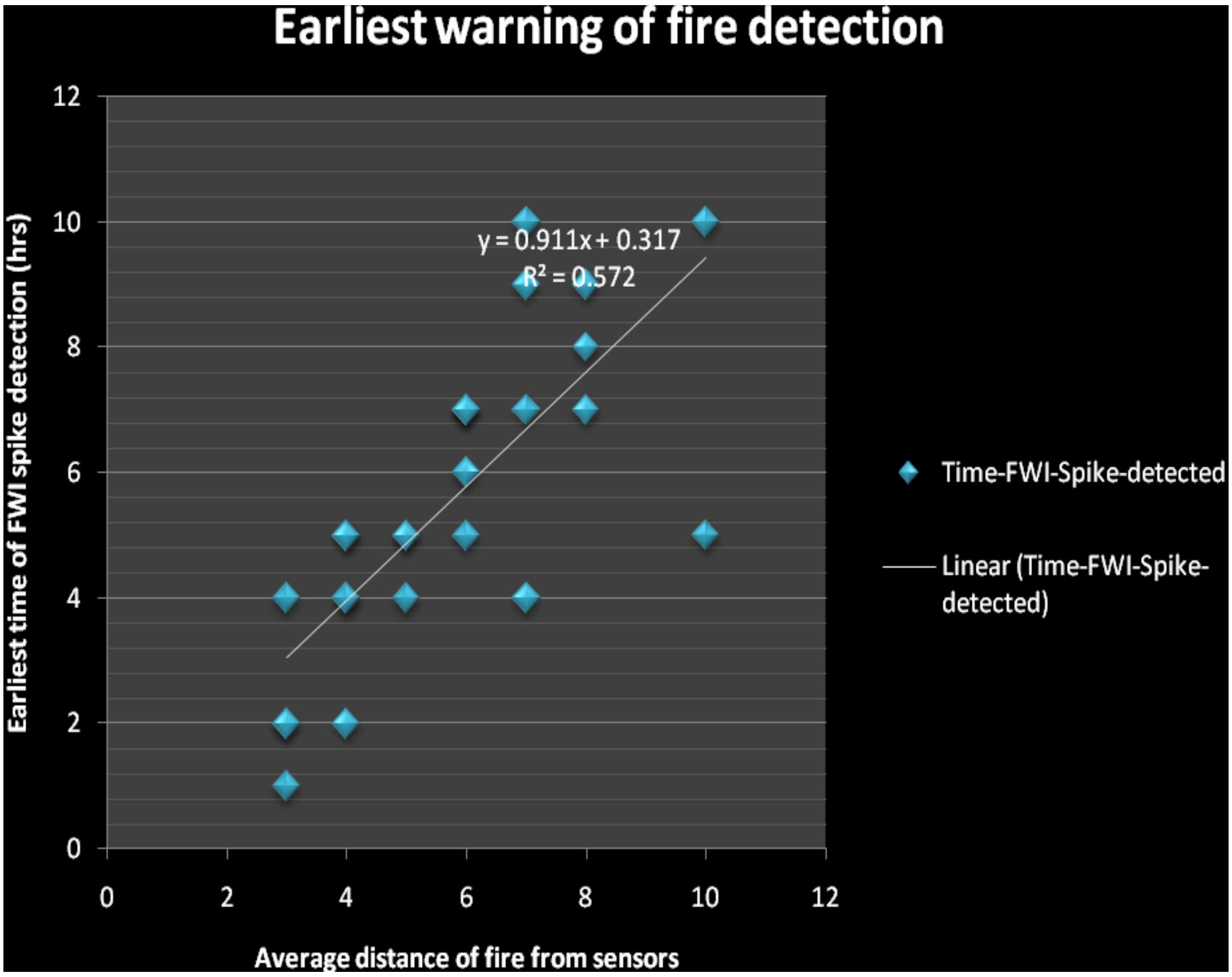

**Figure 88 :Earliest detection of FWI spike vs. average distance of sensors from fire incident**

Here we can clearly note that using the linear interpolation line, the detection follows a trend line of $y = 0.911x + 0.317$ with a reasonable R-squared value of $R^2 = 0.572$. Here the fact that this value is greater than 0 demonstrates the possibility of prediction of one value based on the other value (in linear regression).

Thus, in this case study, we have again demonstrated the effectiveness of using VOMAS to verify and validate the simulation model using the invariant contract. In this particular case, again, our validation was coupled with existing model validation techniques of empirical validation and cross-model validation using multiple cooperative agents in the same simulation model.

### 6.4.3 Case Study III

While the previous two case studies were from ecological modeling and telecommunications respectively, our third case study, which was developed in the chapter previously is from Social Sciences and the two invariant contracts for this case study can be stated as follows:

**Invariant I**: *If for any TCN, the pre-condition that "any random number of researchers are correctly represented" is true, then the post-condition that "the total number of links should be significantly lesser than the number of links in a author-paper network with the same data" is also true.*

**Invariant II**: *If the pre-condition that "researcher evolution data is correctly represented in a TCN" is true, then the post-condition that "the height of each node should correspond correctly to the Hirsch Index (h) of the researcher over time" should also be true.*

#### 6.4.3.1 Simulation parameters

In this case study, the way to understand the experimental results is that for validation of the hypotheses, we have developed essentially two different types of validation VOMAS mechanisms inside the same ABM. The first validation VOMAS is specifically for the first invariant allowing for the creation of a random number of researchers using VOMAS agents based on the TCN hypothesis based complex network model. The second validation VOMAS demonstrates the validation of the second invariant by allowing for the mapping and loading of actual Hirsch index data and demonstrating the evolution of a well-known researcher. The simulation parameters of the two validation simulation experiments are explained below:

Number of researcher nodes $n_{res} \in \{10, 20, ..., 100\}$

Maximum number of initial papers per researcher $\max(n_{papers}) \in \{10, 20, ...40\}$

Average tendency to be cited is a probability which is a uniform random, $P_{cited} = rnd(0,100)$

6.4.3.2 Validation of the representational abilities of TCN models

Researchers are each placed on the Cartesian coordinate system, based on their Hirsch index. Each researcher, shown in black, has two labels. The first label shows the total number of published papers and the second shows the Hirsch index. Each paper, shown in blue, has the total citations as a label. This experiment validates the representational ability of the TCN modeling paradigm for researchers as shown in Figure 89. Notice that this validates the first invariant since we have demonstrated the capability of representation of any number of researchers on the basis of TCN model in simulations.

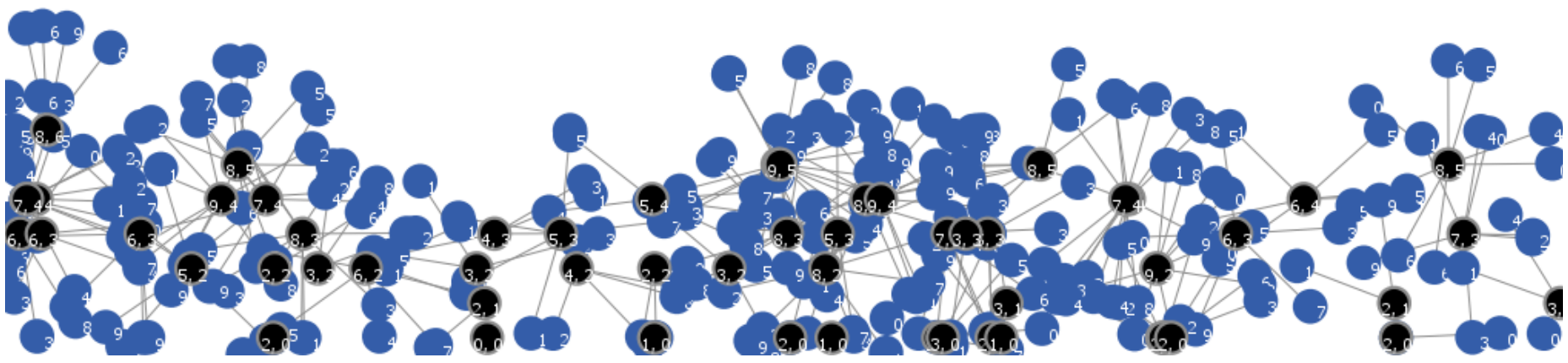

**Figure 89 :**Simulation view of n = 60 random researchers with max-init-papers = 10

6.4.3.3 Validation of Hirsch Index Calculation

For the validation of the second invariant, we shall use concepts from empirical validation by means of calculating evolution of actual Hirsch index values for a renowned researcher in the domain of multiagent systems. The chosen researcher was Prof. Victor Lesser because of his relatively high h-Index as noted in various citation indices. Using "Publish or Perish" software [220], the Google scholar citation index was queried. It was discovered firstly that the index lists a total of 649 papers. However, 546 of these papers have valid and properly indexed years which were therefore used in the subsequent validation exercises. The first indexed paper of the researcher is indexed from the year 1968. The data for the next 20 years for "Victor Lesser" has been plotted in Figure 90.

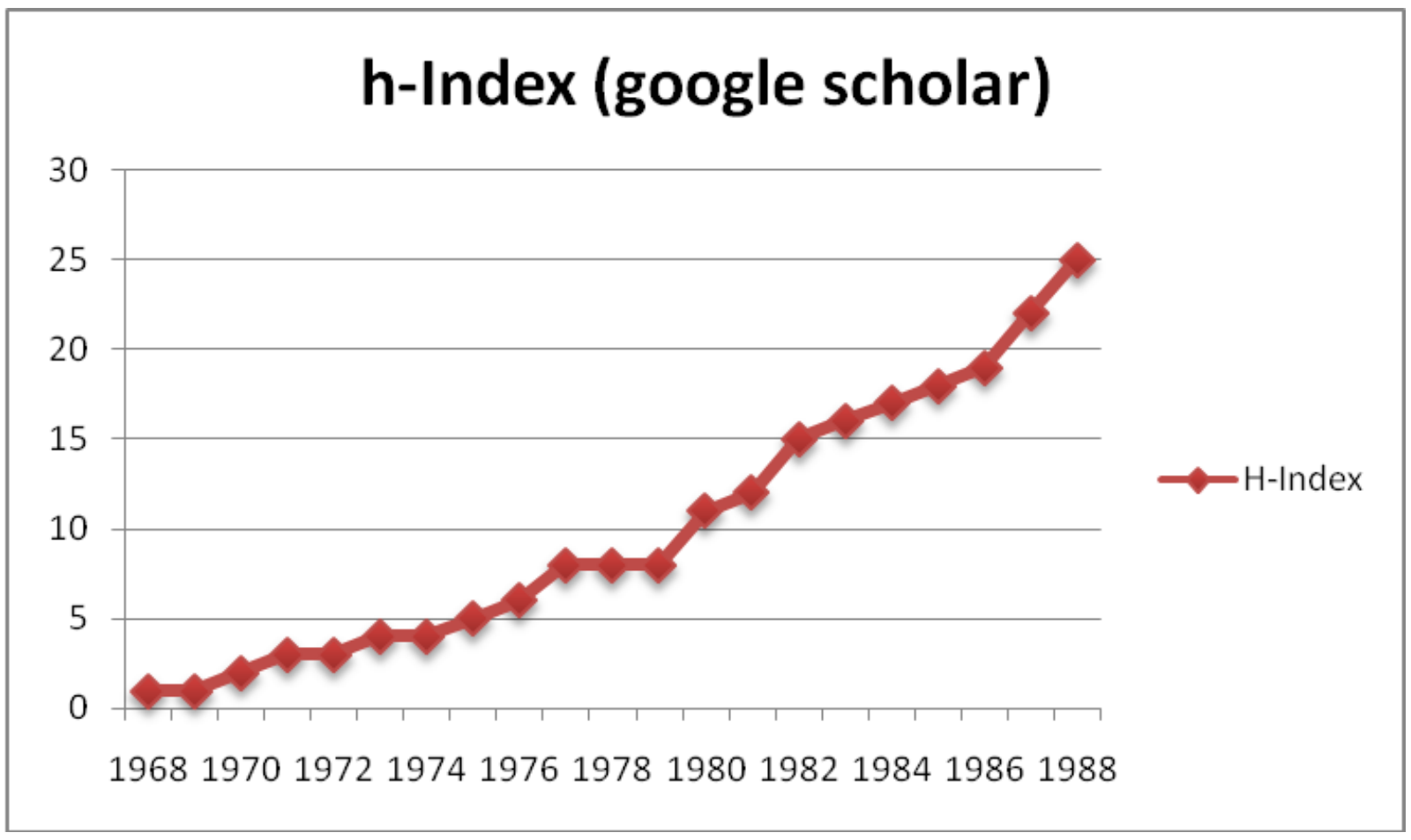

**Figure 90: h-Index plot for twenty years for "Victor Lesser" obtained using Publish or Perish program**

Next, we simulate this using our ABM as shown in Figure 91a and b which show ten years of evolution of h-Index of Victor Lesser as depicted using TCN by means of the simulated agent-based model. The detailed results and the retrieved data (via Google scholar index) are depicted in the experiments are given in Table 23.

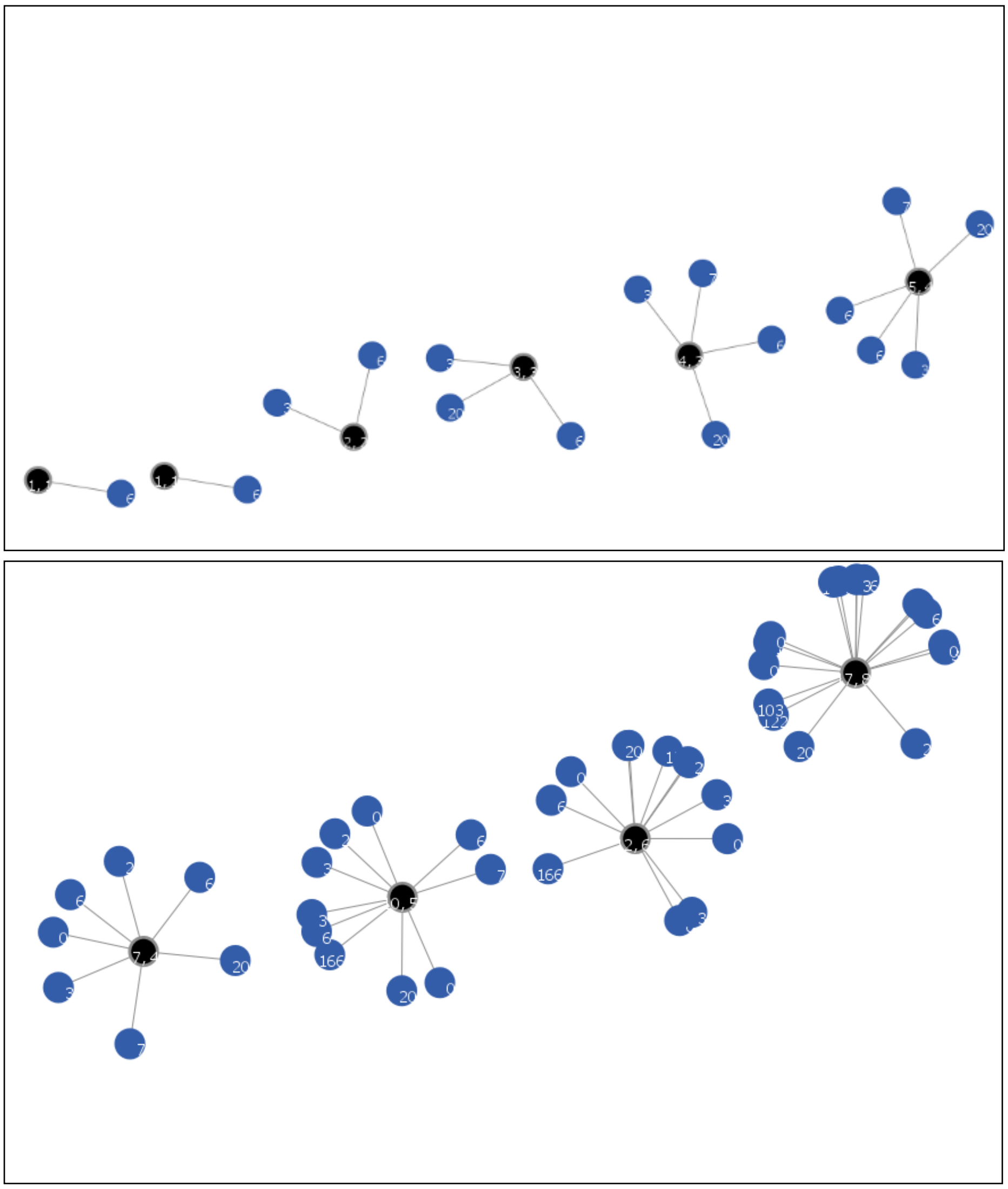

**Figure 91: a, b Validation exercise 1: Evolution of a researcher's Hirsch-index (Victor Lesser)**

**Table 23 Data for validation. Calc hI = Calculated h-Index and h-Index is the h-Index from Publish or Perish software which obtains data via Google Scholar (Current as of July, 2011)**

| Year | H-Index | Citations of papers | Calc hI |
|---|---|---|---|
| 1968 | 1 | 6 | 1 |
| 1969 | 1 | 6 | 1 |
| 1970 | 2 | [6 3] | 2 |
| 1971 | 3 | [20 6 3] | 3 |
| 1972 | 3 | [20 7 6 3] | 3 |
| 1973 | 4 | [20 7 6 6 3] | 4 |
| 1974 | 4 | [20 7 6 6 3 2 0] | 4 |
| 1975 | 5 | [166 20 7 6 6 3 3 2 0 0] | 5 |
| 1976 | 6 | [166 21 20 15 7 6 6 3 3 2 0 0] | 6 |
| 1977 | 8 | [166 122 103 71 21 20 15 8 7 6 6 3 3 2 0 0 0] | 8 |

Table 23 shows two columns for the Hirsch index. One column shows the Hirsch index calculated using Google Scholar and the Publish or Perish program. On the other hand, the "Calc hI" column indicates the Hirsch index calculated by the algorithm alongside the evolution of the researcher. The table also shows the number of cited papers per researcher. Thus, we can see how we have validated both the calculation of the h-Index as well as the effectiveness of TCNs for the modeling of researcher reputation (state) co-evolution with the changes in the topology using our agent-based model.

In terms of the number of nodes required to be displayed in traditional author-paper citation networks versus TCNs, we can observe the results as shown in Figure 92.

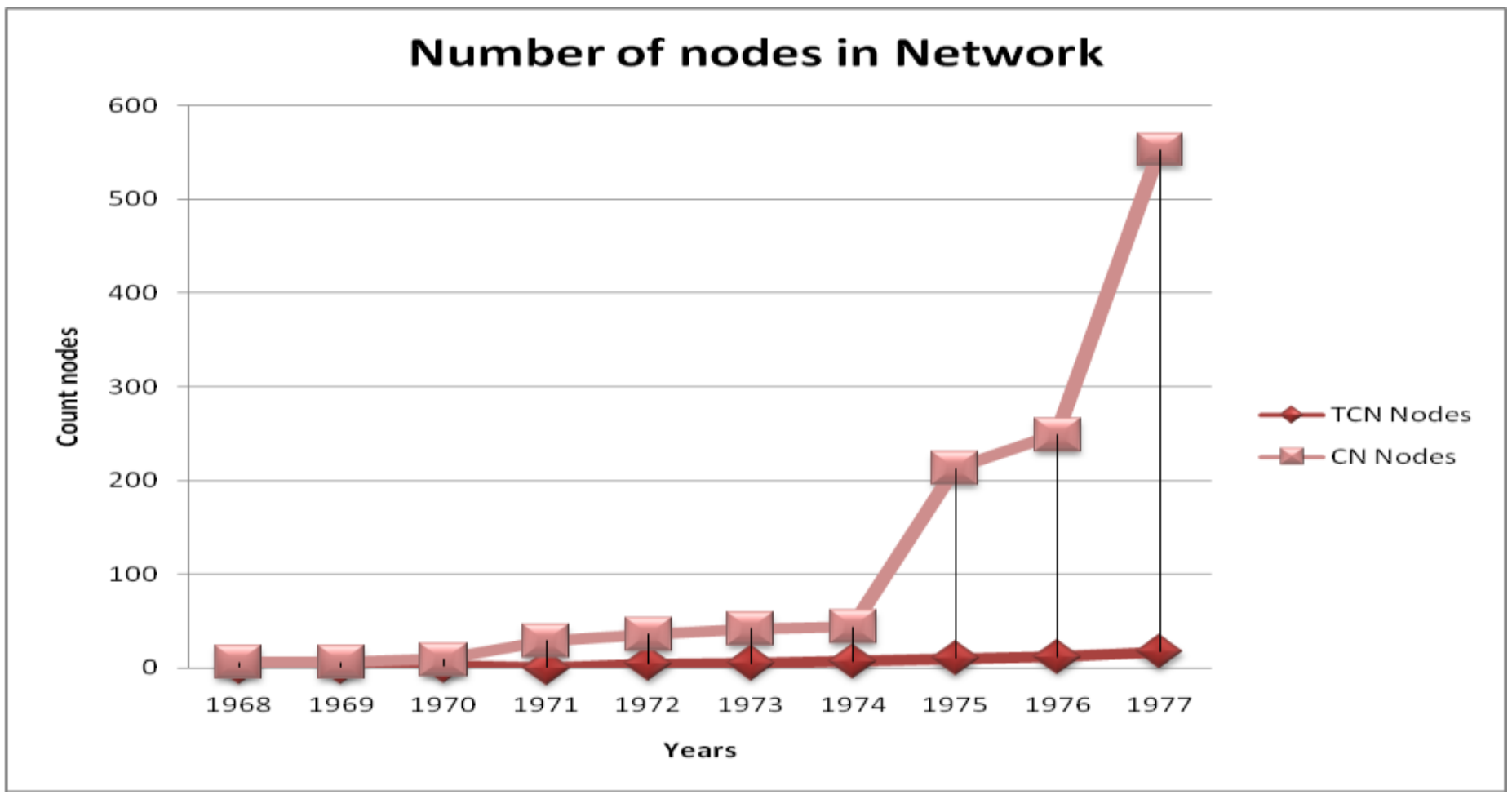

**Figure 92: Validation Exercise 2: Comparison of number of nodes needed to display a author-paper citation network vs. a TCN for Victor Lesser's papers.**

Thus, by means of the proposed methodology, we have demonstrated how to verify and validate the case study. Using two different invariants allowed us to demonstrate different aspects from the earlier two case studies.

### 6.4.4 Critical comparison of VOMAS with other approaches

In the previous section, we have demonstrated the broad applicability of the proposed in-simulation validation scheme for ABMs by application of the methods on three different case studies from different scientific disciplines. In this section, we evaluate existing methods in contrast with our methodological approach and analyze how this validation scheme fits in these methods prevalent in literature for the verification and validation of ABMs.

#### 6.4.4.1 Empirical Validation

Empirical validation is concerned with using empirical data to validate the ABM. Different researchers such as Marks [221] and Fagiolo et al. [202] have proposed the use of empirical validation of agent-based models. The proposed in-simulation validation methodology builds on top of existing empirical validation techniques. It essentially allows for a design of validation techniques using software engineering concepts. While it can still use empirical validation subsequently or else perform conceptual validation depending upon the particular type of simulation model or application case study (as has been shown earlier in the case studies). Thus there is no conflict here with the original related empirical validation methodologies.

#### 6.4.4.2 Validation using Companion Modeling Approach

In Social Sciences, there is an approach to agent-based modeling termed as the "Companion Modeling Approach". In companion modeling, the researcher or modeler are the one and the same and they work together with stakeholders from the authorities to develop models and results. By having stakeholders examine the implications, the models are in essence "face" validated. The process has been generalized by Bousquet and Trébuil [222] as shown in Figure 93. The idea is to go back and forth between different stakeholders and the researchers where the stakeholders are the authorities governing the businesses/economic ecosystems. However, this methodology does not differentiate between the simulation specialists and end-user researchers in any way. As such, while end-user researchers might be experts in their domain of modeling, they can be separate entities from the actual teams who can better develop the simulation models. This is one way in which how our proposed methodology can be considered to extend upon the "companion modeling" concept. In addition, our methodology brings in concepts from Software Engineering and Artificial Intelligence.

Another way in which our proposed methodology is different is the use of descriptive agent-based models (DREAM models as discussed in previous chapter) which are able to describe ABMs across scientific disciplines. In the proposed companion modeling approach, the goal was to develop simulations with only stakeholders from the business ecosystem modeling and not multidisciplinary validation. Still, in this way, the proposed VOMAS-based approach extends this companion modeling set of concepts which were designed for specific application case studies to generalized multidisciplinary approaches. Again, this implies there is no direct conflict between the two approaches and VOMAS intelligently uses and extends existing techniques.

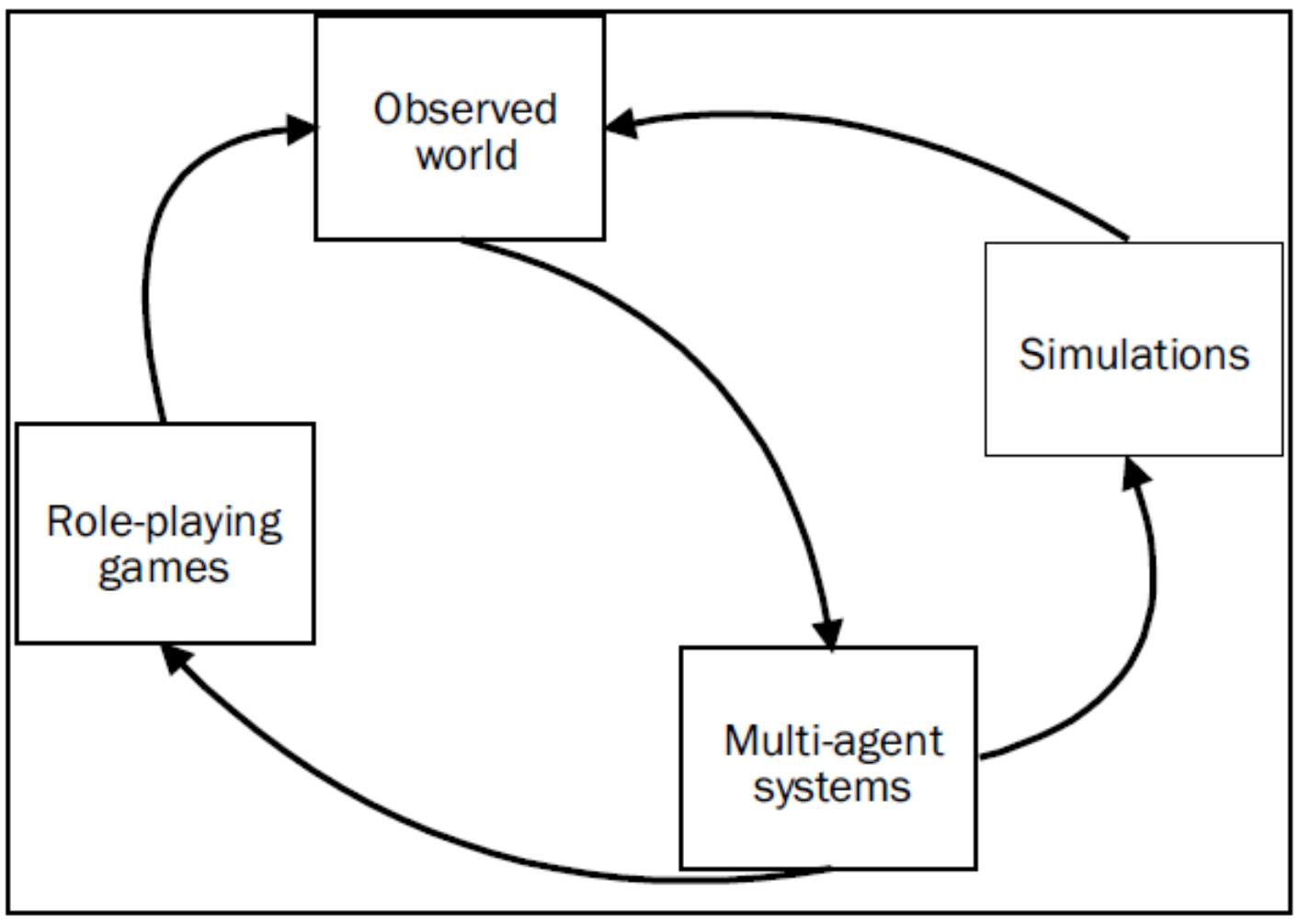

**Figure 93: Companion modeling cycle, figure credit: Bousquet and Trébuil [222]**

## 6.5 Conclusions

In this chapter, we have presented the proposed in-simulation verification and validation methodology of building customized VOMAS inside different ABMs. This validated agent-based modeling level of the framework builds upon previous framework levels such as exploratory agent-based modeling and descriptive agent-based modeling allowing for validated agent-based modeling. As a means of unification of all ideas and concepts, the methods and case studies extensively involve both the use of agent-based models and complex network models and methods in different scientific disciplines. We presented three different case studies from different scientific disciplines including ecological modeling, telecommunications and social sciences demonstrating the broad applicability of the proposed methods involving building customized validation schemes based on the particular case study.

# 7 Conclusions and Future Work

This thesis represents first steps towards the solution of a set of nontrivial problems associated with the colossal task of developing a unified framework applicable to multidisciplinary and inter-disciplinary cas research studies. We propose a combination of agent-based modeling and complex network approaches, two paradigms which are prevalent in cas modeling literature but, to the best of our knowledge, have never previously been combined in the form of a framework enabling comprehensive cas inter-/multi-disciplinary research studies to be carried out based on expected goals and outcomes of the studies.

In this chapter, we first give an overview of the thesis contributions. Next, we critically evaluate the different framework levels and case studies in relation to other state-of-the-art techniques prevalent in cas modeling literature. This is followed by details of the limitations of our work and finally we propose how the different framework levels can be utilized in other cas research case studies in addition to how our proposed framework levels can be further enhanced by other researchers in long term studies.

## 7.1 Overview

To demonstrate the applicability and use of each of the different proposed framework levels, we had to work in different scientific disciplines and nomenclatures. At times, we had to work with domain experts from a variety of backgrounds to develop the different case studies. The application case studies ranged from consumer electronics, agent-based computing, complex adaptive communication networks, ecological modeling, scientometrics and social sciences. While the framework levels propose a systematic set of ideas for modeling and simulation of cas depending on the data types available and the case study objectives, by no means do they reflect a comprehensive framework for modeling all pos-sible types of cas, rather, as stated in the thesis objectives, our work represent a first step towards realizing this ambitious goal.

## 7.2 Research Contributions

The unified framework proposed in this thesis is structured in the form of four different levels based on objectives of the proposed research as well as availability of suitable data. These levels use a combination of agent-based and complex network-based modeling approaches allowing multidisciplinary researchers to better conduct research by adopting a suitable framework level for developing models for their particular case study. Each framework level is substantiated by one or more complete example case studies as a means of demonstration of the generalized applicability and usage.

### 7.2.1 Complex Network Modeling Level for Interaction data

In chapter 3, we proposed the complex network modeling level. This level is based on the scenario where suitable data is some available, categorized and classified. This level is more suitable for use in the case of highly standardized interaction data repositories. To specifically show the broad applicability of the proposed framework level, a standard data repository, the Thomson Reuters web of knowledge, was selected. The initial data retrieved from this source was textual in nature with a large number of tags. Since only a few set of tags were suitable for each type of complex network, these text files were first parsed and then analyzed for tag selection. This was followed by the extraction of suitable complex networks. Subsequently different complex network methods were employed to extract suitable quantitative and qualitative measures about these networks. These measures were used to correlate the scientific implications of the research cas with the structural topological features of each network. Using two separate case studies from two different scientific domains of “Agent-based Computing” and “Consumer Electronics”, we demonstrated the broad applicability of the complex networks modeling methods. This framework level also serves as an encompassing level for existing complex network modeling case studies.

### 7.2.2 Exploratory agent-based modeling level

In chapter 4, we proposed the use of Agent-based Modeling to develop exploratory cas studies. The idea of exploratory ABM is to allow researchers to focus specifically on demonstrating the feasibility of their ABM design by means of developing proof of concept models. Similar to the complex network modeling level, exploratory agent-based modeling level is also an encompassing level and allows different existing agent-based modeling literature to be classified under this level.

To demonstrate the broad applicability of the proposed modeling paradigm, we demonstrated exploratory agent-based modeling in the domain of “Internet of Things”, a sub-area of upcoming research area of Cyber-physical systems where household and commonly devices can use communication networking to retrieve or expose useful information. The case study was used as a proof-of-concept for the use of unstructured P2P search algorithms in a completely different domain by demonstrating how static and mobile devices can dynamically connect together to expose different types of content in their surroundings. In general, simulation results from the case study, demonstrated the effectiveness and benefits of using the proposed exploratory agent-based modeling level for performing feasibility studies and developing proof-of-concept models for aspects such as needed in e.g. funding applications or for planning future research.

### 7.2.3 Descriptive agent-based modeling level

While exploratory ABM can be effective in conducting proof-of-concept feasibility studies, occasionally the goal of modeling is to be able to explore the use of communication of models as a means of inter-disciplinary model comparison for knowledge transfer and learning. Our proposed DescRiptivE Agent-based Modeling (DREAM) approach allows for this type of modeling and simulation. The DREAM model consists of a quantitative complex network model (i.e. a network model along with a quantitative "footprint" based on a centrality analysis of the network) coupled with a pseudocode-based specification model. DREAM model allows for a high level of fidelity with the ABM using a non-code modeling approach. Thus it allows for comparison of different models without expecting end-user cas researchers to be able to program models. Researchers can develop a DREAM model of their cas case study by extending the already-provided baseline complex network model followed by performing a complex network analysis. DREAM models also allow for a quantitative comparison of the centrality- based "footprint" of different ABMs across application case studies in addition to a visual comparison of different models by means of the different complex networks. A case study example demonstrating the use of DREAM in the domain of a heterogeneous ABM consisting of a well-studied "boids" model of flocking coupled with a single-hop WSN model demonstrates the usage of the descriptive agent-based modeling level. Simulation results obtained from the case study experiments also indicated some interesting research outcomes in terms of how sensing can be used to predict and detect group behavior in large and complex mobile agents (such as persons in crowds or airplanes in the sky etc.).

### 7.2.4 Validated agent-based modeling level

The validated agent-based modeling level of the proposed framework is suitable for use in research case studies where the validation of phenomena in the ABM is the central goal of a particular case study. Using a combination of concepts from three different scientific disciplines of multiagent systems, social sciences and software engineering, we propose the use of a VOMAS approach which allows for both the segregation of roles of the Simulation Specialists and the Subject Matter Experts in a team-oriented iterative methodology.

The goal of the VOMAS verification and validation approach is to allow cas researchers to specify validation questions which are transformed to invariants based on pre-conditions and post-conditions. These invariants are enforced in the ABM subsequently by customized VOMAS agents using an in-simulation validation scheme. For the validation and demonstration of the proposed approach, three separate case studies encompassing ecological modeling, WSN monitoring and social simulation of researchers in the form of a Temporal Citation Network (TCN) were presented demonstrating the effectiveness and viability of the proposed validated agent-based modeling level.

## 7.3 Critical review of the proposed levels

The first two levels of the proposed framework can be considered as encompassing levels since they allow the inclusion of existing cas literature under the proposed framework guidelines. Thus complex network models in existing cas literature can be classified under the complex network modeling level of the proposed framework as complex network models while existing agent-based models can be classified under the exploratory agent-based modeling level. The rest of the two framework levels specifically propose descriptive and validated Agent-based modeling respectively. Next, both these levels are critically compared with existing methods in cas literature.

### 7.3.1 Critical Comparison of DREAM with other approaches

While, to the best of our knowledge, there is no other well-known approach in literature which describes agent-based models descriptively and quantitatively similar to the proposed DREAM approach as part of the descriptive agent-based modeling level, here we review the DREAM approach to two somewhat related approaches of knowledge transfer and description of agent-based models (discussed earlier in more details in chapter 5):

#### 7.3.1.1 Template Software approach

The Santa Fe Institute proposes the use of template software approach[130] which essentially entails using software code as a means of model comparison. However, while useful, there are certain obvious limitations of this approach given as follows:

Firstly, software templates are code-based and while helpful (in line) comments can at times augment the readability of code, the fact remains that code is always tied to a single platforms and/or language. As such, this approach does not allow for either a non-code or a quantitative/visual comparison of models as proposed in the descriptive agent-based modeling level of the proposed framework.

Secondly, one key benefit of moving away from pure code –based approaches is the reduction of required programming-related learning curve for researchers to perform inter-disciplinary model comparison without the requirement of a detailed programming knowledge by means of visual or quantitative or pseudocode template based comparison.

#### 7.3.1.2 ODD approach

A textual approach of describing ecological and individual-based (agent-based) models has been proposed by Grimm et al. This approach is termed as the Overview, Design concepts, and Details (ODD) protocol[199]. As compared with the proposed DREAM modeling approach, there are certain key limitations of ODD given as follows:

Firstly, ODD is primarily textual description. A closer examination of the ODD approach reveals that it is basically a way of describing agent-based models using a proposed checklist of headings. Secondly this approach does not allow for a quantitative comparison such as possible by means of the use of complex network methods as described in the proposed modeling methodology. Thirdly, ODD also does not offer any visualization-based approach where researchers can visualize an entire ABM for comparison with another ABM.

Another problem with the ODD approach is that it also does not offer a one-to-one translation of models from the actual model to the source code. Finally, ODD also does not allow for writing detailed technical descriptions of algorithms (since it is textual and does not use pseudocode, flowcharts or UML etc. in any way to describe the algorithms). Thus to summarize, while textual descriptions are effective in giving basic ideas about a model, the proposed DREAM model based descriptive agent-based modeling level allows for a number of other benefits as compared to the ODD approach.

### 7.3.2 Critical comparison of VOMAS with other approaches

Like the previous comparison of DREAM models, here we give the summary of a more detailed comparison, given earlier in Chapter 6, of the proposed in-simulation validation approach with other literature in the domain of verification and validation of ABMs.

#### 7.3.2.1 Empirical Validation

Empirical validation is a technique used primarily in the domain of Agents in Computational Economics (ACE) and social sciences concerned with using empirical data to validate ABM such as by Marks [221] and Fagiolo et al. [202]. While empirical validation is an effective approach, it is not associated with a specific analysis and design approach on how to structure the simulation study or how an agent-based model can be enhanced for performing a systematic verification and validation. As such, the proposed in-simulation validation methodology of the validated agent-based modeling level builds on top of these empirical validation techniques. It essentially allows for the design of validation techniques using software engineering concepts in addition to allowing for performing customized conceptual validation of the case study.

#### 7.3.2.2 Validation using Companion Modeling Approach

In Social Sciences, another approach to validation is termed as the "Companion Modeling Approach". In the companion modeling approach, researchers work together with stakeholders from higher authorities (in the case of Business ecosystems) to develop models and get approval of the results. In other words, by involving the stakeholders, the model is essentially "face" validated as noted by Bousquet and Trébuil [222]. The idea is to go back and forth between different stakeholders and the researchers where the stakeholders are the authorities governing the businesses/economic ecosystems. However, this methodology does not differentiate between the simulation specialists and the end-user researchers in any way. As such, while end-user researchers might be experts in their domain of modeling, they can be separate entities from the actual teams who might be actually developing the simulation models. Secondly, the companion modeling approach does not allow for a step-by-step team-oriented approach to validation. Thirdly this approach is limited to business stakeholder and is domain specific.

Thus, the proposed in-simulation methodology can also be considered to extend existing concepts from the "companion modeling" approach. In addition, the proposed methodology utilizes concepts from Software Engineering and Artificial Intelligence to form a customizable approach to the in-simulation validation of Agent-based simulation models.

## 7.4 Limitation of the Methods

In the previous sections, we have given an outline of our key contributions in this thesis followed by a critically comparison of the different framework levels with existing state-of-the-art approaches in cas literature. Here, we give details of the limitations of the individual framework levels as well as the limitations of case studies and the choice of the selected ABM tool.

### 7.4.1 Limitations and proposed extensions of the framework levels

Here, we outline the limitations of the individual framework levels. Firstly, as mentioned earlier in this chapter, the complex network modeling and the exploratory agent-based modeling levels are specifically designed for encompassing existing literature in cas modeling. As such, they are purposefully designed to allow existing case studies from the domains of complex network and agent-based modeling paradigms.

However, as noted in previous section, while ODD is a completely textual description, none of the framework level currently tie in with the ODD model. As such, one possible extension of the exploratory agent-based modeling level could thus be to tie in ODD with the proposed descriptive agent-based modeling level.

For the VOMAS in-simulation modeling approach, while we have applied it to different case studies, it has not been possible to apply and evaluate the approach to earlier business ecosystem modeling domains, which use empirical validation and companion modeling approaches. As such, a possible iterative improvement in this in-simulation modeling approach can be foreseen if the proposed approach were applied to case studies which have previously been studied by researchers in the companion modeling and the empirical validation areas. This would thus validate the approach further and possibly evolve it further in the directions of these previously well-studied approaches.

### 7.4.2 Selection of Case studies

Our first key limitation arises from the diversity of possible cas that exist in the real world. While we have attempted to cover a large number of representative application case studies, there are simply too many possible domains to be covered in a limited period of time and are hence out of the scope of this thesis. Although we believe we have demonstrated the methods in a number of key target sub-domains of cas such as social sciences, ecological sciences as well as telecommunications, our selected case studies only reflect a limited combination of cas. Thus, extensive further research and exploration of cas case studies need to be performed to fully evaluate our proposed framework in other cas domains.

#### 7.4.3 Selection of Agent-based Modeling Tool

At several times during the research, other agent-based modeling tools were evaluated for possible usage in the different case studies. These included Repast-S, Mason, Swarm, Scratch and other toolkits, a good overview of which can be found in[141]. However, none of these tools appeared to better suit the case studies in comparison with the NetLogo simulation tool. However, there are certain problems associated with NetLogo that limit its usage in certain cases that can only be overcome by advanced programmers capable of developing widgets and extensions to extend NetLogo. Here, we list some of the observed limitations:

A. NetLogo is based on primarily a single file structure. While other files can be imported, it does not offer any proper reuse in the form of object oriented modules. While NetLogo extensions can be built and compiled in Java, the Logo language itself requires code to be re-written or imported instead of being able to make classes and objects for reuse.

B. While the new version (V4.1) of NetLogo allows multi-threaded execution of several experiments simultaneously, it does not directly allow for performing distributed experiments on multiple machines. As a result, it is not very easy to execute large scale realistic models, which might otherwise require resources offered only by a cluster of machines.

C. One major problem with NetLogo is the lack of any strong debugging support. The debugging primitives allowed by NetLogo are essentially the same as printing values, which was the norm in low-level languages such as Assembly and Machine Languages. As such, on the one hand, while NetLogo offers excellent agent-based modeling support, on the other, there is no way of either setting breakpoints or watches as is the norm in most advanced Integrated Development Environments (IDEs).

D. Finally, one other problem with NetLogo is that it does not differentiate between the IDE and the runtime environment. As such, it is not possible to develop NetLogo code in a separate IDE while running it in separate runtime.

## 7.5 Proposed Future directions

The framework and the example case studied allow for a number of possible usages in further research studies. Some of these are given as possible future case studies as follows:

Specifically, at the complex network modeling level, the complex network approach can be further applied to complex adaptive social and biological systems. These can include the extraction and analysis of nucleic acid molecules secondary structures (from genomic data), analysis of social network structures and internet-scale communication networks.

In addition, the exploratory agent-based modeling level can be further applied in different scientific disciplines such as in exploration of how complex networks can be used to study the spread of HIV infections in sub-populations amongst other possible case studies. In addition, the Cyber Physical Systems case study can be further enhanced by practically implementing the study using mobile devices and performing comparative analysis of the results obtained from the simulation study with results obtained from the physical devices.

The proposed descriptive agent-based modeling approach can be further applied to inter-disciplinary case study applications. In addition, the results of sensing in complex environments have numerous possible future applications. Examples of possible applications of sensing of flocking can include the following:

1. Sensing can be used for the identification of collective behavior of people using an image processing or infrared approach. In other words, this technique can be used to detect any group of people such as a group of "shoppers" flocking to a particular set of shops in a mall even though there is a large crowd of thousands of other people moving around.
2. This same approach can be used to detect a group of malicious attackers moving through a crowd thus allowing for the detection of the group "flocking" without cohering too much as that close grouping might have raised suspicions of the security personnel.
3. Another application of this approach is in the domain of detecting a group of stealth aircraft flocking towards a common mission. Stealth aircrafts are known to be invisible to radars in certain conditions. However these aircrafts are physically visible to the naked eye or in the visible light. As such, while it is difficult to infer a coordinated operation based on only the detection of a single aircraft, if there were a randomly deployed group of sensors with appropriate cameras or other effective sensing equipment pointing to the sky, these wirelessly connected sensors could potentially detect a group of stealth aircraft flocking and moving towards a common goal. Using simple shape detection image processing algorithms, each sensor can simply note if a "large" object passes over them. And this detection of "proximity" could be used to quantify flocking at a monitoring station collecting the data from these sensors.

The validated agent-based modeling level of the framework can be used for validation of any agent-based modeling case study. This is particularly useful in physical sciences where validation of hypotheses can be crucial to the study. Possible future application case studies can include the use of agent-based modeling for tumor growth analysis as well as effects of treatment prediction, HIV spread modeling inside the patient's body, use of ABM to study fungal adsorption of metal ions, simulation study of correlation of Helicobacter Pylori spread in patient's gastrointestinal tract and others.

A general future research direction could take the form of development of software tools and components for a possible unified framework. Thus researchers could implement the framework levels in a single tool by, for example, embedding an agent-based modeling tool such as NetLogo.

## 7.6 Final Words

To summarize, our framework design goals have been to develop a set of framework levels allowing inter-disciplinary communication, collaboration and research by means of combining complex network and agent-based modeling methods for multidisciplinary cas researchers. While both complex network models as well as agent-based models have previously been used in different application cas studies, to the best of our knowledge, they have not previously been combined together in the form of a single unified set of framework guidelines. We have undertaken several successful case studies and reported results across a range of selected domains, which demonstrate the effectiveness of our proposed framework. However, our work is still only a first step towards the formulation of the envisaged comprehensive unified framework for the modeling and simulation of cas. It is hoped that continuing collaborative work in this domain would result in further enhancement and applications of the proposed framework.

# Appendix 1

Here we would like to mention the keywords used for searching the ISI web of knowledge in addition to the reasoning behind the selection. Arguably there are numerous ways to classify as sub-domain based on keywords. In this particular case, some of the keywords were even shared with Chemical and Biological Journals (e.g. using the word agent for e.g. Biological agent or Chemical agent even). As such, we had to limit the search to papers with a focus on either agent-based modeling specifically or else in the domain of multiagent systems.

The search was thus performed on titles and the exact search from the ISI web of knowledge was as following:

Title=(agent-based OR individual-based OR multi-agent OR multiagent OR ABM*) AND Title=(model* OR simulat*)

Timespan=All Years. Databases=SCI-EXPANDED, SSCI, A&HCI, CPCI-S.

Date retrieved: 8th September 2010 (1064 records)

# References

[1]C. E. Van Wagner, “Development and structure of the Canadian forest fire weather index system,” *Forestry technical report,* vol. 35, pp. 37, 1987.
[2]M. A. Niazi, and A. Hussain, “Social Network Analysis of trends in the consumer electronics domain,” in Consumer Electronics (ICCE), 2011 IEEE International Conference on, 2011, pp. 219-220.
[3]M. Niazi, and A. Hussain, “Agent-based computing from multi-agent systems to agent-based models: a visual survey,” *Scientometrics*, pp. 1-21, 2011.
[4]M. A. Niazi, and A. Hussain, “Agent-based tools for modeling and simulation of self-organization in peer-to-peer, ad hoc, and other complex networks,” *Communications Magazine, IEEE,* vol. 47, no. 3, pp. 166-173, 2009.
[5]M. A. Niazi, and A. Hussain, “Sensing Emergence in Complex Systems,” *Sensors Journal, IEEE,* vol. 11, no. 10, pp. 2479-2480, 2011.
[6]M. A. Niazi, and A. Hussain, “A Novel Agent-Based Simulation Framework for Sensing in Complex Adaptive Environments,” *Sensors Journal, IEEE,* vol. 11, no. 2, pp. 404-412, 2011.
[7]M. Niazi, A. Hussain, A. R. Baig, and S. Bhatti, “Simulation of the research process,” in 40th Conference on Winter Simulation, Miami, FL, 2008, pp. 1326-1334.
[8]M. Niazi, Q. Siddique, A. Hussain, and M. Kolberg, “Verification and Validation of an Agent-Based Forest Fire Simulation Model,” in SCS Spring Simulation Conference, Orlando, FL, USA, 2010, pp. 142-149.
[9]M. A. Niazi, A. Hussain, and M. Kolberg, “Verification &Validation of Agent Based Simulations using the VOMAS (Virtual Overlay Multi-agent System) approach,” in MAS&S 09 at Multi-Agent Logics, Languages, and Organisations Federated Workshops, Torino, Italy, 2009, pp. 1-7.
[10]J. Holland, “Complex adaptive systems,” *Daedalus,* vol. 121, no. 1, pp. 17-30, 1992.
[11]C. Gershenson, and F. Heylighen, “When Can We Call a System Self-Organizing?,” *ECAL,* vol. Volume 2801/2003, pp. 606-614, 2003.
[12]J. A. Pelesko, "*Self assembly: the science of things that put themselves together*," Chapman & Hall, 2007.
[13]F. Boschetti, and R. Gray, "A Turing Test for Emergence," *Advances in Applied Self-organizing Systems*, Advanced Information and Knowledge Processing M. Prokopenko, ed., pp. 349-364: Springer London, 2008.
[14]H. Kitano, “Systems biology: a brief overview,” *Science,* vol. 295, no. 5560, pp. 1662, 2002.
[15]F. Amadieu, C. Mariné, and C. Laimay, “The attention-guiding effect and cognitive load in the comprehension of animations,” *Computers in Human Behavior,* vol. 27, no. 1, pp. 36-40, 2011.
[16]J. Epstein, “Why model?,” *Journal of Artificial Societies and Social Simulation,* vol. 11, no. 4, pp. 12, 2008.
[17]M. Mitchell, "*Complexity: a guided tour*," Oxford University Press, USA, 2009.
[18]V. Grimm, and S. F. Railsback, "*Individual-based Modeling and Ecology*," Princeton University Press, 2005.
[19]M. J. North, and C. M. Macal, "*Managing business complexity: discovering strategic solutions with agent-based modeling and simulation*," Oxford University Press, USA, 2007.

[20]N. Gilbert, and K. G. Troitzsch, "*Simulation for the social Scientist*," Second ed.: McGraw Hill Education, 2005.
[21]P. L. Lollini, and F. P. Santo Motta, "Discovery of cancer vaccination protocols with a genetic algorithm driving an agent based simulator," *BMC bioinformatics,* vol. 7, no. 1, pp. 352, 2006.
[22]C. Cioffi-Revilla, "Comparing Agent-Based Computational Simulation Models in Cross-Cultural Research," *Cross-Cultural Research,* vol. 45, no. 2, pp. 208-230, May 1, 2011, 2011.
[23]N. Boccara, "*Modeling complex systems*," Second ed.: Springer, 2010.
[24]L. A. N. Amaral, and J. M. Ottino, "Complex networks," *The European Physical Journal B - Condensed Matter and Complex Systems,* vol. 38, no. 2, pp. 147-162, 2004.
[25]R. Axelrod, "*The complexity of cooperation: agent-based models of competition and colloboration.*," Princeton, NJ: Princeton University Press, 1997.
[26]M. E. J. Newman, "The Structure and Function of Complex Networks," *SIAM Review,* vol. 45, no. 2, pp. 167-256, 2003.
[27]U. Nations, *Second generation surveillance for HIV: The next decade*, HASP, Pakistan, Islamabad, 2010.
[28]J. A. Bondy, and U. S. R. Murty, "*Graph theory with applications*," MacMillan, 1976.
[29]Author ed.^eds., "Analysis of biological networks," John Wiley & Sons,, 2008, p.^pp. Pages.
[30]W. d. Nooy, A. Mrvar, and V. Batagelj, "*Exploratory Social Network Analysis With Pajek*," Cambridge, UK: Cambridge University Press, 2005.
[31]O. Balci, "Verification, validation, and accreditation," in Proceedings of the 30th conference on Winter simulation, Washington, D.C., United States, 1998.
[32]S. Forrest, and T. Jones, "Modeling complex adaptive systems with Echo," *Complex systems: Mechanisms of adaptation*, pp. 3–21, 1994.
[33]W. Wayne, "Cyber-physical Systems," *IEEE Computer,* vol. 42, pp. 88-89, 2009.
[34]K. Ashton, "That Internet of Things Thing," *RFID Journal*, 2009.
[35]J. H. Holland, "*Hidden order: How adaptation builds complexity*," Basic Books, 1996.
[36]M. Girvan, and M. Newman, "Community structure in social and biological networks," *Proceedings of the National Academy of Sciences of the United States of America,* vol. 99, no. 12, pp. 7821, 2002.
[37]J. H. Miller, and S. E. Page, "*Complex Adaptive Systems: An Introduction to Computational Models of Social Life*," Princeton University Press, 2007.
[38]K. Kaneko, "*Life: An introduction to complex systems biology*," Springer Heidelberg, Germany:, 2006.
[39]J. A. Thomson, J. Itskovitz-Eldor, S. S. Shapiro *et al.*, "Embryonic stem cell lines derived from human blastocysts," *Science,* vol. 282, no. 5391, pp. 1145, 1998.
[40]K. Nair, "*Tropical forest insect pests: ecology, impact, and management*," Cambridge Univ Pr, 2007.
[41]M. D. Lowman, "Leaf growth dynamics and herbivory in five species of Australian rain-forest canopy trees," *Journal of Ecology*, pp. 433-447, 1992.
[42]S. L. Lima, "Nonlethal effects in the ecology of predator-prey interactions," *Bioscience,* vol. 48, no. 1, pp. 25-34, 1998.
[43]A. M. Hein, and J. F. Gillooly, "Predators, prey, and transient states in the assembly of spatially structured communities," *Ecology,* vol. 92, no. 3, pp. 549-555, 2011.
[44]H. C. Berg, "*E. coli in Motion*," Springer Verlag, 2004.
[45]Y. Bar Yam, "A mathematical theory of strong emergence using multiscale variety," *Complexity,* vol. 9, no. 6, pp. 15-24, 2004.

[46]L. Leydesdorff, "*The challenge of scientometrics: The development, measurement, and self-organization of scientific communications*," Universal-Publishers, 2001.
[47]C. Chen, R. Paul, and B. O'Keefe, “Fitting the jigsaw of citation: Information visualization in domain analysis,” *Journal of the American Society for Information Science and Technology,* vol. 52, no. 4, pp. 315-330, 2001.
[48]S. Card, J. Mackinlay, and B. Shneiderman, "*Readings in information visualization: using vision to think*," Morgan Kaufmann, 1999.
[49]H. White, and K. McCain, “Visualizing a discipline: An author co-citation analysis of information science, 1972-1995,” *Journal of the American Society for Information Science,* vol. 49, no. 4, pp. 327-355, 1998.
[50]A. Pouris, and A. Pouris, “Scientometrics of a pandemic: HIV/AIDS research in South Africa and the World,” *Scientometrics*, pp. 1-12, 2010.
[51]H. Hou, H. Kretschmer, and Z. Liu, “The structure of scientific collaboration networks in Scientometrics,” *Scientometrics,* vol. 75, no. 2, pp. 189-202, 2008.
[52]M. Sierra-Flores, M. Guzmán, A. Raga, and I. Pérez, “The productivity of Mexican astronomers in the field of outflows from young stars,” *Scientometrics,* vol. 81, no. 3, pp. 765-777, 2009.
[53]A. Barabási, H. Jeong, Z. Néda *et al.*, “Evolution of the social network of scientific collaborations,” *Physica A: Statistical Mechanics and its Applications,* vol. 311, no. 3-4, pp. 590-614, 2002.
[54]R. Sooryamoorthy, “Scientific publications of engineers in South Africa, 1975–2005,” *Scientometrics*, pp. 1-16, 2010.
[55]C. Chen, G. Panjwani, J. Proctor *et al.*, “Visualizing the Evolution of HCI,” *People and Computers XIX—The Bigger Picture*, pp. 233-250, 2006.
[56]E. Sandström, and U. Sandström, "CiteSpace Visualization of Proximity Clusters in Dentistry Research." pp. 25–28.
[57]R. Zhao, and J. Wang, “Visualizing the research on pervasive and ubiquitous computing,” *Scientometrics*, pp. 1-20, 2010.
[58]L. Chun-juan, C. Yue, and H. Hai-yan, “Scientometric & Visualization Analysis of Innovation Studies International,” *Technology and Innovation Management,* vol. 1, 2010.
[59]H. White, and B. Griffith, “Author cocitation: A literature measure of intellectual structure,” *Journal of the American Society for Information Science,* vol. 32, no. 3, pp. 163-171, 1981.
[60]H. D. White, and K. W. McCain, “Visualizing a discipline: An author co-citation analysis of information science, 1972–1995,” *Journal of the American Society for Information Science,* vol. 49, no. 4, pp. 327-355, 1998.
[61]H. Small, “Co citation in the scientific literature: A new measure of the relationship between two documents,” *Journal of the American Society for Information Science,* vol. 24, no. 4, pp. 265-269, 1973.
[62]H. Small, and B. Griffith, “The structure of scientific literatures I: Identifying and graphing specialties,” *Science studies*, pp. 17-40, 1974.
[63]B. Griffith, H. Small, J. Stonehill, and S. Dey, “The structure of scientific literatures II: Toward a macro-and microstructure for science,” *Science studies,* vol. 4, no. 4, pp. 339-365, 1974.
[64]H. Small, “Macro-level changes in the structure of co-citation clusters: 1983–1989,” *Scientometrics,* vol. 26, no. 1, pp. 5-20, 1993.
[65]E. Garfield, “Use of Journal Citation Reports and Journal Performance Indicators in measuring short and long term journal impact,” *Croatian medical journal,* vol. 41, no. 4, pp. 368-374, 2000.

[66]M. Amin, and M. Mabe, “Impact factors: use and abuse,” *Medicina [Buenos Aires],* vol. 63, pp. 347-354, 2003.
[67]T. Braun, I. Dióspatonyi, S. Zsindely, and E. Zádor, “Gatekeeper index versus impact factor of science journals,” *Scientometrics,* vol. 71, no. 3, pp. 541-543, 2007.
[68]A. Fersht, “The most influential journals: Impact Factor and Eigenfactor,” *Proceedings of the National Academy of Sciences,* vol. 106, no. 17, pp. 6883, 2009.
[69]E. Garfield, “The history and meaning of the journal impact factor,” *Jama,* vol. 295, no. 1, pp. 90, 2006.
[70]H. Moed, T. Van Leeuwen, and J. Reedijk, “A critical analysis of the journal impact factors of Angewandte Chemie and the journal of the American Chemical Society inaccuracies in published impact factors based on overall citations only,” *Scientometrics,* vol. 37, no. 1, pp. 105-116, 1996.
[71]M. Niazi, H. F. Ahmed, and A. Ali, “Introducing fault-tolerance and responsiveness in web applications using SREFTIA,” in Proceedings of the International Multiconference on Computer Science and Information Technology, Wisla, Poland, 2006, pp. 271-278.
[72]H. S. Gunawi, T. Do, J. M. Hellerstein *et al.*, *Failure as a Service (FaaS): A Cloud Service for Large-Scale, Online Failure Drills*, University of California, Berkeley, Berkeley, 2011.
[73]F. Dressler, "*Self-Organization in Sensor and Actor Networks*," Wiley, 2007.
[74]B. Cohen, "Incentives build robustness in BitTorrent." pp. 68–72.
[75]C. W. Reynolds, “Flocks, herds and schools: A distributed behavioral model,” in Proceedings of the 14th annual conference on Computer graphics and interactive techniques, 1987, pp. 25-34.
[76]E. Fischbein, M. Deri, M. S. Nello, and M. S. Marino, “The role of implicit models in solving verbal problems in multiplication and division,” *Journal for Research in Mathematics Education,* vol. 16, no. 1, pp. 3-17, 1985.
[77]Z. N. O. Albert Laszlo Barabasi, “Network Biology: Understanding the cell's functional organization,” *Nature Reviews,* vol. 5, pp. 101-114, 2004.
[78]V. Batagelj, “Efficient algorithms for citation network analysis,” *Arxiv preprint cs/0309023*, 2003.
[79]K. Thulasiraman, and M. N. S. Swamy, "*Graphs: theory and algorithms*," Wiley Online Library, 1992.
[80]P. Erdős, and A. Rényi, “On the evolution of random graphs,” *Magyar Tud. Akad. Mat. Kutató Int. Közl.,* vol. 5, pp. 17–61, 1960.
[81]D. J. Watts, and S. H. Strogatz, “Collective dynamics of ‘small-world’networks,” *Nature,* vol. 393, no. 6684, pp. 440-442, 1998.
[82]A. L. Barabási, and R. Albert, “Emergence of scaling in random networks,” *Science,* vol. 286, no. 5439, pp. 509, 1999.
[83]F. M. Hulett, “The signal transduction network for Pho regulation in Bacillus subtilis,” *Molecular microbiology,* vol. 19, no. 5, pp. 933-939, 1996.
[84]E. H. Davidson, "*The regulatory genome: gene regulatory networks in development and evolution*," Academic Press, 2006.
[85]E. H. Davidson, “Emerging properties of animal gene regulatory networks,” *Nature,* vol. 468, no. 7326, pp. 911-920, 2010.
[86]Z. N. Oltvai, and A. L. Barabási, “Life's complexity pyramid,” *Science,* vol. 298, no. 5594, pp. 763, 2002.
[87]N. Team, "Network Workbench Tool," Indiana University, Northeastern University, and University of Michigan, 2006.

[88]V. Batagelj, and A. Mrvar, “Pajek-program for large network analysis,” *Connections,* vol. 21, no. 2, pp. 47-57, 1998.
[89]C. Chen, “CiteSpace II: Detecting and visualizing emerging trends and transient patterns in scientific literature,” *Journal of the American Society for Information Science and Technology,* vol. 57, no. 3, pp. 359-377, 2006.
[90]P. Shannon, A. Markiel, O. Ozier *et al.*, “Cytoscape: a software environment for integrated models of biomolecular interaction networks,” *Genome research,* vol. 13, no. 11, pp. 2498, 2003.
[91]M. Baur, M. Benkert, U. Brandes *et al.*, "Visone Software for visual social network analysis." pp. 554-557.
[92]M. Wooldridge, “Agent-based computing,” *Interoperable Communication Networks,* vol. 1, pp. 71-98, 1998.
[93]N. Jennings, “Agent-based computing: Promise and perils,” in 16th Int. Joint Conf. on Artificial Intelligence (IJCAI-99), Stockholm, Sweden, 1999, pp. 1429-1436.
[94]S. R. a. P. Norvig, "*Artificial Intelligence A Modern Approach*," Second ed.: Prentice Hall, 2003.
[95]L. Panait, and S. Luke, “Cooperative multi-agent learning: The state of the art,” *Autonomous Agents and Multi-Agent Systems,* vol. 11, no. 3, pp. 387-434, 2005.
[96]H. James, “Where Are All the Intelligent Agents?,” vol. 22, pp. 2-3, 2007.
[97]T. Finin, R. Fritzson, D. McKay, and R. McEntire, "KQML as an agent communication language." p. 463.
[98]J. Hendler, and D. McGuinness, “The DARPA agent markup language,” *IEEE Intelligent Systems,* vol. 15, no. 6, pp. 67-73, 2000.
[99]D. Connolly, F. van Harmelen, I. Horrocks *et al.*, “Daml+ oil (march 2001) reference description, December 2001,” *Internetquelle: http://www. w3. org/TR/daml+ oil-reference, heruntergeladen am,* vol. 5, 2007.
[100]V. Quera, F. S. Beltran, and R. Dolado, “Flocking Behaviour: Agent-Based Simulation and Hierarchical Leadership,” *Journal of Artificial Societies and Social Simulation,* vol. 13, no. 2, pp. 8, 2010.
[101]V. Grimm, U. Berger, F. Bastiansen *et al.*, “A standard protocol for describing individual-based and agent-based models,” *Ecological Modelling,* vol. 198, no. 1-2, pp. 115-126, 2006.
[102]S. F. Railsback, and V. Grimm, "*Agent-based and Individual-based Modeling: A Practical Introduction*," Princeton University Press, 2011.
[103]S. C. Bankes, "Agent-based modeling: A revolution?," 90003, National Acad Sciences, 2002, pp. 7199-7200.
[104]E. Bonabeau, “Agent-based modeling: Methods and techniques for simulating human systems,” *Proceedings of the National Academy of Sciences of the United States of America,* vol. 99, no. Suppl 3, pp. 7280, 2002.
[105]H. Devillers, J. R. Lobry, and F. Menu, “An agent-based model for predicting the prevalence of Trypanosoma cruzi I and II in their host and vector populations,” *Journal of Theoretical Biology,* vol. 255, no. 3, pp. 307-315, 2008.
[106]G. M. Dancik, D. E. Jones, and K. S. Dorman, “Parameter estimation and sensitivity analysis in an agent-based model of Leishmania major infection,” *J Theor Biol,* vol. 262, no. 3, pp. 398-412, Feb 7, 2010.
[107]V. A. Folcik, G. C. An, and C. G. Orosz, “The Basic Immune Simulator: an agent-based model to study the interactions between innate and adaptive immunity,” *Theor Biol Med Model,* vol. 4, pp. 39, 2007.
[108]V. Galvao, and J. G. Miranda, “A three-dimensional multi-agent-based model for the evolution of Chagas' disease,” *Biosystems,* vol. 100, no. 3, pp. 225-30, Jun, 2010.

[109]V. Galvao, J. G. Miranda, and R. Ribeiro-dos-Santos, “Development of a two-dimensional agent-based model for chronic chagasic cardiomyopathy after stem cell transplantation,” *Bioinformatics,* vol. 24, no. 18, pp. 2051-6, Sep 15, 2008.
[110]Z. Guo, P. M. A. Sloot, and J. C. Tay, “A hybrid agent-based approach for modeling microbiological systems,” *Journal of Theoretical Biology,* vol. 255, no. 2, pp. 163-175, 2008.
[111]C. F. Huang, J. Kaur, A. Maguitman, and L. M. Rocha, “Agent-based model of genotype editing,” *Evol Comput,* vol. 15, no. 3, pp. 253-89, Fall, 2007.
[112]J. Itakura, M. Kurosaki, Y. Itakura *et al.*, “Reproducibility and usability of chronic virus infection model using agent-based simulation; comparing with a mathematical model,” *Biosystems,* vol. 99, no. 1, pp. 70-8, Jan, 2010.
[113]M. Kiran, S. Coakley, N. Walkinshaw, P. McMinn, and M. Holcombe, “Validation and discovery from computational biology models,” *Biosystems,* vol. 93, no. 1-2, pp. 141-150, 2008.
[114]B. J. Lao, and D. T. Kamei, “Investigation of cellular movement in the prostate epithelium using an agent-based model,” *J Theor Biol,* vol. 250, no. 4, pp. 642-54, Feb 21, 2008.
[115]A. M. Bailey, M. B. Lawrence, H. Shang, A. J. Katz, and S. M. Peirce, “Agent-based model of therapeutic adipose-derived stromal cell trafficking during ischemia predicts ability to roll on P-selectin,” *PLoS Comput Biol,* vol. 5, no. 2, pp. e1000294, Feb, 2009.
[116]C. Carpenter, and L. Sattenspiel, “The design and use of an agent-based model to simulate the 1918 influenza epidemic at Norway House, Manitoba,” *Am J Hum Biol,* vol. 21, no. 3, pp. 290-300, May-Jun, 2009.
[117]M. A. Rubin, J. Mayer, T. Greene *et al.*, “An agent-based model for evaluating surveillance methods for catheter-related bloodstream infection,” *AMIA Annu Symp Proc*, pp. 631-5, 2008.
[118]G. M. Odell, and V. E. Foe, “An agent-based model contrasts opposite effects of dynamic and stable microtubules on cleavage furrow positioning,” *J Cell Biol,* vol. 183, no. 3, pp. 471-83, Nov 3, 2008.
[119]E. J. Robinson, F. L. Ratnieks, and M. Holcombe, “An agent-based model to investigate the roles of attractive and repellent pheromones in ant decision making during foraging,” *J Theor Biol,* vol. 255, no. 2, pp. 250-8, Nov 21, 2008.
[120]D. Santoni, M. Pedicini, and F. Castiglione, “Implementation of a regulatory gene network to simulate the TH1/2 differentiation in an agent-based model of hypersensitivity reactions,” *Bioinformatics,* vol. 24, no. 11, pp. 1374-80, Jun 1, 2008.
[121]M. Batty, "*Cities and complexity: understanding cities with cellular automata, agent-based models, and fractals*," The MIT press, 2007.
[122]X. Chen, and F. Zhan, “Agent-based modelling and simulation of urban evacuation: relative effectiveness of simultaneous and staged evacuation strategies,” *Journal of the Operational Research Society,* vol. 59, no. 1, pp. 25-33, 2008.
[123]R. E. Streit, and D. Borenstein, “An agent-based simulation model for analyzing the governance of the Brazilian Financial System,” *Expert Systems with Applications,* vol. 36, no. 9, pp. 11489-11501, 2009.
[124]M. H. F. Zarandi, M. Pourakbar, and I. B. Turksen, “A Fuzzy agent-based model for reduction of bullwhip effect in supply chain systems,” *Expert Systems with Applications,* vol. 34, no. 3, pp. 1680-1691, 2008.

[125]T. Galla, “Independence and interdependence in the nest-site choice by honeybee swarms: Agent-based models, analytical approaches and pattern formation,” *Journal of Theoretical Biology,* vol. 262, no. 1, pp. 186-196, 2010.
[126]C. Gershenson, “Self-organizing traffic lights,” *Arxiv preprint nlin/0411066*, 2004.
[127]T. Carmichael, “Complex adaptive systems and the threshold effect: Towards a general tool for studying dynamic phenomena across diverse domains,” THE UNIVERSITY OF NORTH CAROLINA AT CHARLOTTE.
[128]U. Wilensky, “NetLogo,” *Center for Connected Learning Comp.-Based Modeling, Northwestern University,* vol. Evanston, IL, 1999.
[129]M. Resnick, "StarLogo: an environment for decentralized modeling and decentralized thinking." pp. 11-12.
[130]S. W. Pages. "Code Templates for Agent-based Models," August, 2011; http://www.swarm.org/index.php/Software_templates.
[131]J. M. Vidal, P. Buhler, and H. Goradia, “The Past and Future of Multiagent Systems,” *AAMAS Workshop on Teaching Multi-Agent Systems*, 2004.
[132]J. McCarthy, "*LISP 1.5 programmer's manual*," 1965.
[133]J. Banks, J. S. C. II, B. L. Nelson, and D. M. Nicol, "*Discrete-Event System Simulation*," Fourth ed.: Peason Education, 2005.
[134]A. M. Law, “How to build valid and credible simulation models,” in Winter Simulation Conference Miami, FL, 2008, pp. 39-47.
[135]T. H. Naylor, and J. M. Finger, “Verification of Computer Simulation Models,” *Management Science,* vol. 2, pp. B92-B101, 1967.
[136]J. M. Galán, L. R. Izquierdo, S. S. Izquierdo *et al.*, “Errors and Artefacts in Agent-Based Modelling,” *Journal of Artificial Societies and Social Simulation,* vol. 12, no. 11, 2009.
[137]T. W. Lucas, S. M. Sanchez, F. Martinez, L. R. Sickinger, and J. W. Roginski, “Defense and homeland security applications of multi-agent simulations,” in Proceedings of the 39th conference on Winter simulation: 40 years! The best is yet to come, Washington D.C., 2007.
[138]C. Bianchi, P. Cirillo, M. Gallegati, and P. A. Vagliasindi, “Validating and Calibrating Agent-Based Models: A Case Study,” *Comput. Econ.,* vol. 30, no. 3, pp. 245-264, 2007.
[139]J. S. Hodges, and J. A. Dewar, *Is It You or Your Model Talking? A Framework for Model Validation*, RAND Corporation, Santa Monica, California.
[140]G. Fagiolo, C. Birchenhall, and P. Windrum, “Empirical Validation in Agent-based Models: Introduction to the Special Issue ” *Computational Economics,* vol. 30, no. 3, pp. 189-194, October, 2007, 2007.
[141]C. M. Macal, and M. J. North, “Agent-based modeling and simulation: desktop ABMS,” in Proceedings of the 39th conference on Winter simulation: 40 years! The best is yet to come, Washington D.C., 2007.
[142]R. Axtell, “The emergence of firms in a population of agents: local increasing returns, unstable Nash equilibria, and power law size distributions,” *The Brookings Institution CSED Working Paper,* vol. 3, pp. 138, 2000.
[143]R. Axtell, J. Epstein, and H. Young, “The emergence of classes in a multiagent bargaining model,” *Social dynamics*, pp. 191–211, 1999.
[144]S. Moss, “Alternative Approaches to the Empirical Validation of Agent-Based Models,” *Journal of Artificial Societies and Social Simulation,* vol. 11, no. 15, 2008.

[145]U. Wilensky, and W. Rand, “Making Models Match: Replicating an Agent-Based Model,” *Journal of Artificial Societies and Social Simulation,* vol. 10, no. 4, pp. 2, 10/31, 2007.
[146]A. Schmid, “What is the Truth of Simulation?,” *Journal of Artificial Societies and Social Simulation,* vol. 8, no. 4, pp. 5, 10/31, 2005.
[147]O. Barreteau, and e. al., “Our Companion Modelling Approach ” *Journal of Artificial Societies and Social Simulation,* vol. 6, no. 1, 2003.
[148]M. Makowsky, “An Agent-Based Model of Mortality Shocks, Intergenerational Effects, and Urban Crime,” *Journal of Artificial Societies and Social Simulation,* vol. 9, No. 2, 2006.
[149]L. Cernuzzi, M. Cossentino, and F. Zambonelli, “Process models for agent-based development,” *Engineering Applications of Artificial Intelligence,* vol. 18, no. 2, pp. 205-222, 2005.
[150]P. Garrido, M. Malumbres, and C. Calafate, "ns-2 vs. OPNET: a comparative study of the IEEE 802.11 e technology on MANET environments." p. 37.
[151]A. Sobeih, W. Chen, J. Hou *et al.*, "J-sim: A simulation environment for wireless sensor networks." pp. 175-187.
[152]P. Levis, N. Lee, M. Welsh, and D. Culler, "TOSSIM: Accurate and scalable simulation of entire TinyOS applications." pp. 126-137.
[153]S. Park, A. Savvides, and M. B. Srivastava, “SensorSim: a simulation framework for sensor networks,” in Proceedings of the 3rd ACM international workshop on Modeling, analysis and simulation of wireless and mobile systems, Boston, Massachusetts, United States, 2000, pp. 104-111.
[154]J. Polley, D. Blazakis, J. McGee *et al.*, "Atemu: A fine-grained sensor network simulator."
[155]I. Baumgart, B. Heep, and S. Krause, "OverSim: A flexible overlay network simulation framework." pp. 79-84.
[156]A. Varga, “The OMNeT++ discrete event simulation system,” in Proceedings of the European Simulation Multiconference ESM’2001 (2001), Prague, 2001, pp. 319–324.
[157]M. Jelasity, A. Montresor, G. P. Jesi, and S. Voulgaris, "The Peersim simulator," 2008.
[158]F. Mondada, G. C. Pettinaro, A. Guignard *et al.*, “SWARM-BOT: A new distributed robotic concept,” *Autonomous Robots,* vol. 17, no. 2, pp. 193-221, 2004.
[159]O. Michel, “WebotsTM: Professional Mobile Robot Simulation,” *International Journal of Advanced Robotic Systems,* vol. 1, no. 1, pp. 39-42, 2004.
[160]E. Sahin, S. Girgin, L. Bay nd r, and A. E. Turgut, “Swarm robotics,” *Swarm Intelligence: Introduction and Applications,* vol. 1, pp. 87, 2008.
[161]F. Bellifemine, A. Poggi, and G. Rimassa, “Developing multi-agent systems with JADE,” *Intelligent Agents VII Agent Theories Architectures and Languages*, pp. 42-47, 2001.
[162]M. Pullen, “The Network Workbench: network simulation software for academic investigation of Internet concepts,” *Computer Networks,* vol. 32, no. 3, pp. 365-378, 2000.
[163]A. Barabasi, “The origin of bursts and heavy tails in human dynamics,” *Nature,* vol. 435, no. 7039, pp. 207-211, 2005.
[164]A. Barabási, and A. Gelman, “Bursts: The Hidden Pattern Behind Everything We Do,” *Physics Today,* vol. 63, pp. 46, 2010.
[165]T. Reuters, “Web of science,” *Online factsheet Thomson Reuters, Philadelphia, Pennsylvania (Available from:*

*www.thomsonreuters.com/content/PDF/scientific/Web_of_Science_factsheet.pdf* ), 2008.
[166]R. Sedgewick, and M. Schidlowsky, "*Algorithms in Java, Part 5: Graph Algorithms*," Addison-Wesley Longman Publishing Co., Inc. Boston, MA, USA, 2003.
[167]E. Adar, "Guess: a language and interface for graph exploration." p. 800.
[168]R. W. Keyes, "The evolution of digital electronics towards VLSI," *Solid-State Circuits, IEEE Journal of,* vol. 14, no. 2, pp. 193-201, 1979.
[169]S. Gamm, and R. Haeb-Umbach, "User interface design of voice controlled consumer electronics," *Philips Journal of Research,* vol. 49, no. 4, pp. 439-454, 1995.
[170]J. F. Christensen, M. H. Olesen, and J. S. Kjær, "The industrial dynamics of Open Innovation--Evidence from the transformation of consumer electronics," *Research Policy,* vol. 34, no. 10, pp. 1533-1549, 2005.
[171]R. S. Rosenbloom, and W. J. Abernathy, "The climate for innovation in industry : The role of management attitudes and practices in consumer electronics," *Research Policy,* vol. 11, no. 4, pp. 209-225, 1982.
[172]R. Hisano, and T. Mizuno, "Sales distribution of consumer electronics," *Physica A: Statistical Mechanics and its Applications,* vol. 390, no. 2, pp. 309-318, 2011.
[173]S. F. Railsback, S. L. Lytinen, and S. K. Jackson, "Agent-based simulation platforms: Review and development recommendations," *SIMULATION,* vol. 82, no. 9, pp. 609, 2006.
[174]N. Gershenfeld, R. Krikorian, and D. Cohen, "The Internet of Things," *Scientific American,* vol. 291, no. 4, pp. 76-81, 2004.
[175]R. Groenevelt, E. Altman, and P. Nain, "Relaying in mobile ad hoc networks: the Brownian motion mobility model," *Wireless Networks,* vol. 12, no. 5, pp. 561-571, 2006.
[176]M. C. Gonzalez, C. A. Hidalgo, and A. L. Barabási, "Understanding individual human mobility patterns," *Nature,* vol. 453, no. 7196, pp. 779-782, 2008.
[177]U. Wilensky. "NetLogo Users Discussion Group," August, 2011; http://groups.yahoo.com/group/netlogo-users/.
[178]U. Wilensky. "NetLogo Educators Mailing List," August, 2011; http://groups.yahoo.com/group/netlogo-educators/.
[179]D. L. Parnas, "Precise Documentation: The Key to Better Software," *The Future of Software Engineering*, S. Nanz, ed., pp. 125-148: Springer Berlin Heidelberg, 2011.
[180]Y. Li, "Reengineering a scientific software and lessons learned." pp. 41-45.
[181]C. M. Macal, and M. North. "Repast Interest users Mailing List," August, 2011; https://lists.sourceforge.net/lists/listinfo/repast-interest.
[182]A. I. Concepcion, "DEVS Formalism: A Framework for Hierarchical Model Development," *IEEE Transactions on Software Engineering,* vol. 14, pp. 228-241, 1988.
[183]B. P. Zeigler, "Multifaceted modeling methodology: Grappling with the irreducible complexity of systems," *Behavioral Science,* vol. 29, no. 3, pp. 169-178, 2007.
[184]T. Buzan, and B. Buzan, "*The mind map book*," Pearson Education, 2006.
[185]A. Robins, J. Rountree, and N. Rountree, "Learning and teaching programming: A review and discussion," *Computer Science Education,* vol. 13, no. 2, pp. 137-172, 2003.
[186]J. C. Reynolds, "The essence of Algol." pp. 67-88.
[187]U. Wilensky, "NetLogo Flocking Model," Center for Connected Learning and Computer-Based Modeling, Northwestern University, 1998.
[188]A. Boukerche, "*Algorithms and protocols for wireless sensor networks*," Wiley-IEEE press, 2009.

[189]G. Anastasi, M. Conti, M. Di Francesco, and A. Passarella, “Energy conservation in wireless sensor networks: A survey,” *Ad Hoc Networks,* vol. 7, no. 3, pp. 537-568, 2009.
[190]J. Zheng, and A. Jamalipour, "*Wireless Sensor Networks: A Networking Perspective*," Wiley-IEEE Press, 2009.
[191]J. N. Al-Karaki, R. Ul-Mustafa, and A. E. Kamal, “Data aggregation and routing in Wireless Sensor Networks: Optimal and heuristic algorithms,” *Computer Networks,* vol. 53, no. 7, pp. 945-960, 2009.
[192]M. Younis, and K. Akkaya, “Strategies and techniques for node placement in wireless sensor networks: A survey,” *Ad Hoc Networks,* vol. 6, no. 4, pp. 621-655, 2008.
[193]J. M. Kahn, R. H. Katz, and K. S. J. Pister, "Next century challenges: mobile networking for “Smart Dust”." pp. 271-278.
[194]C. Colombo, and C. Mcinnes, “Orbital dynamics of 'smart dust' devices with solar radiation pressure and drag,” *Journal of Guidance, Control and Dynamics*, 2011.
[195]W. Kim, K. Mechitov, J. Y. Choi, and S. Ham, "On target tracking with binary proximity sensors." pp. 301-308.
[196]A. S. Rao, and M. P. Georgeff, “BDI agents: From theory to practice,” in First International Conference on Multiagent Systems, San Francisco, 1995, pp. 312–319.
[197]R. Cappelli, M. Ferrara, and D. Maltoni, “Fingerprint Indexing Based on Minutia Cylinder-Code,” *Pattern Analysis and Machine Intelligence, IEEE Transactions on*, no. 99, pp. 1-1, 2011.
[198]L. C. Freeman, “A set of measures of centrality based on betweenness,” *Sociometry*, pp. 35-41, 1977.
[199]V. Grimm, U. Berger, D. L. DeAngelis *et al.*, “The ODD protocol: A review and first update,” *Ecological Modelling,* vol. 221, no. 23, pp. 2760-2768, 2010.
[200]B. Edmonds, and D. Hales, “Replication, replication and replication: Some hard lessons from model alignment,” *Journal of Artificial Societies and Social Simulation,* vol. 6, no. 4, 2003.
[201]R. Axelrod, and W. D. Hamilton, “The evolution of cooperation,” *Science,* vol. 211, no. 4489, pp. 1390, 1981.
[202]G. Fagiolo, A. Moneta, and P. Windrum, “A Critical Guide to Empirical Validation of Agent-Based Models in Economics: Methodologies, Procedures, and Open Problems,” *Computational Economics,* vol. 30, no. 3, pp. 195-226, 2007.
[203]R. G. Sargent, “Verification and validation of simulation models,” in Proceedings of the 40th Conference on Winter Simulation, Miami, Florida, 2008, pp. 157-169.
[204]F. P. Brooks, "*The mythical man-month*," Addison-Welsley, 1995.
[205]Y. Le Traon, B. Baudry, and J. M. Jezequel, “Design by Contract to Improve Software Vigilance,” *Software Engineering, IEEE Transactions on,* vol. 32, no. 8, pp. 571-586, 2006.
[206]J. M. Jazequel, and B. Meyer, “Design by contract: the lessons of Ariane,” *Computer,* vol. 30, no. 1, pp. 129-130, 1997.
[207]U. Wilensky, "NetLogo Fire model," Center for Connected Learning and Computer-Based Modeling, Northwestern University 1997.
[208]A. Alexandridis, D. Vakalis, C. I. Siettos, and G. V. Bafas, “A cellular automata model for forest fire spread prediction: The case of the wildfire that swept through Spetses Island,” *Applied Mathematics and Computation,* vol. 204, no. 1, pp. 191-201, 2008.
[209]R. Axtell, R. Axelrod, J. M. Epstein, and M. D. Cohen, “Aligning simulation models: A case study and results,” *Computational & Mathematical Organization Theory,* vol. 1, no. 2, pp. 123-141, 1996.

[210]B. Lawson, P. F. R. Centre, and J. Turner, "*Weather in the Canadian Forest Fire Danger Rating System: a user guide to national standards and practices*," Canadian Forestry Service, 1978.
[211]C. E. Van Wagner, "*Structure of the Canadian forest fire weather index*," Environment Canada, Forestry Service, 1974.
[212]B. J. Stocks, T. J. Lynham, B. D. Lawson *et al.*, “The Canadian Forest Fire Danger Rating System: An Overview,” *The Forestry Chronicle,* vol. 65, no. 6, pp. 450-457, 1989.
[213]W. J. de Groot, “Interpreting the Canadian Forest fire Weather Index (FWI) System,” *In proc of the fourth Central Region Fire Weather Index Committee Scientific and technical Seminar, Edmonton Canada*, 1998.
[214]B. N. Clark, C. J. Colbourn, and D. S. Johnson, “Unit disk graphs,” *Discrete mathematics,* vol. 86, no. 1-3, pp. 165-177, 1990.
[215]Y. Wang, “Topology control for wireless sensor networks,” *Wireless Sensor Networks and Applications*, pp. 113-147, 2008.
[216]F. Kuhn, R. Wattenhofer, and A. Zollinger, “Ad hoc networks beyond unit disk graphs,” *Wireless Networks,* vol. 14, no. 5, pp. 715-729, 2008.
[217]J. Hirsch, “An index to quantify an individual's scientific research output,” *Proceedings of the National Academy of Sciences,* vol. 102, no. 46, pp. 16569, 2005.
[218]L. Egghe, “The Hirsch-index and related impact measures,” *Annual Review of Information Science and Technology,* vol. 44, pp. 65-114, 2010.
[219]A. Harzing, and R. van der Wai, “Google Scholar as a new source for citation analysis,” *Ethics in Science and Environmental Politics(ESEP),* vol. 8, no. 1, pp. 61-73, 2008.
[220]A. Harzing, and R. van der Wal, “A Google Scholar h-index for journals: An alternative metric to measure journal impact in economics and business,” *Journal of the American Society for Information Science and Technology,* vol. 60, no. 1, pp. 41-46, 2009.
[221]R. Marks, “Validating Simulation Models: A General Framework and Four Applied Examples,” *Computational Economics,* vol. 30, no. 3, pp. 265-290, 2007.
[222]F. Bousquet, and G. Trébuil, “Introduction to companion modeling and multi-agent systems for integrated natural resource management in Asia,” *Companion Modeling and Multi-Agent Systems for Integrated Natural Resource Management in Asia*, pp. 1-20, 2005.